\newcommand*{\ATLASLATEXPATH}{}
\pdfoutput=1
\documentclass[atlasdraft=false, texlive=2016, UKenglish, cernpreprint]{\ATLASLATEXPATH atlasdoc}
 
\AtlasJournal{The European Physical Journal C}
\PreprintIdNumber{CERN-EP-2018-192}
 
\usepackage[subfigure]{\ATLASLATEXPATH atlaspackage}
\usepackage{\ATLASLATEXPATH atlasbiblatex}
 
\usepackage{\ATLASLATEXPATH atlasphysics}
 
\graphicspath{{logos/}{figures/}}
 
\usepackage{ANA-JETM-2018-03-PAPER-defs}
\usepackage{\ATLASLATEXPATH atlasjetetmiss}
\usepackage{multirow}
\usepackage[normalem]{ulem}
\usepackage{rotating}
\usepackage{hhline}
\usepackage{subfigure}
 
\AtlasTitle{Performance of top-quark and $\boldsymbol{W}$-boson tagging with ATLAS in Run 2 of the LHC}
 
\AtlasVersion{3.0}
\AtlasAbstract{
The performance of identification algorithms (``taggers'') for hadronically decaying top quarks and $W$ bosons in
$pp$ collisions at $\sqrt{s}$ = 13~TeV recorded by the ATLAS experiment
at the Large Hadron Collider is presented.  A set of techniques based on jet shape observables
are studied to determine a set of
optimal cut-based taggers for use in physics analyses.  The studies are extended
to assess the utility of combinations of  substructure observables as a
multivariate tagger using boosted decision trees or deep neural networks in
comparison with  taggers based on two-variable combinations.  In addition, for
highly boosted top-quark tagging, a deep neural network based on jet constituent inputs as
well as a re-optimisation of the shower deconstruction technique is presented. The
performance of these taggers is studied in data collected during 2015 and
2016 corresponding to 36.1~fb$^{-1}$ for the  $t\bar{t}$ and $\gamma+\text{jet}$
and 36.7~fb$^{-1}$ for the dijet event topologies.
}

\author{The ATLAS Collaboration}
 
\AtlasRefCode{JETM-2018-03}
 
\AtlasJournalRef{Eur. Phys. J. C 79 (2019) 375}
\AtlasDOI{10.1140/epjc/s10052-019-6847-8}

\hypersetup{pdftitle={ATLAS document},pdfauthor={The ATLAS Collaboration}}
 
\begin{document}
 
\maketitle
 
\tableofcontents
\clearpage

\section{Introduction}
With the increase of the Large Hadron Collider (LHC)~\cite{Evans:2008zzb}
centre-of-mass energy to $13$~\TeV\ in Run 2, it is important for
searches for physics phenomena beyond the Standard Model to probe processes
involving highly boosted massive particles, such as \Wboson and \Zboson bosons
and top quarks~\cite{EXOT-2015-04,EXOT-2016-12,EXOT-2016-19}, as well
as Standard Model measurements using these techniques~\cite{TOPQ-2016-09,STDM-2017-04,STDM-2015-23}.  To
fully exploit these final states, it is important to reconstruct and accurately
identify the hadronic decay modes of these massive particles which serve as an
effective tool to reject events produced by background processes and improve
the sensitivity in searches for physics beyond the Standard Model.  Techniques
to achieve this aim were studied by both the ATLAS and CMS collaborations
during the course of Run 1 of the LHC
~\cite{PERF-2015-03,PERF-2015-04,CMS-JME-13-006,CMS-PAS-JME-13-007}.  In this
paper, these studies are performed with Run 2 data with particular attention to
the investigation of multivariate techniques based on both jet shape
observables and an approach using the jet constituents as input observables in
addition to the optimisation of the shower deconstruction technique for highly
boosted top-quark tagging.
 
In Section~\ref{sec:detector} the ATLAS detector is briefly described, followed
by a description of the Monte Carlo and data samples used in the analysis in
Section~\ref{sec:samples}. The set of jet reconstruction and tagging techniques
investigated in this work is described in Section~\ref{sec:TaggerTechniques}.
The optimisation procedure for each tagger, as well as a comprehensive
comparison of the tagging techniques using Monte Carlo simulation are presented
in Section~\ref{sec:TaggerOptimization}. In Section~\ref{sec:DataPerformance},
the $pp$ collision data recorded in 2015 and 2016 are used to evaluate the
performance of these tagging techniques, with the measurement of signal and
background efficiencies using boosted lepton+jet \ttbar, dijet and \gammajet
topologies and the robustness of the various techniques when confronted with varying levels of
event \pileup. Finally, concluding remarks are given in
Section~\ref{sec:conclusion}.
 
\section{ATLAS detector}
\label{sec:detector}
\newcommand{\AtlasCoordFootnote}{ ATLAS uses a right-handed coordinate system
with its origin at the nominal interaction point (IP) in the centre of the
detector and the $z$-axis along the beam pipe. The $x$-axis points from the IP
to the centre of the LHC ring, and the $y$-axis points upwards. Cylindrical
coordinates $(r,\phi)$ are used in the transverse plane, $\phi$ being the
azimuthal angle around the $z$-axis. The pseudorapidity is defined in terms
of the polar angle $\theta$ as $\eta = -\ln \tan(\theta/2)$. Angular distance
is measured in units of $\Delta R \equiv \sqrt{(\Delta\eta)^{2} +
(\Delta\phi)^{2}}$.}
 
The ATLAS detector~\cite{PERF-2007-01,Abbott:2018ikt} at the LHC covers nearly
the entire solid angle around the collision point.\footnote{\AtlasCoordFootnote}
It consists of an inner tracking detector (ID) surrounded by a thin
superconducting solenoid, electromagnetic and hadronic calorimeters, and a muon
spectrometer composed of three large superconducting toroid magnets and
precision tracking chambers. For this study, the most important subsystems are
the calorimeters, which cover the pseudorapidity range $|\eta| < $ 4.9. Within
the region $|\eta| < $ 3.2, electromagnetic calorimetry is provided by barrel
and endcap high-granularity lead/liquid-argon (LAr) sampling calorimeters, with
an additional thin LAr presampler covering $|\eta|< $ 1.8 to correct for energy
loss in material upstream of the calorimeters.  Hadronic calorimetry is provided
by a steel/scintillator-tile calorimeter, segmented into three barrel
structures within $|\eta| < $ 1.7, and two copper/LAr hadronic endcap
calorimeters which instrument the region 1.5$ < |\eta| < $ 3.2. The forward
region 3.1$ < |\eta| < $ 4.9 is instrumented with copper/LAr and tungsten/LAr
calorimeter modules.
 
Inside the calorimeters, the inner tracking detector measures charged-particle trajectories
in a 2~T axial magnetic field produced by the superconducting solenoid.  It covers a
pseudorapidity range $|\eta| < $ 2.5 with pixel and silicon microstrip detectors, and the
region $|\eta| < $ 2.0 with a straw-tube transition radiation tracker.
 
The muon spectrometer (MS) comprises separate trigger and high-precision
tracking chambers measuring the deflection of muons in a magnetic field
generated by superconducting air-core toroid magnets. The precision chamber
system covers the region $|\eta| < $  2.7 with three layers of monitored drift
tubes, complemented by cathode strip chambers in the forward region where the
background is highest. The muon trigger system covers the range $|\eta| < $ 2.4
with resistive plate chambers in the barrel and thin gap chambers in the endcap
regions.
 
A two-level trigger system is used to select  events for offline analysis~\cite{TRIG-2016-01}.
The first step, named the level-1 trigger, is implemented
in hardware and uses a subset of detector information to reduce the event rate
from 40 MHz to 100 kHz.  This is followed by a software-based high-level trigger
which reduces the final event rate to an average of 1 kHz.
 
\section{Data and simulated samples}
\label{sec:samples}
The taggers described in this article were initially designed, as described in Section~\ref{sec:TaggerOptimization}, using Monte Carlo
(MC) simulated samples for two signal processes (i.e. events containing the decay of heavy resonances) and one
background process (i.e. light quark and gluon jets). The dijet
process was used to simulate jets from gluons and non-top quarks.  It was
modelled using the leading-order
\textsc{Pythia8}~(v8.186)~\cite{Sjostrand:2007gs} generator with the
\textsc{NNPDF2.3LO}~\cite{Ball:2013hta} parton distribution function (PDF) set
and a set of tuned parameters called the A14 tune~\cite{ATL-PHYS-PUB-2014-021}.
Events were generated in slices of leading
jet transverse momentum (\pt) to sufficiently populate the kinematic region of
interest (between 200 and 2500~\GeV). Event-by-event weights were applied to
correct for this generation methodology and to produce the expected
smoothly falling jet \pt distribution of the multijet background. The signal
samples containing either high-\pt top-quark or \Wboson-boson jets were obtained
from two physics processes modelling phenomena beyond the Standard Model. For
the \Wboson-boson sample, high-mass sequential standard
model~\cite{Altarelli:1989ff} $W' \rightarrow WZ \rightarrow q\bar{q}q\bar{q}$
events were used. For the top-quark sample, high-mass sequential standard model
$Z' \rightarrow t\bar{t}$  events were used as a source of signal jets.  Both the
\Wboson bosons and top quarks were required to decay hadronically. The two signal
processes were simulated using the \textsc{Pythia8}~\cite{Sjostrand:2007gs}
generator with the \textsc{NNPDF2.3LO} PDF set and A14 tune for multiple values
of the resonance ($W'$ or $Z'$ boson) mass between 400 and 5000~\GeV\ in order to
populate the entire jet \pt range\footnote{As the combination of these signal
samples with different generated heavy resonance masses results in irregular
top-quark and \Wboson-boson \pT distributions, the events are reweighted at
the generator level to either a constant or a falling jet \pT distribution, as
is typical for light jets.  This procedure is described in Section~\ref{sec:optimisation_bdtdnn}} from 200 to 2500~\GeV\ and to reduce the impact of
MC statistical uncertainties on the calculated signal efficiencies.
 
For the study of \Wboson-boson and top-quark jets in data, described in
Section~\ref{sec:DataPerformance}, a number of MC samples are needed to model
both the \ttbar signal and backgrounds. The \textsc{Powheg-Box v2}
generator~\cite{Alioli:2010xd,Frixione:2007vw,Nason:2004rx} was used to simulate
\ttbar and single-top-quark production in the $Wt$- and $s$-channels at
next-to-leading order (NLO), while for the single-top-quark $t$-channel process,
the NLO \textsc{Powheg-Box v1} generator and the \textsc{CT10}
~\cite{Lai:2010vv} NLO PDF set was used. For all processes involving top quarks,
the parton shower, fragmentation, and the underlying event were simulated using
\textsc{Pythia6}~(v6.428) ~\cite{Sjostrand:2006za} with the
\textsc{CTEQ6L1}~\cite{Pumplin:2002vw} PDF set and the corresponding
\textsc{Perugia} 2012 tune (P2012)~\cite{Perugia2012}. The top-quark mass was set
to 172.5~\GeV. The $h_{\text{damp}}$ parameter, which controls the matching of the matrix
element to the parton shower, was set to the mass of the top quark.  The \ttbar
process is normalised to the cross-sections predicted to
next-to-next-to-leading order (NNLO) in $\alphas$ and next-to-next-to-leading
logarithm (NNLL) in soft-gluon terms while the single-top-quark processes are
normalised to the NNLO cross-section predictions~\cite{Czakon:2013goa}.
 
Several additional variations of the \ttbar generator are used for the
estimation of modelling uncertainties. Estimates of the parton showering,
hadronisation modelling and underlying-event uncertainty are derived by
comparing results obtained with the \textsc{Powheg-Box v2} generator interfaced to
\textsc{Herwig}++~(v2.7.1)~\cite{Bahr:2008pv} instead of \textsc{Pythia6}. To
estimate the hard-scattering modelling uncertainty, the NLO
\textsc{MadGraph5\_aMC@NLO}~(v2.2.1) generator~\cite{Alwall:2014hca} (hereafter
referred to as MC@NLO) is used with \textsc{Pythia6}. To estimate the
uncertainty in the modelling of additional radiation, the \textsc{Powheg-Box v2}
generator with \textsc{Pythia6} is used with modified renormalisation and
factorisation scales ($\times 2$ or $\times 0.5$) and a simultaneously modified
$h_{\text{damp}}$ parameter value ($h_{\text{damp}}=m_\text{top}$ or
$h_{\text{damp}}=2 \times m_\text{top}$) as described in
Ref.~\cite{ATL-PHYS-PUB-2016-004}.
 
Samples of $W/Z$+jets and Standard Model diboson ($WW$/$WZ$/$ZZ$) production were
generated with final states that include either one or two charged leptons. The
\textsc{Sherpa}~\cite{Gleisberg:2008ta} generator version 2.1.1 and version
2.2.1 were used to simulate these processes at NLO with the CT10 PDF set to
simulate the diboson and $W/Z$+jets production processes, respectively. The
$W/Z$+jets events are normalised to the NNLO
cross-sections~\cite{Catani:2009sm}.
 
For the study of \gammajet events in data, events containing a photon with
associated jets were simulated using the \textsc{Sherpa}~2.1.1 generator,
requiring a photon transverse momentum above 140~\GeV. Matrix elements were
calculated with up to four partons at LO and merged with the \textsc{Sherpa}
parton shower~\cite{Schumann:2007mg} using the ME+PS@LO
prescription~\cite{Hoeche:2009rj}.  The CT10 PDF set was used in conjunction with
the dedicated parton shower tune developed by the \textsc{Sherpa} authors.
 
The MC samples were processed through the full ATLAS detector
simulation~\cite{SOFT-2010-01} based on
\textsc{Geant4}~\cite{Agostinelli2003250}. Additional simulated proton–proton
collisions generated using \textsc{Pythia8}~(v8.186) with the
\textsc{A2M}~\cite{ATL-PHYS-PUB-2014-021} tune and \textsc{MSTW2008LO} PDF
set~\cite{Martin:2009iq} were overlaid to simulate the effects of additional
collisions from the same and nearby bunch crossings (pile-up), with a mean
number of 24 collisions per bunch crossing. All simulated events were then
processed using the same reconstruction algorithms and analysis chain as is used
for the data.
 
Data were collected in three broad categories to study the signal efficiency and
background rejection. For the signal, a set of observed top-quark and
\Wboson-boson jet candidates is obtained from a sample of \ttbar\ candidate
events in which one top quark decays semileptonically and the other decays
hadronically, the lepton-plus-jets decay signature. The background is studied
using data samples enriched in dijet events and \gammajet events. In addition to
covering different \pT regions, the dijet and \gammajet samples differ in what
partons initiated the jets under study. In the \gammajet topology the jets are
mostly initiated by quarks over the full \pT range studied, while for the
dijet topology the fraction of quarks initiating jets is slightly smaller than
the gluon fraction at low \pT and becomes large at high \pt. The data for the
\ttbar and \gammajet studies were collected during normal operations of the
detector and correspond to an integrated luminosity of $36.1~\ifb$. For the
dijet analysis, additional data where the toroid magnet was turned off are used.
This adds an additional  $0.6~\ifb$.  For both datasets, only data collected
while all relevant detector subsystems were fully functional and in which at
least one primary vertex was reconstructed with at least five associated ID
tracks consistent with the LHC beam spot are used~\cite{PERF-2015-01}.
 
The lepton-plus-jets events were collected with a set of single-electron and
single-muon triggers that became fully efficient for \pt of the reconstructed
lepton greater than 28~\GeV. The dijet events were collected with a single \largeR
jet trigger, where the jet was reconstructed using the same algorithm described
in Section~\ref{sec:reconstruction} and with a radius parameter of $R=1.0$.
This trigger became fully efficient for an offline jet \pt of approximately 450~\GeV.
The \gammajet jet events were collected with a single-photon trigger that
became fully efficient for an offline photon \pT of approximately 155~\GeV.
 
\section{Jet substructure techniques}
\label{sec:TaggerTechniques}
The identification of hadronic jets originating from the decay of boosted
\Wboson and \Zboson bosons and top quarks can broadly, and somewhat arbitrarily,
be divided into two stages: jet reconstruction and jet tagging.  In the first,
the hadronic energy flow of the event is exclusively divided into a number of
jets, composed of constituents, with the primary goal being to most accurately
reconstruct the interesting energy flows in the case of true signal jets while
suppressing contributions from the underlying event and event \pileup.  In the second, the information about
the jet constituents is distilled into a single observable by different means to
obtain a criterion by which to identify a jet as originating from a
hadronically decaying massive particle, such as a \Wboson boson or a top quark.
A number of techniques and observables pertaining to these two categories have
been described and investigated extensively in previous
work~\cite{PERF-2015-03,PERF-2015-04} with only a short summary of the relevant
techniques presented here.  In the case of the identification of \Wboson bosons,
the techniques and conclusions are more broadly applicable to both \Wboson and
\Zboson bosons, with dedicated studies concerning the separation of \Wboson-boson jets
from \Zboson-boson jets performed in Ref.~\cite{PERF-2015-02}.
 
\subsection{Jet reconstruction} \label{sec:reconstruction} In this work, jets
are reconstructed with the intention of capturing the full energy flow resulting
from the decay of a massive particle.  This reconstruction primarily uses inputs
in the form of noise-suppressed topological clusters of calorimeter
cells~\cite{PERF-2014-07} that are individually calibrated to correct for
effects such as the non-compensating response of the calorimeter and inactive
material, and which are assumed to be massless~\cite{PERF-2016-04}.  These
topoclusters are then used as inputs to build two different types of jets.  The
first uses the \akt~algorithm~\cite{Cacciari:2008gp} with a radius parameter of
$R=1.0$ to form jets which are further trimmed\footnote{When using
topoclusters~\cite{PERF-2014-07} as jet inputs, the trimming algorithm, as
opposed to pruning~\cite{Ellis:2009me} or
split-filtering~\cite{Butterworth:2008iy} was found to be optimal in terms of
accurately reconstructing the important aspects of the energy flow as shown in
Ref.~\cite{PERF-2015-03}.} to remove the effects of \pileup and the underlying
event. Trimming~\cite{Krohn:2009th} is a grooming technique in which the
original constituents of the jets are  reclustered using the
\kt~algorithm~\cite{Ellis:1993tq} with a radius parameter $R_{\mathrm{sub}}$
to produce a collection of subjets. These subjets are then discarded if they
have less than a specific fraction ($f_{\mathrm{cut}}$) of the \pt of the
original jet. The trimming parameters used here are $R_{\mathrm{sub}}=$ 0.2 and
$f_{\mathrm{cut}}=0.05$.  These  \largeR jets are then calibrated in a two-step
procedure that first corrects the jet energy scale and then the jet mass
scale~\cite{PERF-2016-04,ATLAS-CONF-2016-035}.  The resulting set of
constituents forms the basis from which further observables are calculated.  The
second type of jet clustering, needed for the \htt
algorithm~\cite{Plehn:2009rk,Plehn:2010st}, makes use of the Cambridge/Aachen
(C/A) jet algorithm~\cite{Dokshitzer:1997in,Wobisch:1998wt} with a radius parameter of
$R=1.5$ which aims to identify top-quark jets across a broad \pt range, in
particular reaching low \pt. These jets, used in conjunction with the \htt
algorithm described in Section~\ref{sec:technique_htt}, are also groomed to
mitigate the effects of \pileup.  Trimming with subjet radius parameter of
$R_{\text{sub}}=0.2$ and momentum fraction $f_{\text{cut}} = 0.05$, the same as
those used in the trimming of the \akt $R=1.0$ jet collection, is found to
produce jet reconstruction and identification performance independent of the
average number of interactions per bunch crossing.
 
In simulation, in addition to jets reconstructed from detector-level
observables, a set of jets based on generator-level information is also used to
characterise the performance of a given tagging algorithm.  These jets are
reconstructed with the \akt algorithm with a radius parameter $R =$ 1.0, using
stable particles from the hard scatter with lifetimes greater than 10 ps,
excluding muons and neutrinos, as constituents. These jets, to which no trimming
algorithm is applied, are referred to as \emph{truth jets}, and the related
observables are denoted by the superscript ``true''.
 
\subsection{Jet labelling}
\label{sec:labelling}
As the aim of this study is the evaluation of the performance of jet tagging
algorithms, the labelling of the particle that initiated the jet is of
particular importance.  For signal jets, this labelling is based on the partonic
decay products of the particle of interest (\Wboson boson or top quark) in a
three-step process.  First, reconstructed jets are matched to truth jets with a
matching criterion of $\Delta R(j_{\text{true}},j_{\text{reco}})<0.75$. Next,
those truth jets are matched to truth \Wboson bosons and top quarks (\Wboson,
$t$) with a matching criterion of $\Delta R(j_{\text{true}},\text{particle})<0.75$. Finally, the partonic decay products of the parent \Wboson boson or
top quark (two quarks for hadronically decaying \Wboson bosons and an additional
$b$-quark) are matched to the reconstructed jet. A reconstructed jet is
labelled as a \Wboson-boson or top-quark jet if the parent particle
and all of its direct decay products are contained within a region in
($\eta$,$\phi$) with $\Delta R < 0.75\times R_{\text{jet}}$, where
$R_{\text{jet}}$ is the jet radius parameter.  In the case of \Wboson bosons, this means
that both of the daughter partons from the $\Wboson\rightarrow q\bar{q'}$ decay
are contained within the jet.  For jets matched to the parent \Wboson boson, at
\pt$\sim$ 200~\GeV\ only 50\% of the jets are fully contained when using this
criterion while for \pt>500~\GeV\ the containment rises to nearly 100\%. In the case of
top-quark jets, the possible final-state topologies for the jet are more
complex, including the possibility of the \largeR jet containing only the $b$-quark
from the top decay, only the two quarks from the \Wboson-boson decay, or a
pairing of a $b$-quark and one of the daughter \Wboson-boson quarks within
$\Delta R < 0.75\times R_{\text{jet}}$ around the jet axis.  As seen in
Figure~\ref{fig:truth_acceptance}, the fraction of \largeR jets falling into
each category depends strongly on the \pt of the parent particle with only 60\%
of jets being fully contained at 600~\GeV\ and with 100\% containment not being
reached even at 1500~\GeV. The value $0.75\times R_{\text{jet}}$ for the jet
labelling criteria is chosen as a compromise between the resulting labelling
efficiency and the resolution of the top-quark and \Wboson-boson jet mass peak.
The jet \pt dependence of the variation in containment, particularly in the case of
top-quark tagging in which a top-quark jet is labelled as such only when the
top parton, the $b$-quark from its decay as well as the two light quarks from
the subsequent \Wboson-boson decay are contained within the region $\Delta R <
0.75\times R_{\text{jet}}$ around the jet axis, serves as a strong motivation
for the various optimisation strategies described in
Section~\ref{sec:TaggerOptimization}.

\begin{figure}[htpb]
\centering
\subfigure[][]{ \label{fig:containment_w}   \includegraphics[width=0.485\textwidth]{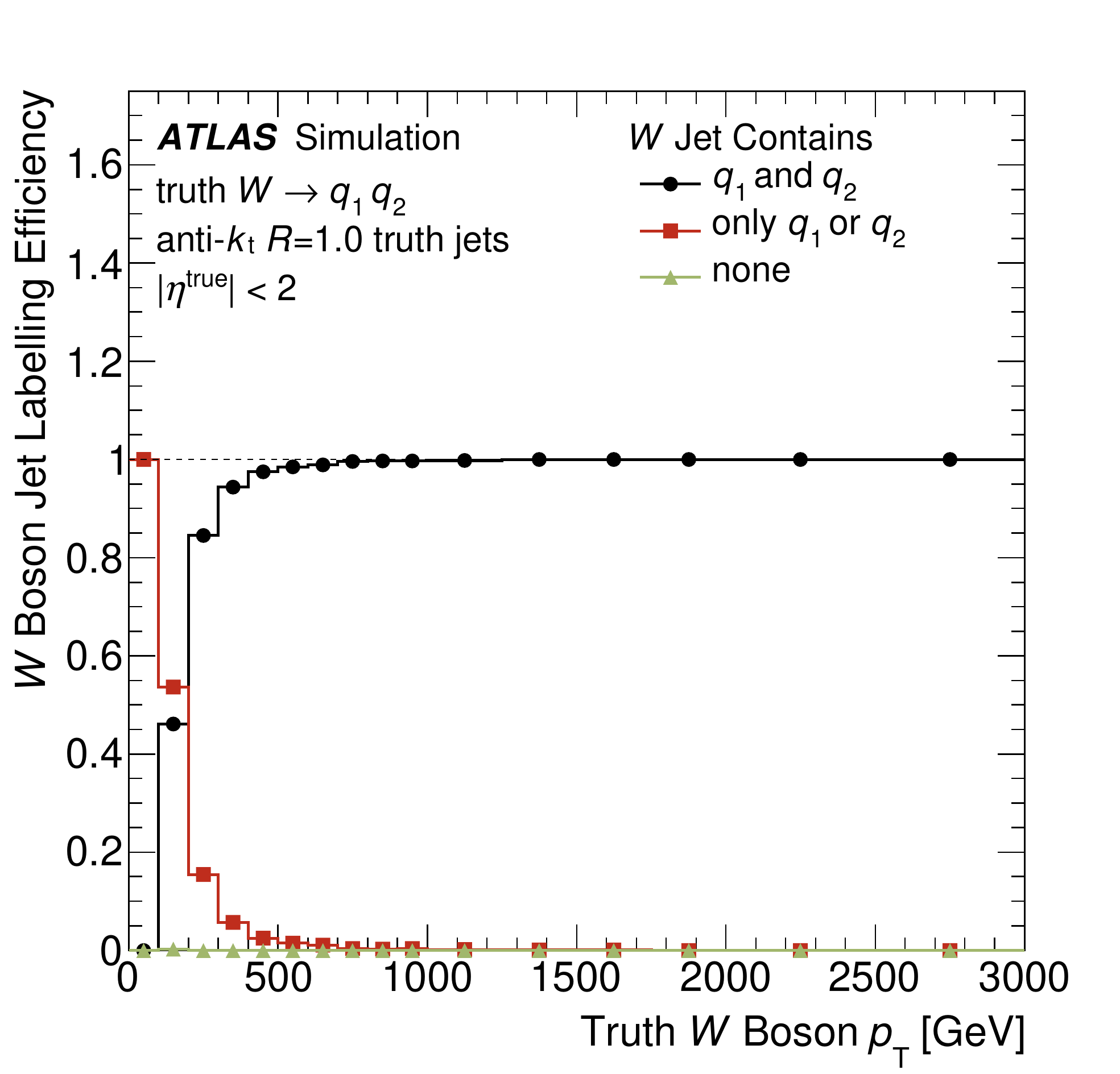} }
\subfigure[][]{ \label{fig:containment_top} \includegraphics[width=0.485\textwidth]{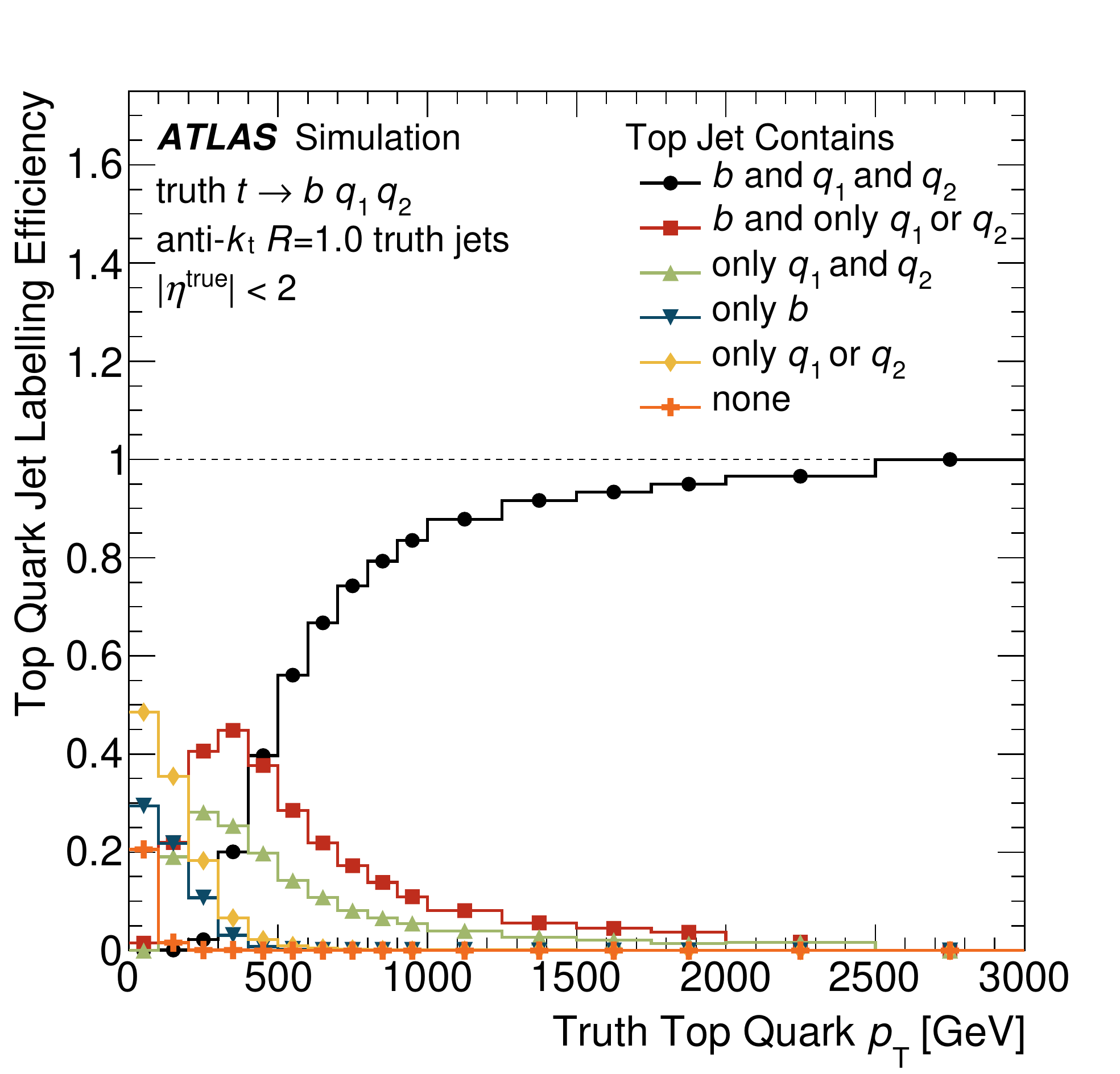} }
\caption{\label{fig:truth_acceptance} Containment of the \Wboson-boson~\subref{fig:containment_w} and
top-quark~\subref{fig:containment_top}decay products in a single truth-level \akt $R=1.0$ jet as a function
of the particle's transverse momentum.}
\end{figure}

\subsection{Tagging techniques}
After reconstructing the jet as a collection of constituents, a number of
methods can be used to classify a jet as originating from a heavy particle
(\Wboson boson or top quark) decay as opposed to a light jet originating from
gluons and quarks of all flavours other than top quarks.  The motivation behind
the various techniques differs, but they all attempt to form a decision
criterion by which to identify a jet as originating from a \Wboson
boson or top quark.
 
\subsubsection{Jet moments}
\label{sec:technique_moments}
The first broad class of observables studied for classification are directly
based on the constituents of the trimmed jet and attempt to quantify a
particular feature of the jet in an analytic way.  Of these features, the most
powerful is the jet mass, which for a jet formed from the decay of a heavy
particle has a scale associated with the mass of the particle, whereas for
light jets high masses are less likely as they need to be generated through
QCD emissions.
Traditionally, the jet mass was calculated as the invariant mass of the
collection of topoclusters of the trimmed jet (\mcalo)~\cite{PERF-2015-03}.
However, at very high \pt, the resolution of this observable decreases when
energy depositions from individual particles begin to merge in clusters.  To
mitigate this effect, the fine spatial granularity of the inner detector is used
to calculate the jet mass as the invariant mass of the
ghost-associated~\cite{Cacciari:2008gn} charged-particle tracks scaled by the ratio
of the transverse momenta of the trimmed jet and the associated tracks to form the
track-assisted mass (\mtaSup).  To achieve good performance across a broad range
of jet transverse momenta, an average of \mcalo and \mtaSup, weighted by the inverse
of their resolutions is calculated to form the combined mass
(\mcomb)~\cite{ATLAS-CONF-2016-035}.
 
In addition to the jet mass, a number of other observables quantify the extent
to which the jet constituents are clustered or uniformly dispersed and can be
used to augment the discrimination power from the jet mass alone.  This can be
done by explicitly using a set of axes (e.g.\ $N$-subjettiness, \tautwoone and
\tauthrtwo), declustering the jet (e.g.\ splitting measures, \Donetwo and
\Dtwothr), or using all jet constituents to quantify the dispersion of the jet
constituents in an axis-independent way (e.g.\ planar flow or energy correlation
functions).  In previous ATLAS studies~\cite{PERF-2015-03,PERF-2015-04}, it was
found that for \Wboson boson tagging, energy correlation variables, in
particular \DTwo, were the best-performing tagging observables while for
top-quark tagging the $N$-subjettiness ratio, \tauthrtwo, was found to be
optimal among the techniques considered.  This can be understood from an
analytical point of view in the context
of \Wboson-boson tagging~\cite{Larkoski:2013eya} and is attributed to additional
wide-angle radiation present in parton jets originating from \Wboson-boson
decays, which is more fully exploited in the energy correlation functions than
in the $N$-subjettiness moments.
 
The full set of jet moments studied in this work is summarised in
Table~\ref{tab:variables} while a more complete description of the observables
under study can be found in Ref.~\cite{PERF-2015-03}.  These moments are studied
individually when paired with the jet mass (\mcomb) as well as in multivariate
combinations, similar to those studied in
Refs.~\cite{Gallicchio:2010dq,CMS-JME-13-006,Adams:2015hiv}, with the intention
of exploiting correlations between the observables and creating a more powerful
single discriminant across a broad \pt range from 200 to 2000~\GeV, the range
commonly probed in searches.
 
\begin{table}
\small
\centering
\renewcommand{\arraystretch}{1.2}
\caption{
Summary of jet moments studied along with an indication of the tagger topology
to which the observable is applicable.  In the case of the energy correlation
observables, the angular exponent $\beta$ is set to 1.0 and for the
$N$-subjettiness observables, the winner-take-all \cite{Larkoski2014}
configuration is used.  A concise description of each jet moment can be found in
Ref.~\cite{PERF-2015-03}.}
\label{tab:variables}
\begin{tabular}{|c|c|c|c|c|}
\hline
Observable  & Variable                      & Used for       & References \\
\hline
\hline
Calibrated jet kinematics                 & \pt, \mcomb                      & top,\Wboson         & \cite{ATLAS-CONF-2016-035} \\
\hline
Energy correlation ratios  & $e_{3}$,\CTwo,\DTwo              & top,\Wboson         & \cite{Larkoski:2013eya,Larkoski:2014gra} \\
 
\hline
\multicolumn{1}{ |c|  }{\multirow{2}{*}{$N$-subjettiness} }
& \tauone,\tautwo,\tautwoone       & top, \Wboson      & \cite{Thaler:2010tr,Thaler:2011gf} \\
& \tauthr,\tauthrtwo               & top          &           \\
\hline
Fox--Wolfram moment         & \FoxWolfRatio                    & \Wboson         & \cite{Fox:1978vu, Chen:2011ah} \\
\hline
\multicolumn{1}{ |c|  }{\multirow{3}{*}{Splitting measures} }
& \zcut                  & \Wboson            &  \cite{Thaler:2008ju, STDM-2012-12} \\
& \Donetwo               & top, \Wboson       &   \\
& \Dtwothr                & top        &  \\
 
\hline
Planar flow & \PlanarFlow & \Wboson & \cite{Almeida:2008tp} \\
\hline
Angularity  & \Angularity & \Wboson & \cite{STDM-2011-38}   \\
\hline
Aplanarity  & \Aplanarity & \Wboson & \cite{Chen:2011ah}   \\
\hline
KtDR        & \KtDR & \Wboson & \cite{Catani:246812} \\
\hline
Qw          & \Qw & top & \cite{Thaler:2008ju} \\
\hline
\end{tabular}
\end{table}
 
\subsubsection{Topocluster-based Tagger}
All of the jet moments presented in Section~\ref{sec:technique_moments} and summarised in Table~\ref{tab:variables} make use
of a specific physical motivation to distil the individual jet constituent
measurements into a single observable.  However, recent simulation-based studies
have found that the more direct use of the jet
constituents~\cite{deOliveira:2015xxd,Baldi:2016fql,Pearkes:2017hku,Kasieczka:2017nvn}
as inputs to a machine-learning algorithm can lead to significant improvements
in discriminating power as compared to more traditional, jet-moment-based
discriminants.  Therefore, in this work, a classifier that makes use of lower-level
input observables is investigated which focuses specifically on the
identification of high-\pt top quarks with $\pt >450~\GeV$.  This
classifier is referred to as ``TopoDNN'' throughout the work.
 
\subsubsection{Shower deconstruction}
\label{sec:technique_sd}
Shower deconstruction (SD)~\cite{Soper:2012pb} is an approach which attempts to
classify jets according to the compatibility of the radiation pattern of the jet with
a predefined set of parton shower hypotheses in a manner similar to the matrix
element method~\cite{Gainer:2013iya}. For a set of input subjets, intended to be
representative of the partonic decay products of the top quark, loose
compatibility with the decay of a top quark is ensured by requiring that the jet
has at least three subjets, that two or more subjets have a mass in a window
centred around the \Wboson-boson mass ($\Delta\MW$), and that at least one more
subjet can be added to obtain a total mass in a window centred around the top-quark
mass ($\Delta\MTop$).  If the jet passes these requirements, then a set of
potential shower histories is constructed for the signal and background models.
Each shower history represents a possible means by which the chosen model could
have resulted in the given subjet configuration.  A probability is assigned to
each shower history based on the parton shower model from which the $\chi$
variable is defined as the likelihood ratio of the signal and background
hypotheses.  The logarithm of this likelihood ratio $\log \chi$ is used as the
final discriminant.  The precise values of the parameters in this algorithms are
described in Section~\ref{sec:optimisation_showerdeconstruction}.
 
\subsubsection{HEPTopTagger}
\label{sec:technique_htt}
An alternative approach to \topquark tagging is the \htt (HTT)
algorithm~\cite{Plehn:2009rk,Plehn:2010st}.  Unlike the previous observables
that are calculated from the constituents of the $R$ = 1.0 trimmed jets, this
technique relies on reconstructing jets using the C/A algorithm with $R=1.5$ to
allow the tagging of fully contained boosted top quarks to be effective at lower
values of \pt ($>200~\GeV$) and to take advantage of the C/A clustering sequence
which attempts to reverse the decay structure of the \topquark decay. The
constituents of the ungroomed uncalibrated C/A jet are analysed with the \htt
algorithm, which identifies the hard jet substructure and
tests it for compatibility with the 3-prong pattern of hadronic \topquark decays
using an algorithm which is designed to mitigate the effects of \pileup
by removing low-\pt portions of the jet.  The \htt studied in this paper is the
original algorithm, from Ref.~\cite{Plehn:2009rk}, not the extended
\textsc{HEPTopTagger2} algorithm~\cite{Kasieczka:2015jma} and is executed with
$\mcut=50~\GeV$, $\Rfilt^{\textrm{max}}$ = 0.25, \Nfilt = 5, $f_W$ = 15\%,
settings found to be optimal in Ref.~\cite{PERF-2015-04}.  The result of the
algorithm is a \topquark-candidate four-vector. The jet is considered to be
tagged if the mass of this resultant \topquark-candidate four-vector
is between $140$ and $210~\GeV$ and its \pt is larger than $200~\GeV$.
 
\section{Tagger optimisation}
\label{sec:TaggerOptimization}
A wide variety of techniques, described in Section~\ref{sec:TaggerTechniques},
exist for identifying \Wboson-boson and top-quark jets. In this section, each of
these techniques is explored and optimised and an inclusive comparison of the
performance of each technique is made based on the \Wboson-boson or top-quark
(signal) efficiency and light-jet (background) rejection, defined as the inverse
of the background efficiency.  This performance is quantified in exclusive
kinematic regimes based on the \pt of the associated \akt $R=1.0$ truth jet
(\pTtrue) to more closely resemble the kinematics of the parent particle and
allow comparison of taggers employing different jet clustering
algorithms.  Finally, to mitigate any bias in the tagging performance due to
differences between the \pt spectra of the signal and background jet samples, the
simulated signal samples described in Section~\ref{sec:samples} are combined and
weighted (separately for \Wboson bosons and top quarks) such that the truth \pt
distribution of the ensemble of signal jets matches that of the light-jet
background.
 
\subsection{Cut-based optimisation}
\label{sec:optimisation_cutbased}
The first approach to tagging is based on selection cuts on jet shape
observables.  This approach was studied in preparation for Run 2
~\cite{ATL-PHYS-PUB-2015-033,ATL-PHYS-PUB-2015-053} to provide a set of guiding
techniques that were used extensively in searches.  The primary goal of these
taggers is to provide a simple set of selections on jet moments that yield a
constant signal efficiency as a function of the transverse momentum of the jet
across a broad \pt range, thus being widely applicable.   In the case of \Wboson-boson
tagging, one of these observables is taken to be \mcomb and the
discrimination power is augmented by a selection on another jet moment defined
in Table~\ref{tab:variables}, while in the case of top tagging, a more inclusive
strategy is explored where all pairwise combinations of jet moments are
investigated.  This optimisation is performed as a function of the \pt of the
associated \akt  $R=1.0$ truth jet for both \Wboson-boson and top-quark tagging.
The tagging strategy resulting from this optimisation provides a benchmark in terms
of tagging performance to which other tagging strategies can be compared.
 
This simple tagger is optimised using a sample of signal \Wboson-boson or top-quark
jets as well as background light jets extracted from the samples described
in Section~\ref{sec:samples}.  In each event the two reco jets matched to the two highest-\pt truth jets
within $|\eta|<2$ are studied.  In the case of signal, \Wboson-boson (top-quark)
jets are retained if they are truth labelled as such according to the procedure
in Section~\ref{sec:labelling} and have a transverse momentum greater than
200~$\GeV$ (350~$\GeV$).  In the case of background, no labelling procedure is
applied and the two highest-\pt jets from the dijet sample are retained.
 
For this study, the general optimisation procedure to determine the two-variable
selection criteria is the same for both \Wboson-boson and top-quark jet tagging.
For each pair of observables, the selection criteria which give the chosen
signal efficiency and the largest background rejection are considered optimal
and taken as the selection criteria in that region of jet \pt.  In the case of
\mcomb, the selection region is two-sided for \Wboson-boson tagging, selecting a
region near \MW, and one-sided in the case of top-quark tagging, selecting an
inclusive region of high jet mass.  In the case of the other jet moments, the
selection criteria are always one-sided, the direction of which depends on the
particular observable in question.  This procedure is repeated for exclusive
bins of jet \pt and a sequence of selection criteria for each of the jet moment
observables is derived.  Finally, this sequence of selection criteria is
parameterised by a smooth function dependent on the jet \pt. All single-sided cuts are
parameterised as a function of \pt with a polynomial function to describe features
which occur due to correlation of the combined-tagger variable.  In the case of the \Wboson-boson
tagging, the \mcomb selection is fit using a four-parameter ($p_{i}$) function
of the form $\sqrt{(p_{0}/\pt + p_{1})^2 + (p_{2}\cdot \pt +p_{3})^{2}}$ chosen
to encapsulate the dominant effects on the jet mass resolution.  Throughout this
work, the targeted signal efficiencies are taken to be constant with respect to
jet \pt with values of 50\% for \Wboson-boson tagging and 80\% for top-quark
tagging.  These signal efficiency working points are largely based on those
commonly used in searches for physics beyond the Standard Model.
In the case of top-quark tagging, a working point with higher efficiency is commonly
used because the dominant backgrounds involve processes including real top
quarks~\cite{EXOT-2015-04,EXOT-2017-34} while in the case of searches for
signals involving \Wboson-boson jets, the backgrounds are largely dominated by
processes involving light-quark jets~\cite{EXOT-2016-01,EXOT-2016-30} thereby
requiring a selection that more effectively rejects background at the expense of
signal efficiency.
 
In Figure \ref{fig:simpletagger_perf}, the resulting background rejections as a
function of the jet \pTtrue are shown for a selection of the most powerful
two-variable combinations.  Based on this study, in the case of \Wboson tagging,
the combination of \mcomb and \DTwo is most powerful in the kinematic range of
interest and is taken as the baseline pairing for \Wboson tagging.  However, at
higher jet \pTtrue, where the power of \DTwo decreases, \Donetwo retains
constant discrimination power.  In the case of top-quark jet tagging, the behaviour
of the most powerful taggers provide a large background rejection at low \pTtrue,
plateauing at a lower value for high jet \pTtrue mostly due to  the migration
of the light-jet mass distribution to higher values and a looser \tauthrtwo cut
to maintain the constant signal efficiency.
The two-variable combinations that do not involve mass perform marginally better
than those with mass across the entire kinematic range studied.
As a consequence, the specific cut-based top-quark jet tagger used in an analysis may
depend on the context of the analysis and not on the performance alone. Therefore,
the baseline two-variable cut-based top-quark jet tagger is selected to be the
one composed of one-sided selections on \mcomb and \tauthrtwo, as it has
been commonly used in ATLAS.
 
\begin{figure}[htpb]
\centering
\subfigure[][]{ \label{fig:simpletagger_perf_1} \includegraphics[width=0.49\textwidth]{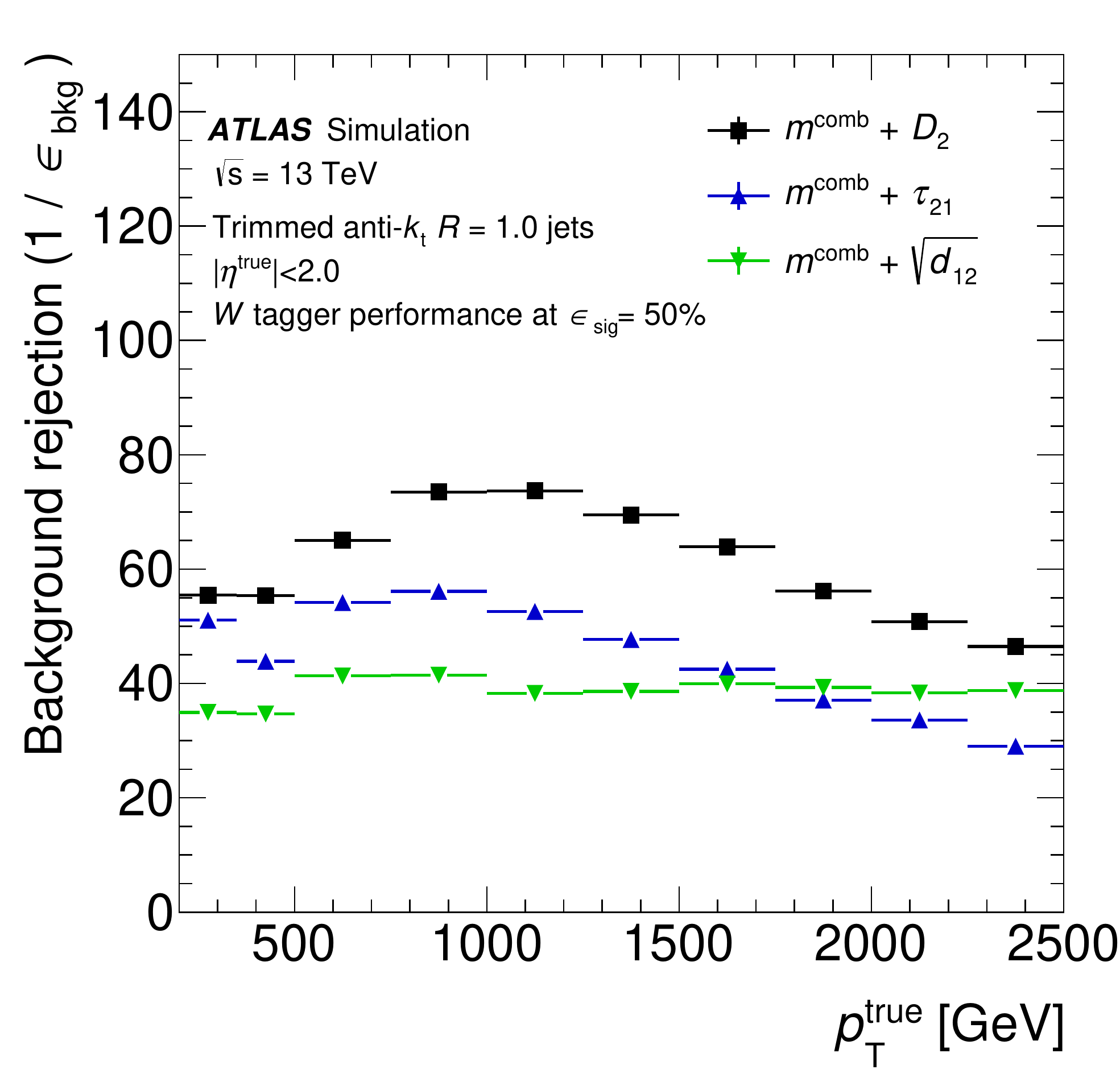}}
\subfigure[][]{ \label{fig:simpletagger_perf_2} \includegraphics[width=0.49\textwidth]{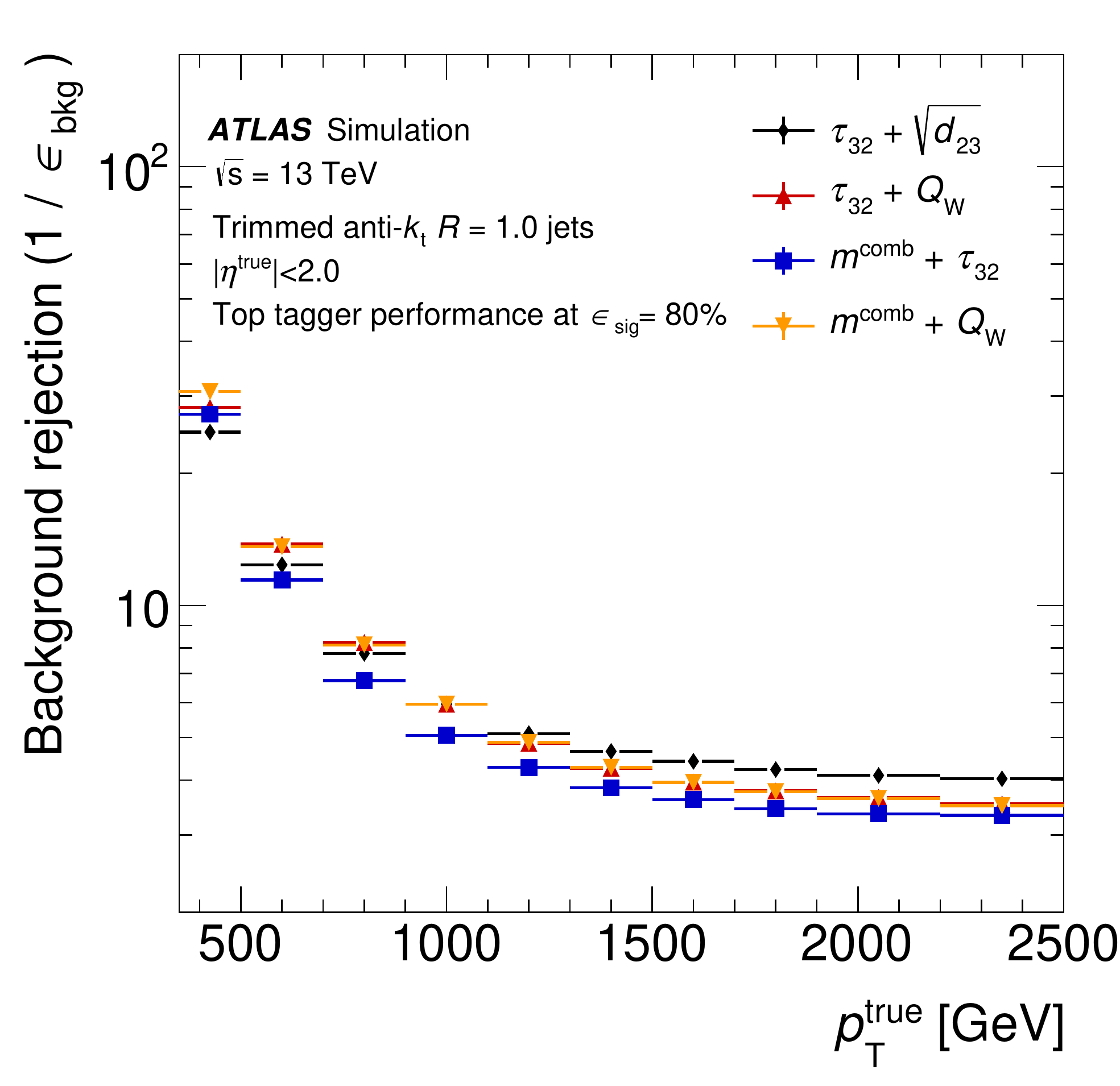}}
\caption{
\label{fig:simpletagger_perf} The \Wboson-boson~\subref{fig:simpletagger_perf_1}
and top-quark tagging~\subref{fig:simpletagger_perf_2} background rejection as a
function of jet \pTtrue for the best performing two-variable combinations at
fixed signal efficiency.}
\end{figure}
 
\subsection{Jet-moment-based multivariate taggers}
\label{sec:optimisation_bdtdnn}
Some of the moments presented in Section~\ref{sec:technique_moments} contain
complementary information and it has been shown that combining these observables
by creating a multivariate \Wboson-boson or top-quark classifier provides higher
discrimination, albeit to differing
degrees~\cite{CMS-JME-13-006,CMS-PAS-JME-15-002,ATL-PHYS-PUB-2017-004}. In this
work boosted decision tree (BDT) and deep neural network (DNN) algorithms are
investigated following a procedure similar to the one in Ref.~\cite{ATL-PHYS-PUB-2017-004}.
The goal is to discriminate \Wboson-boson and top-quark jets from light jets and
to provide a single jet-tagging discriminant that is widely applicable in place
of the single jet moment, described in Section~\ref{sec:optimisation_cutbased},
to augment the discrimination of \mcomb alone across a broad \pt range,
providing another widely applicable and more powerful tagger.
 
The two algorithmic classes used here, BDTs and DNNs, are explored in parallel
to determine if one of the architectures is better suited to exploit
differences between the input observables and their correlations among
high-level variables in signal and background. The DNN used here is a
fully-connected feed-forward network. Given that both algorithms have access
to the same set of input features, of which there are approximately ten, it is
expected that the discrimination power will be approximately the same. The
internal settings, so called hyper-parameters, used for the BDTs and DNNs are
summarized in Appendix~\ref{sec:bdt_&_dnn_hyperparam}.
For the design of all multivariate
discriminants, exclusive subsamples of signal and background jets are derived
from the more inclusive sample selected as in
Section~\ref{sec:optimisation_cutbased} to be used separately for the training
and testing of the discriminant.  To ensure that all jet substructure features
are well-defined for the training, two additional selection criteria are applied
to the jet mass ($\mcomb > 40$~\GeV) and number of constituents (\NConst$\geq$3).
The jets which fail to meet these criteria are not used in the training.  However, in
the evaluation of the performance of the tagger, such jets are classified as
background jets only if they fail the \mcomb requirement, taking this auxiliary
selection into account in the calculation of the signal efficiency and background
rejection.  The chosen input observables used for either \Wboson-boson or
top-quark tagging are the full set of observables summarised in
Table~\ref{tab:variables}, noting that both the jet mass (\mcomb) and transverse
momentum are directly used as inputs.  Therefore, when defining a final working
point for this tagger, unlike in the case of the cut-based taggers in
Section~\ref{sec:optimisation_cutbased}, no additional direct selection beyond
the $\mcomb>40$~\GeV\ requirement is imposed on the mass.  Finally, in the design of the
classifiers, all studies are performed in a wide \pTtrue bin\footnote{These
bins are taken to be $[200,2000]$~\GeV{} for W boson tagging and
$[350,2000]$~\GeV{} for top quark tagging.} and jets are given weights to
create a constant \pTtrue spectra so as to not bias the training.
However, the performance comparison of these taggers with the cut-based ones, as
well as the full comparison of all tagging techniques in
Section~\ref{sec:optimisation_summary}, is made with \pTtrue distributions for
signal jets weighted to match that of the multijet background sample.
 
The set of observables used in the BDT classifiers is determined using a
procedure in which the observables applicable to each topology, specified in
Table~\ref{tab:variables}, which give the largest increase in relative
performance are sequentially added to the network.  For each successive
observable that is to be added to the classifier, the BDT classifier is trained
with jets from the training set and the relative performance is evaluated using
jets from the testing sample and the variable which gives the greatest increase
in relative background rejection at a fixed relative signal efficiency of 50\%
(\Wboson-boson tagging) and 80\% (top-quark tagging) is retained.
\emph{Relative signal efficiency} and \emph{relative background rejection} take
into account only the jets that satisfy the training criteria, where \emph{relative
signal efficiency} is defined as
\begin{equation*}
\epsilon_{\mathrm{sig}}^{\mathrm{rel}} = \frac{N^{\mathrm{tagged}}_{\mathrm{signal, \mcomb > 40~\GeV, N^{\text{const}} > 2}}}{N^{\mathrm{tagged \& untagged}}_{\mathrm{signal, \mcomb > 40~\GeV, \NConst > 2}}}
\end{equation*}
and in a similar manner, \emph{relative background rejection} is defined as $1/
\epsilon_{\mathrm{bkg}}^{\mathrm{rel}}$.  The smallest set of variables which
reaches the highest relative background rejection within statistical
uncertainties is selected.  The minimum number of selected variables is 11 for
\Wboson-boson tagging and 10 for top-quark tagging. The relative background
rejection achieved at each stage for both classifiers is shown in
Figure~\ref{fig:bdt_variable_addition}.
 
\begin{figure}[htpb]
\centering
\subfigure[][]{  \label{fig:bdt_variable_addition_1} \includegraphics[width=0.49\textwidth]{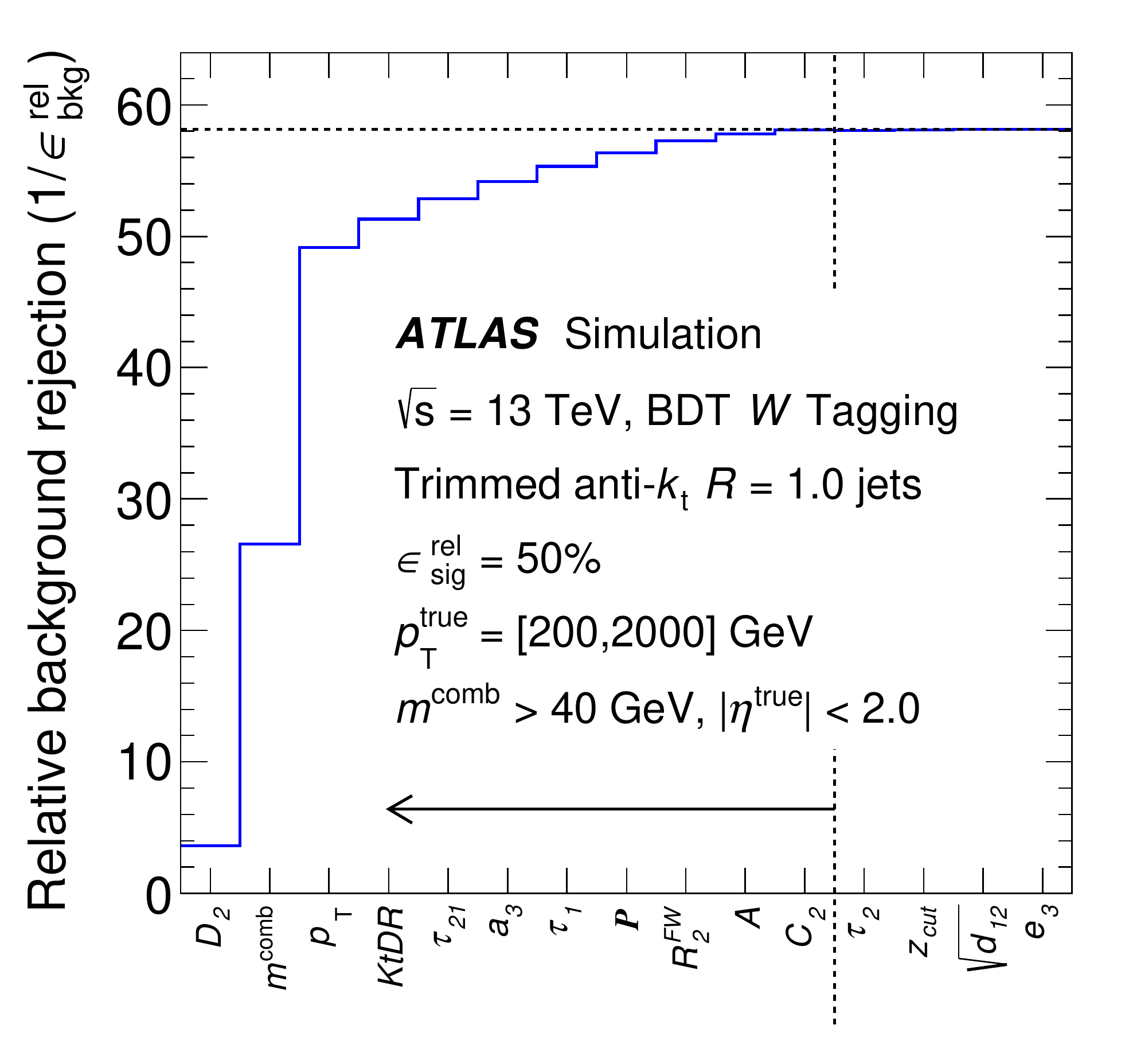}}
\subfigure[][]{  \label{fig:bdt_variable_addition_2} \includegraphics[width=0.49\textwidth]{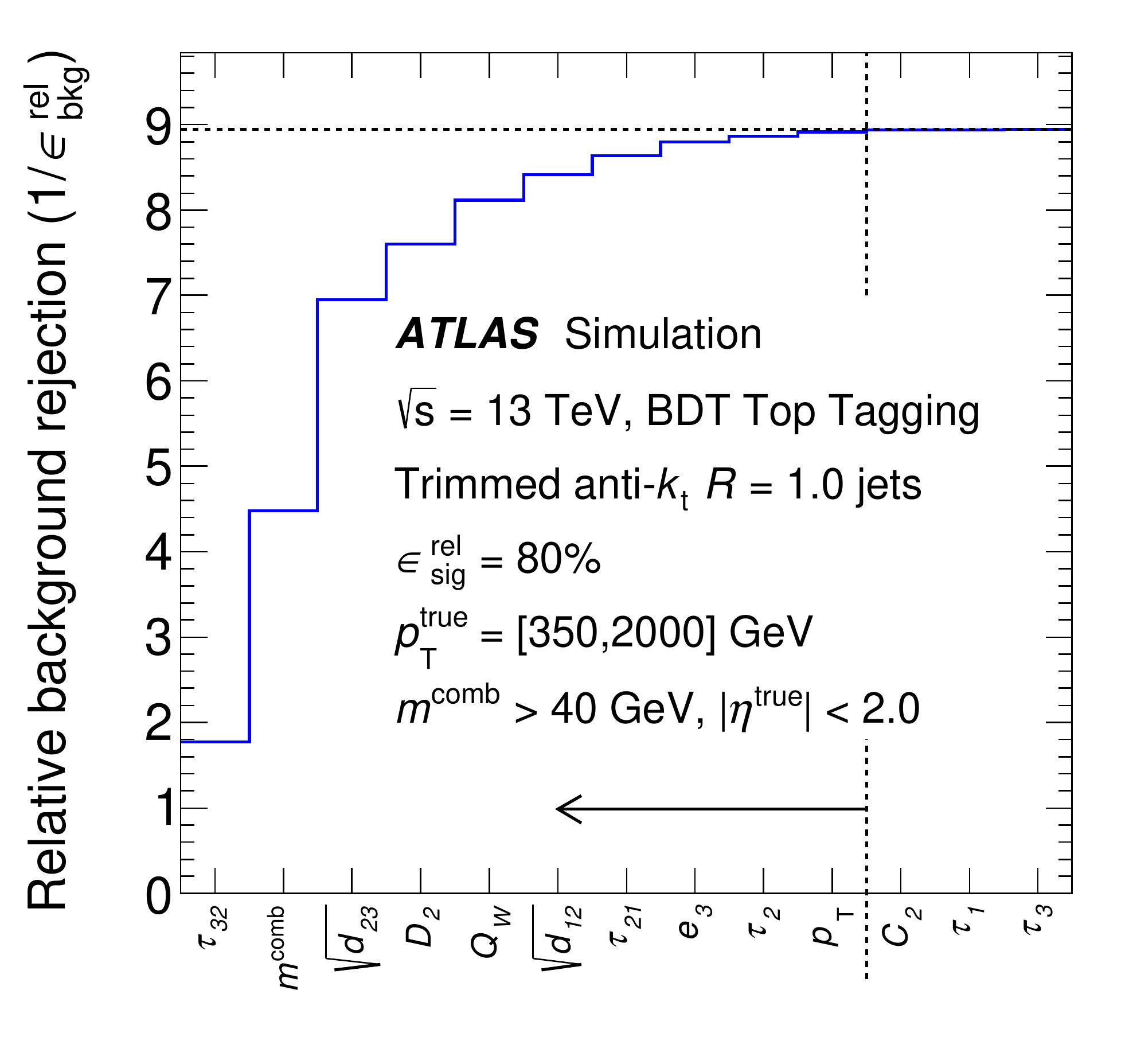}}
\caption{
\label{fig:bdt_variable_addition} The relative background rejection of the
jet-shape-based BDT discriminant for different sets of variables, with more variables
added successively at the 50\% (\Wboson-boson tagging) and 80\%
(top-quark tagging) relative signal efficiency working point for
\Wboson-boson~\subref{fig:bdt_variable_addition_1} and
top-quark~\subref{fig:bdt_variable_addition_2} tagging. Only jets which satisfy the
training criteria are considered when calculating the relative signal
efficiency and relative background rejection. The performance is evaluated with
constant \pTtrue spectra. Uncertainties are not presented.  The horizontal
dashed lines indicate the level of performance saturation, while the vertical
dashed lines and solid arrow represent the set of jet moments used in the final
construction of the discriminant.}
\end{figure}

In a similar manner, the observables used in the DNN classifier are chosen by
comparing the performance when using different sets of input variables to find
the set of observables which gives the largest relative background rejection at
a fixed relative signal efficiency. In this case, variables are not added in
succession due to the time requirements to train the large number of networks.
Instead, groups of observables are chosen by selecting variables according to their
dependence on the momentum scale of the jet substructure objects, what
features of the substructure they describe and their dependence on other
substructure variables. A summary of all the variables tested for the DNN is
shown in Table~\ref{tab:bdtdnn_variable_summary}.  For each group, the DNN
classifier is constructed using the training set of jets and the relative
performance is evaluated using the jets in a testing set.  The relative
background rejection achieved inclusively in jet \pTtrue is shown in
Figure~\ref{fig:dnn_variable_addition}.  The performance of the DNN tagger
depends on both the number of variables and the information content in the
group.  The chosen groups of inputs for \Wboson-boson tagging and top-quark
tagging are listed in Table \ref{tab:bdtdnn_variable_summary}.  Within
statistical uncertainties, the number of variables necessary for maximum
rejection at a fixed relative signal efficiency of 50\% (\Wboson-boson tagging)
and 80\% (top-quark tagging) is found to be 12 variables for \Wboson-boson
tagging (Group 8 in Table~\ref{tab:bdtdnn_variable_summary}) and 13 variables
for top-quark tagging (Group 9 in Table~\ref{tab:bdtdnn_variable_summary}).
 
\begin{figure}[hp]
\centering
\subfigure[][]{  \label{fig:dnn_variable_addition_1} \includegraphics[width=0.49\textwidth]{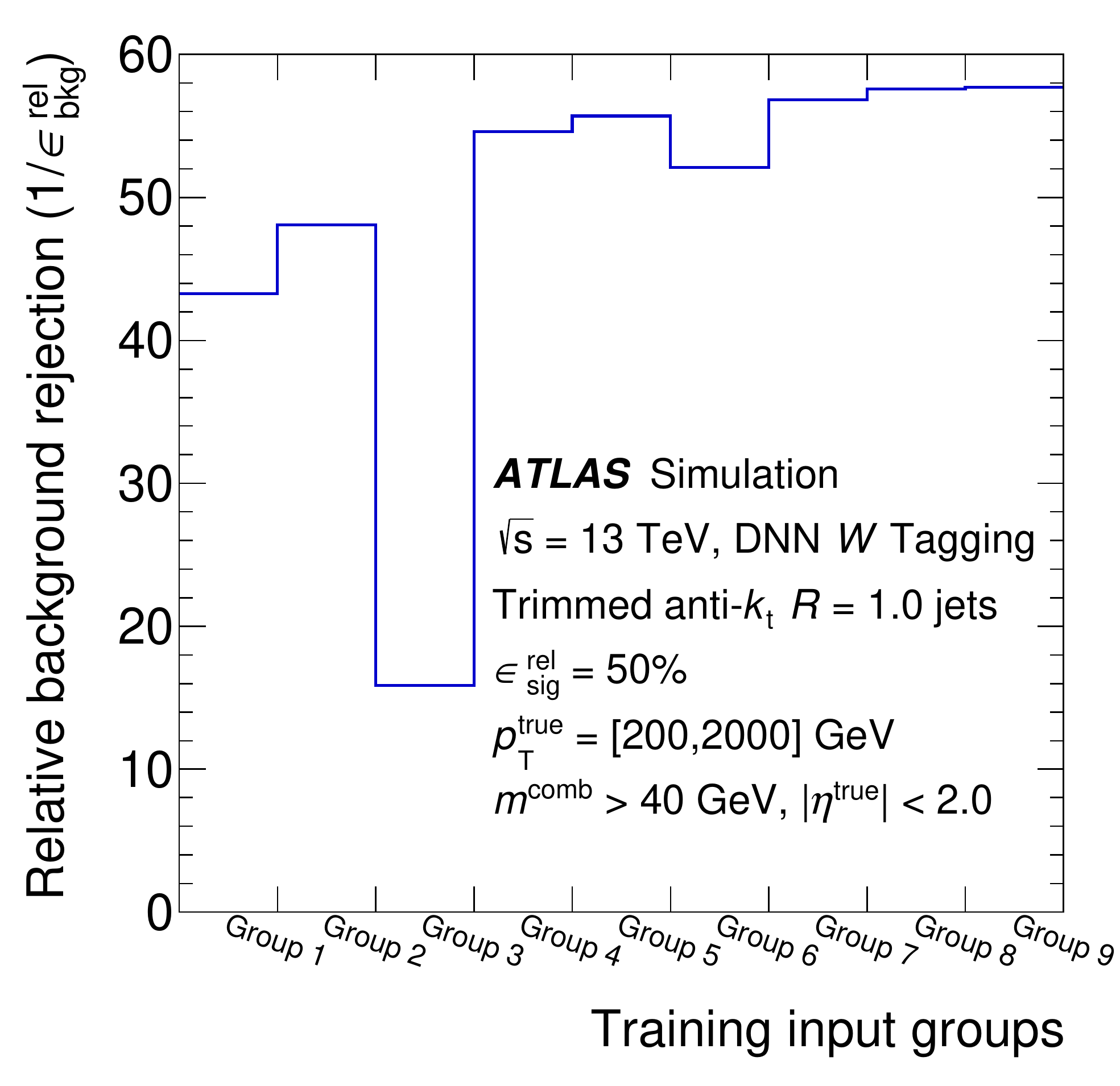}}
\subfigure[][]{  \label{fig:dnn_variable_addition_2} \includegraphics[width=0.49\textwidth]{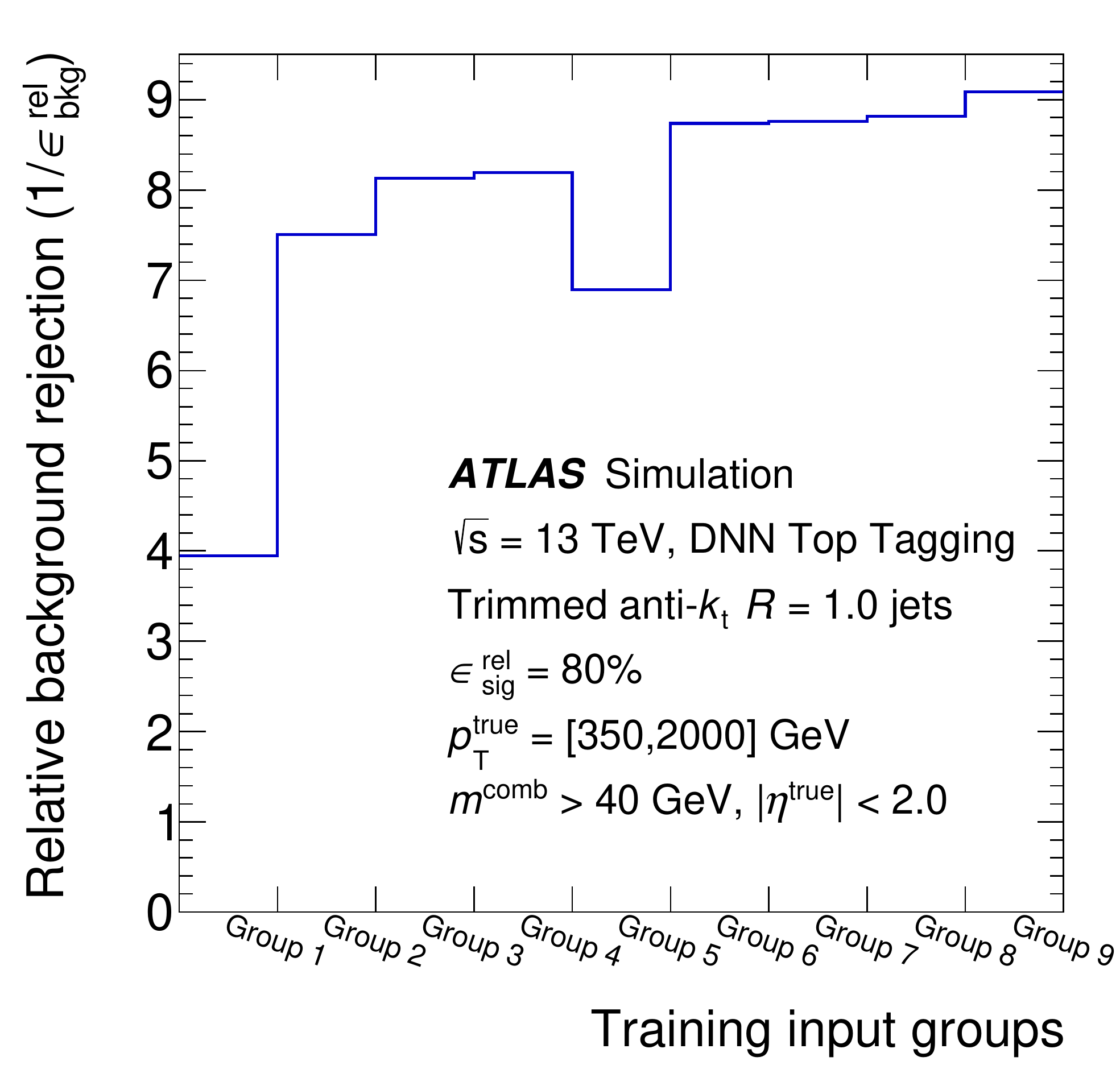}}
\caption{
\label{fig:dnn_variable_addition} Distributions showing the training with
different set of variables and relative improvement in performance for the DNN
\Wboson-boson~\subref{fig:dnn_variable_addition_1} and
top-quark~\subref{fig:dnn_variable_addition_2} taggers at the 50\% and 80\%
relative signal efficiency working point, respectively. The grouping of
observables was decided prior to training and discriminator performance
evaluation. Only jets which satisfy the training criteria are considered when
calculating the relative signal
efficiency and relative background rejection. The performance is evaluated with
constant \pTtrue spectra. Uncertainties are not presented.}
\end{figure}

\begin{table}[hp]
\centering
\scriptsize
\caption{
\label{tab:bdtdnn_variable_summary}A summary of the set of observables that were
tested for \Wboson-boson and top-quark tagging for the various DNN input
observable groups as well as the final set of DNN and BDT input observables as
chosen using Figures~\ref{fig:bdt_variable_addition}
and~\ref{fig:dnn_variable_addition}.}
\begin{tabular}{ c| c c c c c c c c c |c c|| c c c c c c c c c |c c| }    
\hhline{~|-|-|-|-|-|-|-|-|-|-|-||-|-|-|-|-|-|-|-|-|-|-}
& \multicolumn{11}{c||}{\Wboson Boson Tagging}                                                            & \multicolumn{11}{c|}{Top Quark Tagging}            \\
\hhline{~|-|-|-|-|-|-|-|-|-|-|-||-|-|-|-|-|-|-|-|-|-|-}
& \multicolumn{9}{c|}{DNN Test Groups}                                                            & \multicolumn{2}{c||}{Chosen Inputs} & \multicolumn{9}{c|}{DNN Test Groups}                                                            & \multicolumn{2}{c|}{Chosen Inputs} \\
\hhline{|-||-|-|-|-|-|-|-|-|-|-|-||-|-|-|-|-|-|-|-|-|-|-}
\hhline{=::=:=:=:=:=:=:=:=:=:=:=::=:=:=:=:=:=:=:=:=:=:=}
\multicolumn{1}{|c||}{Observable}    & 1   & 2   & 3   & 4   & 5   & 6   & 7   & 8   & 9   & BDT & DNN & 1   & 2   & 3   & 4   & 5   & 6   & 7   & 8   & 9   & BDT & DNN \\
\multicolumn{1}{|c||}{\mcomb}        & \ci & \ci &     & \ci & \ci & \ci & \ci & \ci & \ci & \ci & \ci &     & \ci & \ci & \ci &     & \ci & \ci & \ci & \ci & \ci & \ci   \\
\multicolumn{1}{|c||}{\pt}           & \ci & \ci &     &     & \ci & \ci &     & \ci & \ci & \ci & \ci &     &     & \ci & \ci &     &     & \ci & \ci & \ci & \ci & \ci   \\
\multicolumn{1}{|c||}{\ECFThrNorm}   & \ci & \ci &     &     &     & \ci &     &     & \ci &     &     &     &     &     & \ci &     &     & \ci &     & \ci & \ci & \ci   \\
\multicolumn{1}{|c||}{\CTwo}         &     &     & \ci & \ci & \ci &     & \ci & \ci & \ci &     & \ci & \ci & \ci & \ci &     & \ci & \ci &     & \ci & \ci &     & \ci   \\
\multicolumn{1}{|c||}{\DTwo}         &     &     & \ci & \ci & \ci &     & \ci & \ci & \ci & \ci & \ci & \ci & \ci & \ci &     & \ci & \ci &     & \ci & \ci & \ci & \ci   \\
\multicolumn{1}{|c||}{\tauone}       & \ci & \ci &     &     &     & \ci &     &     & \ci & \ci &     &     &     &     & \ci &     &     & \ci &     & \ci &     & \ci   \\
\multicolumn{1}{|c||}{\tautwo}       & \ci & \ci &     &     &     & \ci &     &     & \ci &     &     &     &     &     & \ci &     &     & \ci &     & \ci & \ci & \ci   \\
\multicolumn{1}{|c||}{\tauthr}       &     &     &     &     &     &     &     &     &     &     &     &     &     &     & \ci &     &     & \ci &     & \ci &     & \ci   \\
\multicolumn{1}{|c||}{\tautwoone}    &     &     & \ci & \ci & \ci &     & \ci & \ci & \ci & \ci & \ci & \ci & \ci & \ci &     & \ci & \ci &     & \ci & \ci & \ci & \ci   \\
\multicolumn{1}{|c||}{\tauthrtwo}    &     &     &     &     &     &     &     &     &     &     &     & \ci & \ci & \ci &     & \ci & \ci &     & \ci & \ci & \ci & \ci   \\
\multicolumn{1}{|c||}{\FoxWolfRatio} &     &     & \ci & \ci & \ci & \ci & \ci & \ci & \ci & \ci & \ci &     &     &     &     &     &     &     &     &     &     & \\
\multicolumn{1}{|c||}{\PlanarFlow}   &     &     & \ci & \ci & \ci & \ci & \ci & \ci & \ci & \ci & \ci &     &     &     &     &     &     &     &     &     &     & \\
\multicolumn{1}{|c||}{\Angularity}   &     &     & \ci & \ci & \ci & \ci & \ci & \ci & \ci & \ci & \ci &     &     &     &     &     &     &     &     &     &     & \\
\multicolumn{1}{|c||}{\Aplanarity}   &     &     & \ci & \ci & \ci & \ci & \ci & \ci & \ci & \ci & \ci &     &     &     &     &     &     &     &     &     &     & \\
\multicolumn{1}{|c||}{\zcut}         &     &     & \ci & \ci & \ci &     & \ci & \ci & \ci &     & \ci &     &     &     &     &     &     &     &     &     &     & \\
\multicolumn{1}{|c||}{\Donetwo}      &     & \ci &     &     &     & \ci & \ci & \ci & \ci & \ci & \ci &     &     &     &     & \ci & \ci & \ci & \ci & \ci & \ci & \ci   \\
\multicolumn{1}{|c||}{\Dtwothr}      &     &     &     &     &     &     &     &     &     &     &     &     &     &     &     & \ci & \ci & \ci & \ci & \ci & \ci & \ci   \\
\multicolumn{1}{|c||}{\KtDR}         &     & \ci &     &     &     & \ci & \ci & \ci & \ci & \ci & \ci &     &     &     &     &     &     &     &     &     &     & \\
\multicolumn{1}{|c||}{\Qw}           &     &     &     &     &     &     &     &     &     &     &     &     &     &     &     & \ci & \ci & \ci & \ci & \ci & \ci & \ci   \\
\hhline{|-||-|-|-|-|-|-|-|-|-|-|-||-|-|-|-|-|-|-|-|-|-|-}
\end{tabular}
\end{table}
 
Similarly to the cut-based two-variable optimised taggers, for the chosen BDT and
DNN taggers the working points are defined as a function of the reconstructed
jet \pt so that they yield constant signal efficiencies versus \pt.  In both
cases, the target signal efficiency working point is obtained by the fixed jet
mass requirement of $\mcomb > 40~\GeV$, relevant $\NConst$ criteria and a
single-sided selection on the relevant discriminant. The performance of the
resulting BDT and DNN discriminants is characterised by the background
rejection, evaluated as a function of jet \pTtrue, for a fixed signal efficiency
of 50\% (\Wboson-boson tagging) and 80\% (top-quark tagging), where the relative variation
of the signal efficiency for the fixed-efficiency taggers is less than 5\%.  It can be seen in
Figure~\ref{fig:WPs_W_top} that in the case of \Wboson-boson tagging, the
performance improvements beyond the cut-based taggers are highest at low jet \pt
and decrease at higher \pTtrue, presumably due to the merging of calorimeter
energy depositions and subsequent loss of granularity in discerning substructure
information.  However, in the case of top-quark tagging, the improvements in
performance are more sizeable, showing increases in background rejection of
roughly a factor of two over the entire kinematic range studied.  This is
presumably due to the greater complexity of the top-quark decay in contrast to that
of the isolated \Wboson boson, indicating that among the observables studied
here, excluding the multivariate classifiers, no single observable adequately
captures the full set of features that provide ability to discriminate signal
from background. There are richer correlations between the observables
that can be further exploited by the multivariate classification algorithms.
A common feature of both tagging topologies is that the
particular algorithm (i.e.\ BDT and DNN) used to construct the discriminant does
not influence the performance that can be obtained.  This is somewhat expected
due to the relatively small number of inputs found to be useful for the DNN and
helps to put a ceiling on the performance achievable using the combination of
those jet moments examined in this work~\cite{Baldi:2016fql}.

\begin{figure}[htpb]
\centering
\subfigure[][]{  \label{fig:WPs_W_top_1} \includegraphics[width=0.49\textwidth]{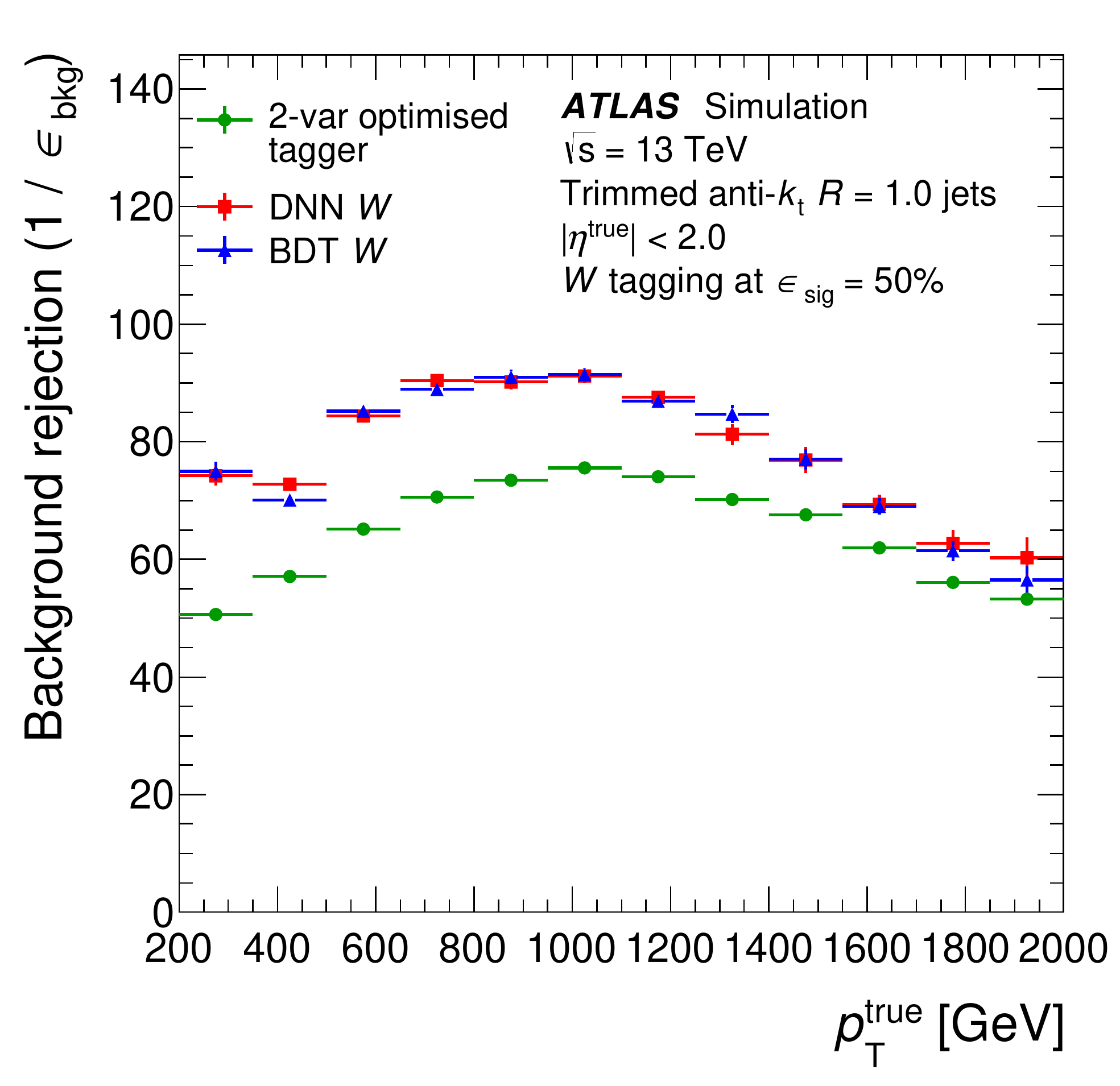}}
\subfigure[][]{  \label{fig:WPs_W_top_2} \includegraphics[width=0.49\textwidth]{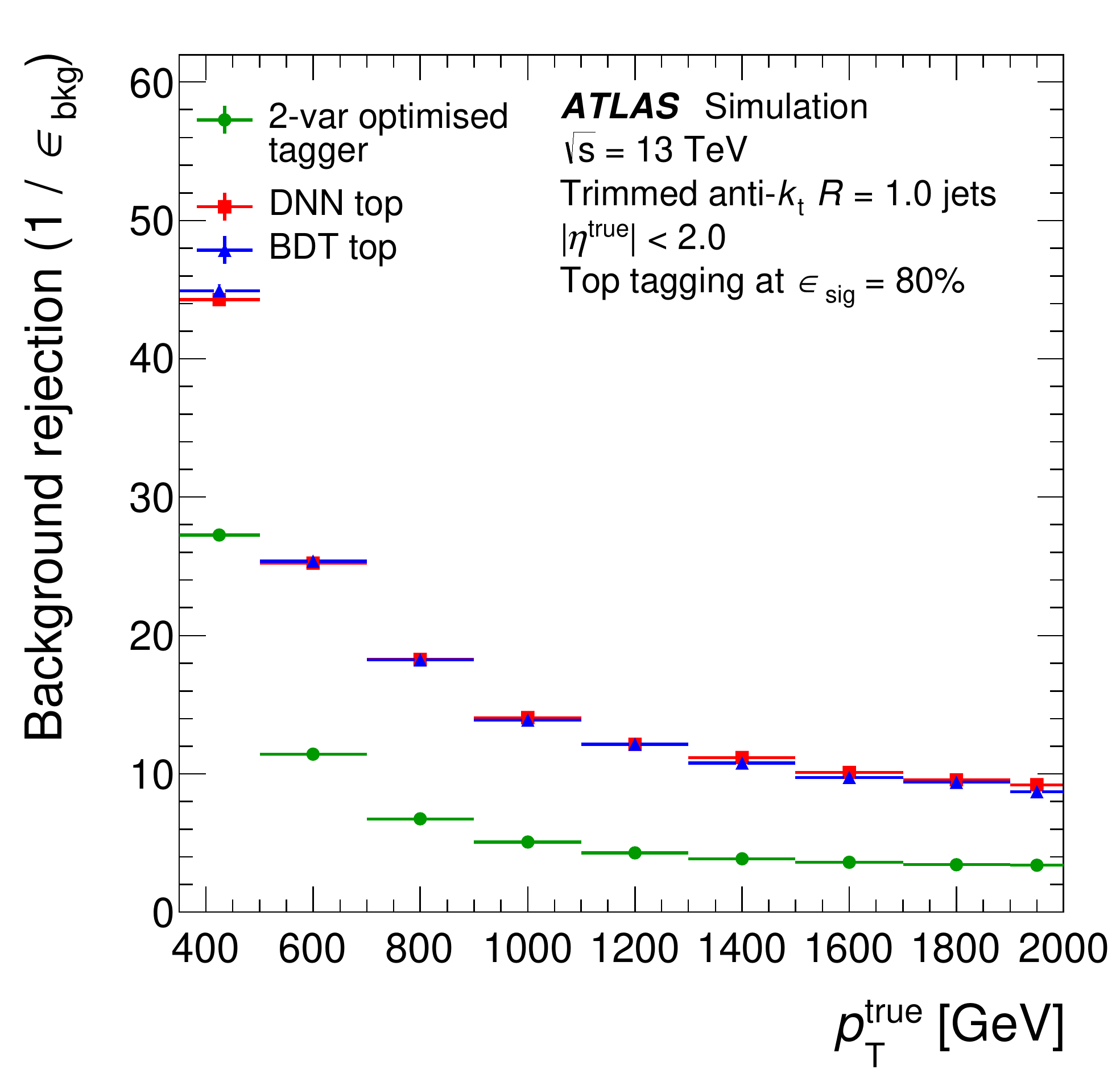}}
 
\caption{
\label{fig:WPs_W_top} The background rejection comparison of \Wboson-boson
taggers at fixed 50\% signal efficiency working point~\subref{fig:WPs_W_top_1}
and top-quark taggers at fixed 80\% signal efficiency working
point~\subref{fig:WPs_W_top_2} for the multivariate jet-shape-based taggers as
well as the two-variable optimised taggers, which are composed of a selection on
\mcomb and \DTwo in the case of \Wboson-boson jet tagging and \mcomb and
\tauthrtwo for top-quark jet tagging. The performance is evaluated with the
\pTtrue distribution of the signal jets weighted to match that of the multijet
background samples. Statistical uncertainties of the background rejection are
presented. }
\end{figure}
 
\clearpage
 
\subsection{Topocluster-based deep neural network tagger}
\label{sec:optimisation_ttt}
Recently, a number of jet phenomenology studies have found that using lower-level
information more directly pertaining to the jet energy flow
can lead to further improvements in the ability
to distinguish signal \Wbosonboson and \topquark jets from light
jets~\cite{deOliveira:2015xxd,Baldi:2016fql,Pearkes:2017hku,Kasieczka:2017nvn,Louppe:2017ipp,Egan:2017ojy,Butter:2017cot,Macaluso:2018tck,Moore:2018lsr}.
Furthermore, it was seen in Figure~\ref{fig:WPs_W_top} that the performance
gains for the high-level variables BDT and DNN combination are significantly
larger for \topquark tagging than for \Wbosonboson tagging. Consequently, a \topquark
jet tagger based directly on the jet constituents, focusing on the high-\pt top quarks with
$\pt >450~\GeV$, is designed.
 
The jet tagger based on low-level jet input information studied in this work
closely follows that described in Ref.~\cite{Pearkes:2017hku} and the reader is
referred there for a more in-depth review of the optimisation of the techniques
used; only a brief summary is provided in the following.  The first aspect of
note which sets this tagger apart from those studied in
Refs.~\cite{deOliveira:2015xxd,Baldi:2016fql,Kasieczka:2017nvn} is that there
is no use of pixelation in this tagger, similar to the taggers studied in Ref.~\cite{Louppe:2017ipp}.
Compared to the taggers studied in Ref.~\cite{Louppe:2017ipp}, the architecture of this tagger
does not employ sequenced, variable-length inputs and the input features used in
this tagger are the four-vectors of fixed-number of topoclusters in the individual \largeR \akt
trimmed jet in the $(\pt,\eta,\phi)$ representation, noting that topoclusters
are taken as massless by convention.  As a preprocessing step, the \pt of each
constituent four-vector is normalised by $1/1700$ to bring the scale of the
input network features within the same magnitude  between approximately 0 and
1.  The $(\eta,\phi)$ location of the set of constituents is then transformed
by a process that involves a translation, a rotation, and a flip based on the
assumed three-subjet topology of a \topquark decay. Of the full set of
constituents, only the 10 highest-\pt constituents are used as input to the
neural network.  This was found to provide optimal background rejection for
this network architecture as compared to using more or fewer clusters and can
be qualitatively understood by examining the fraction of the jet \pt carried by
each of the clusters, shown in Figure~\ref{fig:topoclustertoptagger_pt_frac}
where the distribution of the \pt-fraction for a subset of the 10 highest-\pt clusters is
shown along with the mean value of each of the 20 highest-\pt cluster distributions.  It is
seen that the first 10 clusters, on average carry more than 99\% of the \pt of
the jet.  Therefore, including further clusters saturates the information for
the network to disentangle when discriminating signal from background.  If a
jet has fewer than 10 constituents, the remaining inputs to the neural network
are taken to be null vectors.  The three components of each four-vector are
used as input to a fully connected neural network with four hidden layers
composed of 300, 102, 12 and 6 nodes, respectively. This network architecture
was determined through manual hyper-parameter tuning, exploring configurations
with between 4-6 layers and 40-1000 nodes per layer, and where the used
architecture and hyper-parameters are exactly the same as the one used
in~\cite{Pearkes:2017hku}. The network is trained on jets where only the
initial top parton is required to be matched to the reconstructed jet obtained
from the $\Zboson^{'}$ (signal) and light jets (background) in the high-\pt
region from 450~\GeV\ to 2400~\GeV\ in \pt.  To remove bias in the training due
to the difference in kinematics between the signal and background samples, a subset
of the background ensemble of jets is selected in a random fashion such that the
jet \pt distribution is the same in both signal and background, as opposed to
the BDT and DNN taggers described in Section~\ref{sec:optimisation_bdtdnn},
which use event-by-event reweighting.
 
\begin{figure}[htpb]
\centering
\subfigure[][]{   \label{fig:topoclustertoptagger_pt_frac_a} \includegraphics[width=0.45\textwidth]{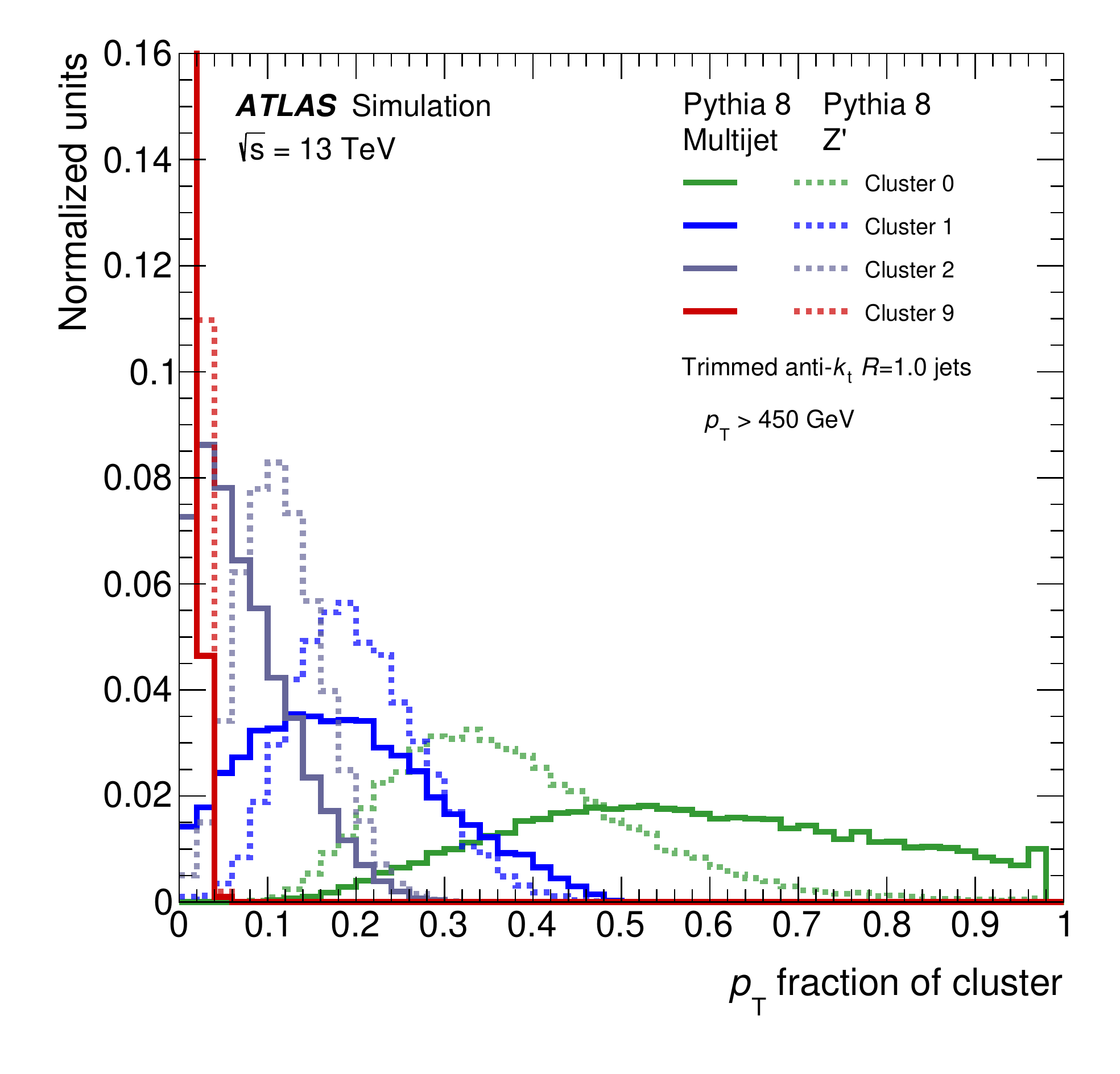}}
\subfigure[][]{   \label{fig:topoclustertoptagger_pt_frac_b} \includegraphics[width=0.45\textwidth]{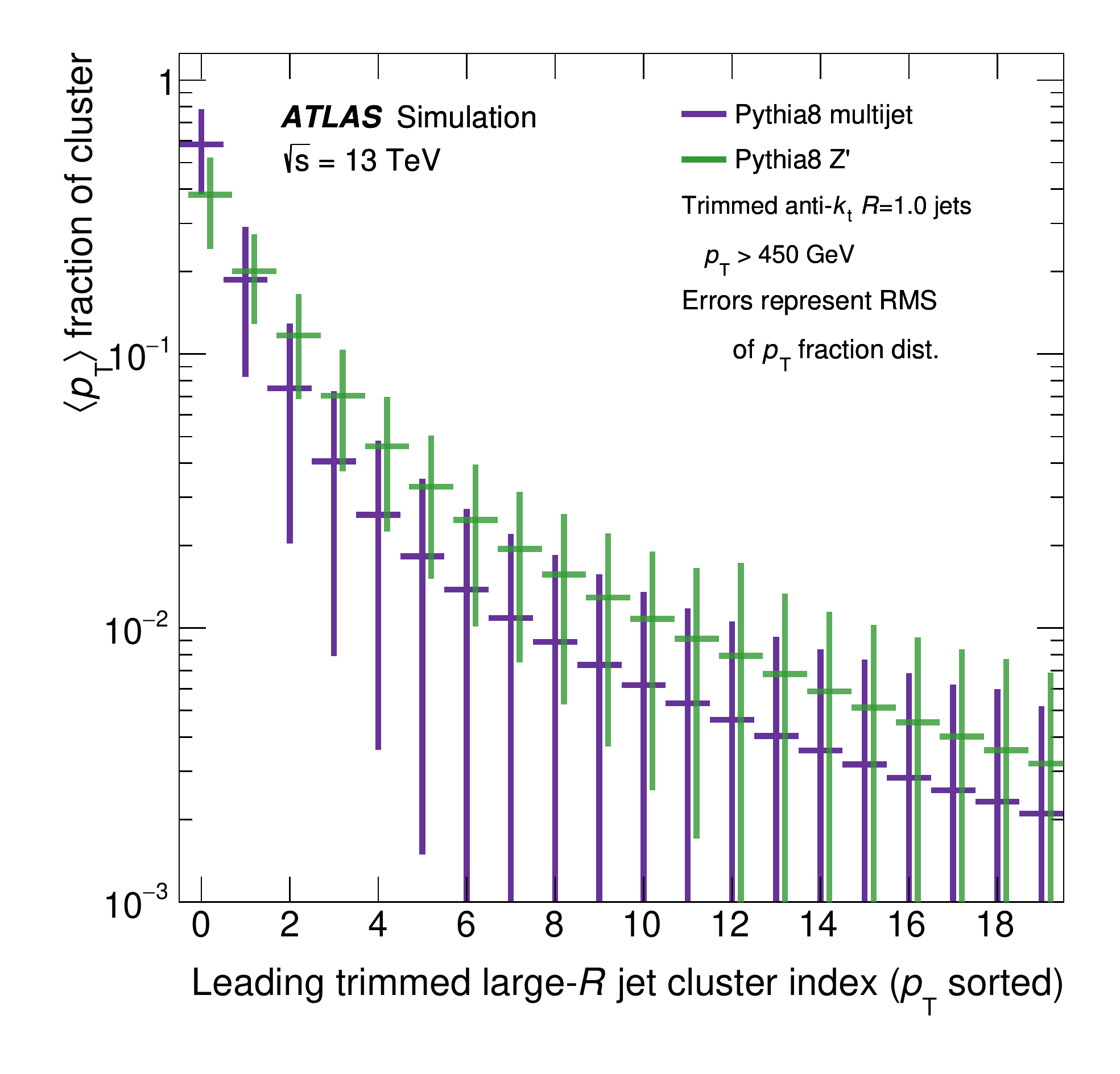}}
 
\caption{
\label{fig:topoclustertoptagger_pt_frac}The distribution of the
fraction of \pt carried by the highest-\pt cluster (Cluster 0) along with the
next-highest (Cluster 1), third-highest (Cluster 2), and tenth-highest-\pt
(Cluster 9) clusters~\subref{fig:topoclustertoptagger_pt_frac_a} along with the average value of the ratio of the cluster
\pt to the jet \pt for the 20 highest-\pt
clusters~\subref{fig:topoclustertoptagger_pt_frac_b}.  The dashed lines
in~\subref{fig:topoclustertoptagger_pt_frac_a} show distributions for signal
jets, and the full lines show distributions for background jets. The vertical lines on
each point in~\subref{fig:topoclustertoptagger_pt_frac_b} represent the RMS of
the corresponding distribution of the fraction of \pt of a given cluster
in~\subref{fig:topoclustertoptagger_pt_frac_a}.
In~\subref{fig:topoclustertoptagger_pt_frac_a}, the distribution for the
tenth-highest-\pt cluster (Cluster 9) extends beyond the maximum value of the vertical axis.
The light-quark jet sample is taken from jets that pass the multijet selection
as described in Section~\ref{subsubsec:background_analysis} while the top-quark
jet sample is taken from jets that pass the semileptonic selection as described
in Section~\ref{subsubsec:leptonjets_analysis}. }
\end{figure}
 
\subsection{Shower deconstruction tagger}
\label{sec:optimisation_showerdeconstruction}
The shower deconstruction tagging method was studied extensively in Run
1~\cite{PERF-2015-04}.  The aim of the method, described in
Section~\ref{sec:technique_sd}, is to determine whether the subjet pattern is
compatible with a parton shower profile typical of a top-quark decay.  In
previous ATLAS studies, the subjets were defined by forming C/A subjets with $R$
= 0.2 using the ungroomed \largeR jet constituents as inputs.  However, in
Run 2, shower deconstruction was recommissioned in the context of the search for
a heavy $\Wboson^{'}$ boson decaying to a top quark and a bottom quark where the
mass-splitting between the $\Wboson^{'}$ and the top quark was large enough to
produce top quarks with momenta of roughly 1~\TeV\ and above~\cite{EXOT-2017-02}.
The approach taken in Run 1 to reconstruct the subjet inputs to the
shower deconstruction algorithm was found to have a low signal efficiency,
largely due to the subjet multiplicity falling below three and therefore
producing a set of subjets that are unable to fulfil the initial consistency
checks between the subjet pairings and triplets with \Wboson-boson and top-quark
masses, respectively.  This drop in efficiency was recovered by altering the
manner in which subjets are constructed to instead use the exclusive-\kt jet clustering
algorithm, run on the constituents of the trimmed \largeR jet.  Since splitting
scales are less dependent on the \largeR jet \pT than the geometric distance
between the jet and its constituents, a stopping criterion is imposed to halt
clustering if \kt splitting scales larger than 15~\GeV\ are found. At that
stage, the resulting set of subjets are used as subjet inputs to shower
deconstruction.  Because the computation time of shower deconstruction scales
exponentially with the number of input subjets, the total number of subjets is
limited to at most the six highest-\pt subjets, compared to a limit of nine
in Run 1, with no loss in performance.  Finally, the parameters controlling the
top-quark topology check using subjet pairings and triplets, \MW and
\MTop respectively, were fixed to 20~\GeV\ and 40~\GeV, the same
as in Run 1.
 
\subsection{Summary of tagger performance studies in simulation}
\label{sec:optimisation_summary}
A direct comparison of the performance of all of the tagging techniques,
described in Section~\ref{sec:TaggerTechniques} and individually optimised in
Section~\ref{sec:TaggerOptimization}, is important in providing guidance as to
which technique can be most beneficial when applied in an analysis.  The primary
metric used to assess the performance of the taggers is the background rejection
as a function of the signal efficiency, characterised in the form of a receiver
operating characteristic (ROC) curve, shown in Figures~\ref{fig:roccurves_w}
and~\ref{fig:roccurves_top} for \Wboson-boson and top-quark tagging,
respectively, for both a low-  and high-\pt kinematic region.  For comparison,
two relatively simple cut-based taggers composed of selections on \mcomb and a
single substructure observable are shown.  In the case of \Wboson-boson tagging,
a fixed mass window requirement of $60 < \mcomb < 100~\GeV$ is applied and a
cut on the \DTwo observable is used for the ROC curve.  In the case of top-quark
tagging, the mass selection is one-sided, requiring $\mcomb > 60~\GeV$, and a
requirement on \tauthrtwo is varied to obtain the ROC curve.  These simple
taggers, along with the specific working points tuned to give constant signal
efficiency and maximal background rejection, are provided as a point of
reference for subsequent optimisations that were performed for studies of the
more advanced techniques.
 
When examining Figures~\ref{fig:roccurves_w} and~\ref{fig:roccurves_top}, it can
be seen that a careful tuning of the simple two-variable cut-based taggers can
lead to sizeable gains due to taking into consideration the correlation
between \mcomb and the auxiliary jet moment observable.  The gains are
significantly larger in the case of the BDT- and DNN-based high-level
observable discriminants, and lead to larger improvements for
top-quark tagging than for \Wboson-boson tagging.  However, the BDT and DNN
algorithms perform similarly to each other for all signal efficiencies,
indicating that they are leveraging the correlations of the input jet moment
observables equally well.  Therefore, when studying the performance of these
tagging techniques in data in Section~\ref{sec:DataPerformance}, only the
DNN-based taggers are included.  The performance of the BDT-based taggers was
studied and found to be similar.  Finally, in the case of top-quark tagging,
where more dedicated tagging techniques are studied, the conclusion is similar.
Dedicated approaches, including shower deconstruction and \htt, are more
performant than a simple cut-based approach on \mcomb and \tauthrtwo, but the
combination of many jet-moment observables in a BDT or DNN yields the best
overall performance out of the techniques tested in this study.  Of particular
note, however, is the comparison of the BDT
and the fully-connected feed-forward DNN taggers using high-level observables and
those using lower-level inputs,
namely the jet constituents, here taken to be topoclusters.  The performance of
these two approaches is similar, with the TopoDNN tagger having slightly higher background rejection at high jet \pt,
resulting in conclusions qualitatively similar to
those found in Ref.~\cite{Kasieczka:2017nvn}, particularly at high jet \pt where
the details of the signal sample used for training are less relevant.
 
\begin{figure}[htpb]
\centering
\subfigure[][]{ \label{fig:roccurves_w_lowpt}
\includegraphics[width=0.49\textwidth]{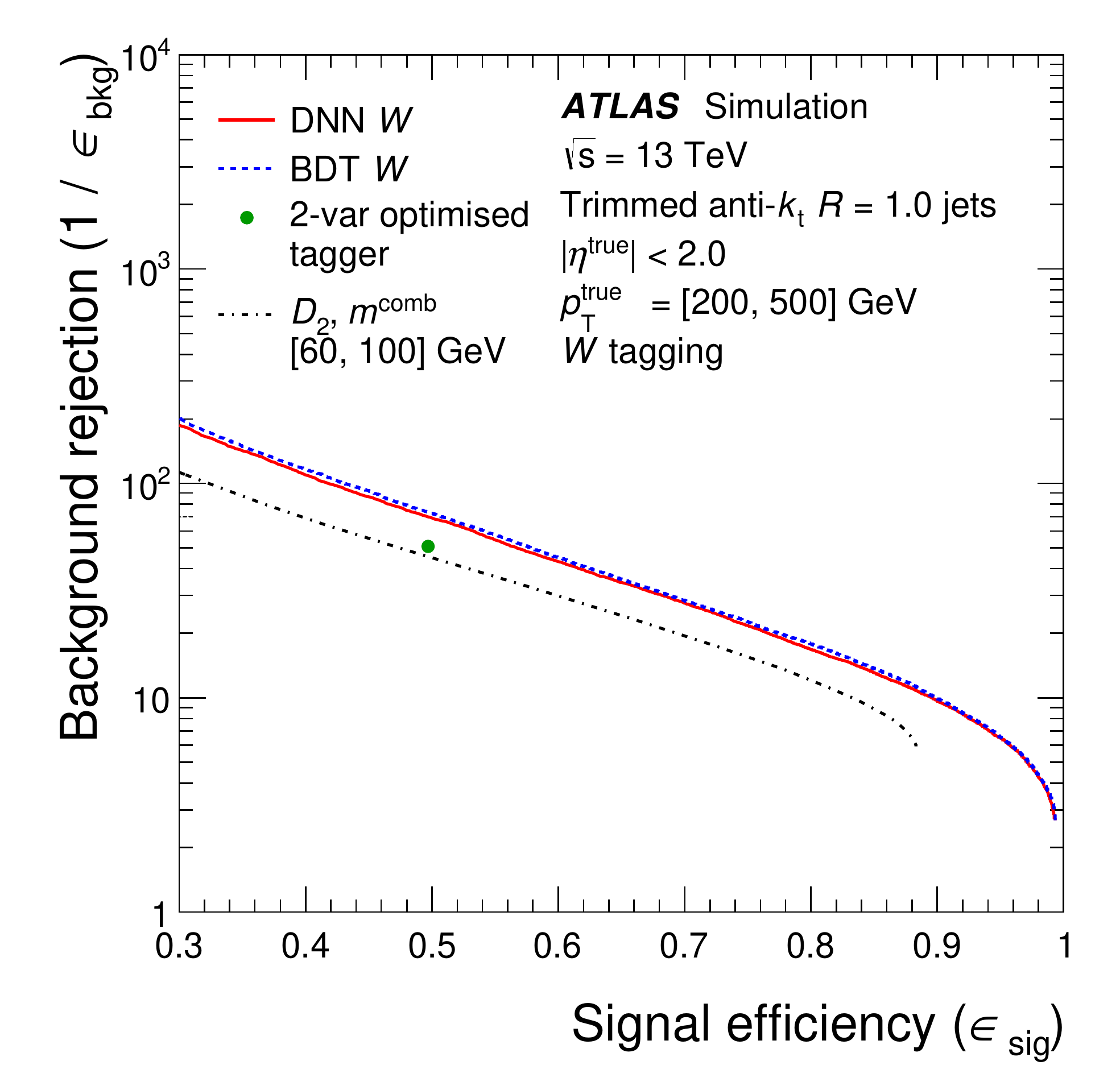}}
\subfigure[][]{ \label{fig:roccurves_w_highpt}
\includegraphics[width=0.49\textwidth]{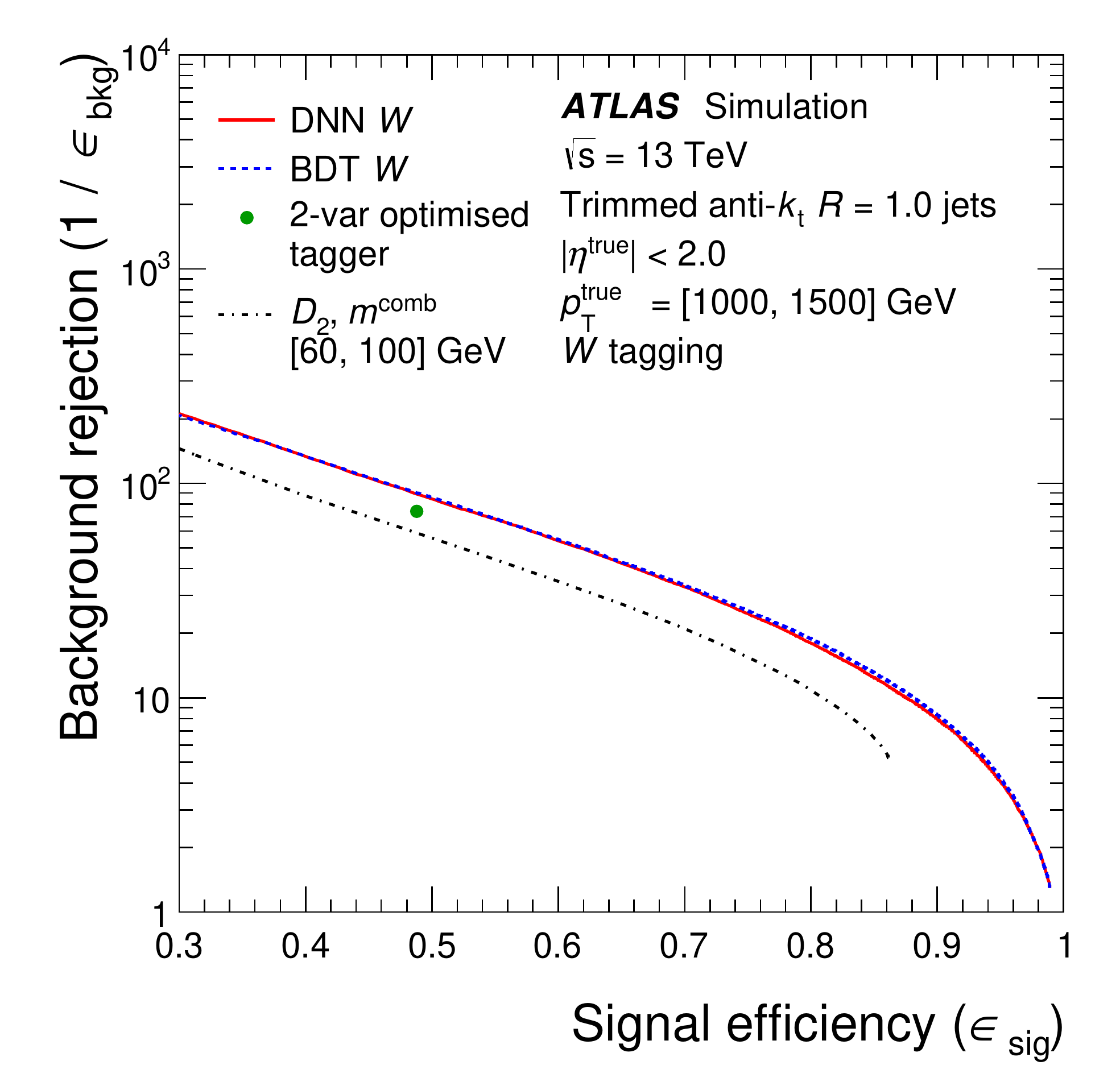}}
\caption{
\label{fig:roccurves_w} The performance comparison of the \Wboson-boson taggers
in a low-\pTtrue~\subref{fig:roccurves_w_lowpt} and high-\pTtrue~\subref{fig:roccurves_w_highpt} bin.  The performance is
evaluated with the \pTtrue distribution of the signal jets weighted to match
that of the dijet background samples.}
\end{figure}
 
\begin{figure}[htpb]
\centering
\subfigure[][]{ \label{fig:roccurves_top_lowpt}
\includegraphics[width=0.49\textwidth]{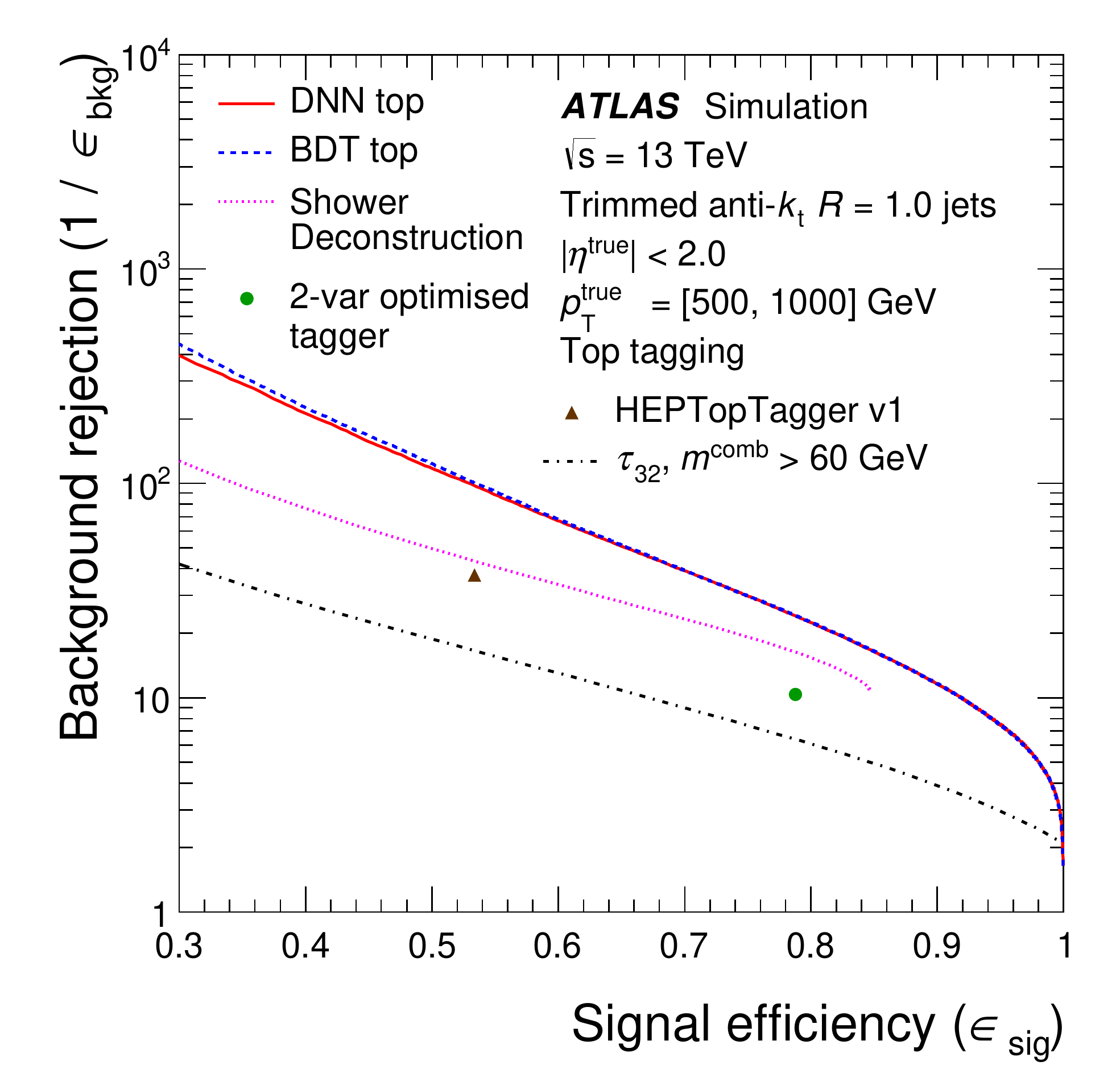}}
\subfigure[][]{ \label{fig:roccurves_top_highpt}
\includegraphics[width=0.49\textwidth]{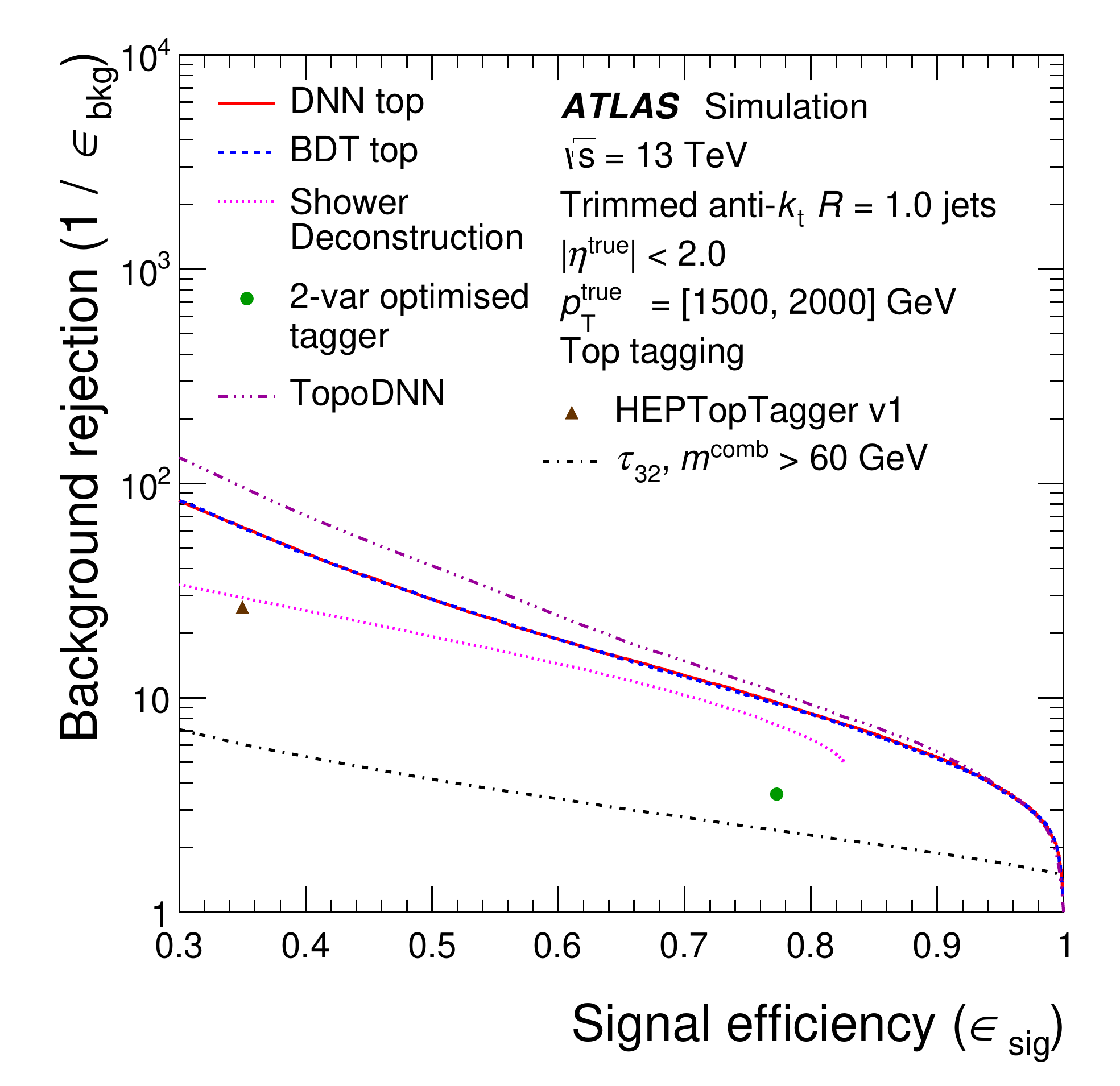}}
\caption{
\label{fig:roccurves_top} The performance comparison of the top-quark taggers in
a low-\pTtrue~\subref{fig:roccurves_top_lowpt} and high-\pTtrue~\subref{fig:roccurves_top_highpt} bin.  The performance is evaluated
with the \pTtrue distribution of the signal jets weighted to match that of the
dijet background samples.}
\end{figure}
\clearpage
 
\section{Performance in data}
\label{sec:DataPerformance}
The taggers studied in the previous sections are validated using signal and
background-enriched data samples collected during 2015 and 2016 at a
centre-of-mass energy of $\sqrt{s} = 13~\TeV$ and corresponding to an
integrated luminosity of $36.1$~\ifb.  In the case of \Wboson-boson and
top-quark jets, the lepton-plus-jets \ttbar signature is used, which provides a
sample of signal jets in a \pt range of approximately 200~\GeV\ to
1000~\GeV.  In the case of background light jets, two topologies are studied: a
\gammajet sample enriched in light-quark jets and spanning a \pt range
of approximately 200~\GeV\ to 2000~\GeV\ and a multijet sample which probes a
mixture of light-quark and gluon jets in a \pt range of approximately
500~\GeV\ to 3500~\GeV.  The primary aim of these studies is to validate the
modelling of the Monte Carlo simulation in data for the techniques studied in
Section~\ref{sec:TaggerTechniques}.  This is achieved by directly studying
the full spectrum of a subset of important observables used in the tagging as well
as directly measuring both the signal efficiency and background efficiency of
the various techniques in the phase space accessible in this data sample.  In
the case of the measured signal efficiency and background rejection, the
performance is evaluated differentially as a function of the jet transverse
momentum as well as the average number of interactions per bunch crossing
($\mu$).
 
\subsection{Signal efficiency in boosted \ttbar events}
\label{subsec:datamc_ttbar}
To study the modelling of signal \Wboson-boson and top-quark \largeR jet
tagging, a sample of data enriched in \ttbar events where one top quark decays
hadronically and the other semileptonically in both the electron and the muon
decay channel is selected in a similar manner to
Refs.~\cite{PERF-2015-03,PERF-2015-04}.  The inclusive sample of events is
decomposed into two exclusive subsamples, enriched in \Wboson-boson jets and
top-quark jets, based on the proximity of a \bjet to the \largeR jet.  The inclusive
distributions of the key observables used in each tagging method are examined
and the signal efficiency is measured for a set of fixed signal efficiency
working points, for which systematic uncertainties can be derived and associated
with a particular tagging method.
 
\subsubsection{Analysis and selection}
\label{subsubsec:leptonjets_analysis}
To select the inclusive set of lepton-plus-jets \ttbar events, both the data and
Monte Carlo simulated events are required to pass either an inclusive electron
trigger or an inclusive muon trigger, where the thresholds were varied between
the 2015 and 2016 datasets due to increases in instantaneous luminosity.
In the electron channel,
events from the 2015 data-taking period are required to pass at least one of
three triggers: one isolated electron with $\pt > 24~\GeV$, one electron with
$\pt > 60~\GeV$ without any isolation requirement, or one electron with $\pt >
120~\GeV$ without any isolation requirement and relaxed identification criteria.
In the 2016 data-taking period, the thresholds of these electron triggers required
$\pt > 26~\GeV$, $\pt > 60~\GeV$ and $\pt > 140~\GeV$, respectively.
In the muon channel, events from the 2015 data-taking period are required to
pass at least one of two muon triggers: one isolated muon with $\pT > 20~\GeV$
or one muon with $\pT > 50~\GeV$ and no isolation requirement. In the 2016
data-taking period, the thresholds of these triggers required $\pt > 26~\GeV$
and $\pt > 50~\GeV$, respectively.
 
Events are then required to contain exactly one electron or muon candidate with
$\pt > 30~\GeV$ that is matched to the trigger-level counterpart associated
with the appropriate trigger. Electron candidates are reconstructed as ID tracks
that are matched to a cluster of energy in the electromagnetic calorimeter.
Electron candidates are required to be within $|\eta| < 2.47$, excluding the
calorimeter transition region from $1.37<|\eta|<1.52$, and satisfy the ``tight''
likelihood-based identification criterion based on shower shape and track
selection requirements~\cite{PERF-2016-01,ATL-PHYS-PUB-2015-041}. Muons are
reconstructed as tracks found in the ID that are matched to tracks reconstructed
in the muon spectrometer.  They are required to be within $|\eta| < 2.5$ and are
required to satisfy the ``medium'' muon identification quality criteria defined
in Ref.~\cite{PERF-2015-10}. For both electrons and muons, the reconstructed
lepton candidate is required to be isolated from additional activity in the
event by imposing isolation criterion defined by a sum of \pt{} of tracks in an
isolation cone with variable radius depending on the lepton
\pt{}~\cite{PERF-2015-10,ATLAS-CONF-2016-024}.

In addition to identified leptons, small-radius jets are used to reconstruct the
missing transverse momentum and identify the signal topology.  These jets are
reconstructed from topoclusters calibrated to the electromagnetic scale using
the \akt algorithm with a radius parameter of $R$ = 0.4. The energy of these
jets is corrected for the effects of \pileup by using a technique based on jet
area~\cite{Cacciari:2008gn} and the jet energy is further corrected using a jet
energy scale calibration based on both Monte Carlo simulation and
data~\cite{PERF-2016-04}.
To ensure that the reconstructed jets are well-measured, they are required to have
$\pt>20~\GeV$, $|\eta|<2.5$ and to satisfy  ``loose'' quality criteria to prevent
mismeasurements due to calorimeter noise spikes and non-collision
backgrounds~\cite{ATLAS-CONF-2015-029}. For jets with $\pT < 60~\GeV$ and
$|\eta|$ < 2.4, a requirement that the jets arise from the primary vertex, using
the ID tracks associated with the jet, is imposed to suppress \pileup
jets~\cite{PERF-2014-03}.
 
For the identification of $b$-quark candidate jets, jets reconstructed from ID
tracks with the \akt algorithm with radius parameter $R$ = 0.2 are used.  These
jets are $b$-tagged using a multivariate discriminant based on impact parameter
and secondary vertex information ~\cite{ATL-PHYS-PUB-2016-012}.  The 70\% signal
efficiency point selection is used.  Event-by-event scale factors, evaluated in
\ttbar events~\cite{PERF-2016-05}, are applied to account for mismodelling of
the selection efficiency.
 
The missing transverse momentum is reconstructed as the negative vectorial sum
of the momenta of all reconstructed physics objects in the plane transverse to
the beamline~\cite{PERF-2016-07}.  In this case, the sum consists of the single
identified lepton and the full set of reconstructed and fully calibrated
small-$R$ calorimeter jets as well as ID tracks not associated with the lepton or
jets. These ID tracks are included to account for the soft hadronic energy flow
in the event.  In the following the magnitude of the missing transverse momentum
vector is denoted by \met.
 
First, events containing a leptonically decaying \Wboson boson are preselected
by requiring one electron or muon candidate with \pt > 30~\GeV\ and rejecting
events that contain additional electrons or muons with \pt > 25~\GeV. The
missing transverse momentum is required to be greater than 20~\GeV\ and the
scalar sum of \met and the transverse mass of the leptonically decaying \Wboson
boson candidate\footnote{$\MTW$ = $\sqrt{2\pt^\ell\met(1-\cos\Delta\phi)}$ is
calculated from the transverse momentum of the lepton, $\pt^\ell$, and \met in
the event. $\Delta\phi$ is the azimuthal angle between the lepton momentum and
the \met direction.} must satisfy $\met+\MTW>60~\GeV$.  To ensure the topology
is consistent with a \ttbar event, at least one small-$R$ jet is required to have
$\pt>25$~\GeV\ and to be close to the lepton ($\Delta R(\text{lepton},
\text{jet}) < 1.5$).  To study \Wboson-boson and top-quark tagging, the highest-\pt
\largeR jet is studied, which is either a trimmed \akt $R =$ 1.0 jet or a
C/A $R =$ 1.5 jet in the case of \htt, with $\pt>200~\GeV$ and $|\eta|<2.0$.
The C/A jets are also trimmed using the same trimming parameters as for the \akt
jets, such that their kinematics are robust against \pileup. Since \htt is
designed to tag ungroomed jets, the constituents of the C/A jet before trimming
are used as inputs to the tagging algorithm.
The signal \topquark jet candidate is required to be well-separated from the
semileptonic top-quark decay by requiring $\Delta R>$1.5 between the large-$R$
jet and the small-$R$ jet close to lepton.  Additionally, the angular separation in the
transverse plane between the lepton and the \largeR jet is required to be
$\Delta\phi>2.3$.
 
Finally, the sample preselected as above is divided into two subsamples,
intended to be representative of a fully contained top-quark decay or an isolated and
fully contained \Wboson-boson decay based on the proximity of a $b$-tagged track
jet to the highest-\pt \largeR jet. The track jets are clustered from at least
two  tracks using the \akt algorithm with a radius parameter of $R=0.2$.  All
tracks must fulfil $|\eta| < 2.5$ and  $\pT > 10~\GeV$. The sample enriched
in top quarks (``top-quark selection'') is defined by requiring a $b$-tagged
track jet to have an angular separation of $\Delta R(b\text{-jet},\
\text{large-}R\ \text{jet})<1.0$ ($\Delta R(b\text{-jet},\ \text{large-}R\
\text{jet})<1.5$) from the \largeR \akt trimmed jet (C/A jet).  In order to
enhance the fraction of fully contained top quarks, an additional requirement of
$\pt>350~\GeV$ ($\pt>200~\GeV$) is also applied.  The sample enriched in
\Wboson-boson jets (``\Wboson-boson selection'') is defined by requiring a $b$-tagged
track jet to have angular separation $\Delta R(b\text{-jet},\ \text{large-}R\
\text{jet})>1.0$ from the \largeR \akt trimmed jet.  Because the geometrical
separation of the daughter $b$-quark and the top parton decreases with
increasing \pt, this requirement limits the efficiency of the \Wboson-boson
selection at high jet \pt, which limits the kinematic reach to approximately 600~\GeV.
These requirements result in relatively pure samples of \Wboson-boson and
top-quark jets as shown in Figure~\ref{fig:DMC_signal_mass_1} for the \akt $R =$
1.0 trimmed jet mass, including the full set of systematic uncertainties
summarised in Section~\ref{subsec:datamc_systematics}, while
Figure~\ref{fig:DMC_signal_mass_2} shows the C/A $R =$ 1.5 trimmed jet mass.
The disagreement between the peak positions in Monte Carlo simulation and
data observed near \MW and \MTop is attributed to a mismodelling of the jet mass
scale as studied in Ref.~\cite{JETM-2018-02}. In this paper, the \ttbar
and single-top Monte Carlo samples are divided into three subsamples based on
the jet labelling criteria outlined in Section~\ref{sec:labelling} to highlight
the fraction of events in each sample of interest (``$\ttbar$ (top)'' and
``$\ttbar$ (\Wboson)''), with all other events in these samples being grouped
together in a single subsample (``$\ttbar$ (other)''). The backgrounds are
derived from the Monte Carlo simulations described in Section~\ref{sec:samples},
with the exception of the multijet background, which is estimated using a
data-driven method based on looser lepton selection criteria with a dedicated
evaluation of the probability of prompt lepton reconstruction and the
probability of fake/non-prompt lepton reconstruction, as was performed in
Ref.~\cite{EXOT-2015-04}.  The event yield in the simulation is normalised to that
in the data at this stage of the selection throughout Section~\ref{subsubsec:leptonjets_analysis}.
 
\begin{figure}[!h]
\begin{centering}
\subfigure[][]{   \label{fig:DMC_signal_mass_1_a} \includegraphics[width=0.45\textwidth]{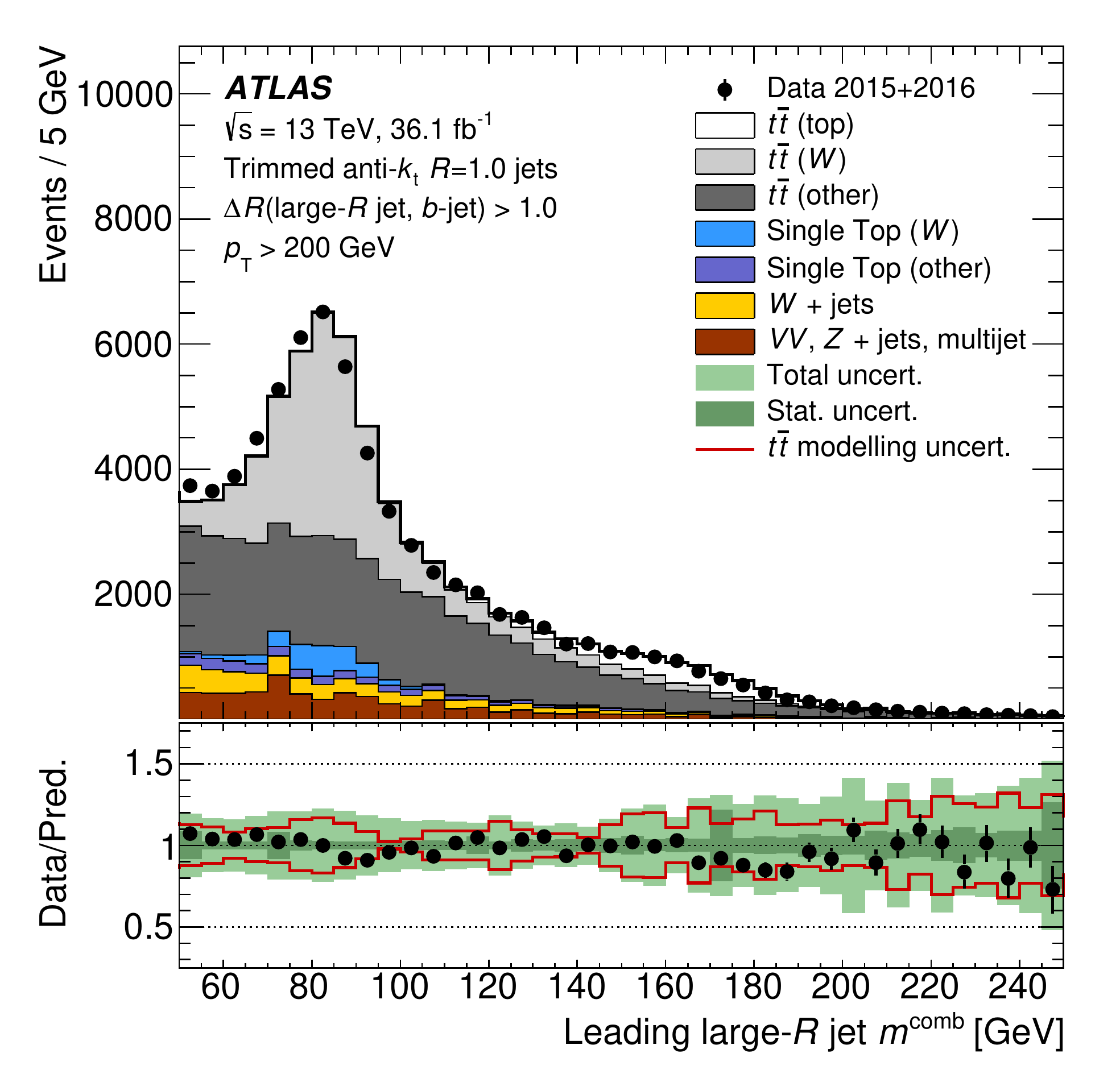}}
\subfigure[][]{       \label{fig:DMC_signal_mass_1_b} \includegraphics[width=0.45\textwidth]{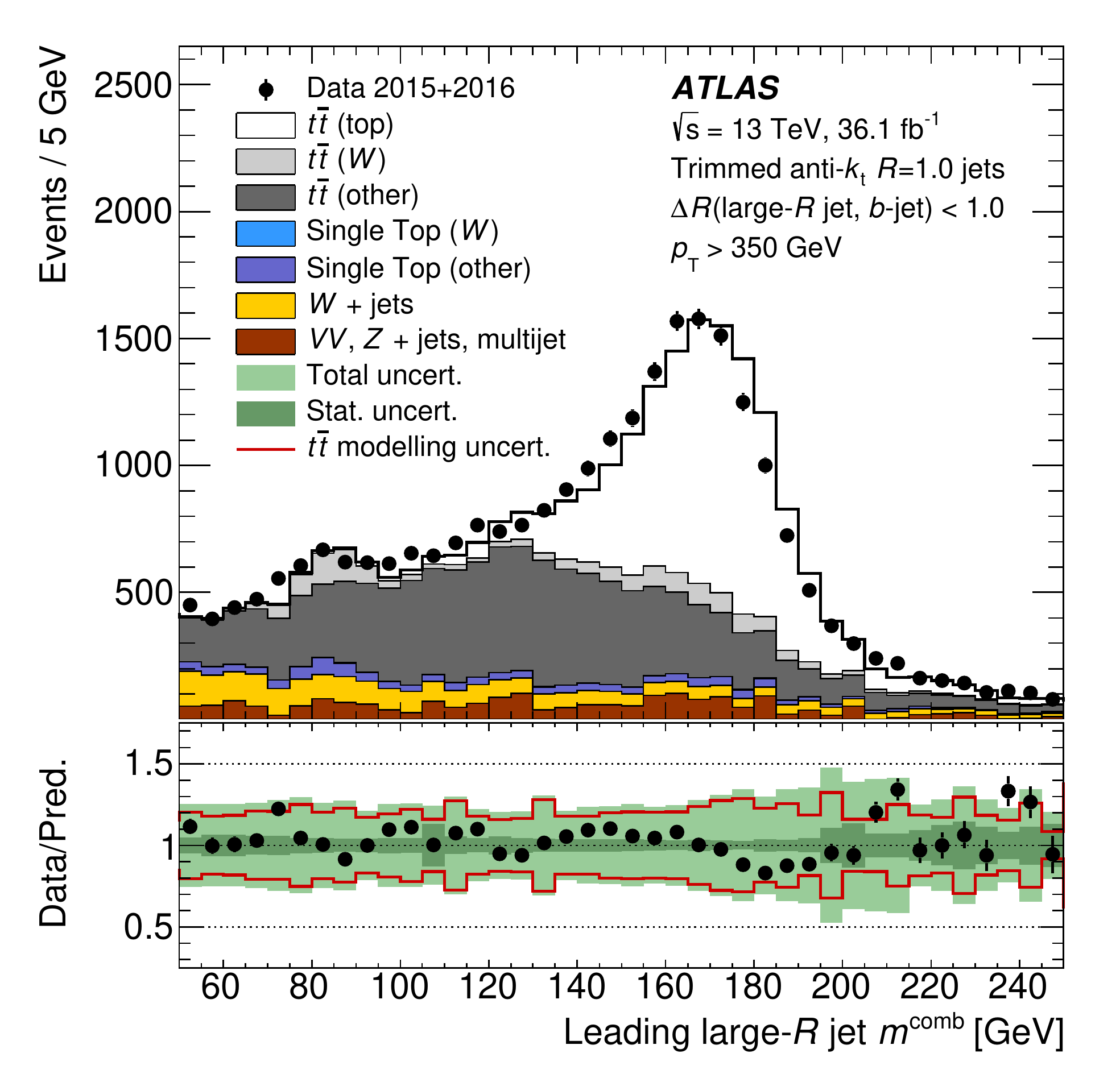}}
\caption{
\label{fig:DMC_signal_mass_1} A comparison of the observed data and predicted MC
distributions of the mass of the leading \pt \akt trimmed jet in the event for
the \Wboson boson \subref{fig:DMC_signal_mass_1_a} and top quark
\subref{fig:DMC_signal_mass_1_b} selections in a sample enriched in lepton+jets
\ttbar events. Simulated distributions are normalised to data.  The \ttbar
sample is divided into a set of subsamples (e.g.\ $\ttbar$ (top)) based on
criteria described in Section~\ref{sec:labelling}.  The statistical uncertainty
of the background prediction (Stat. uncert.) results from limited Monte
Carlo statistics as well as the limited size of the data sample used in the
data-driven estimation of the multijet background.}
\end{centering}
\end{figure}
 
\begin{figure}[!h]
\begin{centering}
\includegraphics[width=0.45\textwidth]{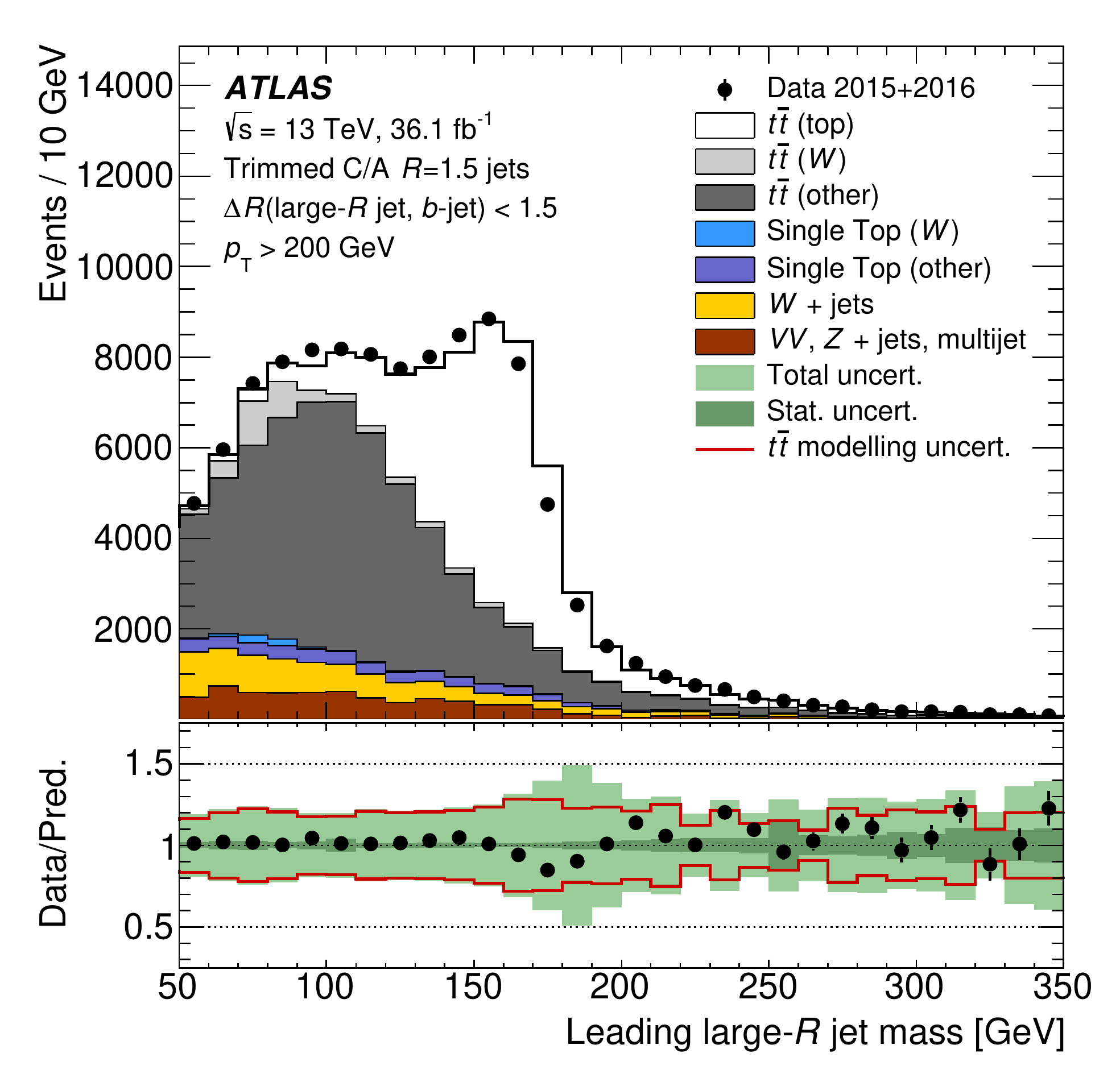}
\caption{
\label{fig:DMC_signal_mass_2} A comparison of the observed data and predicted MC
distributions of the mass of the leading \pt C/A $R =$ 1.5 trimmed jet in events
passing the top-quark selection in a sample enriched in lepton+jets \ttbar
events. Simulated distributions are normalised to data.  The \ttbar sample is
divided into a set of subsamples (e.g.\ $\ttbar$ (top)) based on criteria
described in Section~\ref{sec:labelling}.  The statistical uncertainty of the
background prediction (Stat. uncert.) results from limited Monte Carlo
statistics as well as the limited size of the data sample used in the
data-driven estimation of the multijet background.}
\end{centering}
\end{figure}
 
The primary tagging observables used by the other tagging techniques described
in Section~\ref{sec:TaggerTechniques} are examined in
Figures~\ref{fig:DMC_signal_jss1}--\ref{fig:DMC_signal_jss5}.  For these spectra,
the full set of systematic uncertainties described in
Section~\ref{subsec:datamc_systematics} are included for the \DTwo and
\tauthrtwo observables, whereas for the other spectra, no dedicated experimental
systematic uncertainty in the scale or resolution of the observable itself is
included.  Instead, the mismodelling of the simulation relative to data is
taken into account as a derived uncertainty in the \textit{in situ} measurement of the
signal efficiency of the tagger itself, in a manner similar to that commonly
used to evaluate mismodelling in the detector response in the context of the
identification of heavy-flavour jets~\cite{PERF-2012-04}.  However, for nearly
all regions of phase space, the overall relative yield of data is well-described
by the Monte Carlo prediction within the theoretical uncertainties, derived from
the comparison of various \ttbar Monte Carlo generators.
 
\begin{figure}[!h]
\begin{centering}
\subfigure[][]{   \label{fig:DMC_signal_jss1_a} \includegraphics[width=0.45\textwidth]{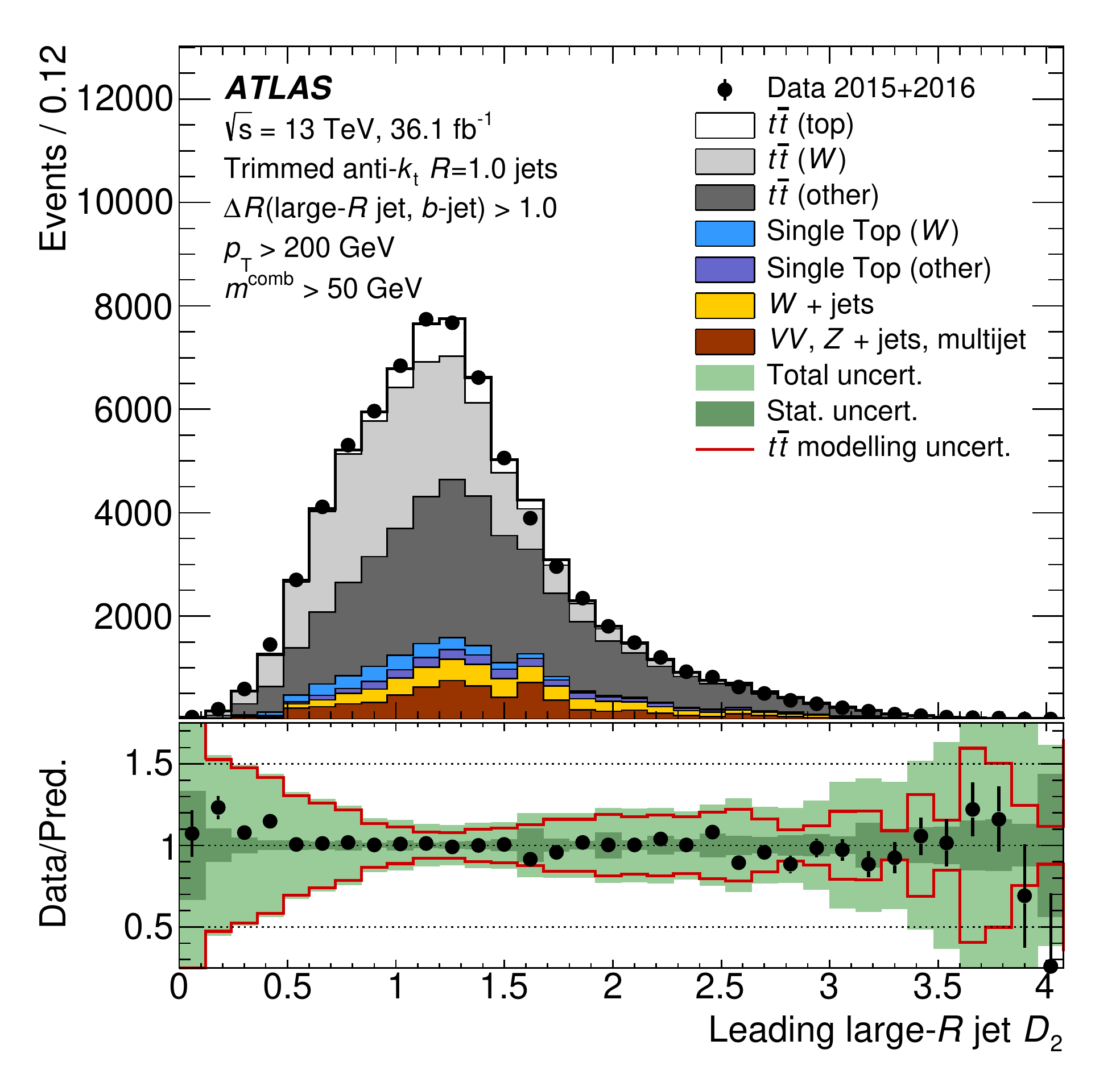}}
\subfigure[][]{       \label{fig:DMC_signal_jss1_b} \includegraphics[width=0.45\textwidth]{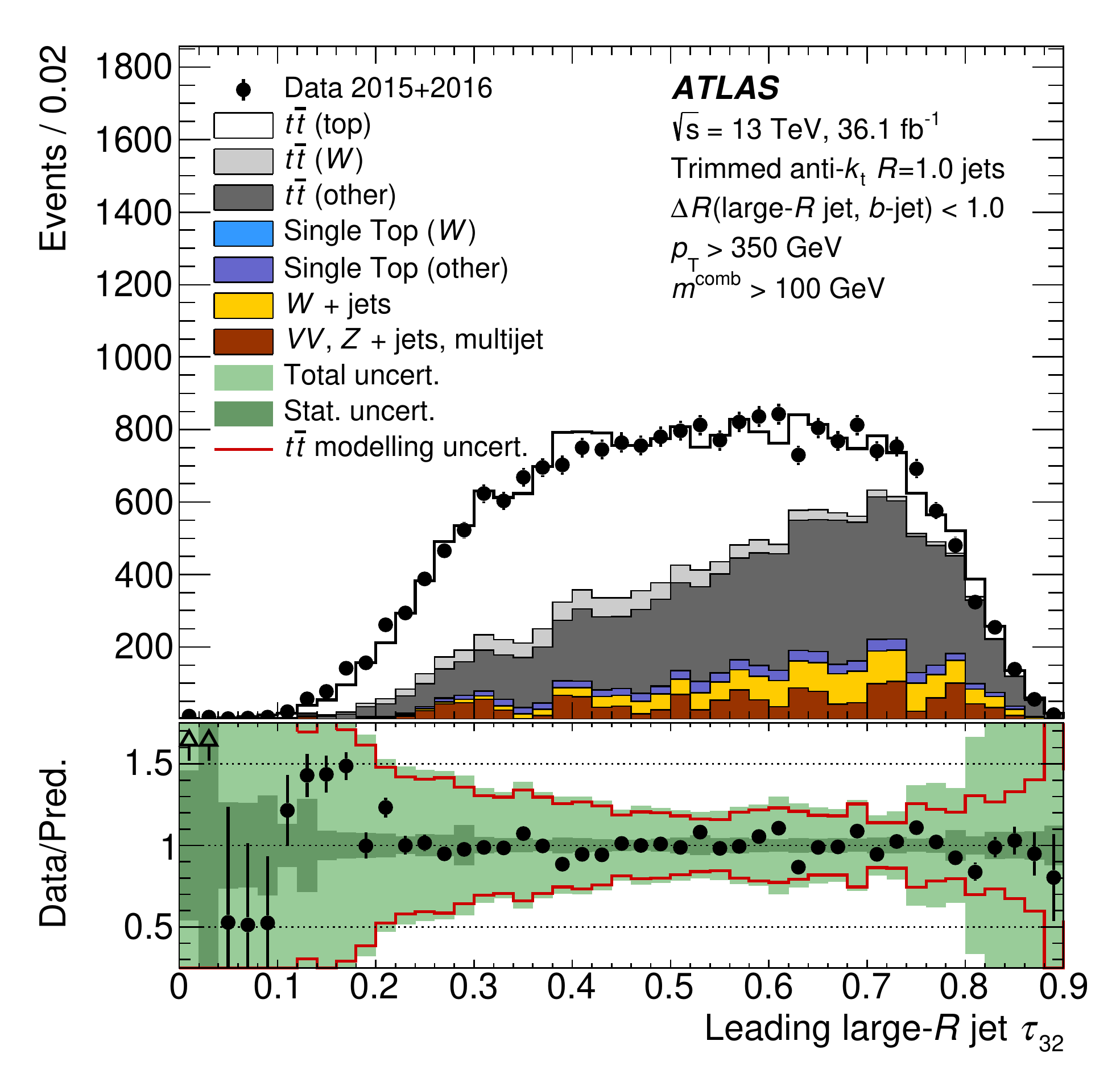}}
\caption{
\label{fig:DMC_signal_jss1} A comparison of the observed data and predicted MC
distributions of the \akt $R =$ 1.0 trimmed jet \DTwo
\subref{fig:DMC_signal_jss1_a} and \tauthrtwo \subref{fig:DMC_signal_jss1_b} for
the \Wboson-boson and top-quark selections, respectively, in a sample enriched
in lepton+jets \ttbar events. Simulated distributions are normalised to data.
The \ttbar sample is divided into a set of subsamples (e.g.\ $\ttbar$
(top)) based on criteria described in Section~\ref{sec:labelling}.  The
statistical uncertainty of the background prediction (Stat. uncert.) results
from limited Monte Carlo statistics as well as the limited size of the data
sample used in the data-driven estimation of the multijet background.}
\end{centering}
\end{figure}
 
\begin{figure}[!h]
\begin{centering}
\subfigure[][]{   \label{fig:DMC_signal_jss2_a} \includegraphics[width=0.45\textwidth]{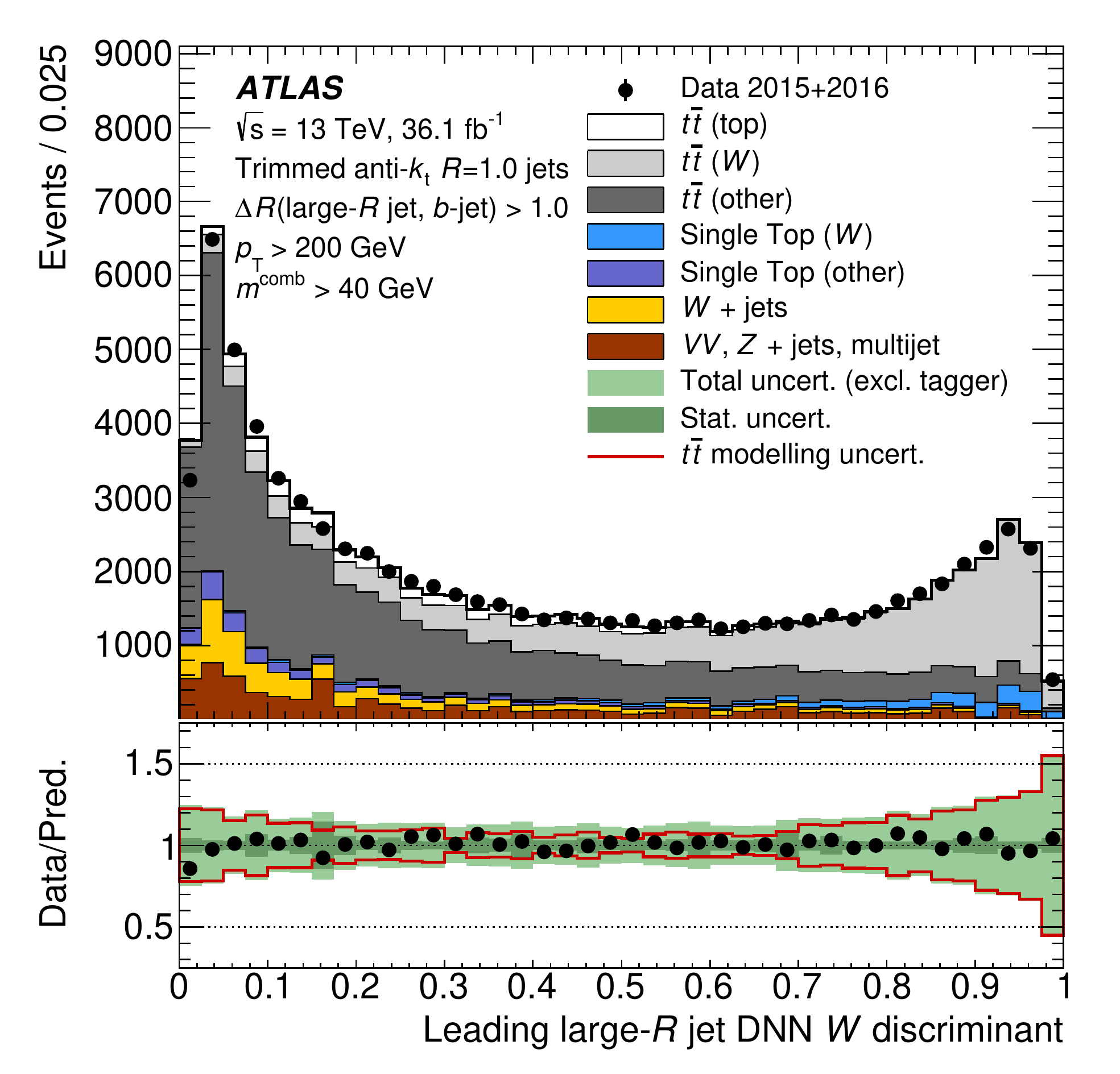}}
\subfigure[][]{       \label{fig:DMC_signal_jss2_b} \includegraphics[width=0.45\textwidth]{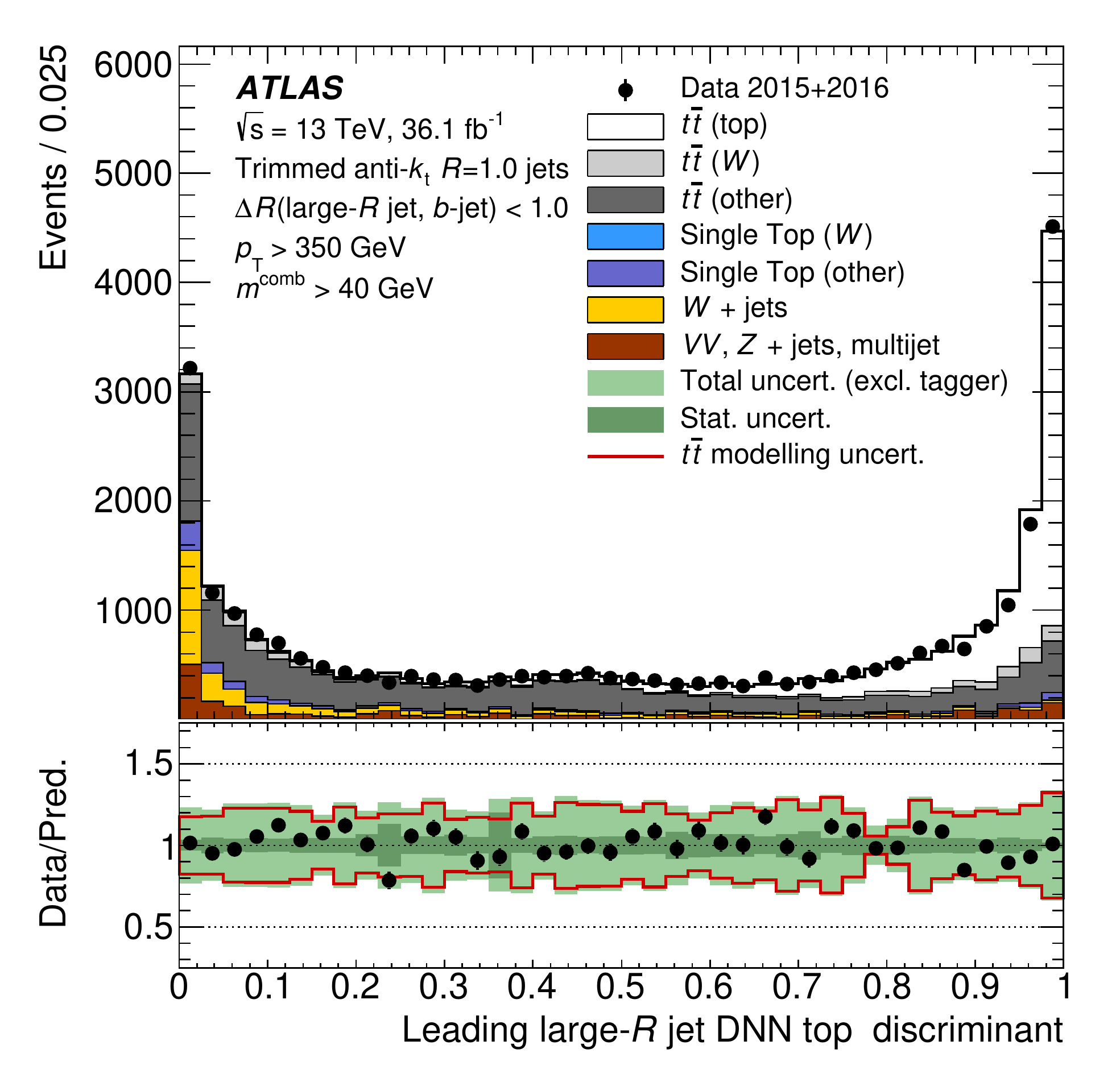}}
\caption{\label{fig:DMC_signal_jss2} A comparison of the observed data and predicted MC
distributions of the \akt $R =$ 1.0 trimmed jet DNN discriminant for \Wboson
boson \subref{fig:DMC_signal_jss2_a} and top quark
\subref{fig:DMC_signal_jss2_b} tagging for the respective event selections in a
sample enriched in lepton+jets \ttbar events. Simulated distributions are
normalised to data.  The \ttbar sample is divided into a set of subsamples
(e.g.\ $\ttbar$ (top)) based on criteria described in
Section~\ref{sec:labelling}.  The statistical uncertainty of the background
prediction (Stat. uncert.) results from limited Monte Carlo statistics as
well as the limited size of the data sample used in the data-driven estimation
of the multijet background.} \end{centering}
\end{figure}
 
\begin{figure}[!h]
\begin{centering}
\includegraphics[width=0.45\textwidth]{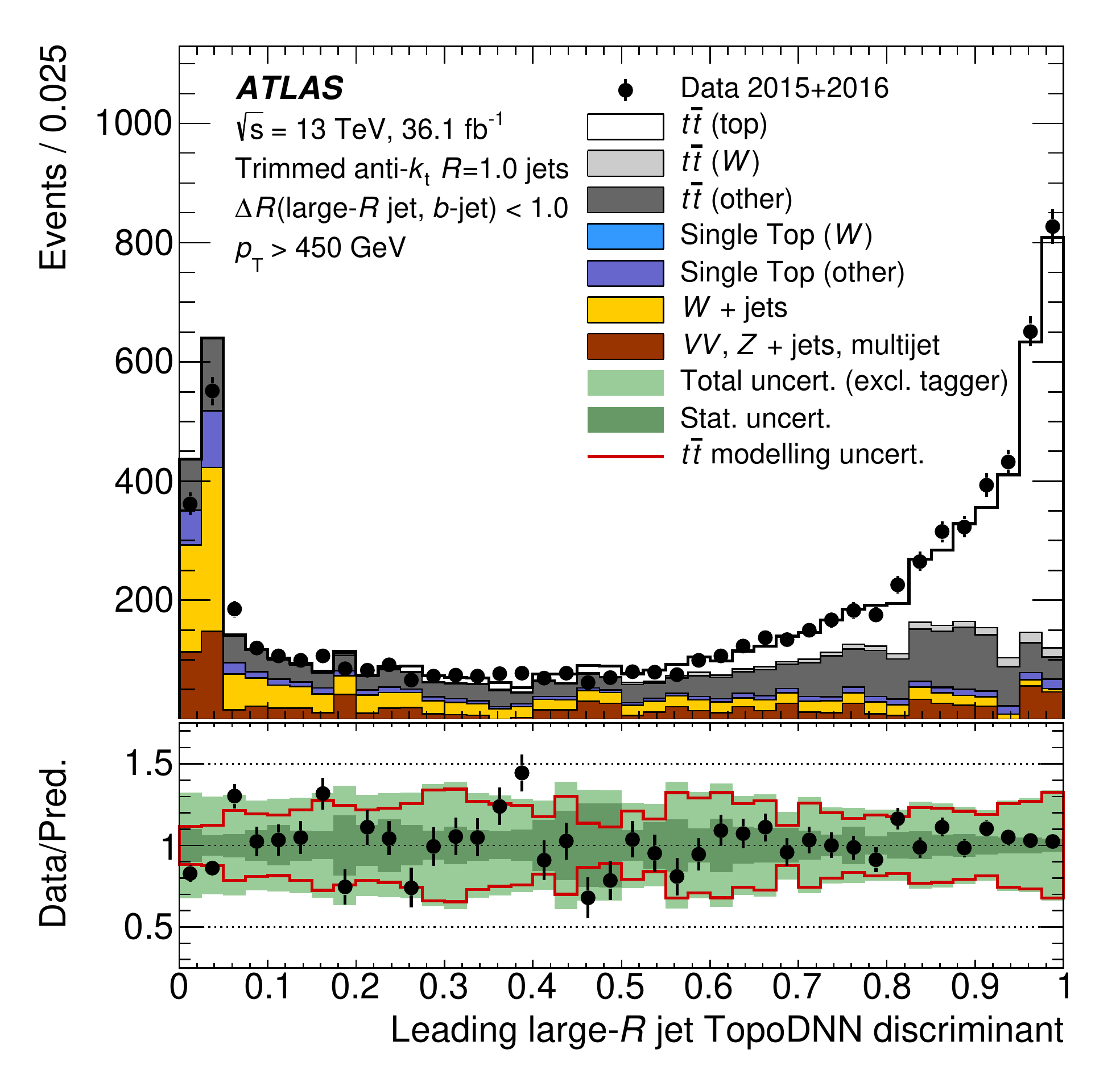}
\caption{\label{fig:DMC_signal_jss3}A comparison of the observed data and predicted MC
distributions of the TopoDNN top tagger discriminant for the top-quark
event selection in a sample enriched in lepton+jets \ttbar events. Simulated
distributions are normalised to data.  The \ttbar sample is divided into a
set of subsamples (e.g.\ $\ttbar$ (top)) based on criteria described in
Section~\ref{sec:labelling}.  In this case, a \pt > 450~\GeV\ selection is
applied to the large-$R$ jet to specifically focus on the kinematic region of
interest for which this tagging algorithm was designed, as described in
Section~\ref{sec:optimisation_ttt}.  The statistical uncertainty of the
background prediction (Stat. uncert.) results from limited Monte Carlo
statistics as well as the limited size of the data sample used in the
data-driven estimation of the multijet background.} \end{centering}
\end{figure}
 
\begin{figure}[!h]
\begin{centering}
\includegraphics[width=0.45\textwidth]{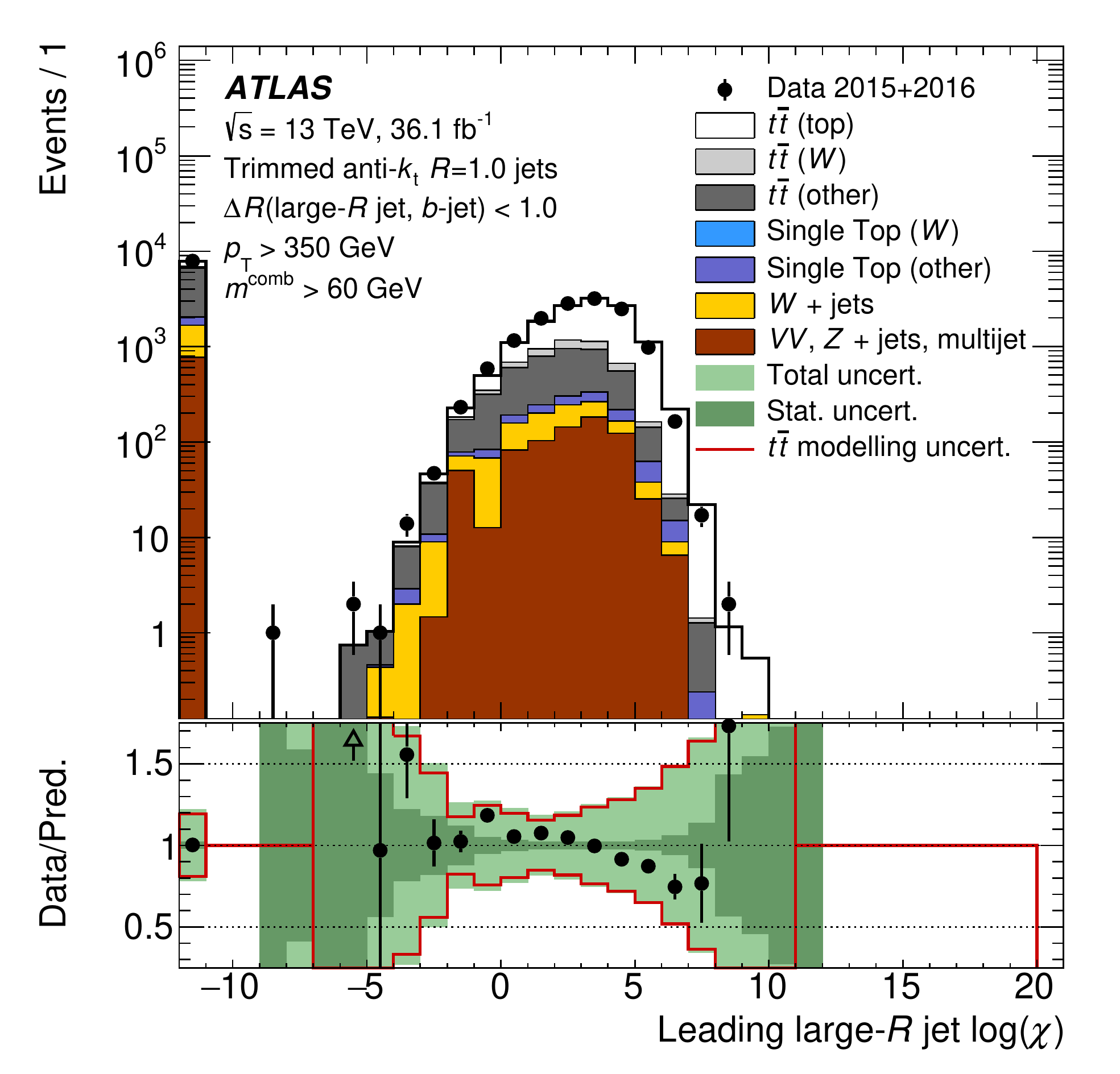}
\caption{
A comparison of the observed data and predicted MC
distributions of the $\log\chi$ shower deconstruction discriminant for the top-quark
event selection in a sample enriched in lepton+jets \ttbar events.
Simulated distributions are normalised to data.  The \ttbar sample is
divided into a set of subsamples (e.g.\ $\ttbar$ (top)) based on criteria
described in Section~\ref{sec:labelling}.  The ensemble of jets with a large
negative $\log\chi$ value correspond to the set of jets where no subjet
configuration is roughly consistent with a top-quark jet topology, as described
in Section~\ref{sec:optimisation_showerdeconstruction}.  The statistical
uncertainty of the background prediction (Stat. uncert.) results from
limited Monte Carlo statistics as well as the limited size of the data sample
used in the data-driven estimation of the multijet background.} \end{centering}
\end{figure}
 
\begin{figure}[!h]
\begin{centering}
\includegraphics[width=0.45\textwidth]{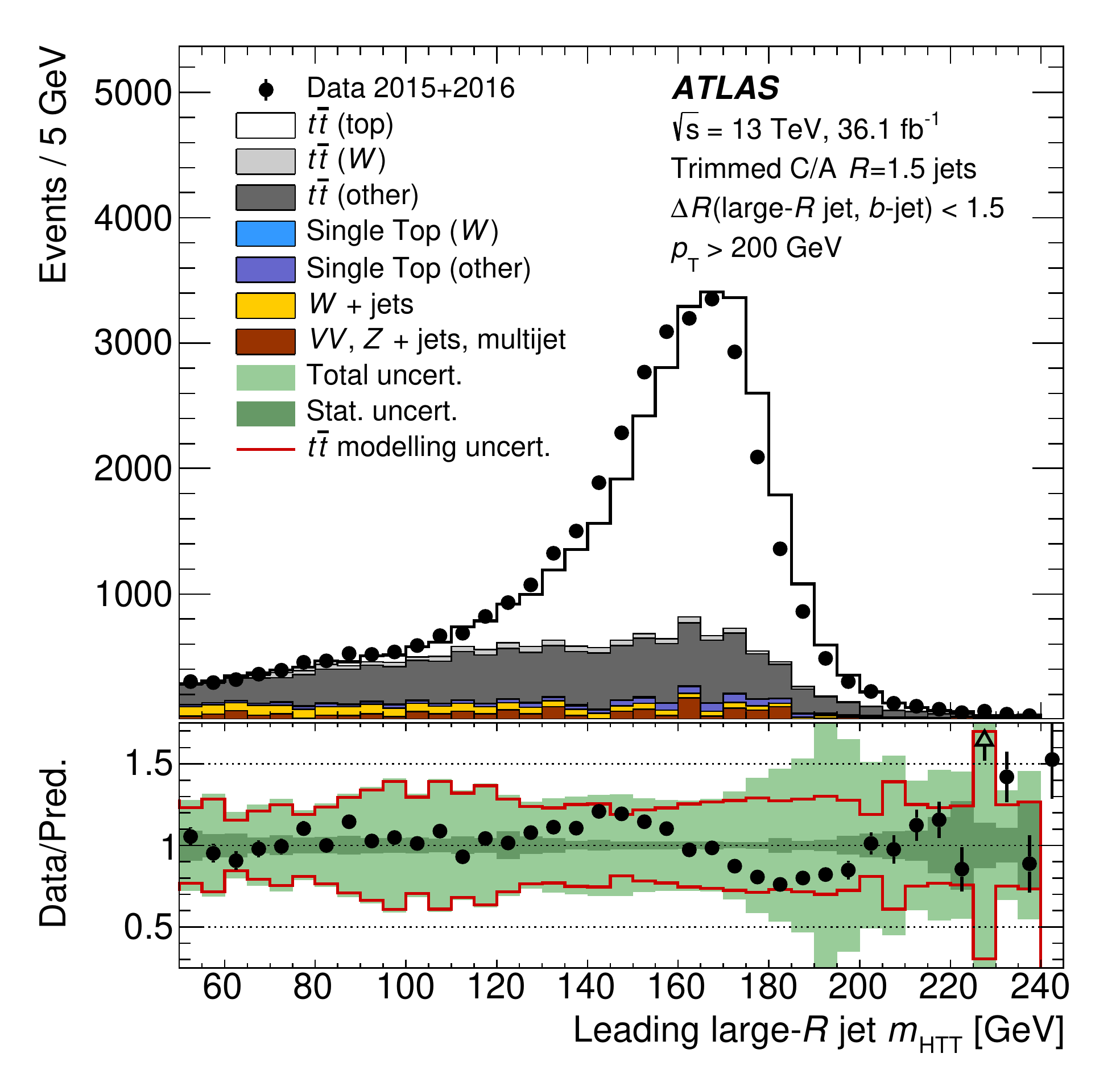}
\caption{
\label{fig:DMC_signal_jss5} A comparison of the observed data and predicted MC
distributions of the \htt mass for the top-quark event selection in a sample
enriched in lepton-plus-jets \ttbar events. Simulated distributions are
normalised to data.  The \ttbar sample is divided into a set of subsamples
(e.g.\ $\ttbar$ (top)) based on criteria described in
Section~\ref{sec:labelling}.  The statistical uncertainty of the background
prediction (Stat. uncert.) results from limited Monte Carlo statistics as
well as the limited size of the data sample used in the data-driven estimation
of the multijet background.} \end{centering}
\end{figure}
\clearpage
\newpage
 
\subsubsection{Signal efficiencies}
\label{sec:measured_signal_efficiency}
Due to the relatively high purity of the samples of \Wboson-boson and \topquark jets
that result from the selection described in
Section~\ref{subsubsec:leptonjets_analysis}, it is possible to measure the
signal efficiency in data.  This measurement, when compared with the Monte Carlo
prediction, can be used to estimate the systematic uncertainty of a particular
tagging method when applied in the context of an independent analysis.  It can
also be used to provide an \textit{in situ} correction in the form of a
jet-by-jet efficiency scale factor~\cite{PERF-2016-05,PERF-2012-04}.  Because
the aim is to provide an efficiency measurement for a particular tagging method,
it is necessary to define selection criteria based on the particular tagging
discriminants described in Section~\ref{sec:TaggerTechniques} for which the
comparison of the Monte Carlo prediction to data was shown in a rather inclusive
selection of signal-like events in Section~\ref{subsubsec:leptonjets_analysis}.
In particular, the seven tagger working points for which the signal efficiency
is measured here are:
 
\begin{itemize}
\item \DTwo+\mcomb (\Wboson boson): A pair of selections on \mcomb and \DTwo,
tuned as a function of \pt, that give the largest background rejection for a
fixed 50\% signal efficiency for fully contained \Wbosonboson jets;
\item \topsimple (top quark):  A pair of selections on \mcomb and \tauthrtwo,
tuned as a function of \pt, that give the largest background rejection for a
fixed 80\% signal efficiency for fully contained \topquark jets;
\item DNN (\Wboson boson): A single-sided selection of $\mcomb >$ 40~\GeV\ and a
selection on the DNN discriminant, tuned to give a fixed 50\% signal
efficiency as a function of \pt for fully contained \Wbosonboson jets;
\item DNN (top quark): A single-sided selection of $\mcomb >$ 40~\GeV\ and a
selection on the DNN discriminant, tuned to give a fixed 80\% signal
efficiency as a function of \pt for fully contained \topquark jets;
\item TopoDNN (top quark):  A selection on the DNN discriminant, tuned to
give a fixed 80\% signal efficiency as a function of \pt for fully contained
\topquark jets;
\item \SDlong (top quark): A single-sided selection of $\mcomb >$ 60~\GeV\ and a
selection on  $\log\chi$, tuned to give a fixed 80\% signal efficiency as a
function of \pt for fully contained \topquark jets;
\item \htt (top quark): A requirement on the \htt candidate trimmed jet
kinematics to have a mass between $140$ and $210~\GeV$ and a \pt larger than
$200~\GeV$.
\end{itemize}
 
The numbers of signal-like events in data that pass and fail each of these
requirements are obtained from a chi-square template fit of ``signal'' and ``background''
distributions predicted by Monte Carlo simulations to the data to correct for
mismodelling of the cross-section of the various processes contributing to the
phase space of interest.  The labelling of ``signal'' events follows
Section~\ref{sec:labelling} and is based on Monte Carlo simulations of \ttbar
and single-top-quark events.  To increase the stability of the fit,
background templates whose shapes are similar are merged.  This procedure
results in a signal ($t\bar{t}(W)$ and $\text{single top}(W)$) and background
($t\bar{t}(\text{top})$+$t\bar{t}(\text{other})$+$\text{single
top}(\text{other})$+$\text{non-}t\bar{t}$) component template in the case of
\Wbosonboson tagging and a signal ($t\bar{t}(\text{top})$) and two background
($t\bar{t}(W)$+$t\bar{t}(\text{other})$ and non-$t\bar{t}$) component
templates in the \topquark efficiency measurement, and the normalisation of each
template is allowed to float freely in the fit.  The fit is performed using
distributions of the mass of the leading \akt trimmed jet, thus separating
signal and background events, as demonstrated in Figure~\ref{fig:eff_fit_mass}
in the case of the simple \topsimple \topquark tagger. For the measurement of
the \htt signal efficiency, the fit is performed using distributions of the mass
of the leading C/A trimmed jet instead.  Distributions of events that either pass or
fail the tagger under study are fit simultaneously.  The total normalisation of
each grouped background component is allowed to float and is extracted in the
fit, while the efficiency of the tagger on background events is fixed to the
value in Monte Carlo simulation.  Normalisations of signal distributions in the
pass and fail categories ($N^{\mathrm{tagged}}_{\mathrm{fitted\ signal}}$  and
$N^{\mathrm{not\ tagged}}_{\mathrm{fitted\ signal}}$) are extracted from the
fit.  Therefore, the tagger efficiency for signal events in data can be
extracted as
\begin{equation*}
\epsilon_{\mathrm{data}} =
\frac{N^{\mathrm{tagged}}_{\mathrm{fitted\ signal}}}
{N^{\mathrm{tagged}}_{\mathrm{fitted\ signal}} + N^{\mathrm{not\ tagged}}_{\mathrm{fitted\ signal}}}.
\end{equation*}
This can be compared to the tagger efficiency in Monte Carlo simulation, which
is based on the numbers of predicted signal events that pass,
$N^{\mathrm{tagged}}_{\mathrm{signal}}$, and fail, $N^{\mathrm{not\
tagged}}_{\mathrm{signal}}$, the tagger under study:
\begin{equation*}
\epsilon_{\mathrm{MC}} =
\frac{N^{\mathrm{tagged}}_{\mathrm{signal}}}
{N^{\mathrm{tagged}}_{\mathrm{signal}} + N^{\mathrm{not\ tagged}}_{\mathrm{signal}}}.
\end{equation*}
 
\begin{figure}[bh]
\begin{centering}
\subfigure[][]{   \label{fig:eff_fit_mass_1} \includegraphics[width=0.45\textwidth]{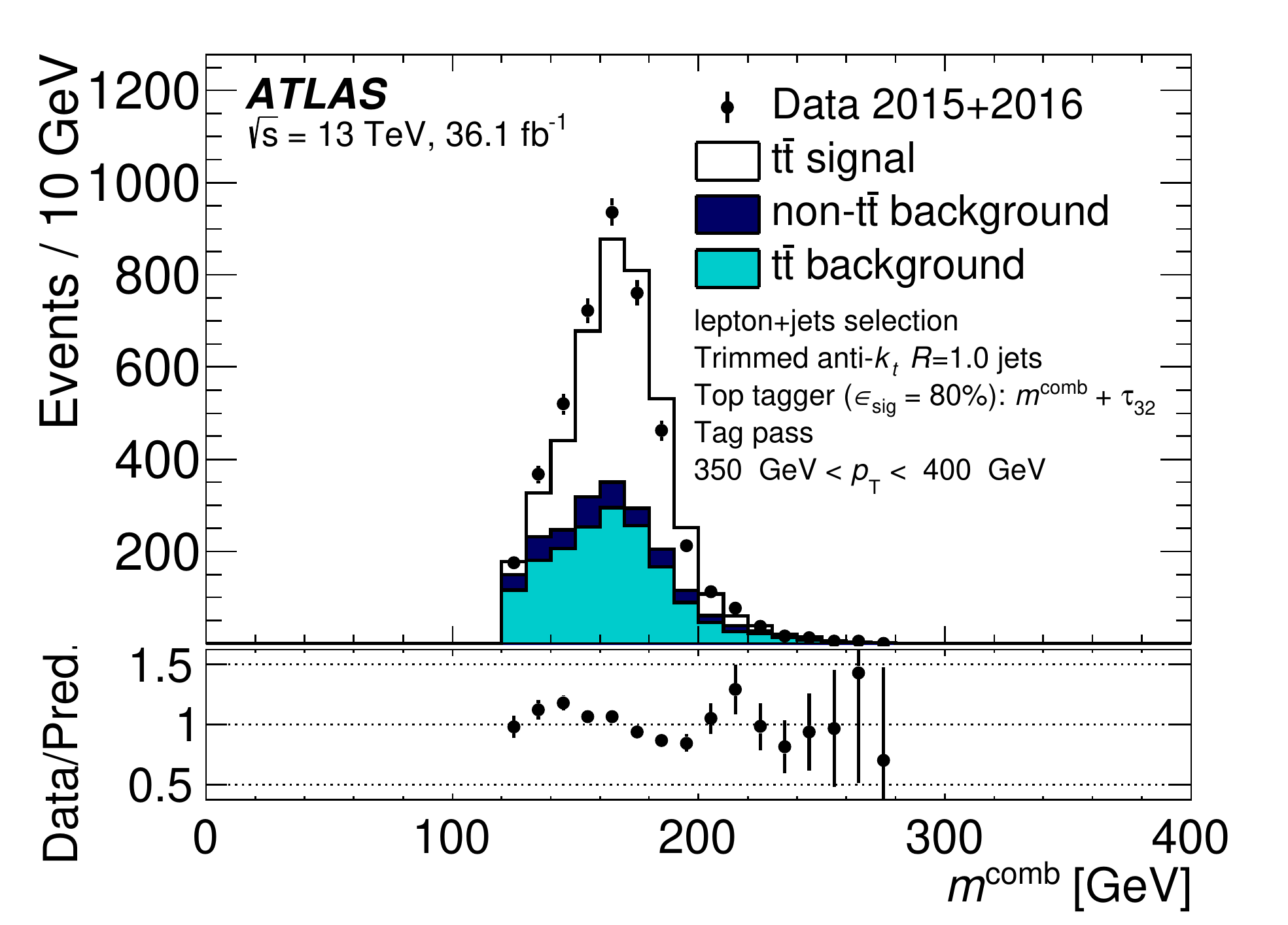}}
\subfigure[][]{   \label{fig:eff_fit_mass_2} \includegraphics[width=0.45\textwidth]{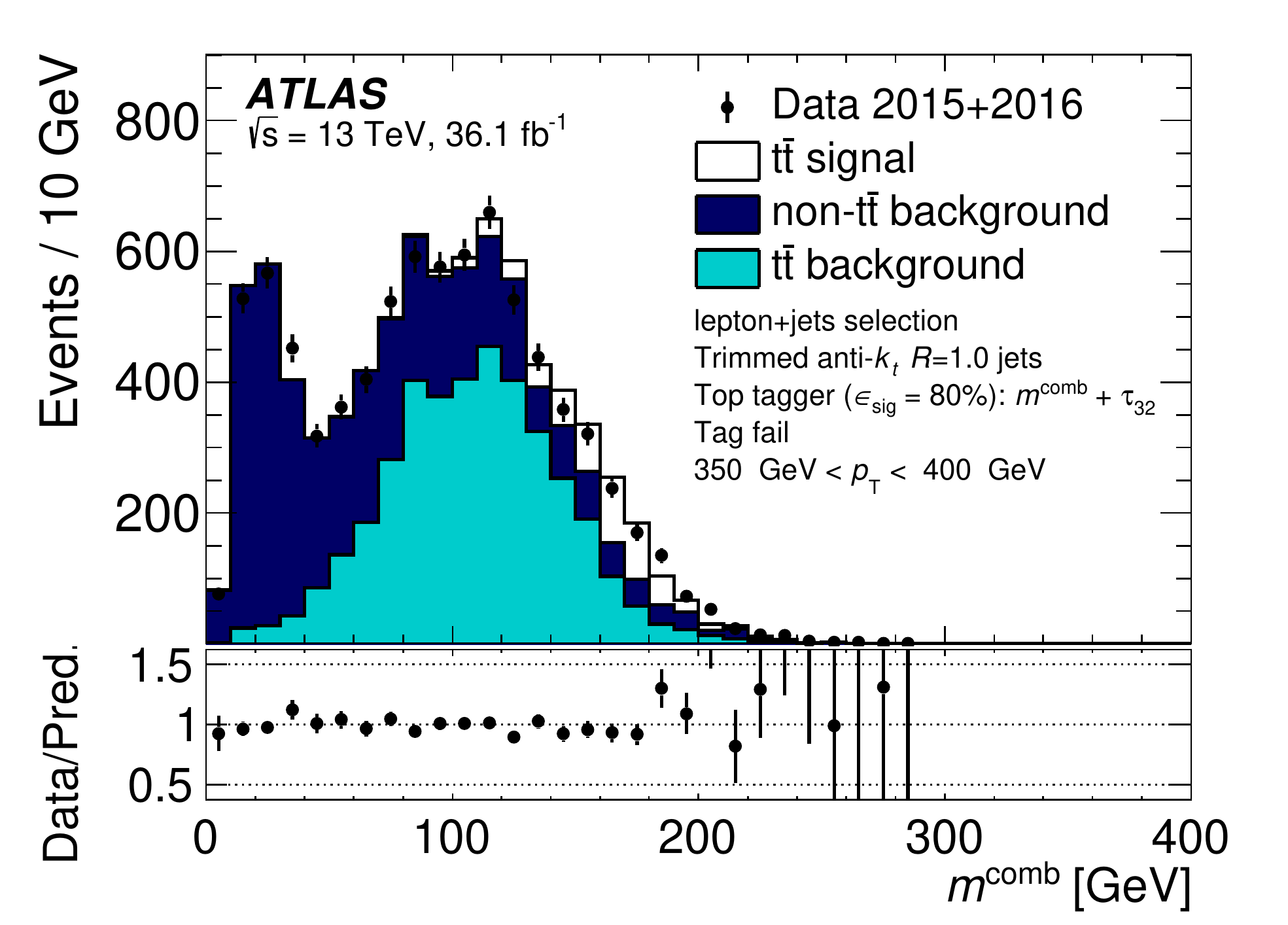}}
\caption{
The \akt trimmed jet mass distribution in the pass~\subref{fig:eff_fit_mass_1}
and fail~\subref{fig:eff_fit_mass_2} categories for the \topsimple \topquark
tagger working point after the chi-square fit has been performed. The templates
shown here are those used in the chi-square fit for the extraction of the three
normalisation factors.  The first, \textit{$t\bar{t}$
signal}, includes only the $t\bar{t}(\text{top})$ contribution, while
\textit{$t\bar{t}$ background} includes contributions from $t\bar{t}(W)$ and
$t\bar{t}(\text{other})$ and the \textit{non-$t\bar{t}$ background} component
includes all other backgrounds.  Only statistical uncertainties are shown.}
\label{fig:eff_fit_mass}
\end{centering}
\end{figure}
 
The signal efficiency is measured in data and obtained in simulations as a
function of the \pt\ of the \largeR jet as well as the average number of
interactions per bunch crossing ($\mu$).  The results are shown in
Figures~\ref{fig:Wtag_sigeff_1} and~\ref{fig:Wtag_sigeff_2} for the \Wbosonboson
taggers and in Figures~\ref{fig:toptag_sigeff_1}--\ref{fig:toptag_sigeff_5} for
the \topquark taggers.
 
The signal efficiency for the \Wbosonboson and \topquark taggers in Monte Carlo
simulation is compatible with the measured efficiency in data within
uncertainties.  In the case of the \Wbosonboson tagger working points, there is
a systematic difference between the target 50\% signal efficiency and that
measured in data due to event topology differences between \Wbosonboson jets
from these two samples, as was investigated in Ref.~\cite{PERF-2015-03}.  The
total uncertainty of the measured signal efficiency is typically about $50\%$
and $15\%$ for the \Wbosonboson and \topquark tagger efficiencies, respectively,
and is largely dominated by the subtraction of the non-contained \topquark
contribution.  In most of the kinematic phase space, these uncertainties are
dominated by systematic uncertainties, described in
Section~\ref{subsec:datamc_systematics}, specifically by the theoretical
uncertainties in \ttbar modelling, largely coming from the subtraction of the
component of the \ttbar Monte Carlo prediction that consists of either
non-\Wboson-boson jets or non-contained \topquark jets.
 
When examining the measured signal efficiency as a function of the average
number of interactions per bunch crossing, it is found to be quite
robust against increasing levels of event \pileup, even when considering only
the statistical uncertainties due to the size of the data sample, noting that the
systematic uncertainties are correlated between bins.
 
\begin{figure}[bh]
\begin{centering}
\subfigure[][]{   \label{fig:Wtag_sigeff_1_a} \includegraphics[width=0.45\textwidth]{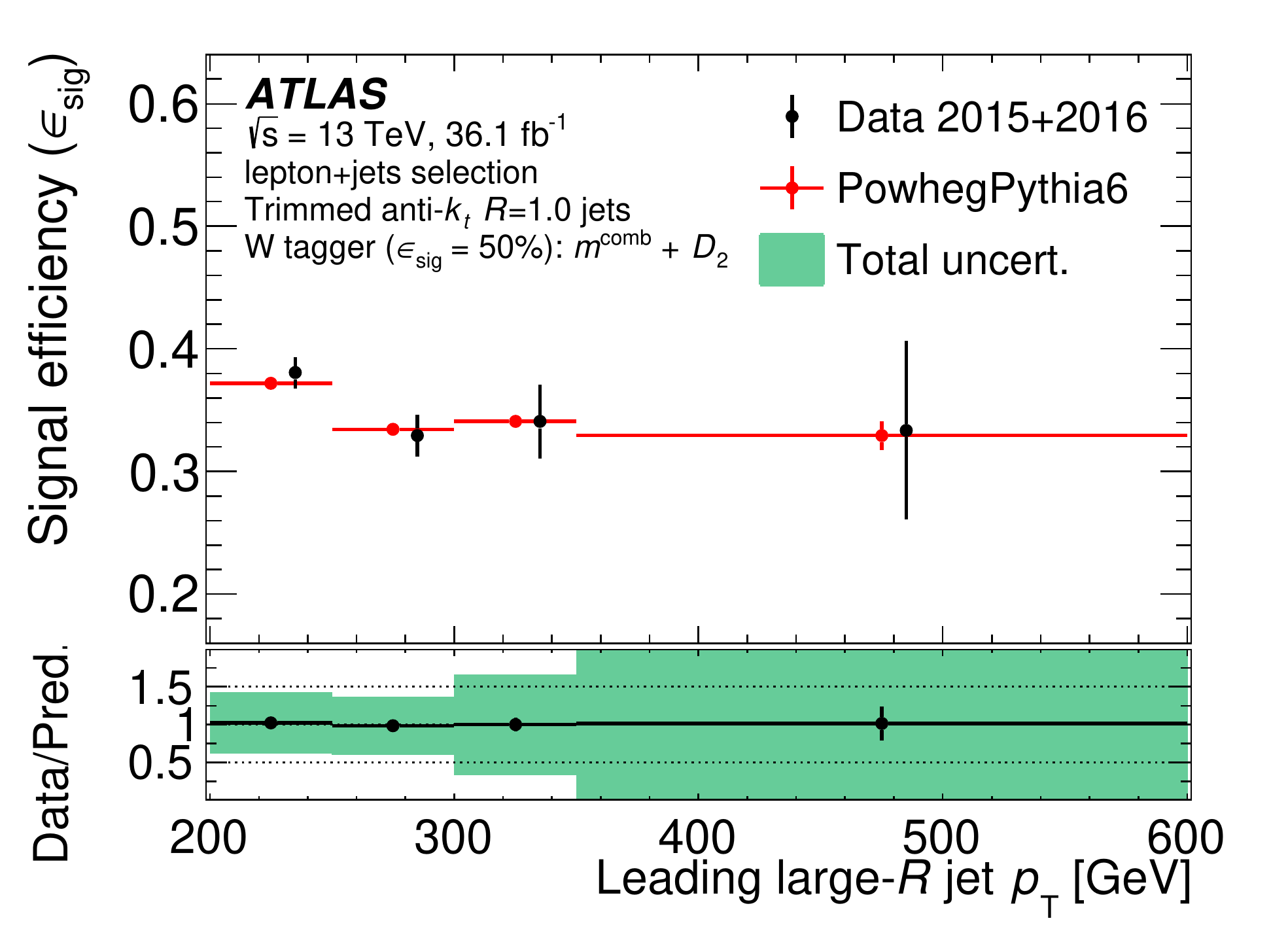}}
\subfigure[][]{   \label{fig:Wtag_sigeff_1_b} \includegraphics[width=0.45\textwidth]{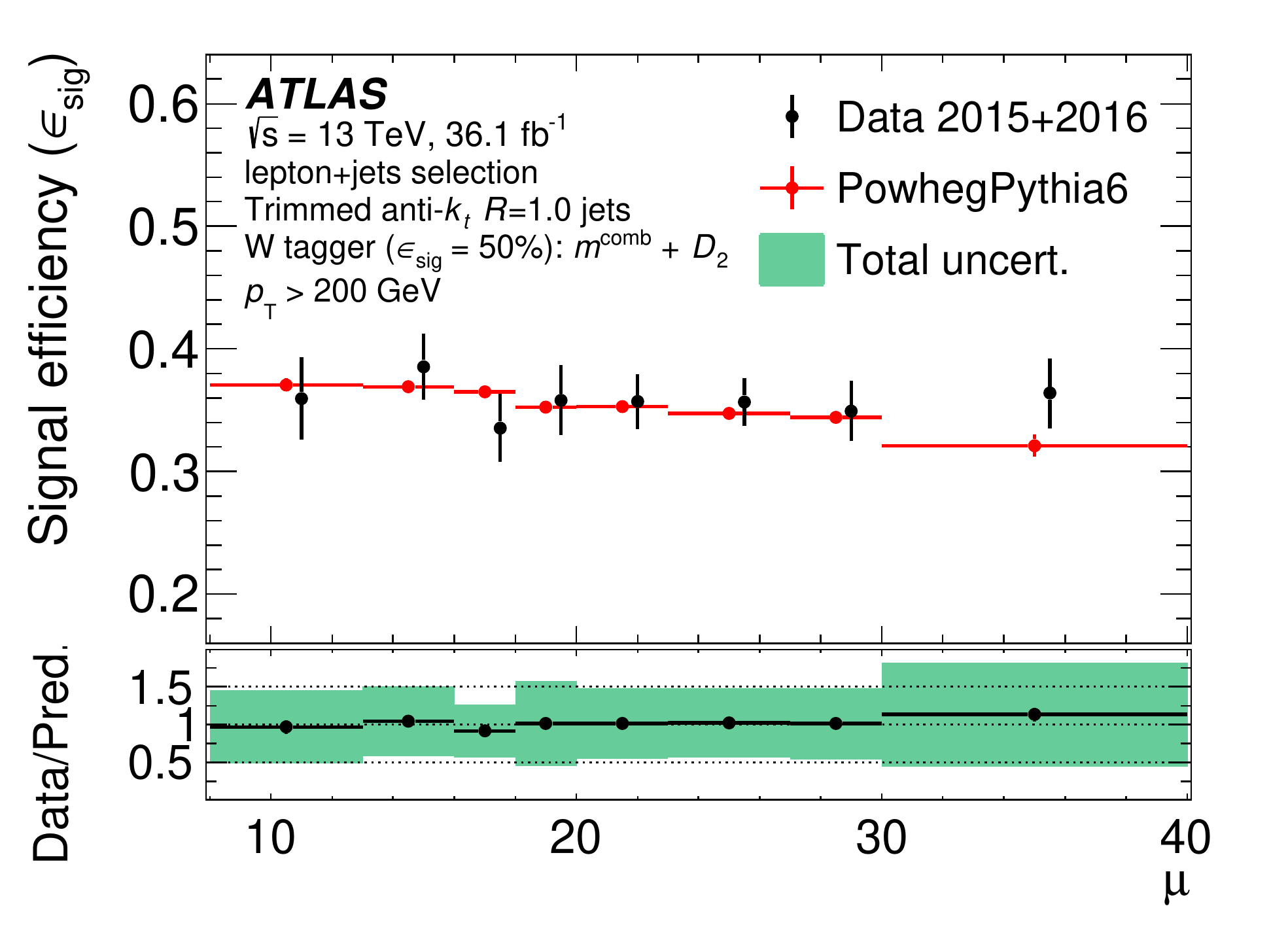}}
\caption{
The signal efficiency on contained \Wbosonboson jets for the two-variable
\wsimple \Wbosonboson tagger as a function of the \largeR jet
\pt~\subref{fig:Wtag_sigeff_1_a} and the average number of interactions per
bunch crossing $\mu$~\subref{fig:Wtag_sigeff_1_b} in data and simulation.  Statistical uncertainties of the signal efficiency
measurement in data and simulation are shown as error bars in the top panel.  In the bottom panel, the ratio of the
measured signal efficiency in data to that estimated in Monte Carlo simulation
is shown with statistical uncertainties as error bars on the data points and
the sum in quadrature of statistical and systematic uncertainties as a shaded
band.  When considering experimental uncertainties arising from the \largeR
jet, only those coming from the jet energy scale and resolution are
considered.}
\label{fig:Wtag_sigeff_1}
\end{centering}
\end{figure}
 
\begin{figure}[bh]
\begin{centering}
\subfigure[][]{   \label{fig:Wtag_sigeff_2_a} \includegraphics[width=0.45\textwidth]{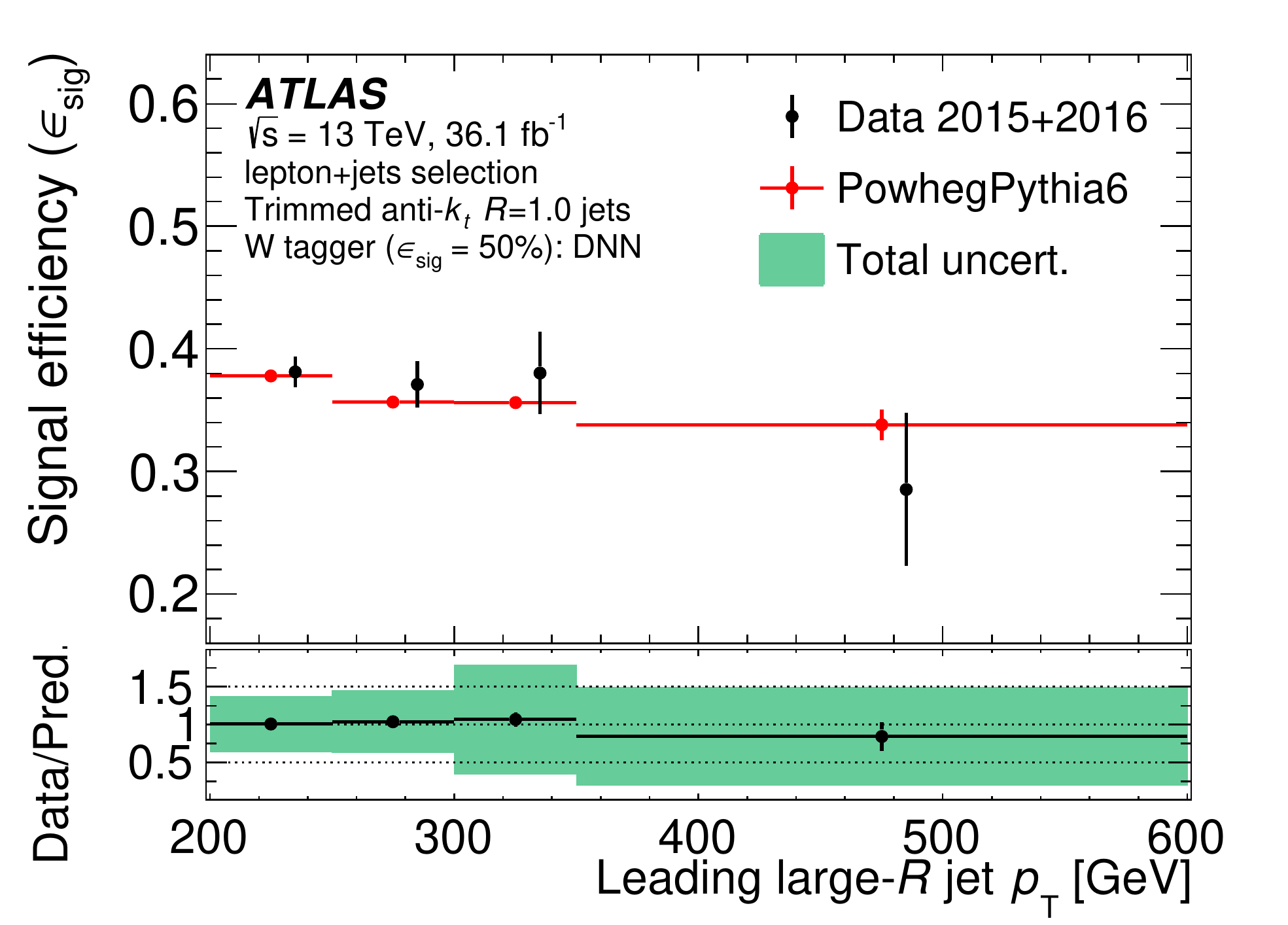}}
\subfigure[][]{   \label{fig:Wtag_sigeff_2_b} \includegraphics[width=0.45\textwidth]{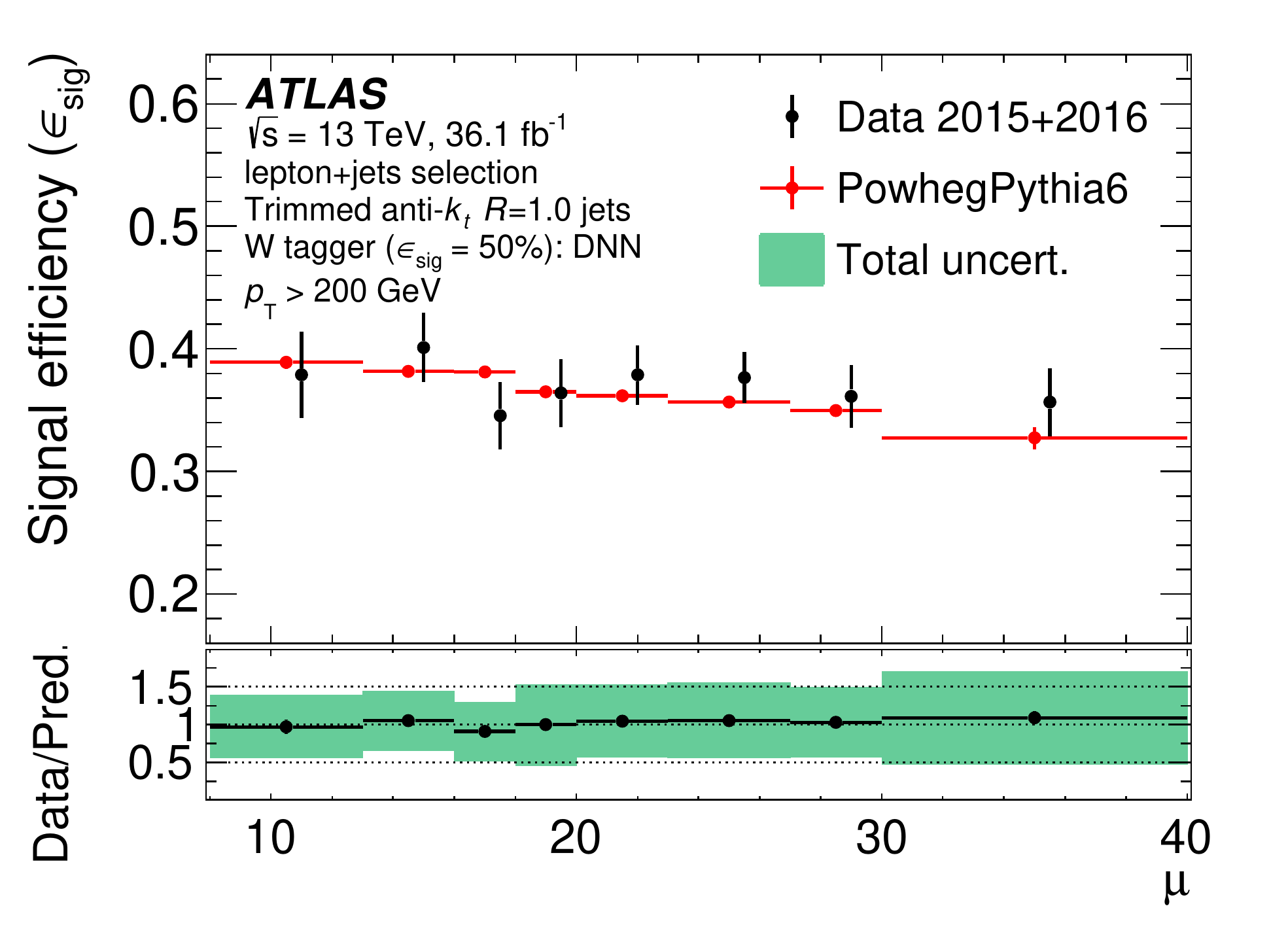}}
\caption{
The signal efficiency on contained \Wbosonboson jets for the jet \DNNlong
\Wbosonboson tagger as a function of the \largeR jet
\pt~\subref{fig:Wtag_sigeff_2_a} and the average number of interactions per
bunch crossing $\mu$~\subref{fig:Wtag_sigeff_2_b} in data and simulation.  Statistical uncertainties of the signal efficiency
measurement in data and simulation are shown as error bars in the top panel.  In the bottom panel, the ratio of the
measured signal efficiency in data to that estimated in Monte Carlo is shown
with statistical uncertainties as error bars on the data points and the sum in
quadrature of statistical and systematic uncertainties as a shaded band.  When
considering experimental uncertainties arising from the \largeR jet, only those
coming from the jet energy scale and resolution are considered.}
\label{fig:Wtag_sigeff_2}
\end{centering}
\end{figure}

\begin{figure}[bh]
\begin{centering}
\subfigure[][]{   \label{fig:toptag_sigeff_1_a} \includegraphics[width=0.45\textwidth]{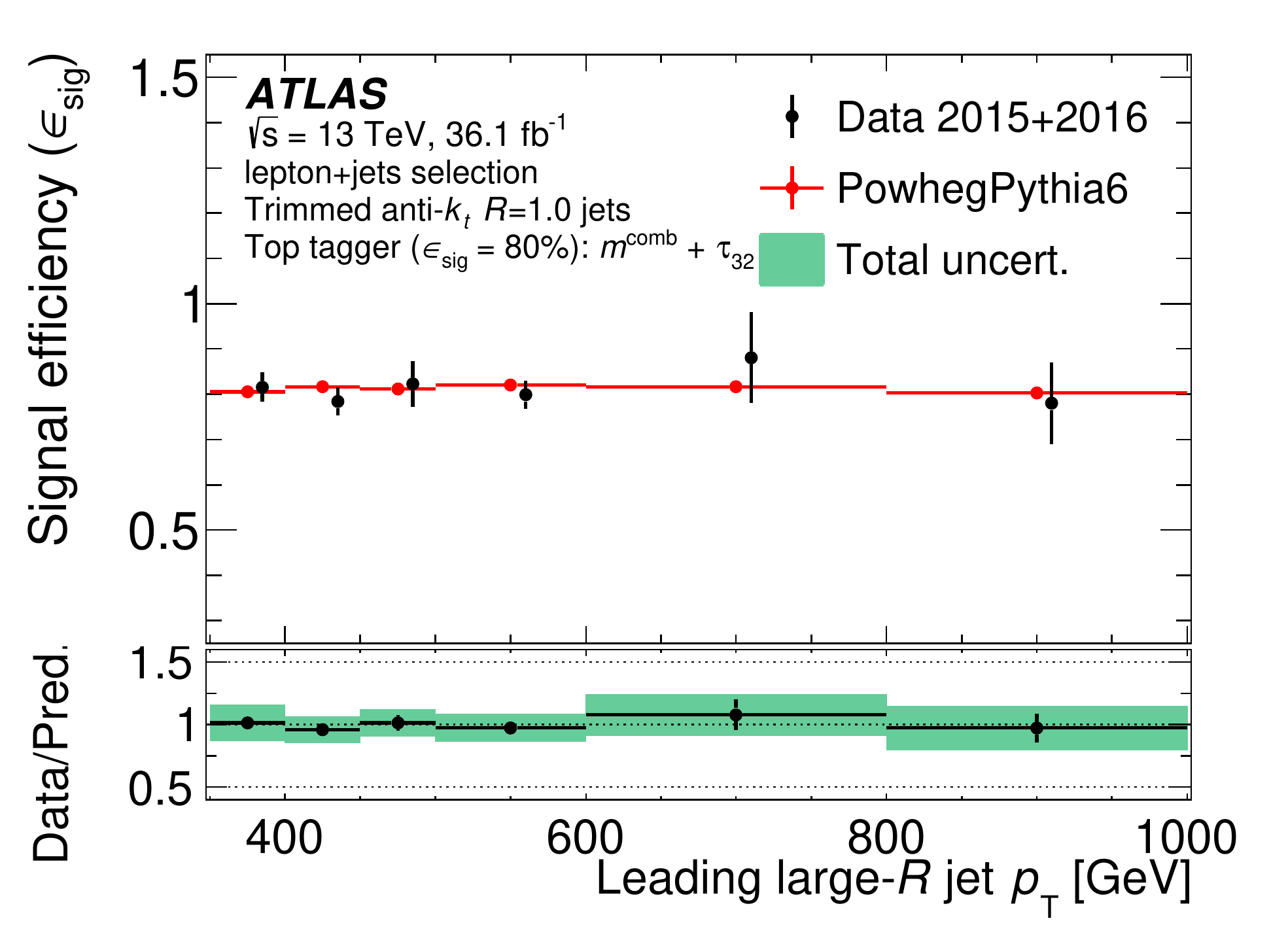}}
\subfigure[][]{   \label{fig:toptag_sigeff_1_b} \includegraphics[width=0.45\textwidth]{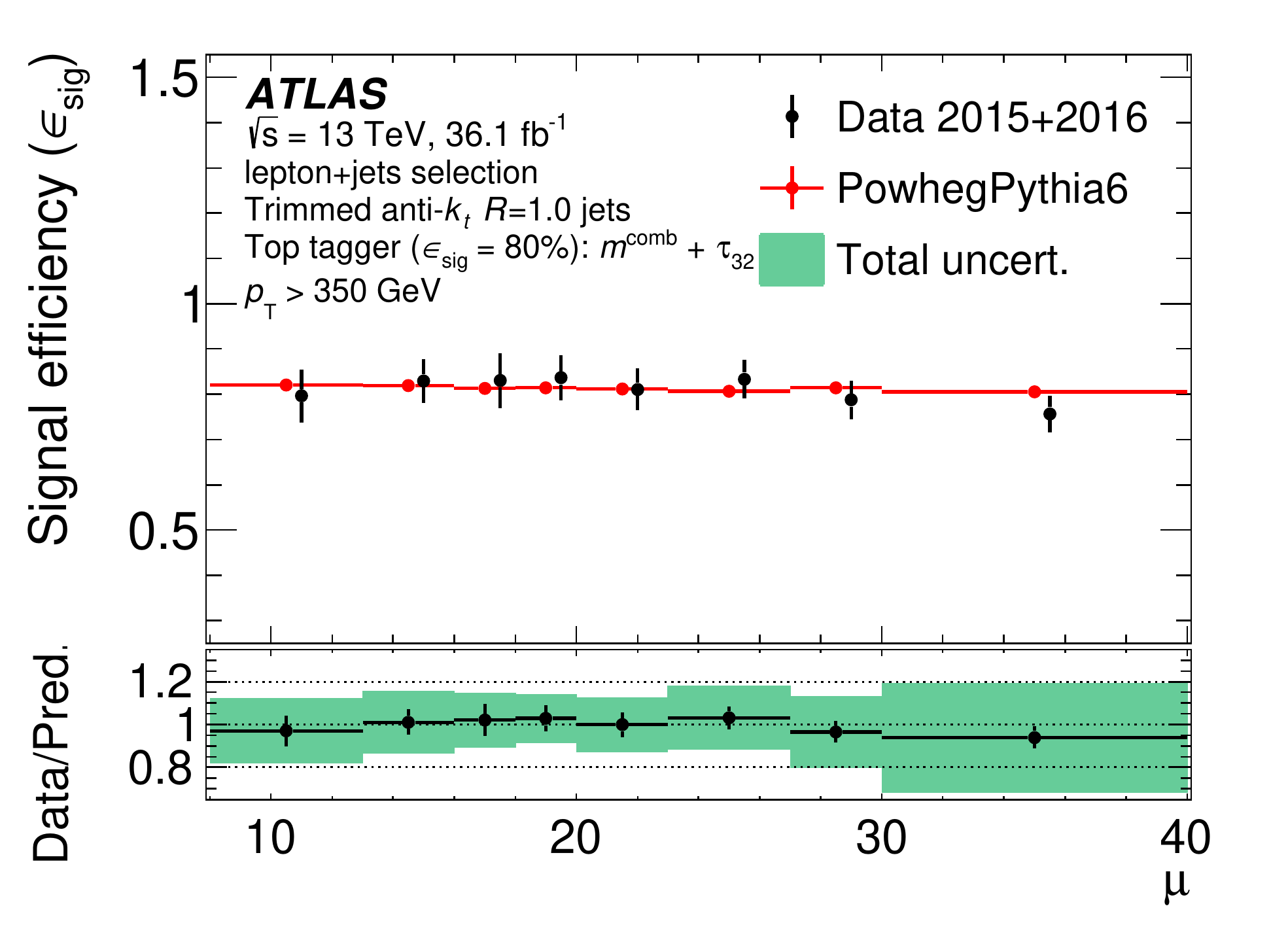}}
\caption{
The signal efficiency on contained \topquark jets for the two-variable
\topsimple \topquark tagger as a function of the \largeR jet
\pt~\subref{fig:toptag_sigeff_1_a} and the average number of interactions per
bunch crossing $\mu$~\subref{fig:toptag_sigeff_1_b} in data and simulation.
Statistical uncertainties of the signal efficiency
measurement in data and simulation are shown as error bars in the top panel.  In the bottom panel, the ratio of the
measured signal efficiency in data to that estimated in Monte Carlo is shown
with statistical uncertainties as error bars on the data points and the sum in
quadratre of statistical and systematic uncertainties as a shaded  band.  When
considering experimental uncertainties arising from the \largeR jet, only those
coming from the jet energy scale and resolution are considered.}
\label{fig:toptag_sigeff_1}
\end{centering}
\end{figure}
 
\begin{figure}[bh]
\begin{centering}
\subfigure[][]{   \label{fig:toptag_sigeff_2_a} \includegraphics[width=0.45\textwidth]{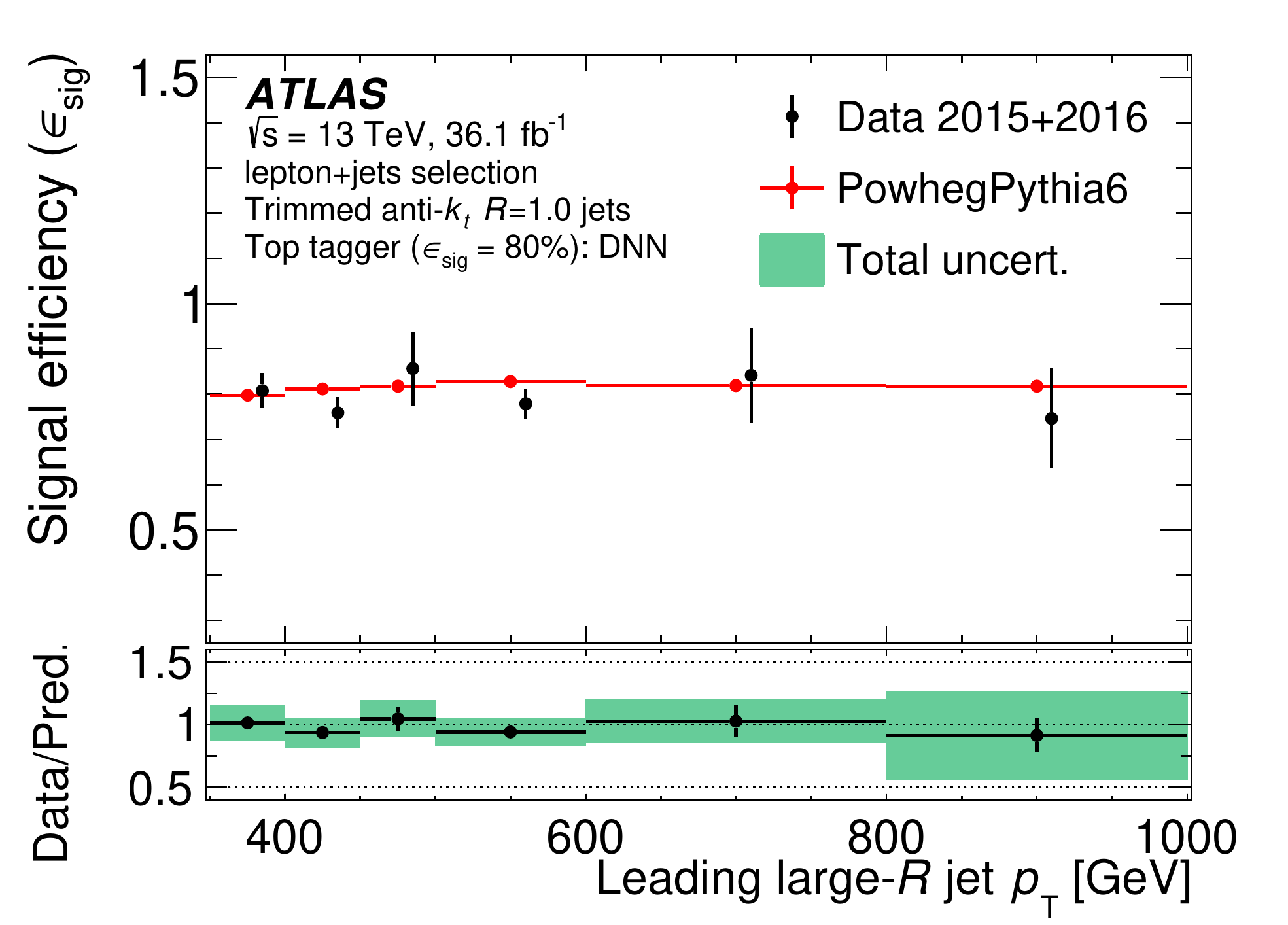}}
\subfigure[][]{   \label{fig:toptag_sigeff_2_b} \includegraphics[width=0.45\textwidth]{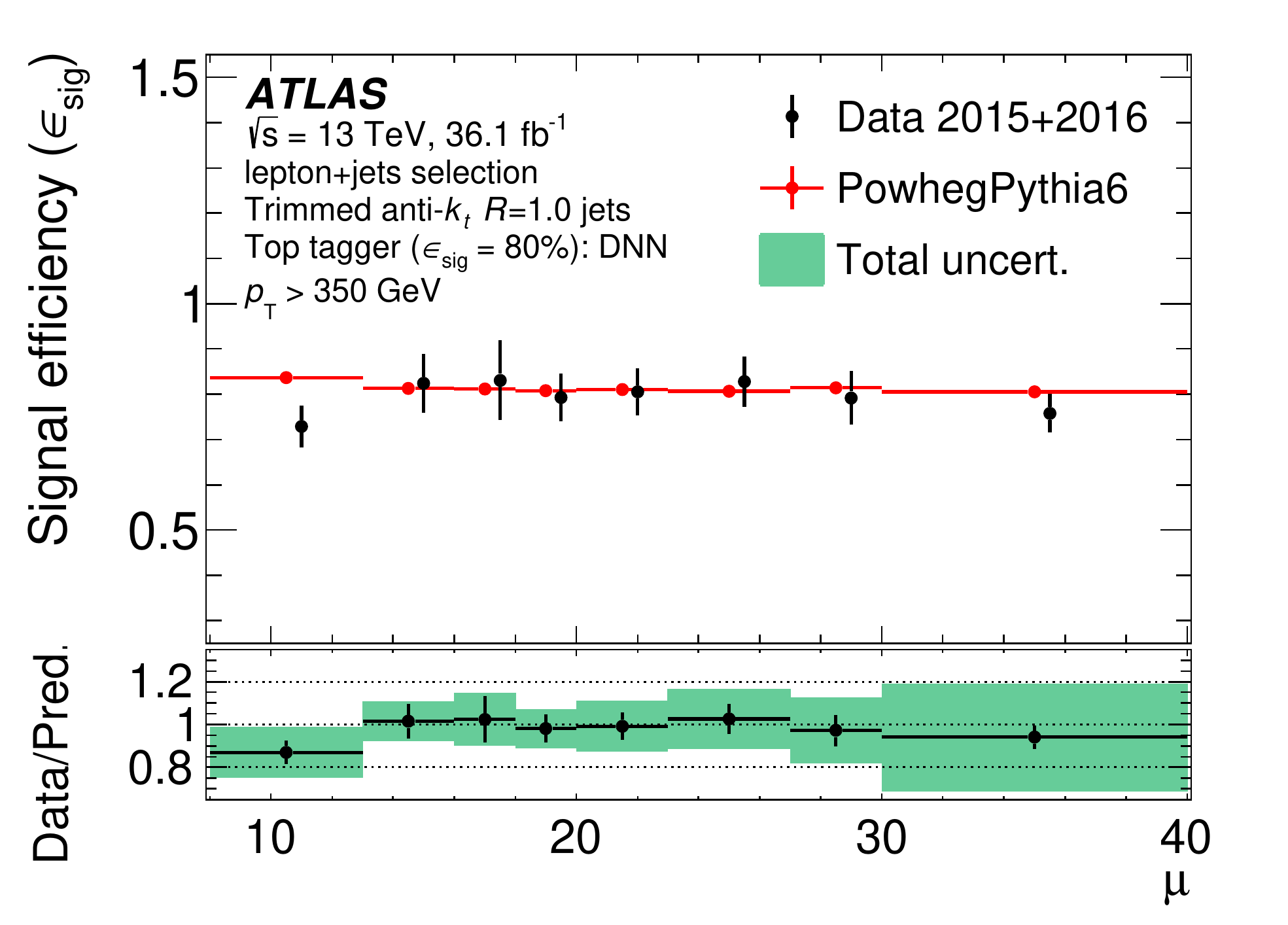}}
\caption{
The signal efficiency on contained \topquark jets for the jet \DNNlong
\topquark tagger as a function of the \largeR jet
\pt~\subref{fig:toptag_sigeff_2_a} and the average number of interactions per
bunch crossing $\mu$~\subref{fig:toptag_sigeff_2_b} in data and simulation.
Statistical uncertainties of the signal efficiency
measurement in data and simulation are shown as error bars in the top panel.  In the bottom panel, the ratio of the
measured signal efficiency in data to that estimated in Monte Carlo is shown
with statistical uncertainties as error bars on the data points and the sum in
quadrature of statistical and systematic uncertainties as a shaded band.  When
considering experimental uncertainties arising from the \largeR jet, only those
coming from the jet energy scale and resolution
are considered.}
\end{centering}
\end{figure}
 
\begin{figure}[bh]
\begin{centering}
\subfigure[][]{   \label{fig:toptag_sigeff_3_a} \includegraphics[width=0.45\textwidth]{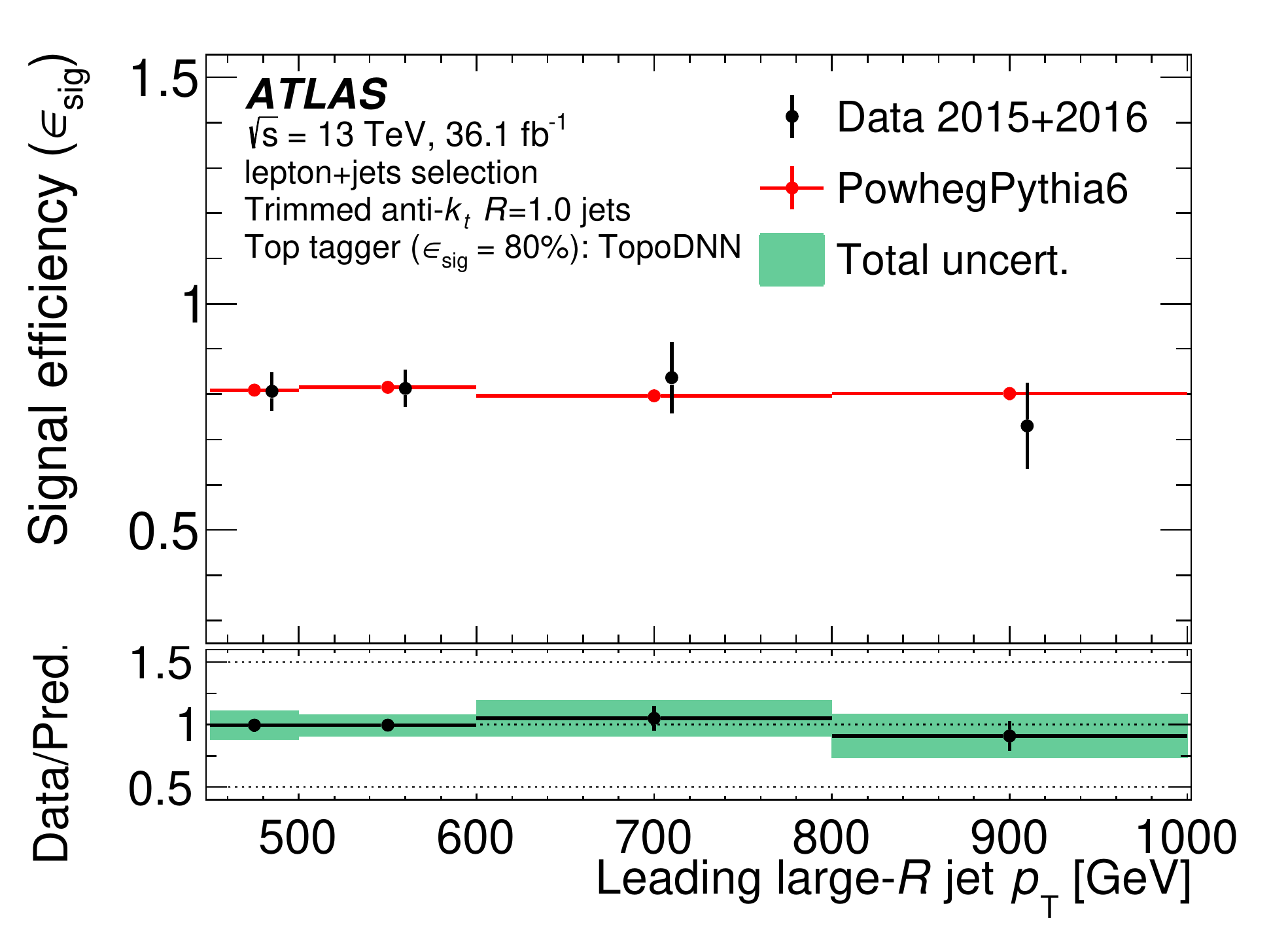}}
\subfigure[][]{   \label{fig:toptag_sigeff_3_b} \includegraphics[width=0.45\textwidth]{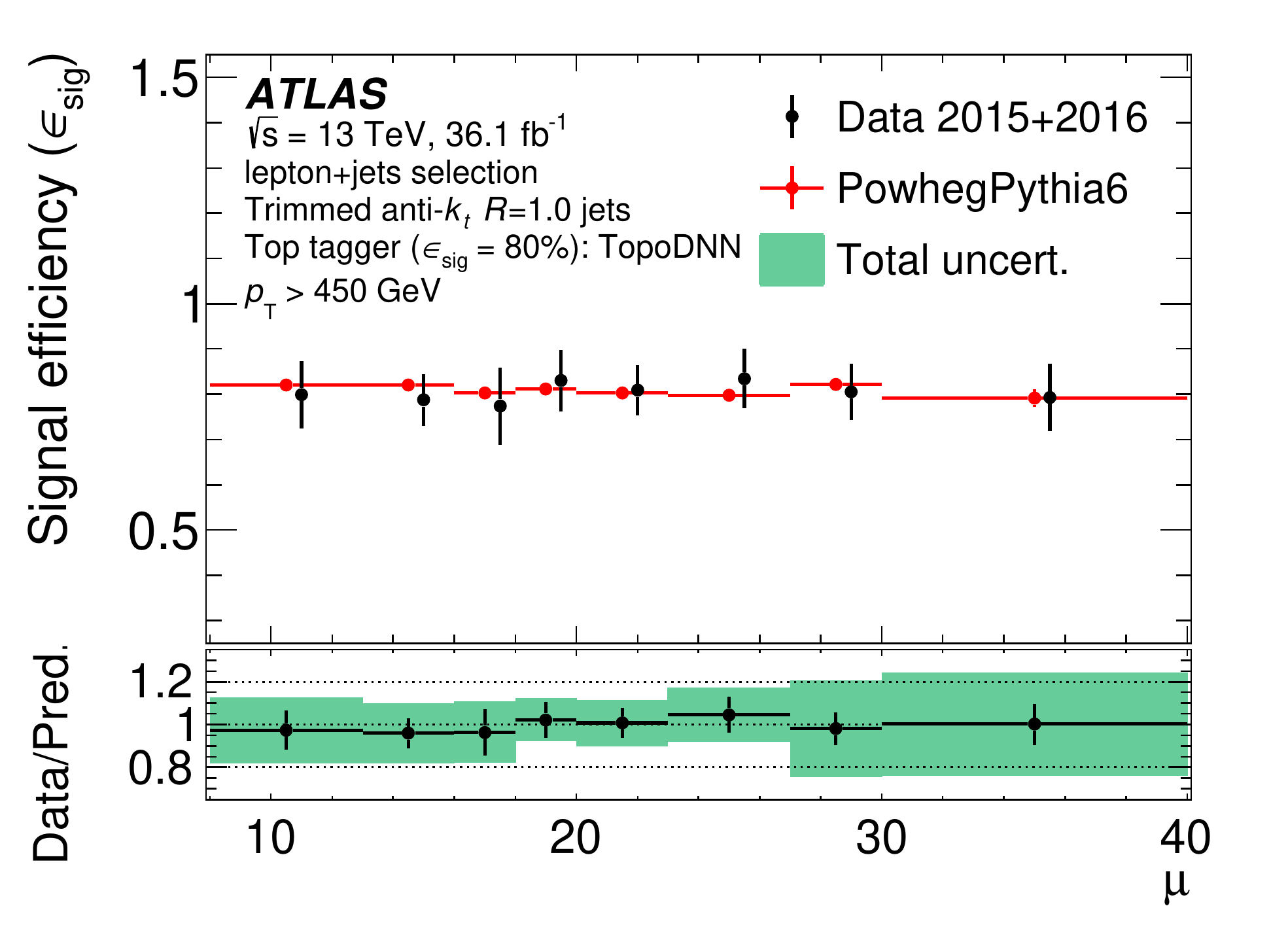}}
\caption{
The signal efficiency on contained \topquark jets for the TopoDNN
\topquark tagger as a function of the \largeR jet
\pt~\subref{fig:toptag_sigeff_3_a} and the average number of interactions per
bunch crossing $\mu$~\subref{fig:toptag_sigeff_3_b} in data and simulation.
Statistical uncertainties of the signal efficiency
measurement in data and simulation are shown as error bars in the top panel.  In the bottom panel, the ratio of the
measured signal efficiency in data to that estimated in Monte Carlo is shown
with statistical uncertainties as error bars on the data points and the sum in
quadrature of statistical and systematic uncertainties as a shaded band.  When
considering experimental uncertainties arising from the \largeR jet, only those
coming from the jet energy scale and resolution are considered.}
\end{centering}
\end{figure}
 
\begin{figure}[bh]
\begin{centering}
\subfigure[][]{   \label{fig:toptag_sigeff_4_a} \includegraphics[width=0.45\textwidth]{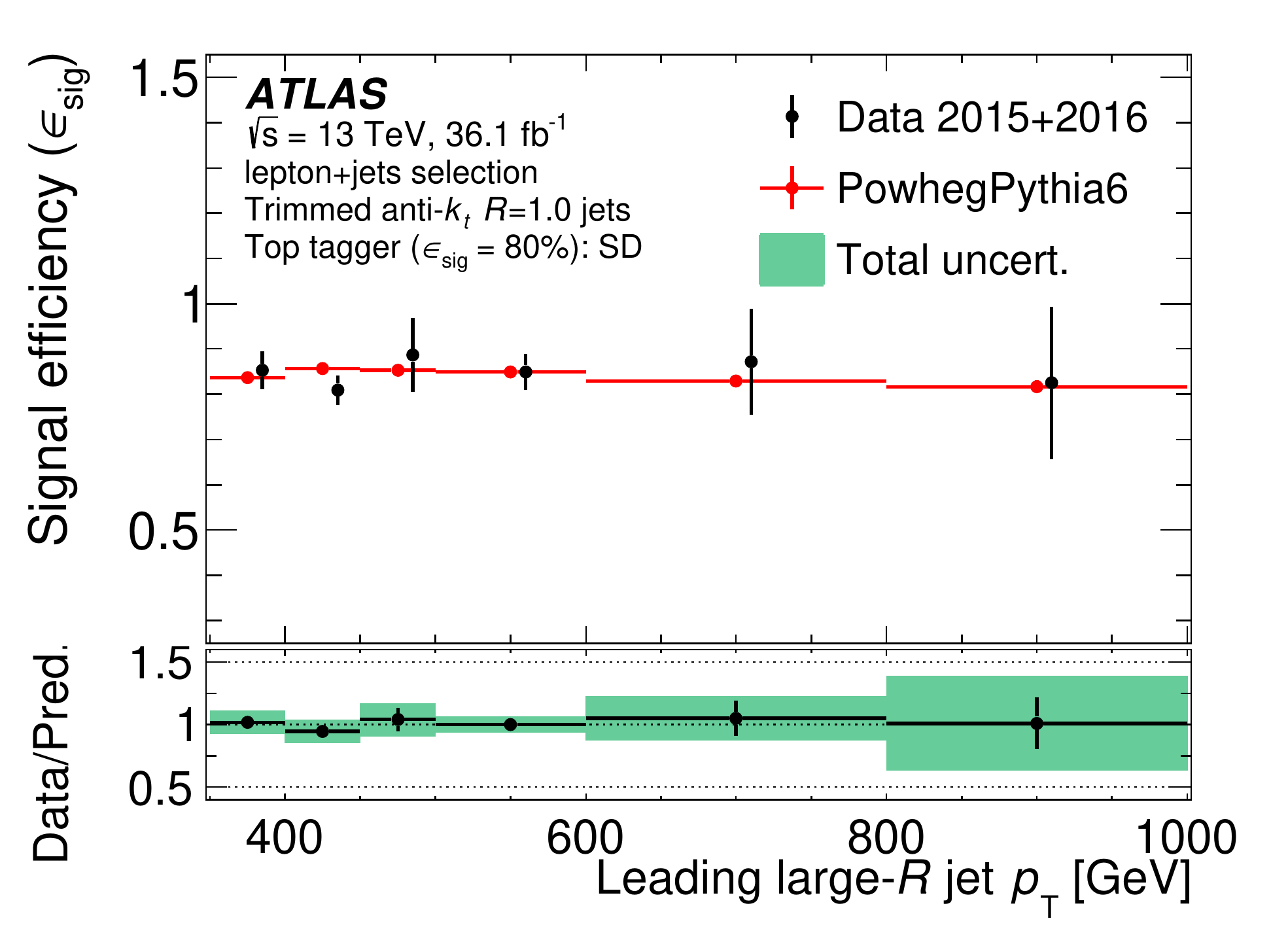}}
\subfigure[][]{   \label{fig:toptag_sigeff_4_b} \includegraphics[width=0.45\textwidth]{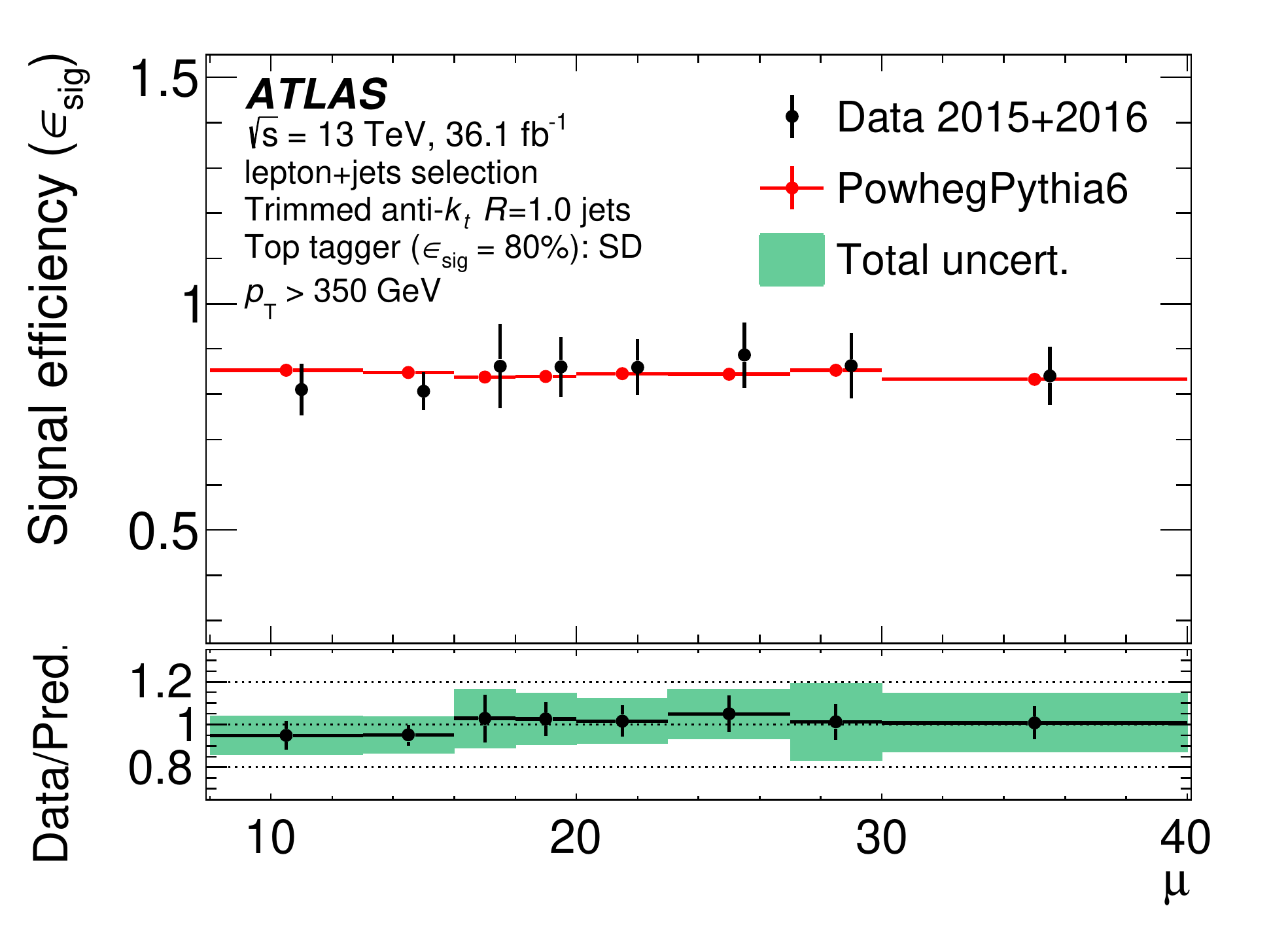}}
\caption{
The signal efficiency on contained \topquark jets for the \SDlong \topquark
tagger as a function of the \largeR jet \pt~\subref{fig:toptag_sigeff_4_a} and
the average interactions per bunch crossing
$\mu$~\subref{fig:toptag_sigeff_4_b} in data and simulation.  Statistical uncertainties of the signal efficiency
measurement in data and simulation are shown as error bars in the top panel.  In the bottom panel, the ratio of the measured signal
efficiency in data to that estimated in Monte Carlo is shown with statistical
uncertainties as error bars on the data points and the sum in quadrature of
statistical and systematic uncertainties as a shaded band.  When considering
experimental uncertainties arising from the \largeR jet, only those coming from
the jet energy scale and resolution are considered.}
\end{centering}
\end{figure}
 
\begin{figure}[bh]
\begin{centering}
\subfigure[][]{   \label{fig:toptag_sigeff_5_a} \includegraphics[width=0.45\textwidth]{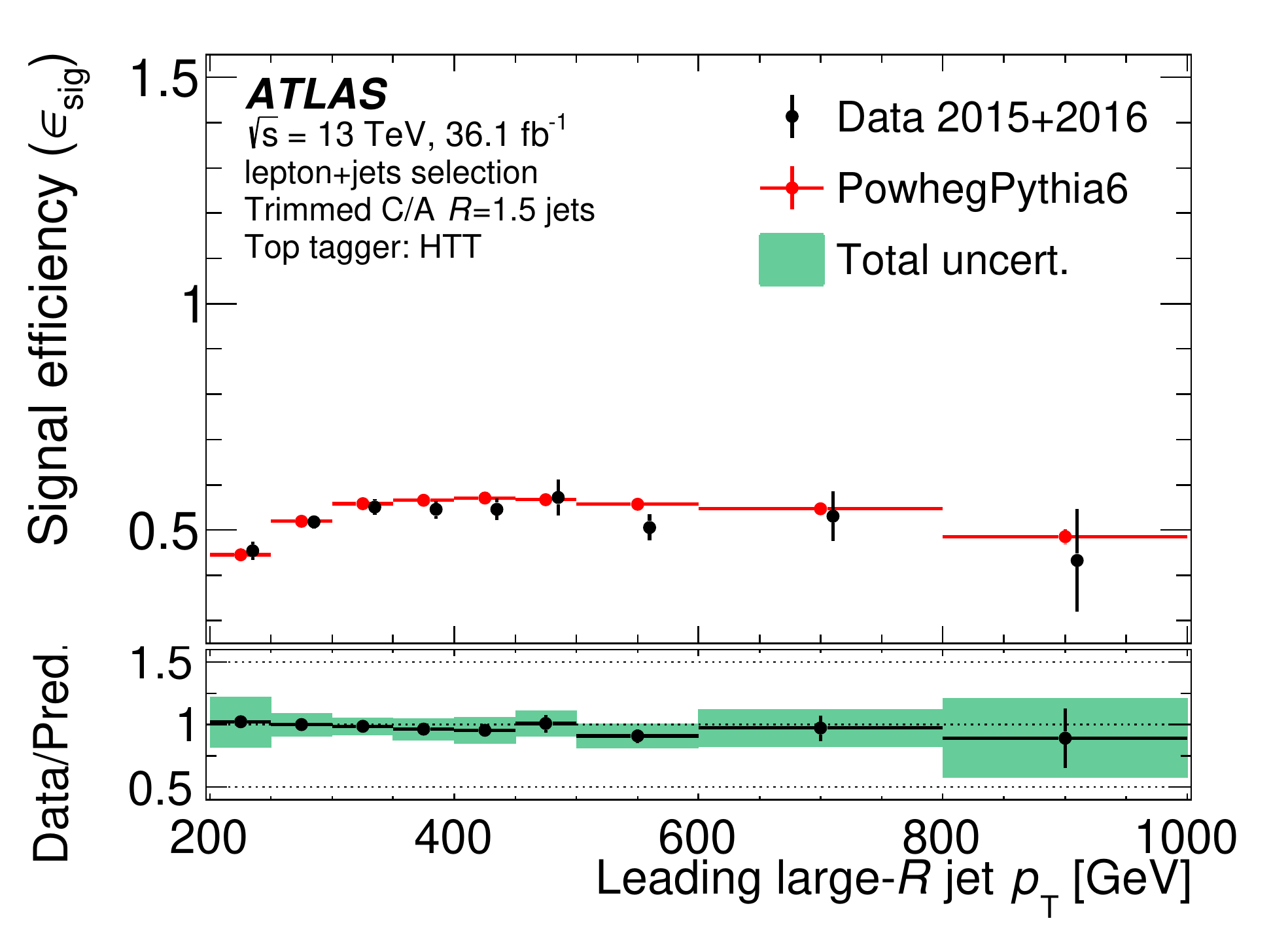}}
\subfigure[][]{   \label{fig:toptag_sigeff_5_b} \includegraphics[width=0.45\textwidth]{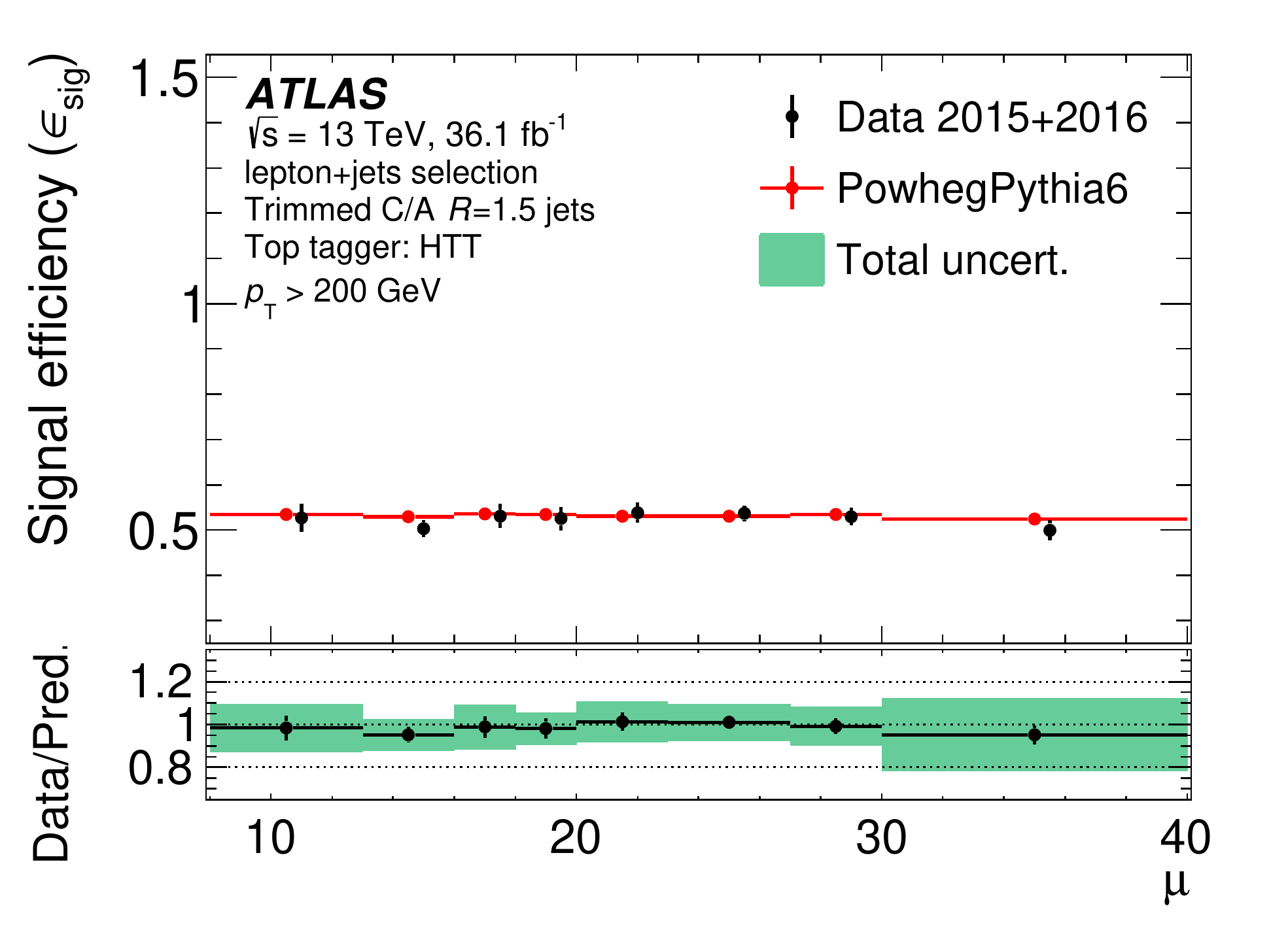}}
\caption{
The signal efficiency on contained \topquark jets for the \htt \topquark tagger
as a function of the \largeR jet \pt~\subref{fig:toptag_sigeff_5_a} and the
average number of interactions per bunch crossing
$\mu$~\subref{fig:toptag_sigeff_5_b} in data and simulation.  Statistical uncertainties of the signal efficiency
measurement in data and simulation are shown as error bars in the top panel.  In the bottom panel, the ratio of the measured signal
efficiency in data to that estimated in Monte Carlo is shown with statistical
uncertainties as error bars on the data points and the sum in quadrature of
statistical and systematic uncertainties as a shaded band.  When considering
experimental uncertainties arising from the \largeR jet, only those coming from
the jet energy scale and resolution are considered.  The signal efficiency on
contained \topquark jets for the \htt is not constant with respect to jet \pt
as the tagger was not re-optimised after the Run-1
analysis~\cite{PERF-2015-04}.}
\label{fig:toptag_sigeff_5}
\end{centering}
\end{figure}
\clearpage
\newpage
\clearpage
\newpage
 
\subsection{Background rejection from multijet and \gammajet events}
\label{subsec:datamc_dijetgammajet}
In addition to studying the modelling of signal \Wboson-boson and top-quark jets
using a sample of \ttbar events, the behaviour of background light jets is
studied in two sets of events (enriched in multijet and \gammajet processes) to
cover a broad kinematic range and probe the behaviour of quark- and
gluon-enriched regions of phase space separately~\cite{Gras:2017jty}.  The first
sample, multijet events, provides a means to study a mixture of light-quark and
gluon jets in the kinematic range from \pt of approximately 450~\GeV\ to
3000~\GeV\ while the \gammajet sample is greatly enhanced in the fraction of
quark jets produced and provides a means to study jets with \pt from $\sim$
200~\GeV\ to 2000~\GeV. As in the case of the study of signal \Wboson-boson and
top-quark jets in Section~\ref{subsec:datamc_ttbar}, the distributions of
important tagging observables are examined and the background rejection is
quantified in both data and Monte Carlo simulation.
 
\subsubsection{Analysis and selection}
\label{subsubsec:background_analysis}
To select the multijet sample, events are selected in both data and Monte Carlo simulation
using a single-jet trigger based on a single \largeR \akt trimmed jet with $R =$
1.0 with an online requirement of \et > 360~\GeV\ during 2015 data taking and
420~\GeV\ in 2016.  Events are then required to have at least one fully-calibrated \largeR
\akt trimmed jet with radius 1.0 with \pt > 450~\GeV\ so that the trigger is
fully efficient.  After this selection, the modelling of the highest-\pt \largeR
jet (both \akt $R=$1.0 trimmed and C/A $R =$ 1.5) in the event is examined with
respect to both the \textsc{Pythia} and \textsc{Herwig++} generators described
in Section~\ref{sec:samples}.
 
In the case of the \gammajet sample, events are selected in both data and Monte
Carlo simulation with a single-photon trigger which selects photons satisfying ``loose''
quality criteria and which pass an online requirement of \ET > 120~\GeV\ in 2015
and 140~\GeV\ in 2016.  Photon candidates are required to be within $|\eta| <
2.5$ and satisfy a likelihood-based identification criterion based on shower
shape observables in the electromagnetic calorimeter as well as the relative
amount of energy in the hadronic and electromagnetic calorimeters, and are
required to be isolated from other activity in the event.  Both the
identification and isolation criteria are required to satisfy the ``tight'' working
point described in Ref.~\cite{ATL-PHYS-PUB-2016-014}.  In addition, \largeR jets
are required to have  $\pT >$ 200~\GeV, $|\eta| <$ 2.0 and to be well-separated
from the reconstructed photon with $\Delta\phi(\mathrm{jet,\gamma}) >
\frac{\pi}{2}$.  Finally, events with at least one photon with \ET > 155~\GeV\
are selected to ensure that the trigger is fully efficient.
 
In both selections, the normalisation of the simulated multijet and \gammajet
predictions is derived directly from data after the initial inclusive selection,
taking into account the small contribution from hadronically decaying \Wboson-boson,
\Zboson-boson and \ttbar events. First the predicted contribution from
processes containing real hadronically decaying \Wboson bosons and top quarks is
subtracted from data. The remaining Monte Carlo samples are then normalised to
reproduce the same yield as the background-subtracted data.
 
Figures~\ref{fig:DMC_bkg_mass_1} and~\ref{fig:DMC_bkg_mass_2} show a comparison
of the distributions of the leading \akt $R=$1.0 and C/A $R=$1.5 jet mass in the
inclusive multijet and \gammajet selections.  In addition, the primary tagging
observables used to perform \Wboson-boson and top-quark tagging described in
Section~\ref{sec:TaggerTechniques} are shown in
Figures~\ref{fig:DMC_bkg_jss1}--\ref{fig:DMC_bkg_jss5}.  In general, the
modelling of the shape of the tagging discriminants in data by the Monte Carlo simulation
agrees at the 20\% level, with non-negligible differences observed when comparing
\textsc{Pythia8} to \textsc{Herwig} Monte Carlo predictions.  Finally, the jet
mass distribution for jets that are positively tagged using the jet-shape-based
DNN discriminant optimised in Section~\ref{sec:optimisation_bdtdnn} is shown in
Figures~\ref{fig:DMC_bkg_tagmass_1} and~\ref{fig:DMC_bkg_tagmass_2} for the
multijet and \gammajet topologies.  Good agreement between data and Monte Carlo
simulation is observed within uncertainties, which are dominated by Monte Carlo
modelling.  It is further observed that the jet mass distribution is strongly
distorted after the application of the tagger, a feature which is shared by all
tagging techniques described in Section~\ref{sec:TaggerTechniques}.
 
\begin{figure}[!h]
\begin{centering}
\subfigure[][]{   \label{fig:DMC_bkg_mass_1_a} \includegraphics[width=0.45\textwidth]{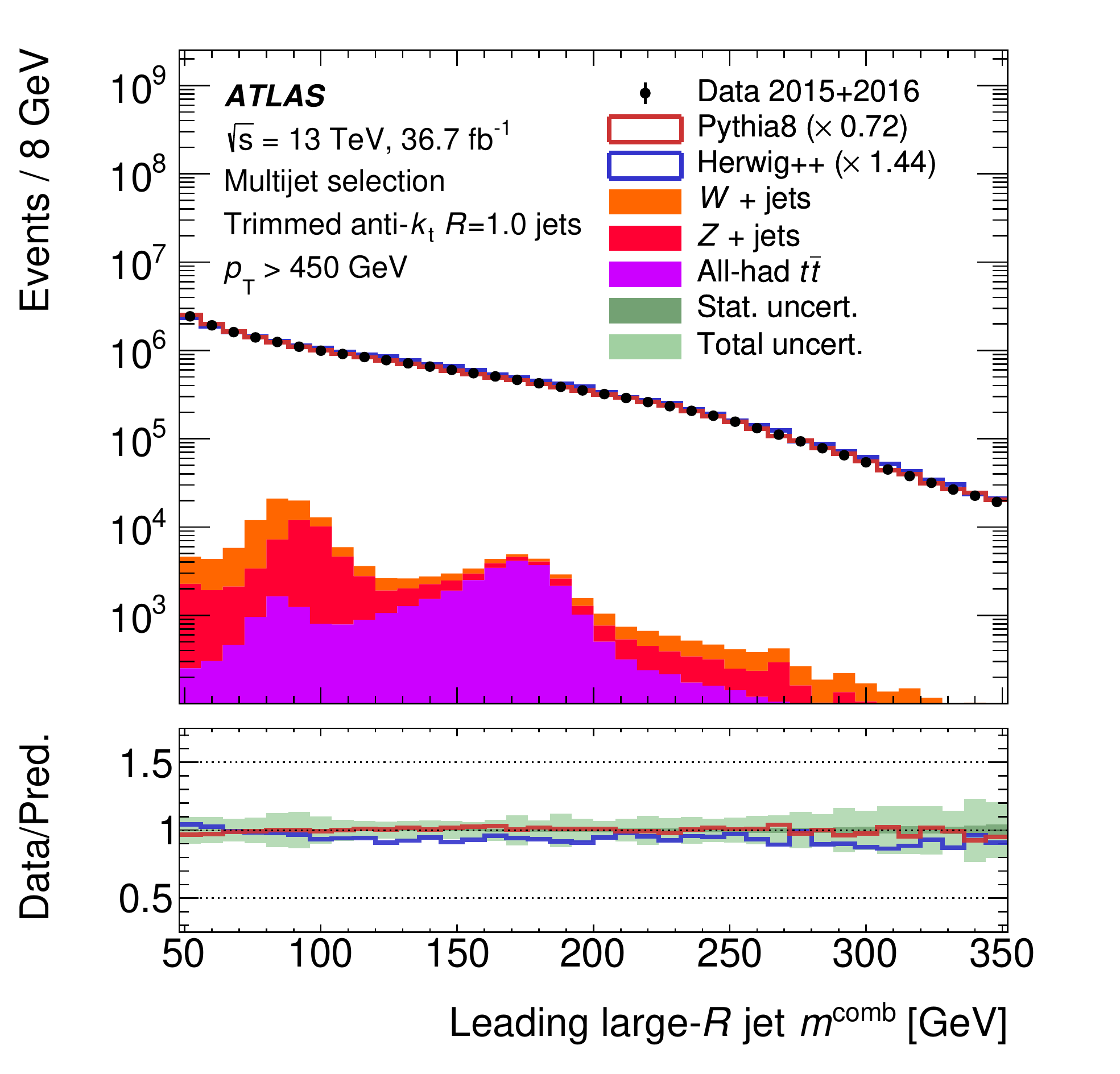}}
\subfigure[][]{  \label{fig:DMC_bkg_mass_1_b} \includegraphics[width=0.45\textwidth]{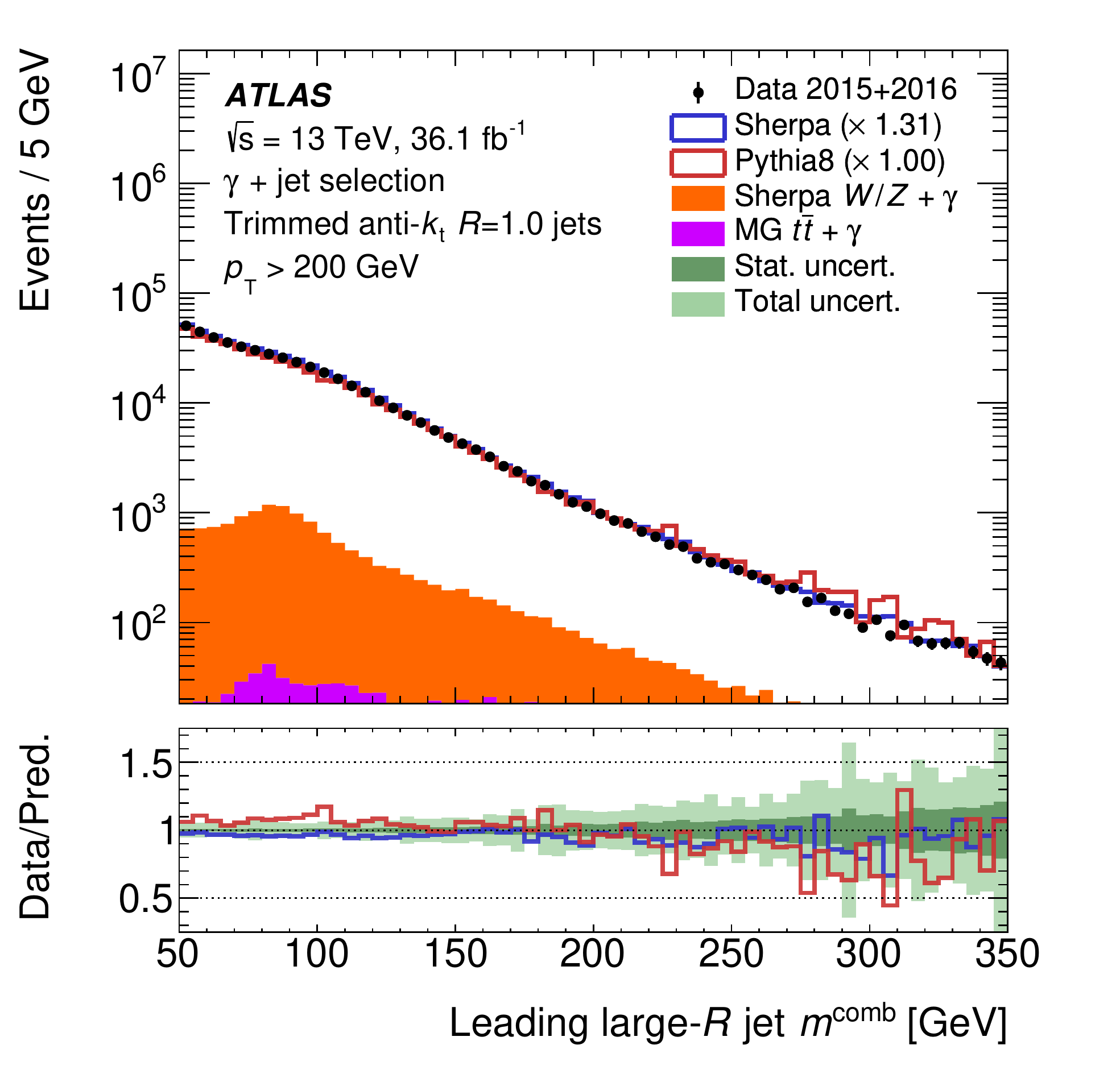}}
\caption{
\label{fig:DMC_bkg_mass_1} A comparison of the observed data and predicted MC
distributions of the mass of the leading \pt \akt trimmed jet in events for the
multijet \subref{fig:DMC_bkg_mass_1_a} and \gammajet
\subref{fig:DMC_bkg_mass_1_b} selections.  The data-driven normalisation
correction, described in Section~\ref{subsubsec:background_analysis}, is
shown in the legend beside the specific sample to which it applies.
Systematic uncertainties are indicated as a band in the lower panel and include all
experimental uncertainties related to the selection of events, as well as the
reconstruction and calibration of the \largeR jet.}
\end{centering}
\end{figure}
 
\begin{figure}[!h]
\begin{centering}
\subfigure[][]{   \label{fig:DMC_bkg_mass_2_a} \includegraphics[width=0.45\textwidth]{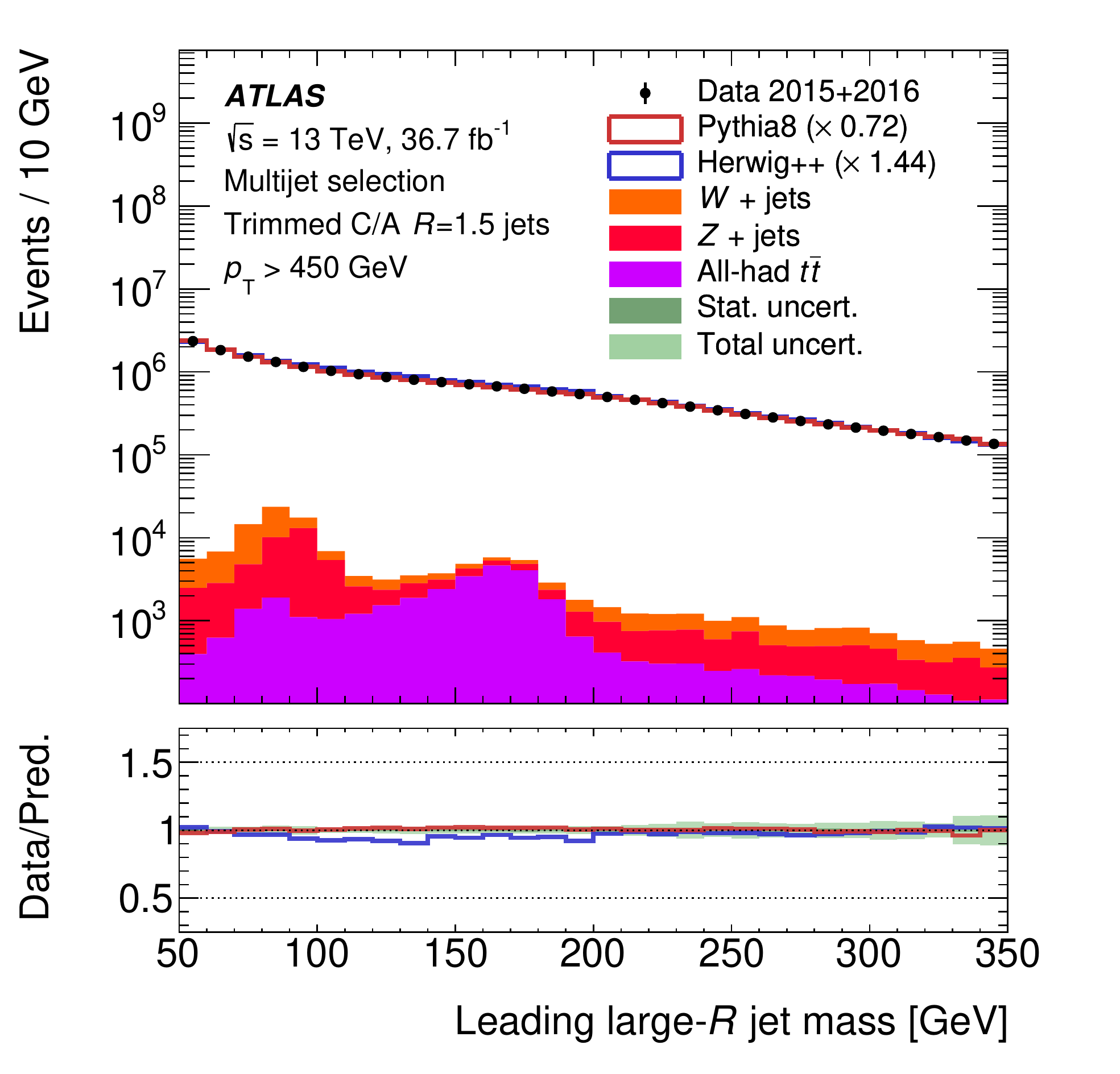}}
\subfigure[][]{  \label{fig:DMC_bkg_mass_2_b} \includegraphics[width=0.45\textwidth]{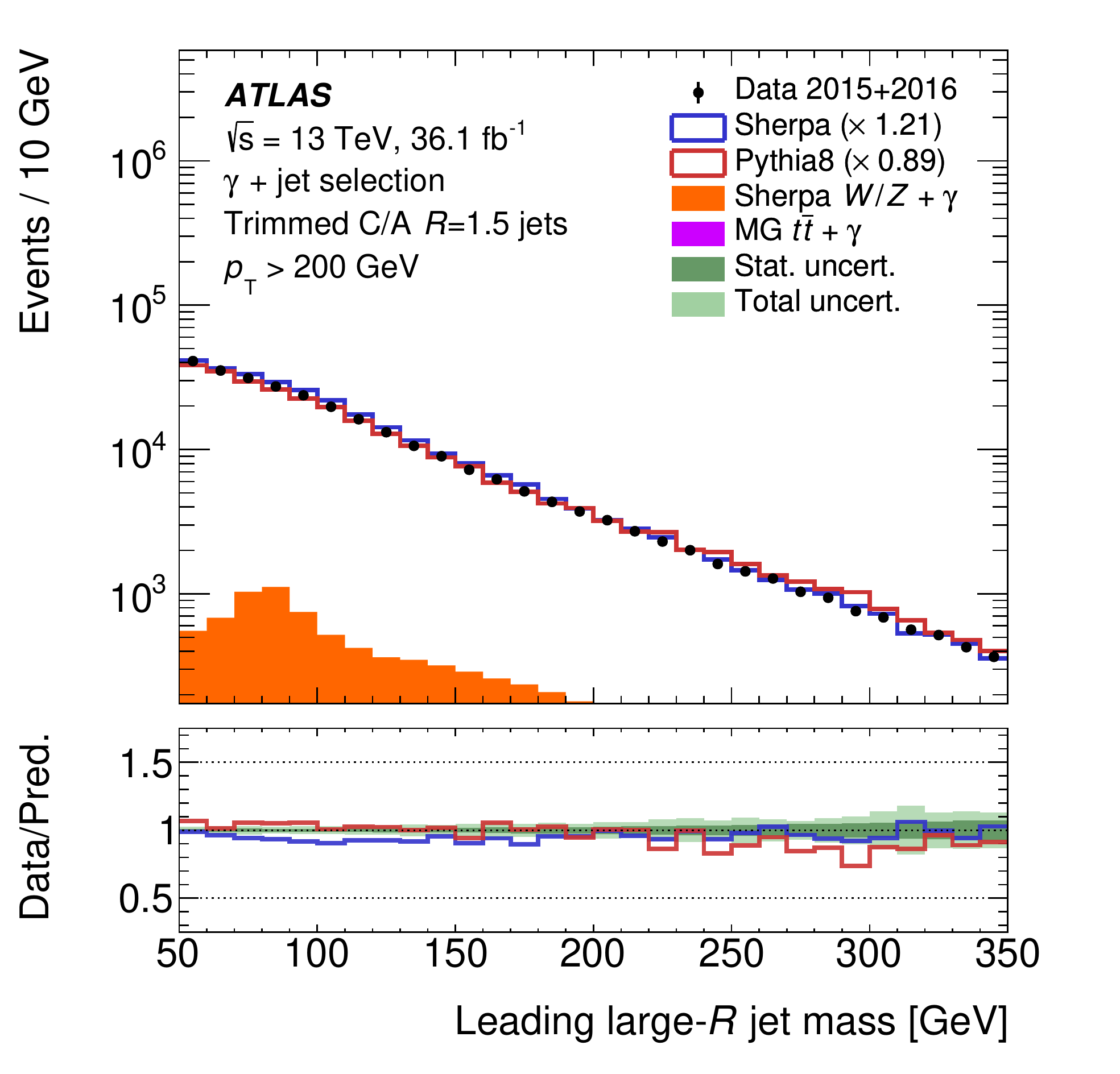}}
\caption{
\label{fig:DMC_bkg_mass_2} A comparison of the observed data and predicted MC
distributions of the mass of the leading \pt C/A $R=$1.5 trimmed jet in events
for the multijet \subref{fig:DMC_bkg_mass_2_a} and \gammajet
\subref{fig:DMC_bkg_mass_2_b} selections.  The data-driven normalisation
correction, described in Section~\ref{subsubsec:background_analysis}, is
shown in the legend beside the specific sample to which it applies.
Systematic uncertainties are indicated as a band in the lower panel and include all
experimental uncertainties related to the selection of events, as well as the
reconstruction and calibration of the \largeR jet.}
\end{centering}
\end{figure}

\begin{figure}[!h]
\begin{centering}
\subfigure[][]{    \label{fig:DMC_bkg_jss1_a} \includegraphics[width=0.45\textwidth]{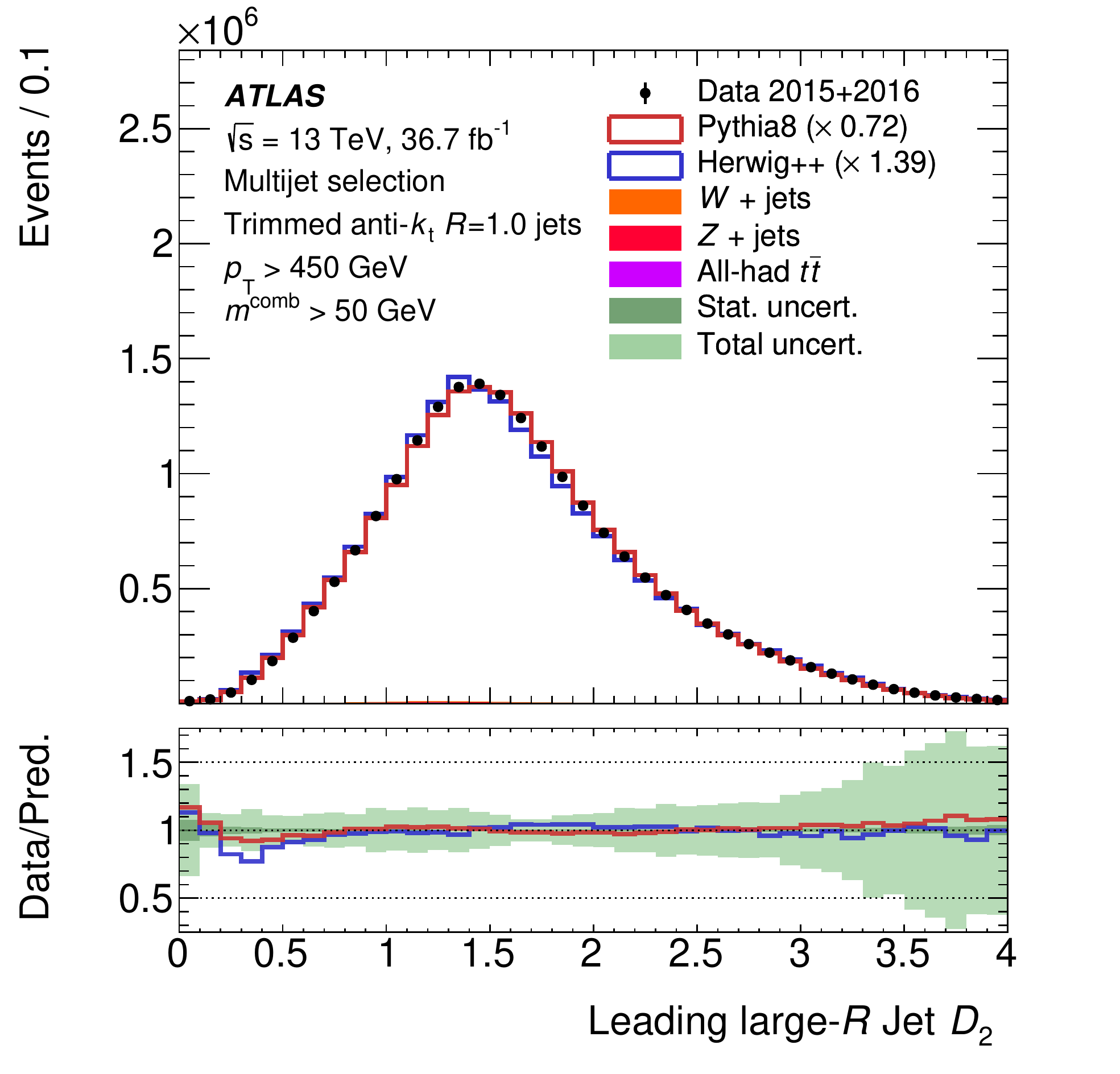}}
\subfigure[][]{   \label{fig:DMC_bkg_jss1_b} \includegraphics[width=0.45\textwidth]{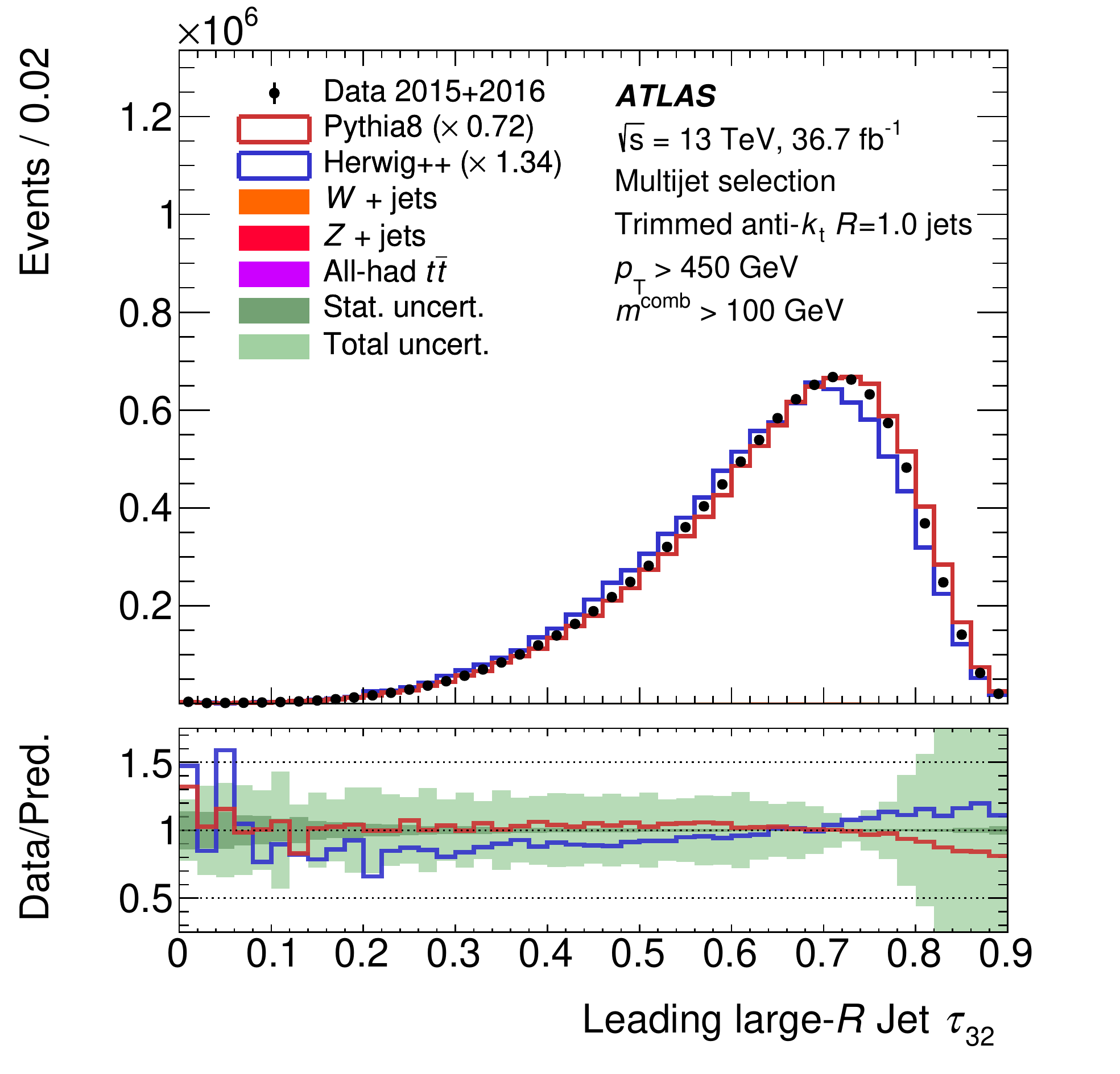}} \\
\subfigure[][]{    \label{fig:DMC_bkg_jss1_c} \includegraphics[width=0.45\textwidth]{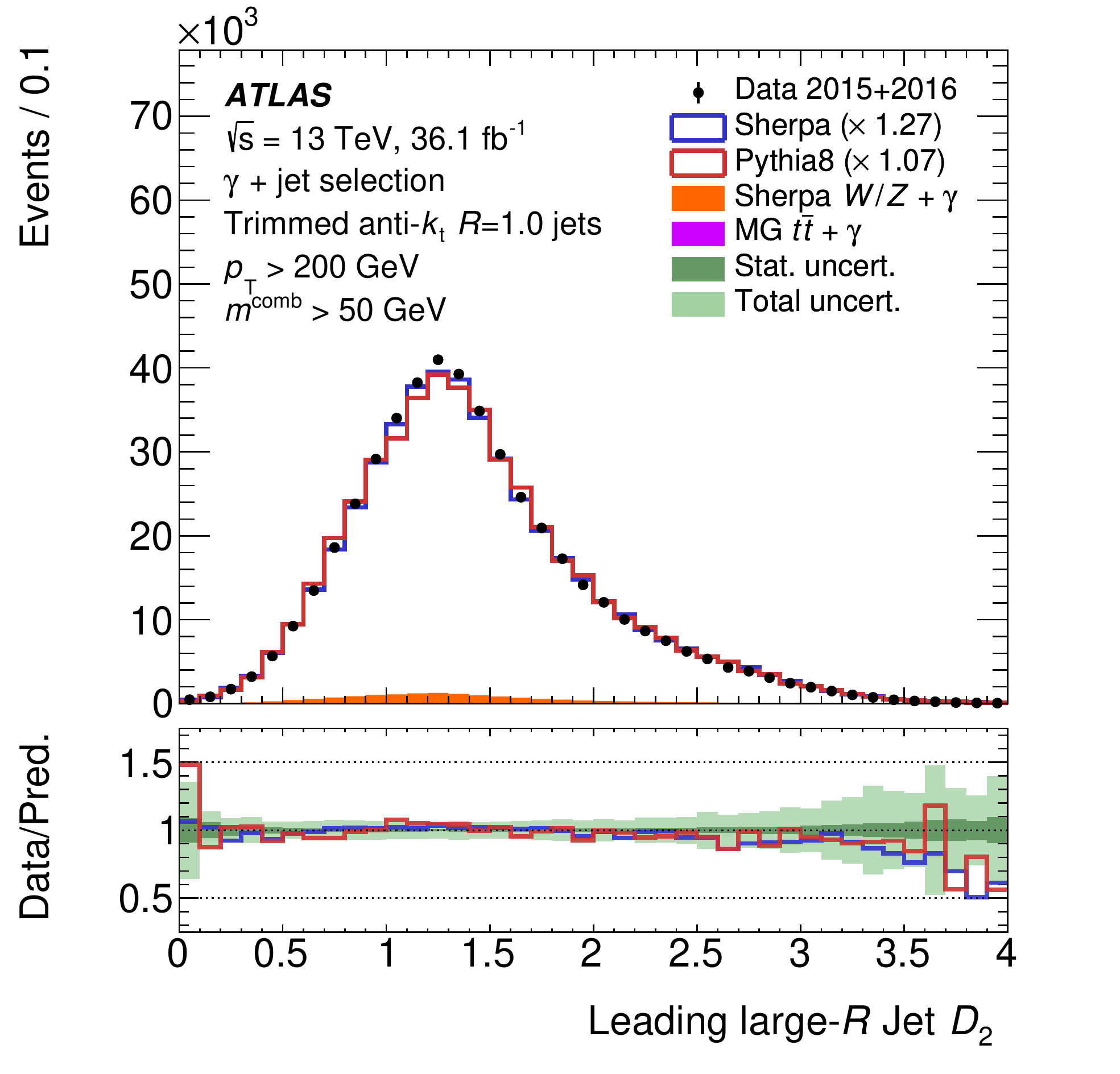}}
\subfigure[][]{   \label{fig:DMC_bkg_jss1_d} \includegraphics[width=0.45\textwidth]{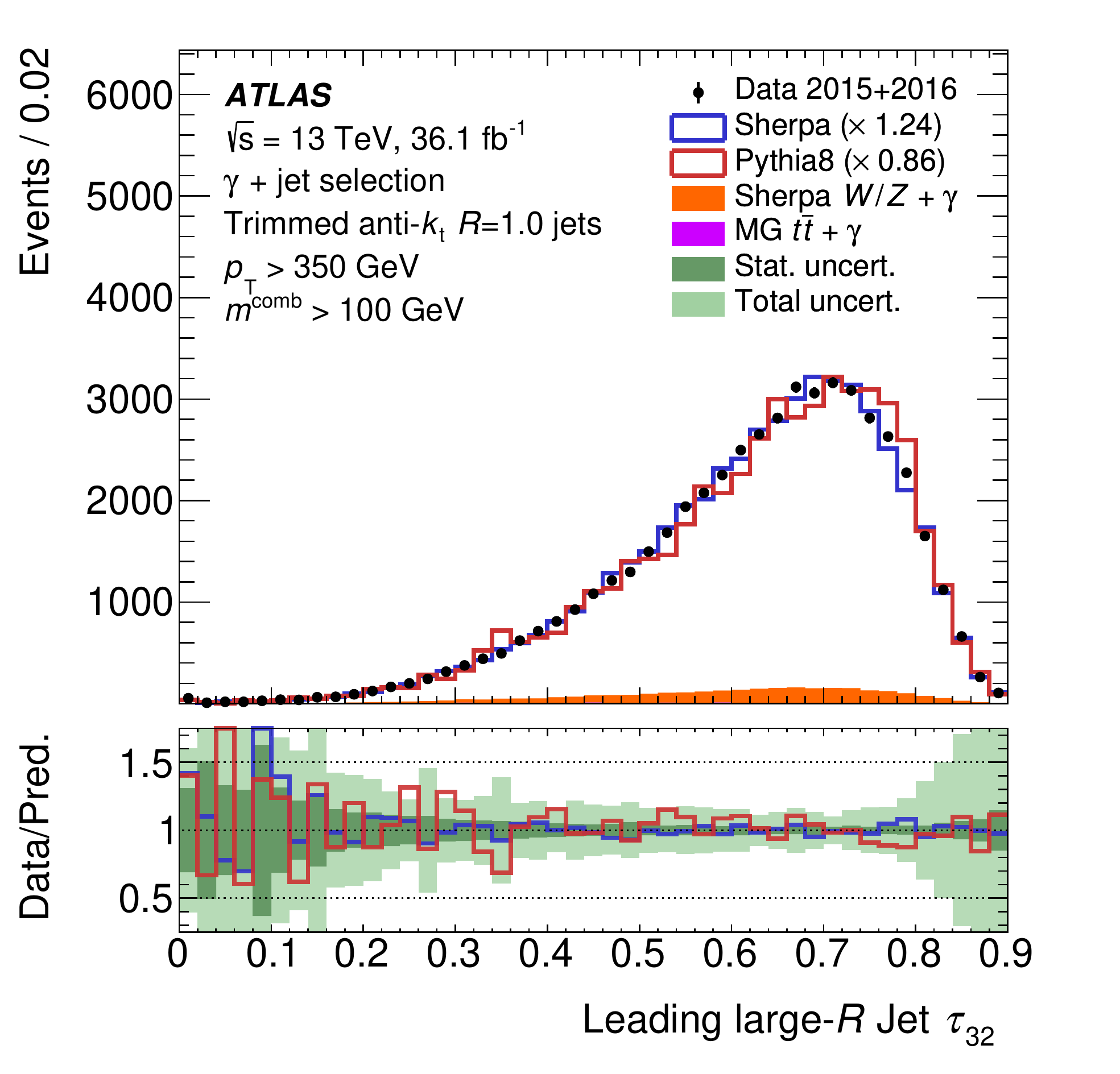}}
\caption{
\label{fig:DMC_bkg_jss1} A comparison of the observed data and MC predictions in
the multijet and \gammajet event samples for the \akt $R=$1.0 trimmed jet \DTwo
\subref{fig:DMC_bkg_jss1_a}\subref{fig:DMC_bkg_jss1_c} and \tauthrtwo
\subref{fig:DMC_bkg_jss1_b}\subref{fig:DMC_bkg_jss1_d} spectra.  The data-driven
normalisation correction, described in
Section~\ref{subsubsec:background_analysis}, is shown in the legend
beside the specific sample to which it applies.  Systematic uncertainties are
indicated as a band in the lower panel and include all experimental uncertainties related to
the selection of events, as well as the reconstruction and calibration of the
\largeR jet.}
\end{centering}
\end{figure}
 
\begin{figure}[!h]
\begin{centering}
\subfigure[][]{    \label{fig:DMC_bkg_jss2_a} \includegraphics[width=0.45\textwidth]{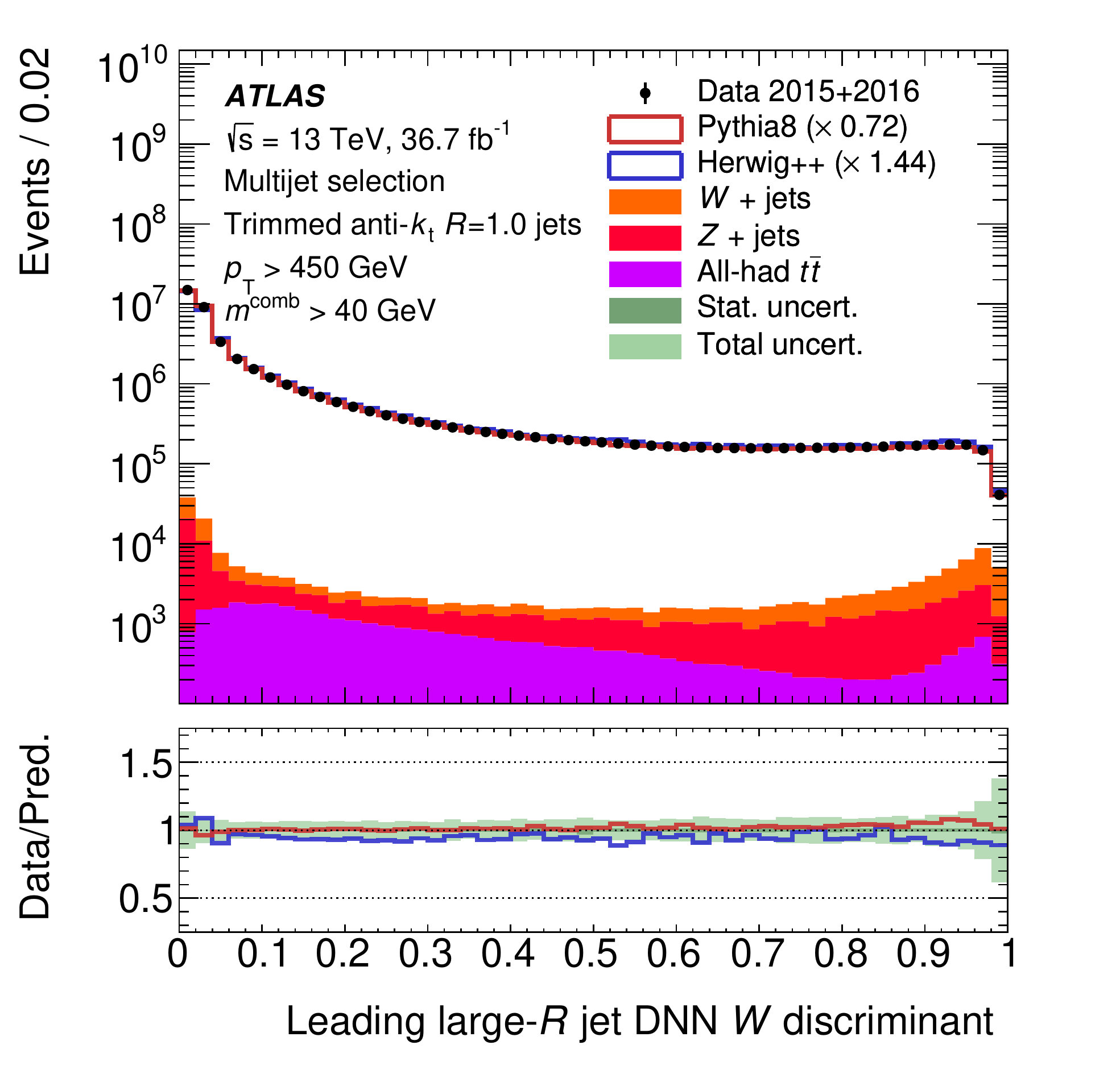}}
\subfigure[][]{   \label{fig:DMC_bkg_jss2_b} \includegraphics[width=0.45\textwidth]{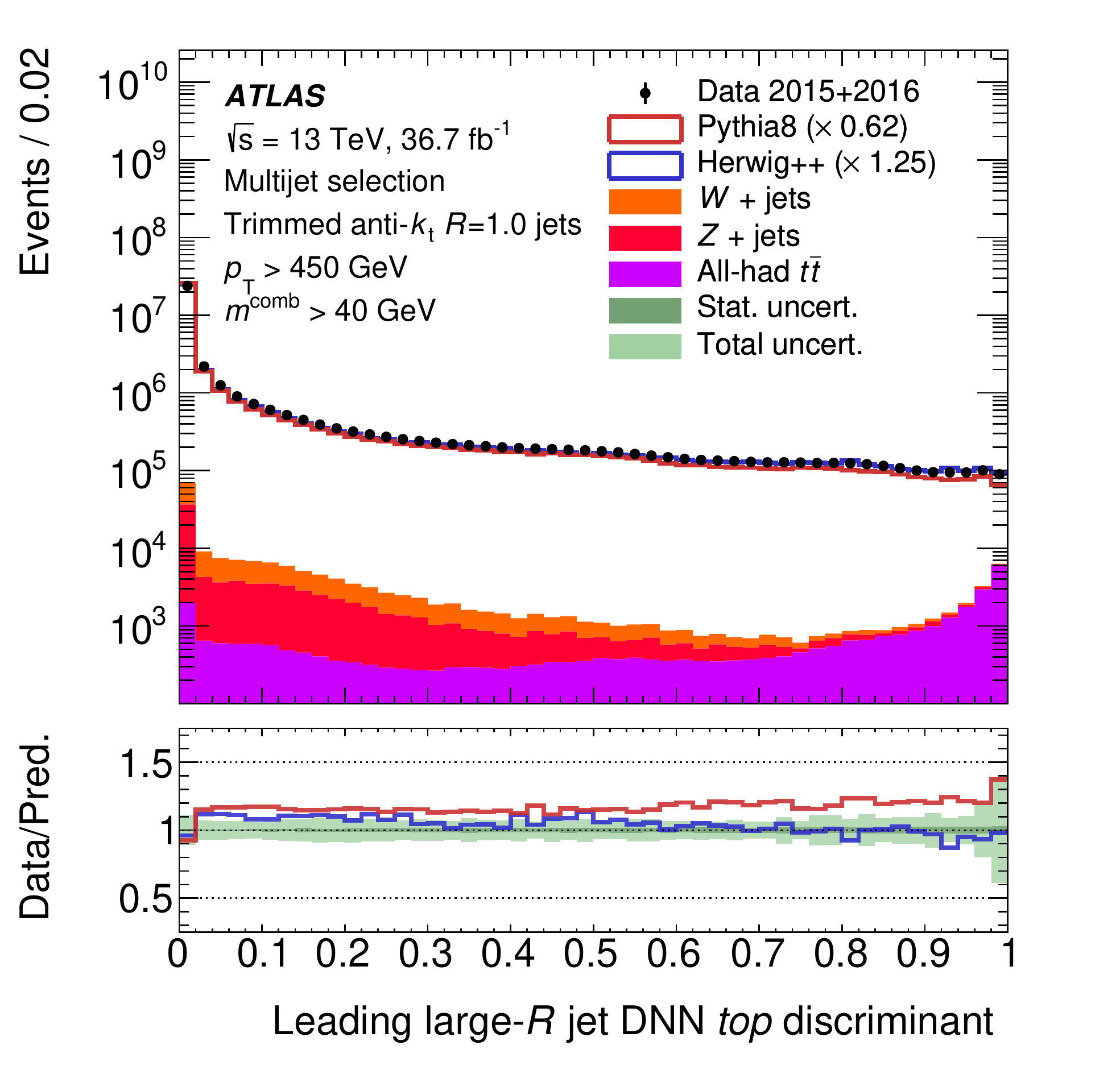}} \\
\subfigure[][]{    \label{fig:DMC_bkg_jss2_c} \includegraphics[width=0.45\textwidth]{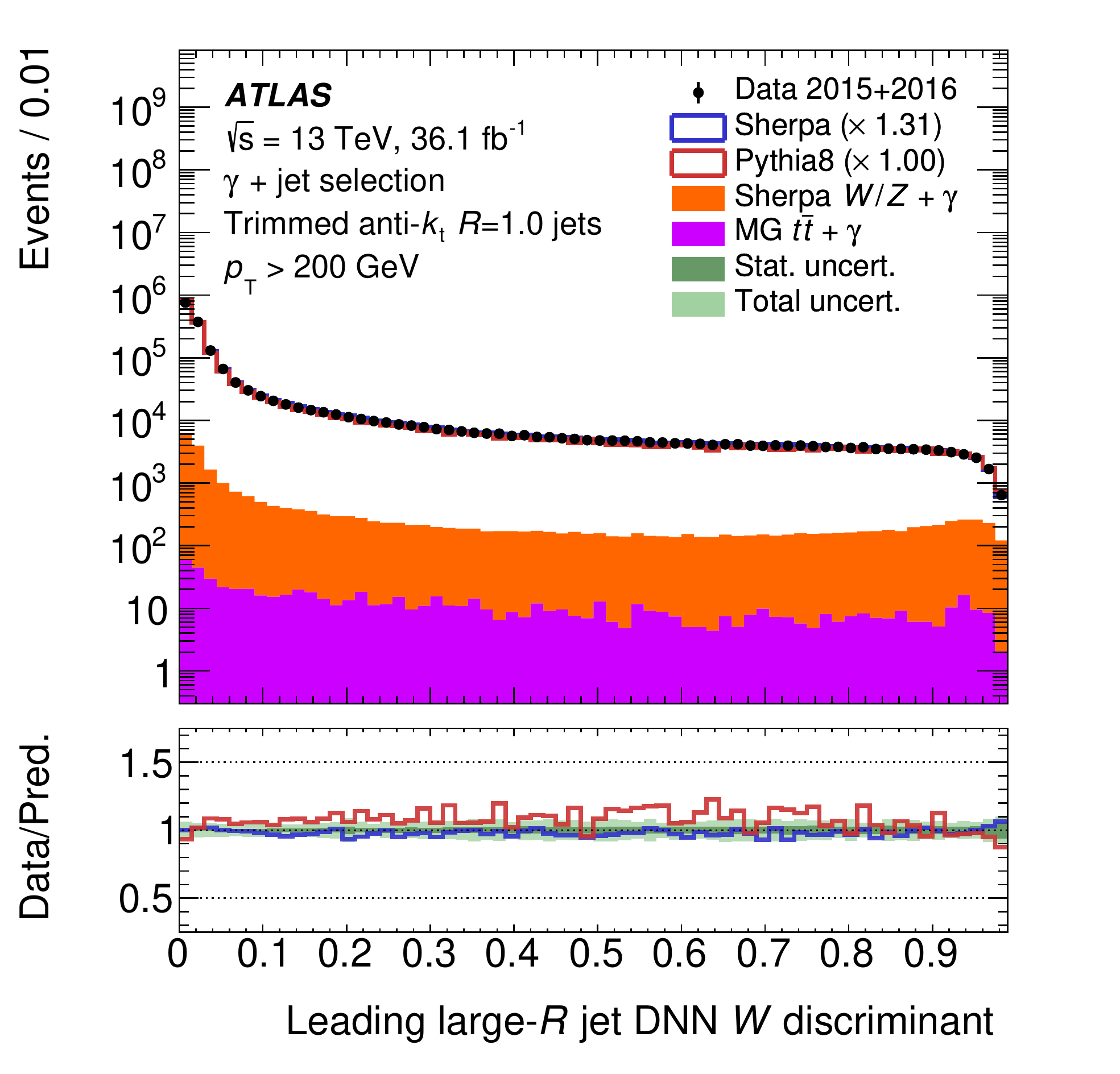}}
\subfigure[][]{   \label{fig:DMC_bkg_jss2_d} \includegraphics[width=0.45\textwidth]{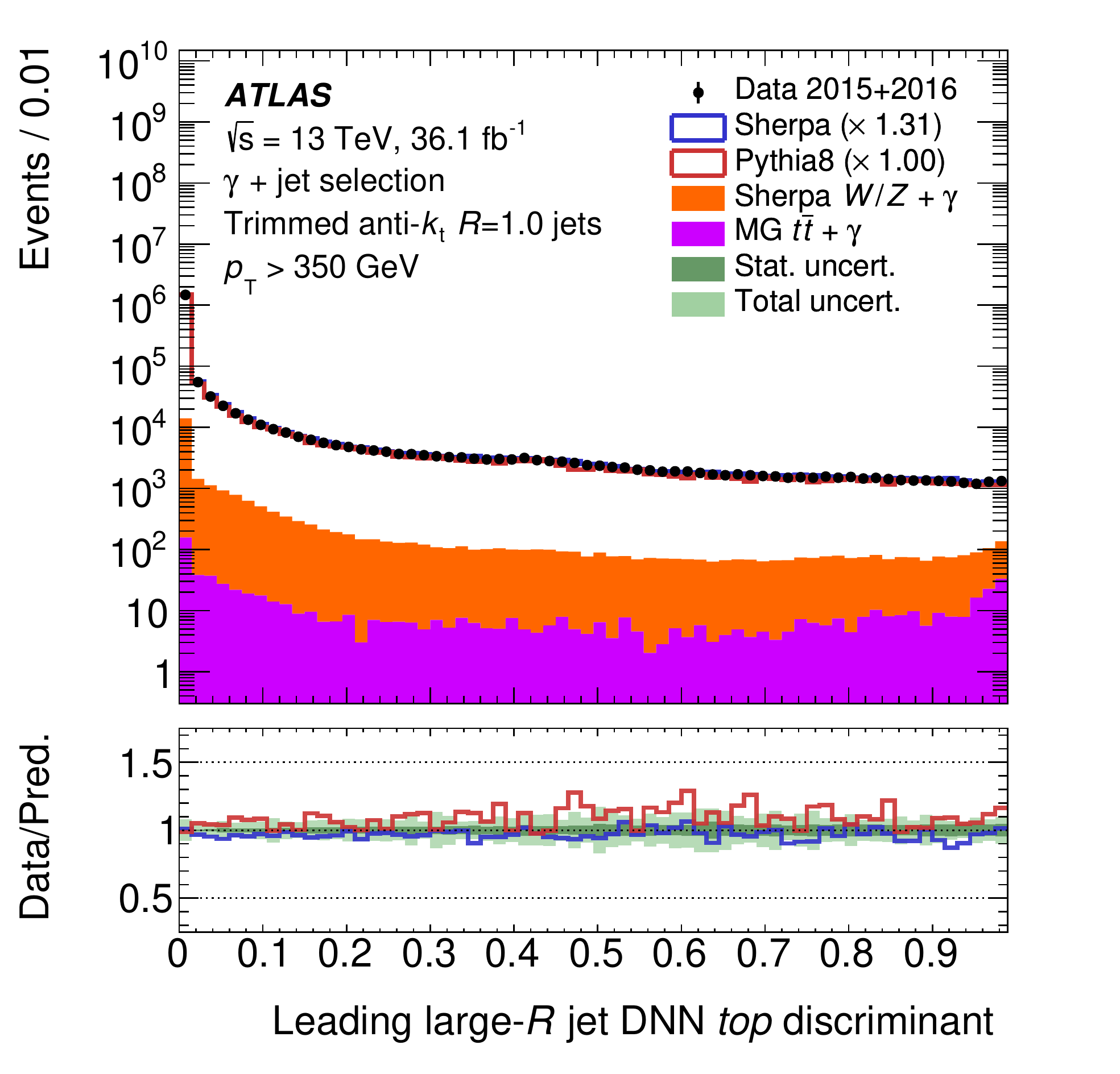}}
\caption{
A comparison of the observed data and MC predictions in
the multijet and \gammajet event samples for the \akt $R=$1.0 trimmed jet
spectra of the \Wboson-boson
\subref{fig:DMC_bkg_jss2_a}\subref{fig:DMC_bkg_jss2_c} and top-quark
\subref{fig:DMC_bkg_jss2_b}\subref{fig:DMC_bkg_jss2_d} DNN discriminants.  The
data-driven normalisation correction, described in
Section~\ref{subsubsec:background_analysis}, is shown in the legend
beside the specific sample to which it applies.  Systematic uncertainties are
indicated as a band in the lower panel and include all experimental uncertainties related to
the selection of events, as well as the reconstruction and calibration of the
\largeR jet.}
\end{centering}
\end{figure}

\begin{figure}[!h]
\begin{centering}
\subfigure[][]{    \label{fig:DMC_bkg_jss3_a} \includegraphics[width=0.45\textwidth]{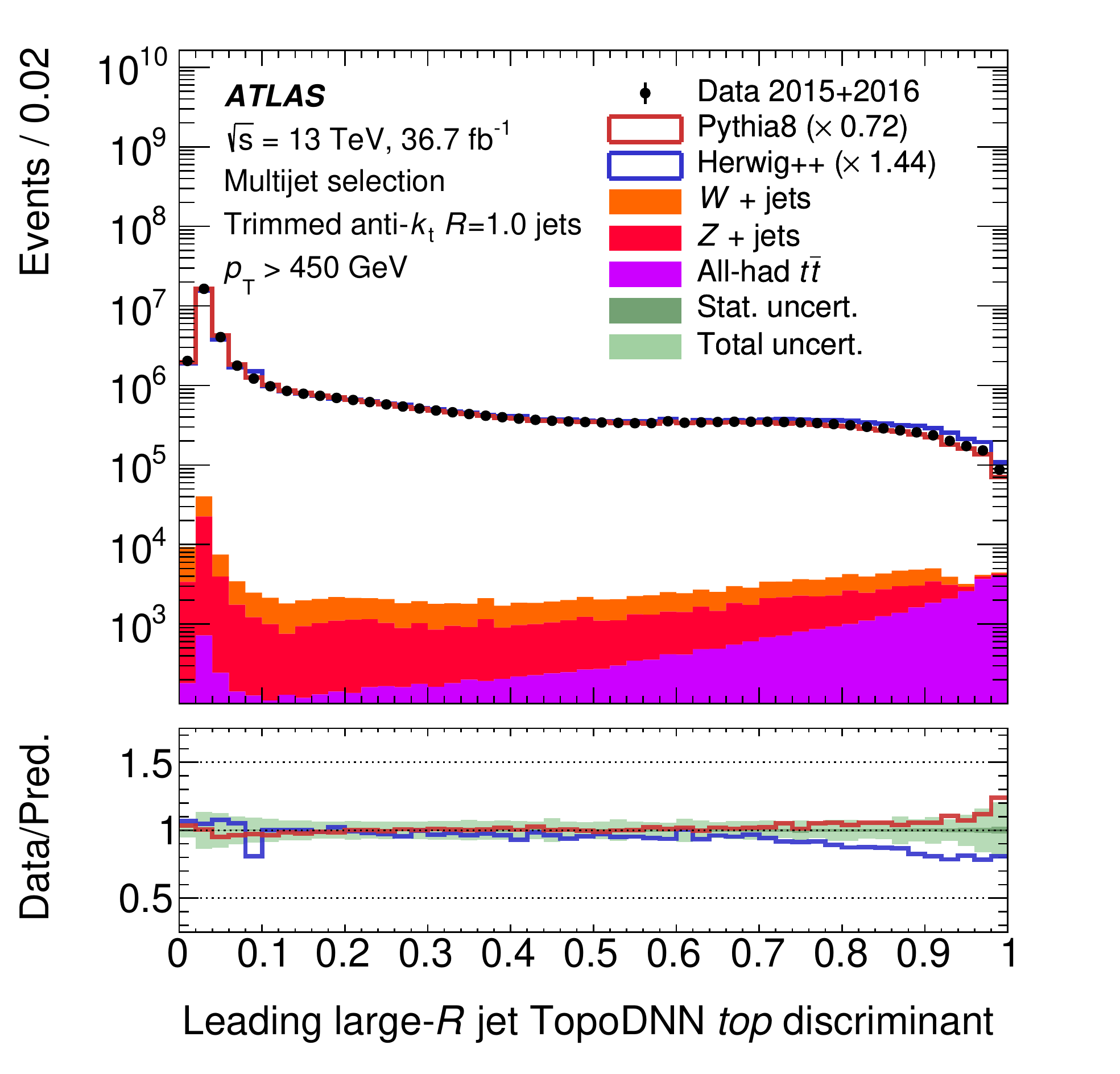}}
\subfigure[][]{   \label{fig:DMC_bkg_jss3_b} \includegraphics[width=0.45\textwidth]{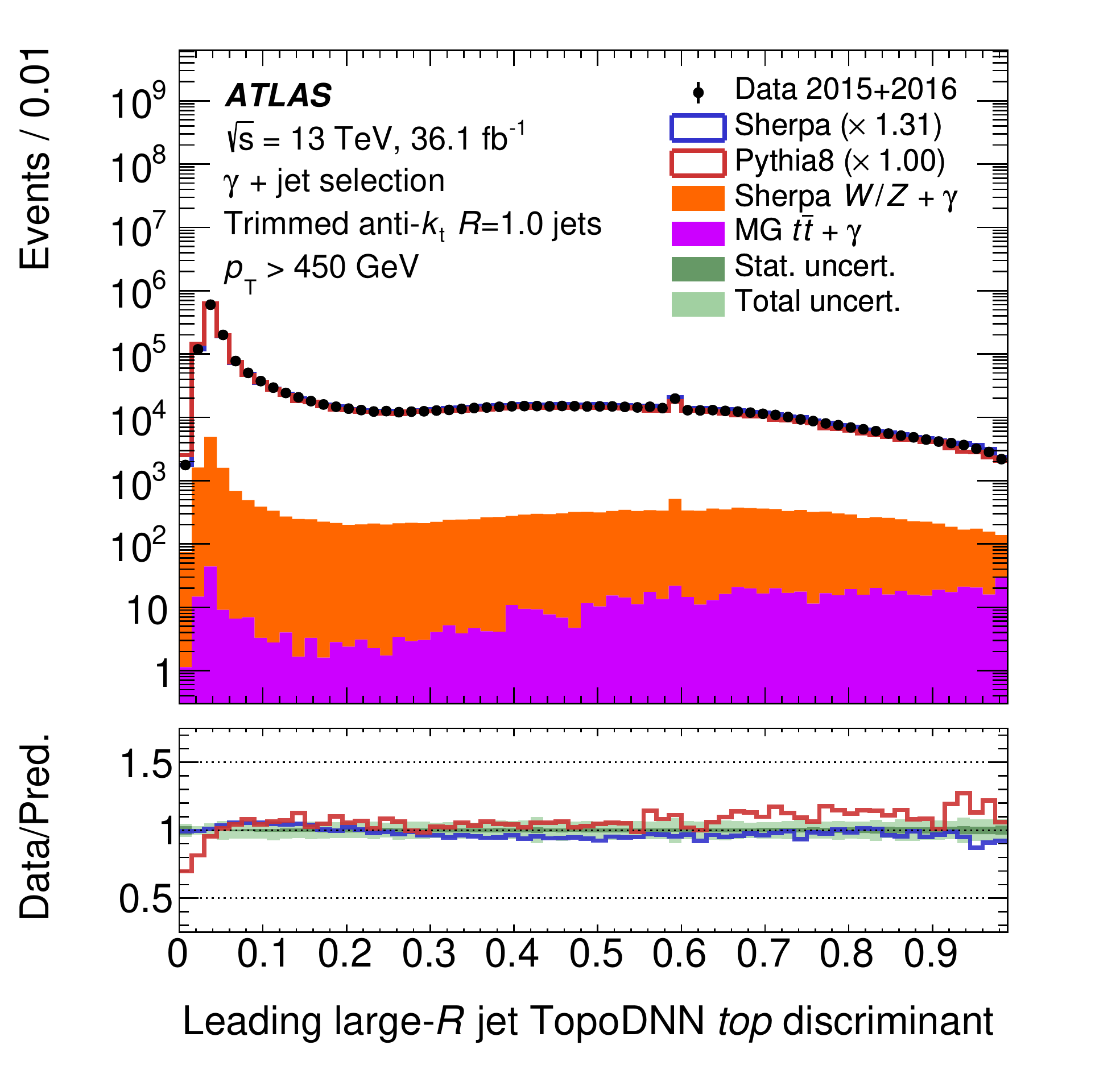}}
\caption{
A comparison of the observed data and MC predictions in
the multijet \subref{fig:DMC_bkg_jss3_a} and \gammajet
\subref{fig:DMC_bkg_jss3_b} event samples for the \akt $R=$1.0 trimmed jet
spectra of the TopoDNN top tagger discriminant.  The data-driven
normalisation correction, described in
Section~\ref{subsubsec:background_analysis}, is shown in the legend
beside the specific sample to which it applies.  Systematic uncertainties are
indicated as a band in the lower panel and include all experimental uncertainties related to
the selection of events, as well as the reconstruction and calibration of the
\largeR jet.}
\end{centering}
\end{figure}
 
\begin{figure}[!h]
\begin{centering}
\subfigure[][]{    \label{fig:DMC_bkg_jss4_a} \includegraphics[width=0.45\textwidth]{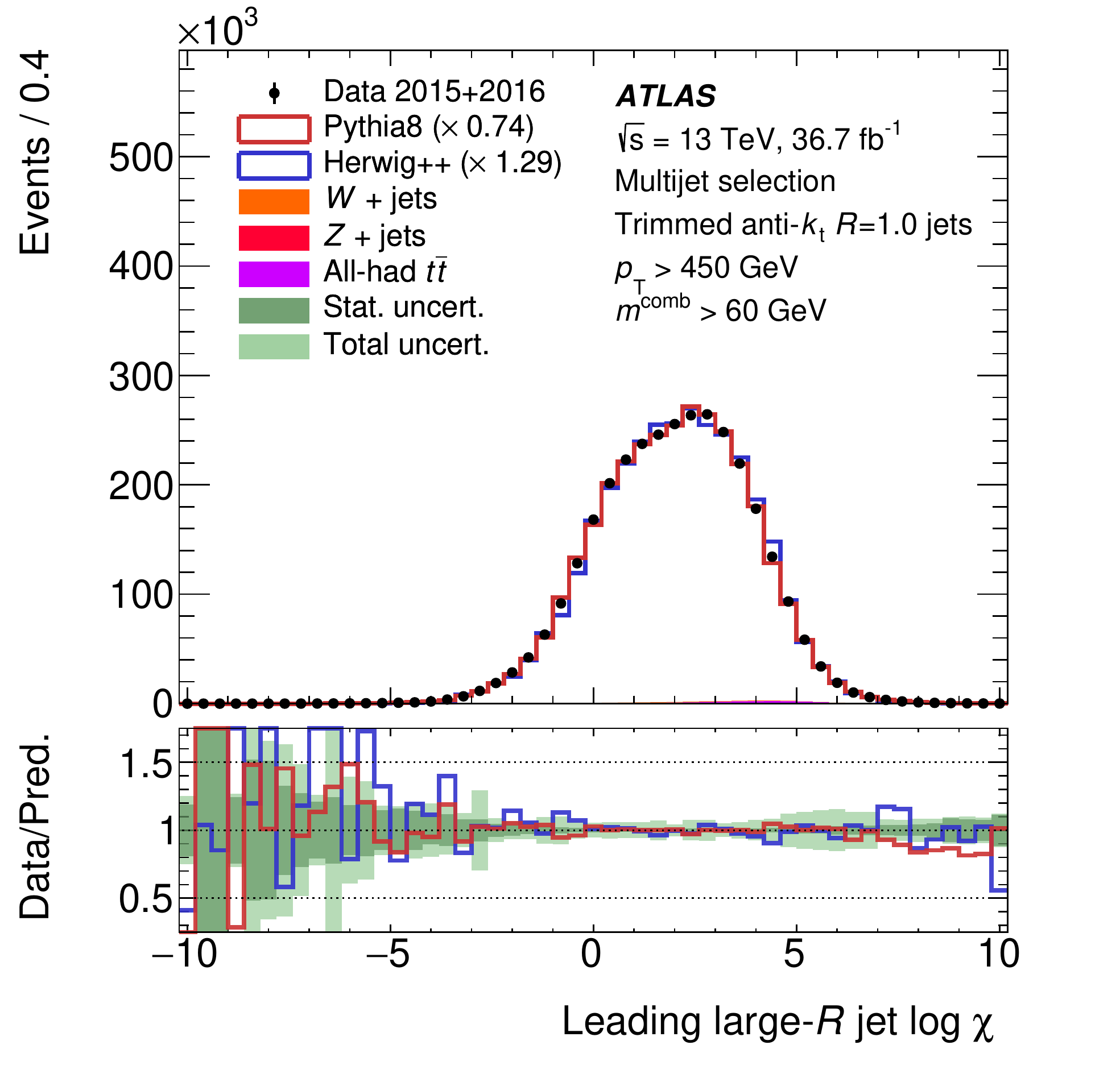}}
\subfigure[][]{   \label{fig:DMC_bkg_jss4_b} \includegraphics[width=0.45\textwidth]{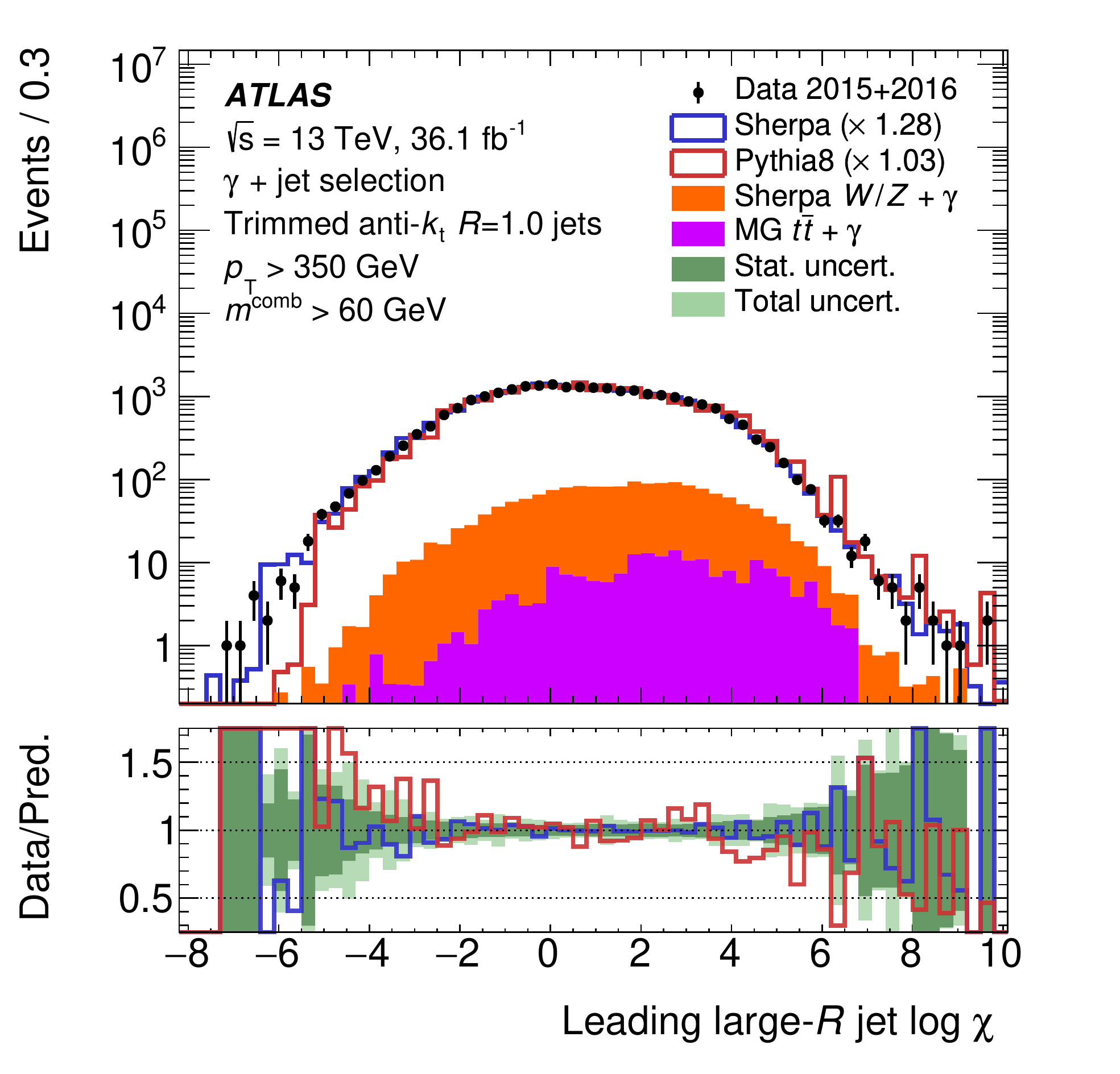}}
\caption{
A comparison of the observed data and MC predictions in
the multijet \subref{fig:DMC_bkg_jss4_a} and \gammajet
\subref{fig:DMC_bkg_jss4_b} event samples for the \akt $R=$1.0 trimmed jet
spectra of the $\log\chi$ shower deconstruction discriminant.  The data-driven
normalisation correction, described in
Section~\ref{subsubsec:background_analysis}, is shown in the legend
beside the specific sample to which it applies.  Systematic uncertainties are
indicated as a band in the lower panel and include all experimental uncertainties related to
the selection of events, as well as the reconstruction and calibration of the
\largeR jet.}
\end{centering}
\end{figure}
 
\begin{figure}[!h]
\begin{centering}
\subfigure[][]{    \label{fig:DMC_bkg_jss5_a} \includegraphics[width=0.45\textwidth]{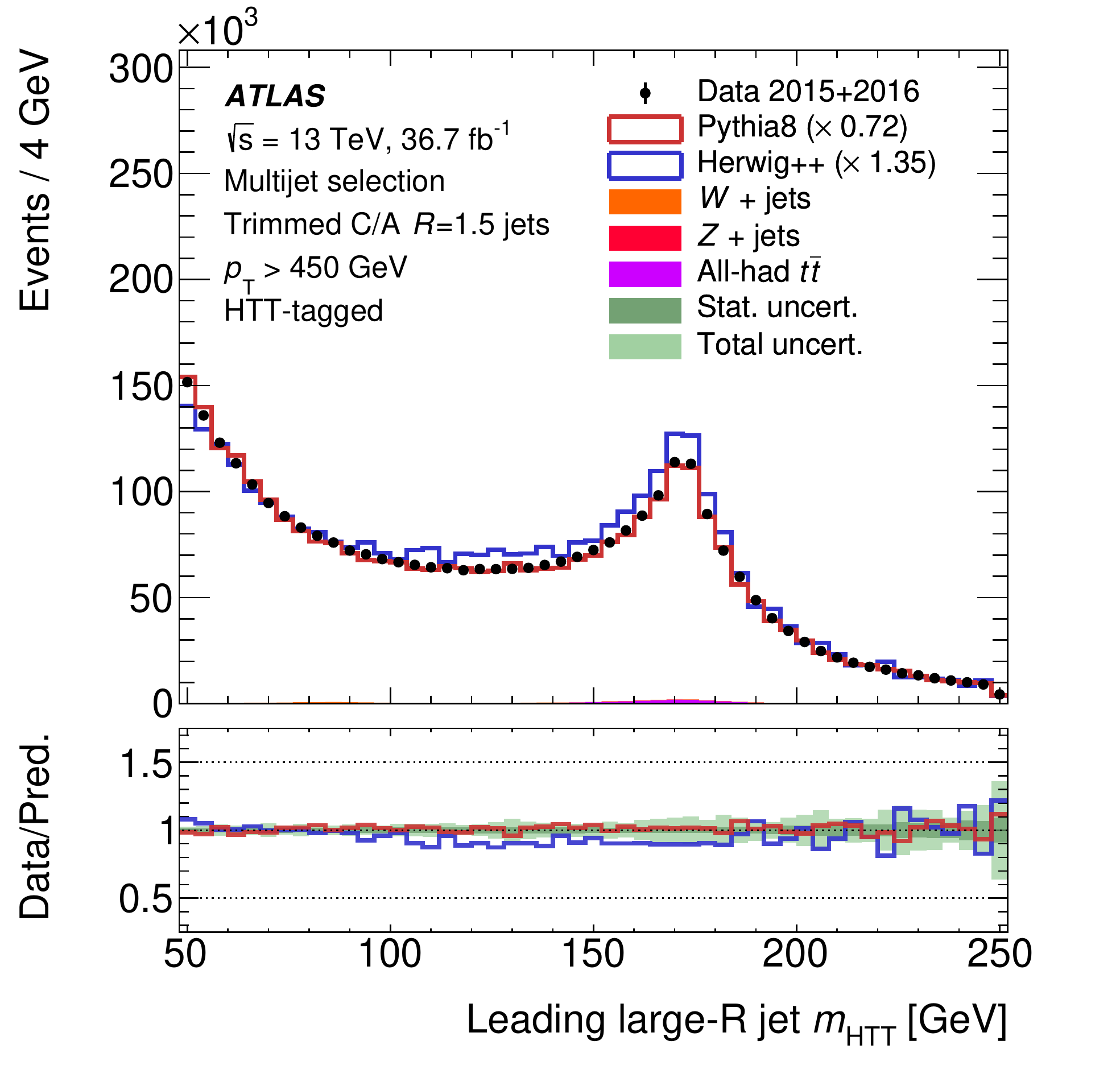}}
\subfigure[][]{   \label{fig:DMC_bkg_jss5_b} \includegraphics[width=0.45\textwidth]{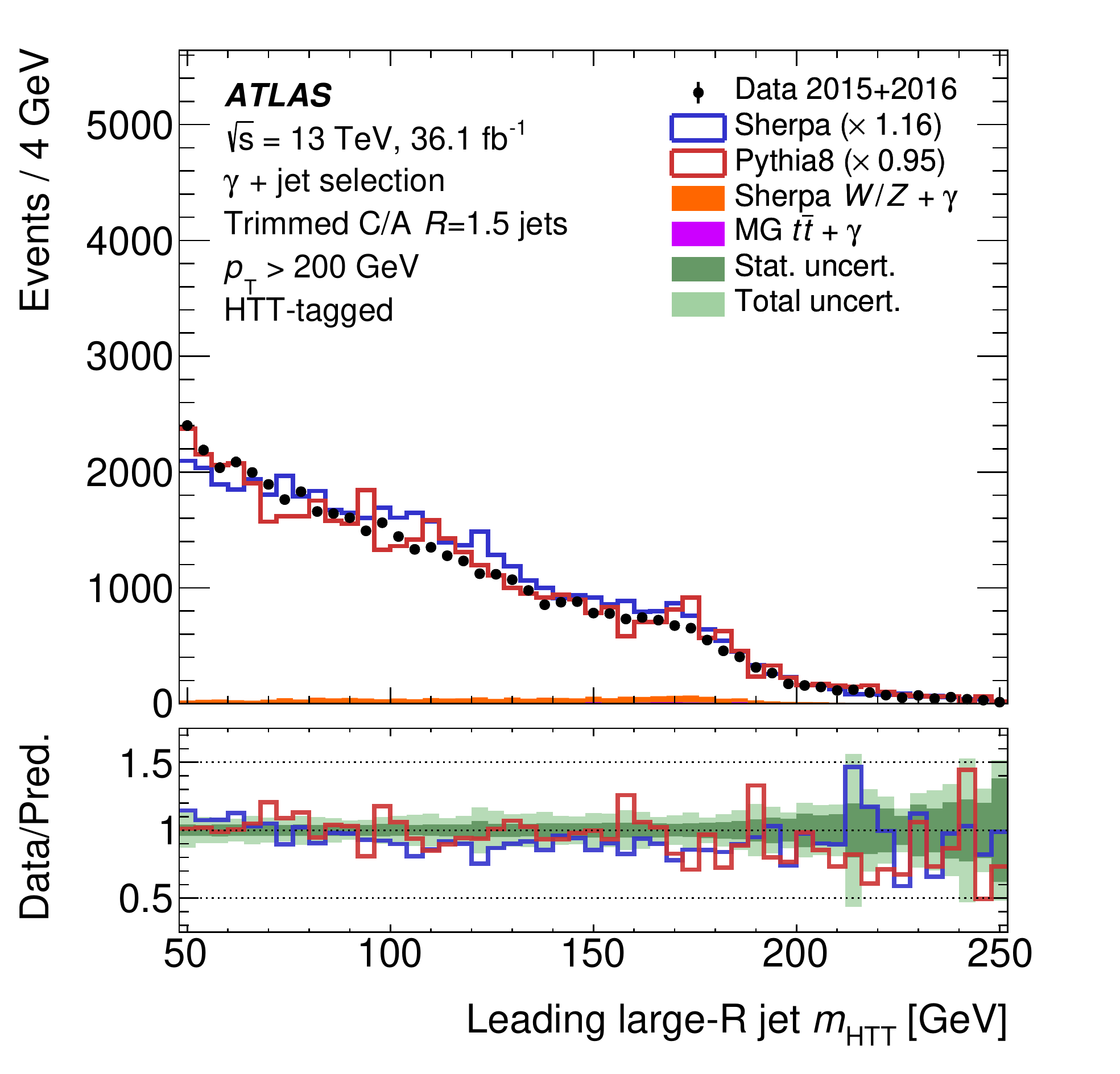}}
\caption{
\label{fig:DMC_bkg_jss5} A comparison of the observed data and predicted MC
distributions in the multijet \subref{fig:DMC_bkg_jss5_a} and \gammajet
\subref{fig:DMC_bkg_jss5_b} event samples for the \htt mass.  The data-driven
normalisation correction, described in
Section~\ref{subsubsec:background_analysis}, is shown in the legend beside the
specific sample to which it applies.  Systematic uncertainties are indicated as a band in
the lower panel and include all experimental uncertainties related to the
selection of events, as well as the reconstruction and calibration of the
\largeR jet. The difference in the shape of the \htt mass distribution between
the multijet and the \gammajet selections, in particular the absence of a
pronounced top-mass peak in the \gammajet selection, is caused by the
difference in the jet \pt thresholds.}
\end{centering}
\end{figure}

\begin{figure}[!h]
\begin{centering}
\subfigure[][]{    \label{fig:DMC_bkg_tagmass_1_a} \includegraphics[width=0.45\textwidth]{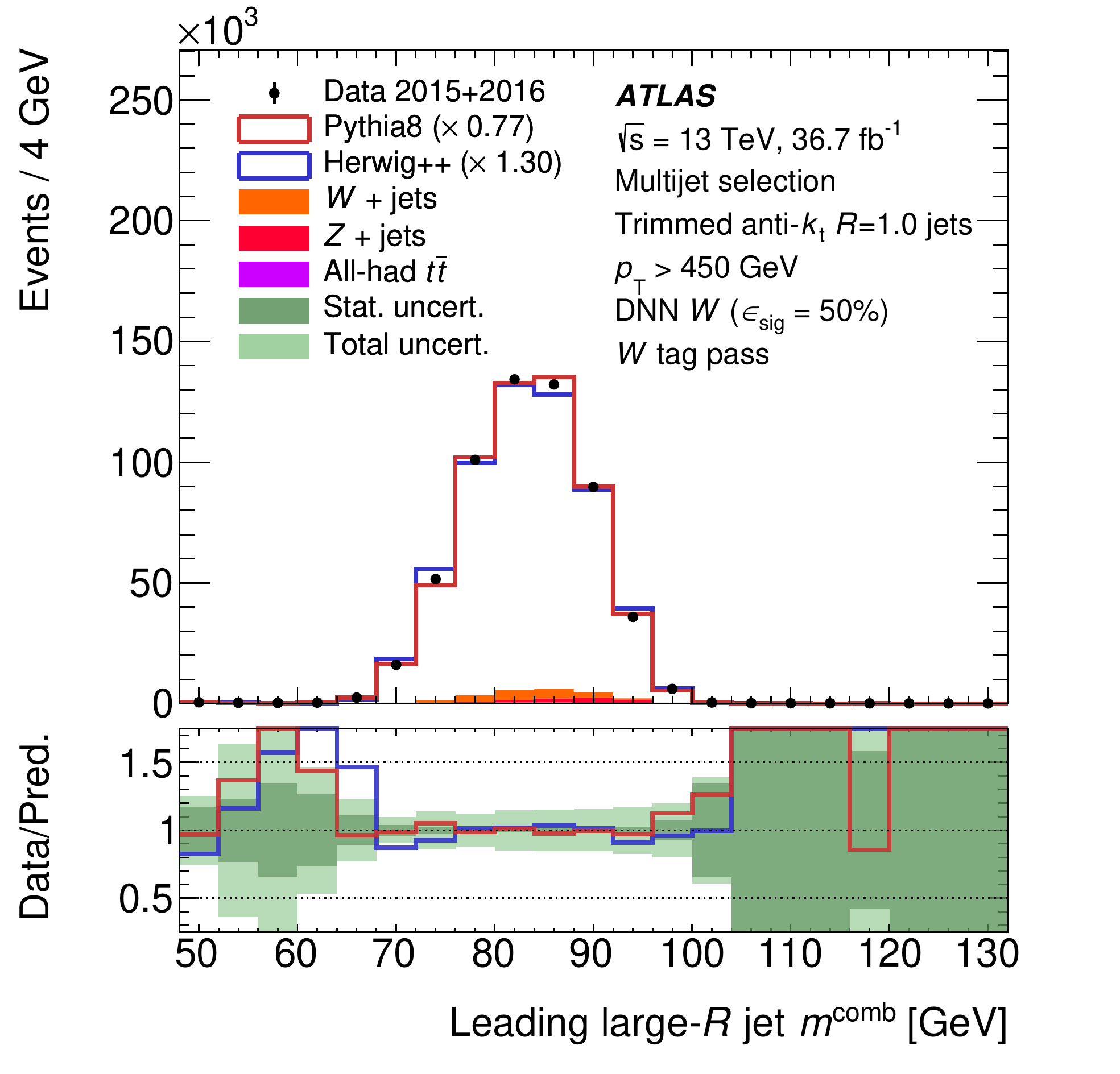}}
\subfigure[][]{   \label{fig:DMC_bkg_tagmass_1_b} \includegraphics[width=0.45\textwidth]{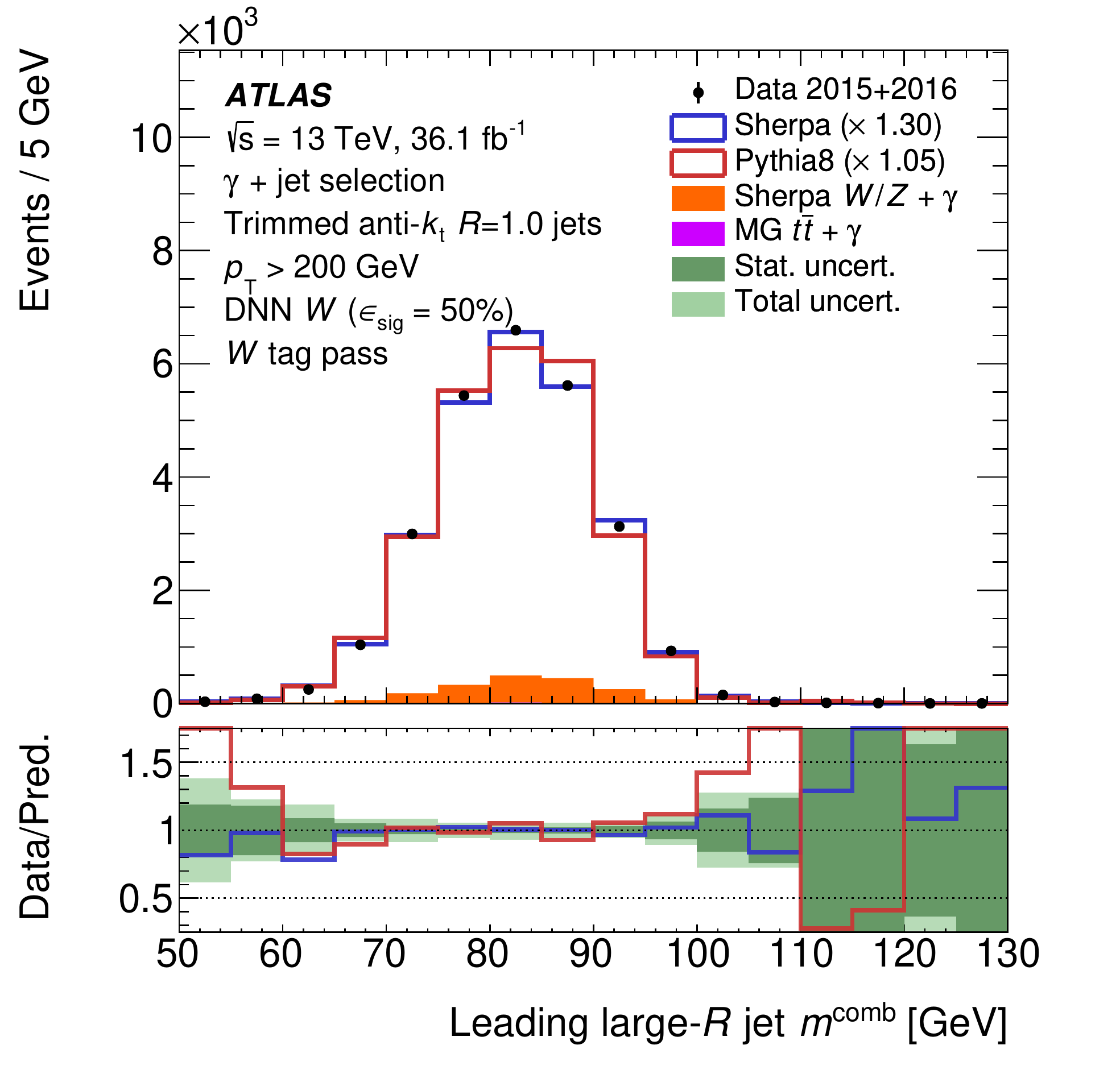}}
\caption{
\label{fig:DMC_bkg_tagmass_1} A comparison of the observed data and predicted MC
distributions of the \akt $R=$1.0 trimmed jet \mcomb observable for events from
the multijet \subref{fig:DMC_bkg_tagmass_1_a} or \gammajet
\subref{fig:DMC_bkg_tagmass_1_b} selections that pass the selection on the
jet-shape-based \Wboson-boson DNN tagger.  The data-driven normalisation correction,
described in Section~\ref{subsubsec:background_analysis}, is shown
in the legend beside the specific sample to which it applies. Systematic
uncertainties are indicated as a band in the lower panel and include all experimental
uncertainties related to the selection of events, as well as the reconstruction
and calibration of the \largeR jet.}
\end{centering}
\end{figure}
 
\begin{figure}[!h]
\begin{centering}
\subfigure[][]{    \label{fig:DMC_bkg_tagmass_2_a} \includegraphics[width=0.45\textwidth]{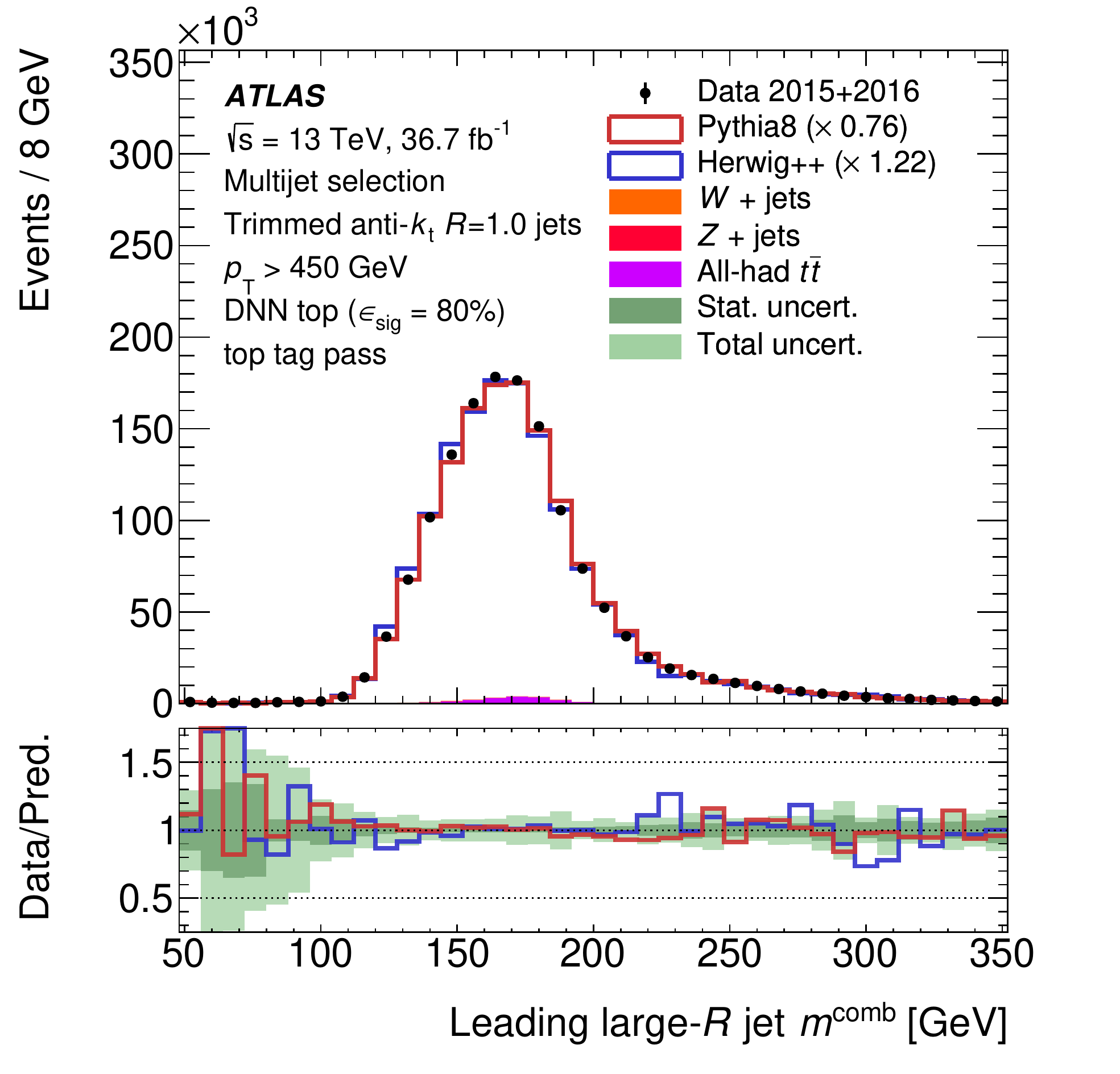}}
\subfigure[][]{   \label{fig:DMC_bkg_tagmass_2_b} \includegraphics[width=0.45\textwidth]{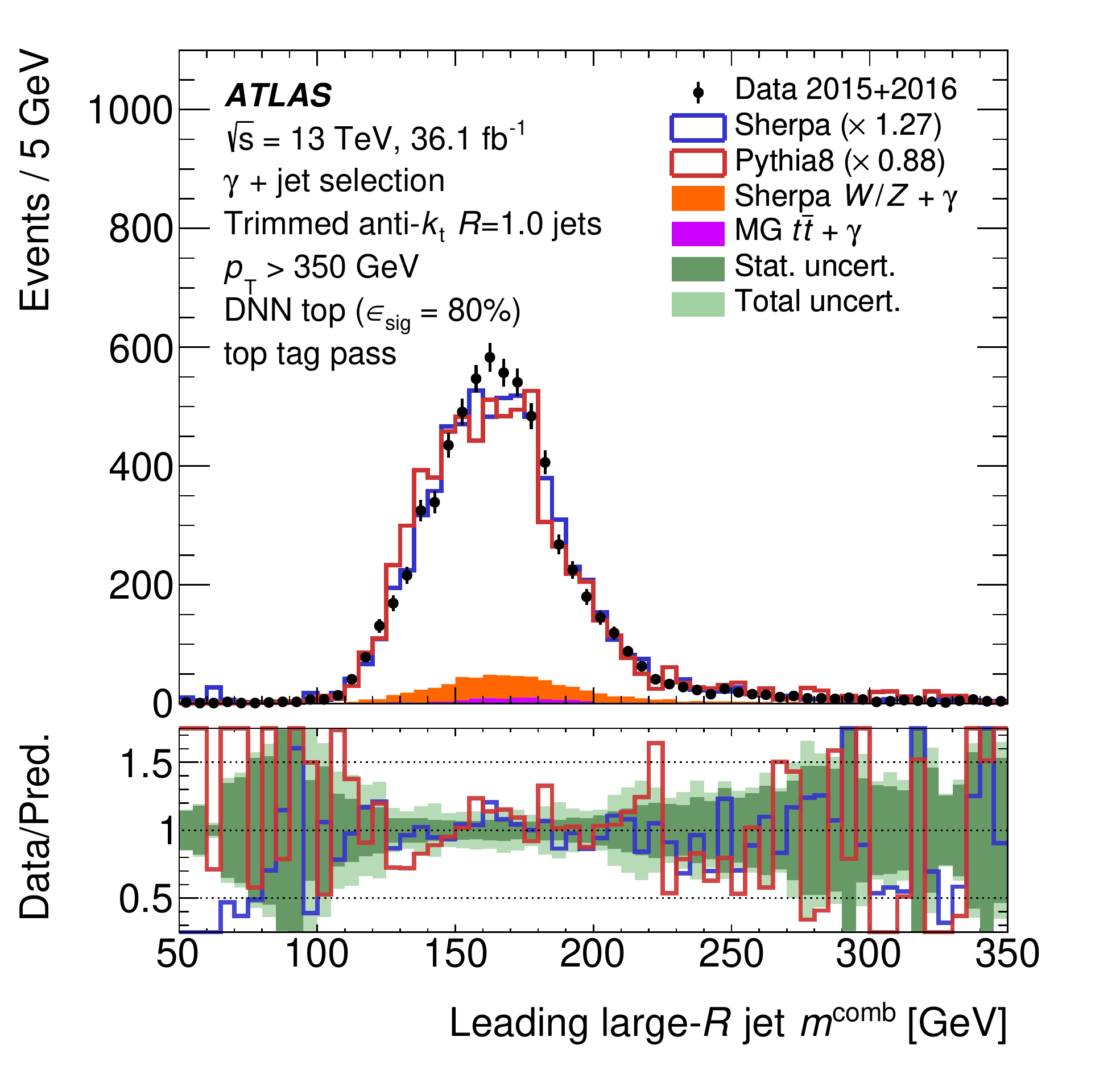}}
\caption{
\label{fig:DMC_bkg_tagmass_2} A comparison of the observed data and predicted MC
distributions of the \akt $R=$1.0 trimmed jet \mcomb observable for events from
the multijet \subref{fig:DMC_bkg_tagmass_2_a} or \gammajet
\subref{fig:DMC_bkg_tagmass_2_b} selections that pass the selection on the
jet-shape-based top quark DNN tagger.  The data-driven normalisation correction,
described in Section~\ref{subsubsec:background_analysis}, is shown
in the legend beside the specific sample to which it applies. Systematic
uncertainties are indicated as a band in the lower panel and include all experimental
uncertainties related to the selection of events, as well as the reconstruction
and calibration of the \largeR jet.}
\end{centering}
\end{figure}
\clearpage
\newpage
 
\subsubsection{Background rejection measurements}
\label{sec:measured_background_rejection}
In a similar manner to the measurement of the signal efficiency in
Section~\ref{sec:measured_signal_efficiency}, the background rejection
$1/\epsilon_{\mathrm{bkg}}$ is measured for the \Wbosonboson and \topquark
tagging working points described in
Section~\ref{sec:measured_signal_efficiency}.  This measurement is performed
in both the multijet and \gammajet topologies as a function of the transverse
momentum of the highest-\pt jet in the event, taken to be the leading jet
studied in Section~\ref{subsubsec:background_analysis}, as well as $\mu$.
 
The approach in this measurement is simpler than the chi-square fit
approach used in Section~\ref{sec:measured_signal_efficiency} due to the purity
of these samples.  In particular, after subtracting the signal contamination from
data and performing the normalisation of the multijet and \gammajet samples in
the inclusive selection described in
Section~\ref{subsubsec:background_analysis}, the background efficiency is
calculated directly as the fraction of events that satisfy the full set of tagging
criteria in data and in Monte Carlo simulation.  The results are shown in
Figures~\ref{fig:DMC_bkg_rejection_1}--\ref{fig:DMC_bkg_rejection_7}, for the full
set of tagging techniques.  In the case of \Wbosonboson tagging
(Figures~\ref{fig:DMC_bkg_rejection_1} and~\ref{fig:DMC_bkg_rejection_2}), the
dependence of the background rejection on jet \pt arises from the requirement of
a fixed signal efficiency.  At low jet \pt, there is a non-negligible fraction
of signal \Wbosonboson jets which are not sufficiently collimated due to
radiation from parton shower outside of the jet area despite the signal
labelling requirement on the \DeltaR between the quarks from \Wbosonboson decay
and the jet axis. As a result, a broader jet-mass selection is required to
maintain the 50\% signal
efficiency.  As the jet \pt increases, the sample of signal jets becomes better
contained within a radius of 1.0, thereby allowing a stricter mass
requirement, and the background rejection increases.  However, the \Wbosonboson
signal jets become fully contained at  \pt$\sim$ 800~\GeV, and with increasing
jet \pt the experimental resolution worsens and the Sudakov peak of the light-jet
mass migrates into the signal region, thereby leading to a degradation of
the background rejection.
 
Good agreement is generally observed between the predicted and measured
rejections.  For the multijet topology, the \textsc{Pythia8} prediction of the
background rejection describes the observed one, while the \textsc{Herwig++}
prediction is lower than the rejection in data.  Although the rejections for the
two topologies are similar, there are relatively large uncertainties at higher
jet \pT, with clear differences observed between the generators examined for the
dominant samples in each topology.  In particular, in the case of \Wbosonboson
tagging, it is observed that these generator differences are larger for the more
complex jet-shape DNN tagger, shown in Figure~\ref{fig:DMC_bkg_rejection_2},
than for the cut-based tagger, shown in Figure~\ref{fig:DMC_bkg_rejection_1}.
In the case of \topquark tagging, in
addition to the trend between the jet-shape DNN and cut-based taggers,
a similar trend can be seen in which a more algorithmically
involved classifier, namely the TopoDNN tagger, shown in
Figure~\ref{fig:DMC_bkg_rejection_5}, shows larger differences between
generators than the jet-shape DNN tagger, shown in
Figure~\ref{fig:DMC_bkg_rejection_4}.
 
When examining the background rejection with respect to $\mu$,
in the case of \Wbosonboson tagging, a trend of increasing background
rejection for higher $\mu$ exists.  This is
observed in both the multijet and \gammajet topologies and found to be the
same size for both the \wsimple \Wbosonboson tagger and the jet \DNNlong \Wbosonboson
tagger.  In the case of \topquark tagging, the \topsimple
\topquark tagger, the jet \DNNlong \topquark tagger, and the TopoDNN tagger show no clear
trend as a function of \pileup, likely due to the high-\pt regime selected by
the top-quark taggers.  However, the \SDlong
\topquark tagger shows minor trends with the background rejection decreasing as
the level of \pileup increases.  The background rejection of the \htt shows
little dependence on $\mu$.  In all cases, this trend is well-described by the
Monte Carlo simulation.
 
\begin{figure}[!h]
\centering
\subfigure[][]{   \label{fig:DMC_bkg_rejection_1_a} \includegraphics[width=0.45\textwidth]{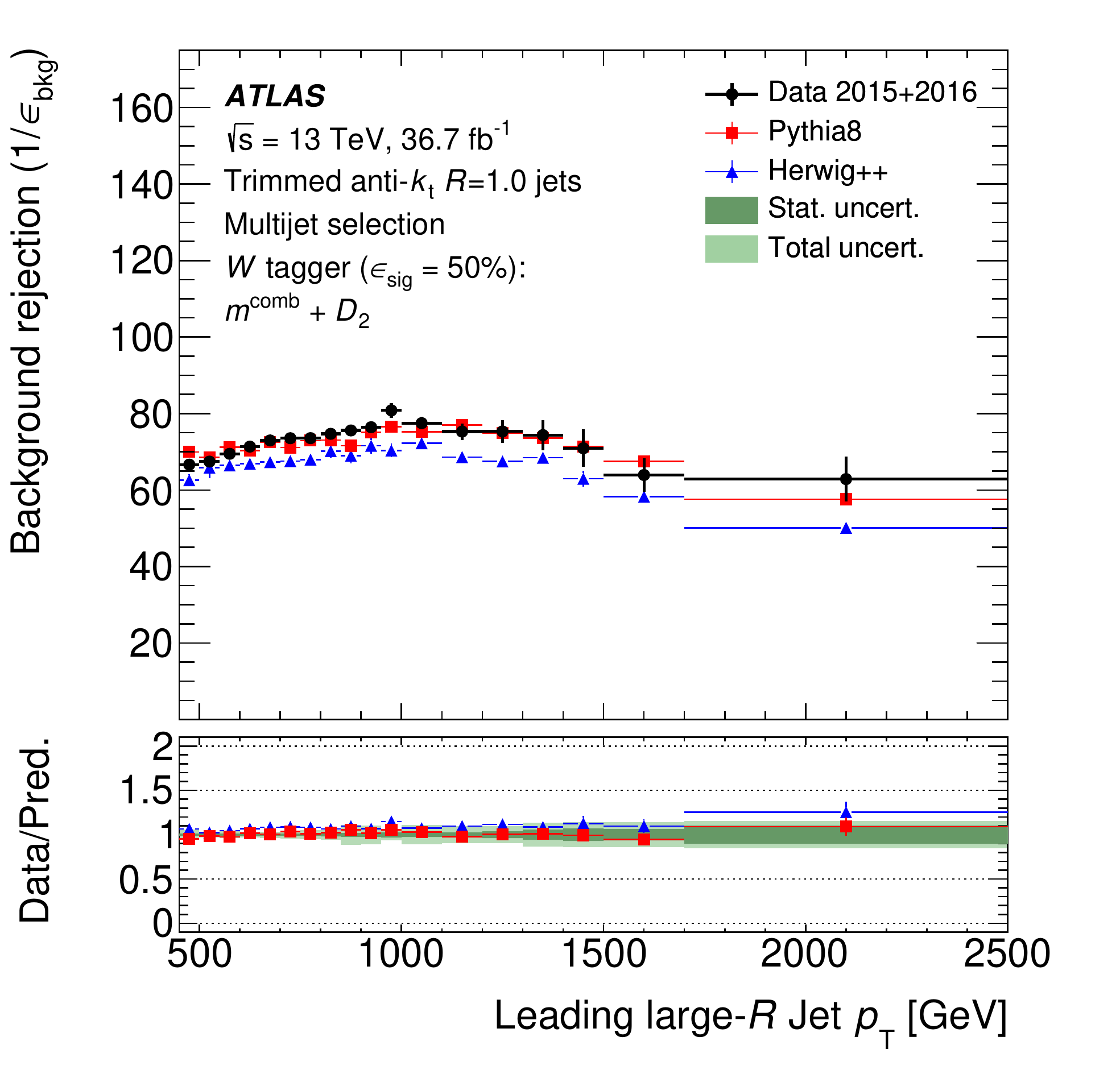}}
\subfigure[][]{  \label{fig:DMC_bkg_rejection_1_b} \includegraphics[width=0.45\textwidth]{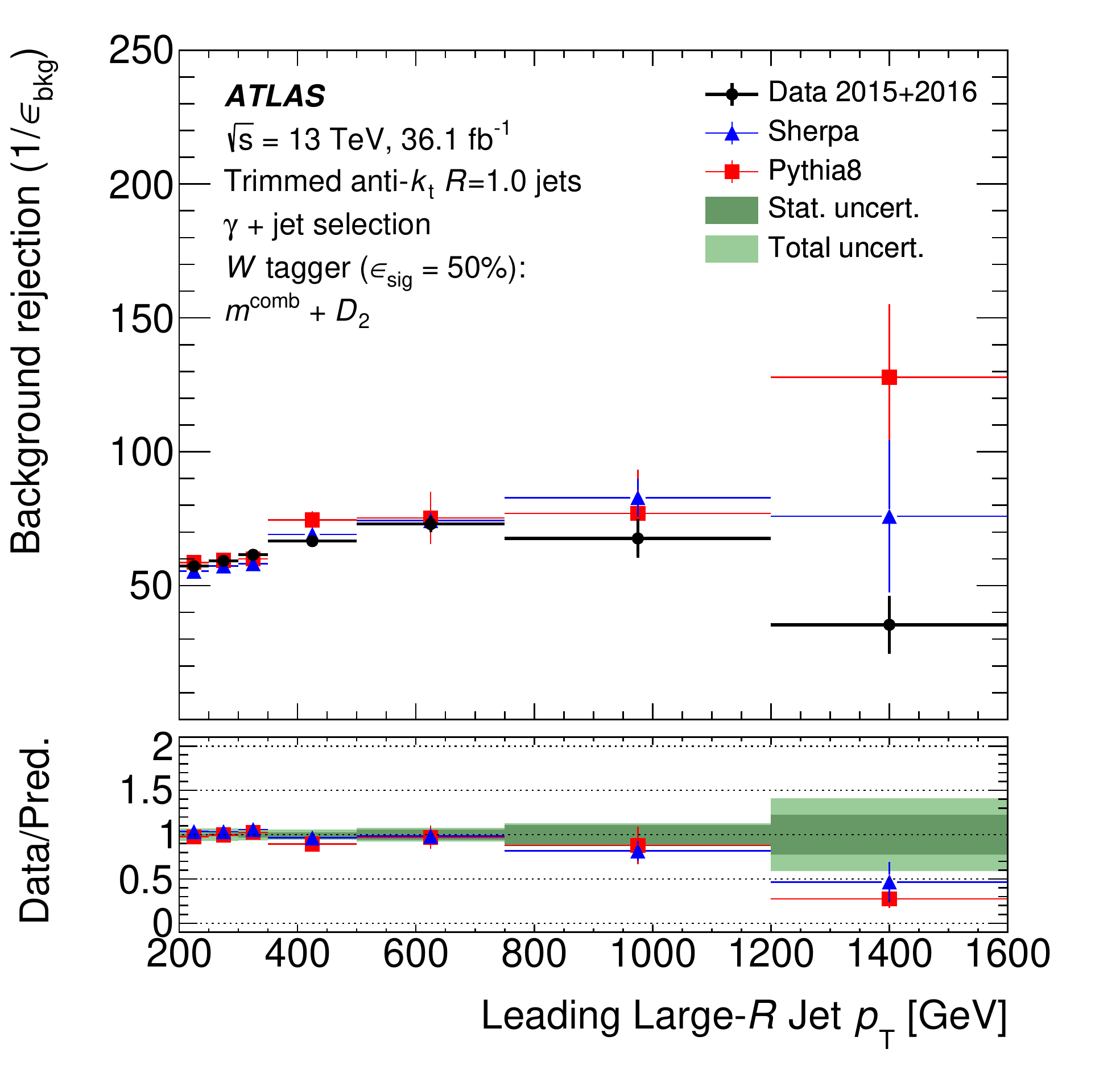}} \\
\subfigure[][]{   \label{fig:DMC_bkg_rejection_1_c} \includegraphics[width=0.45\textwidth]{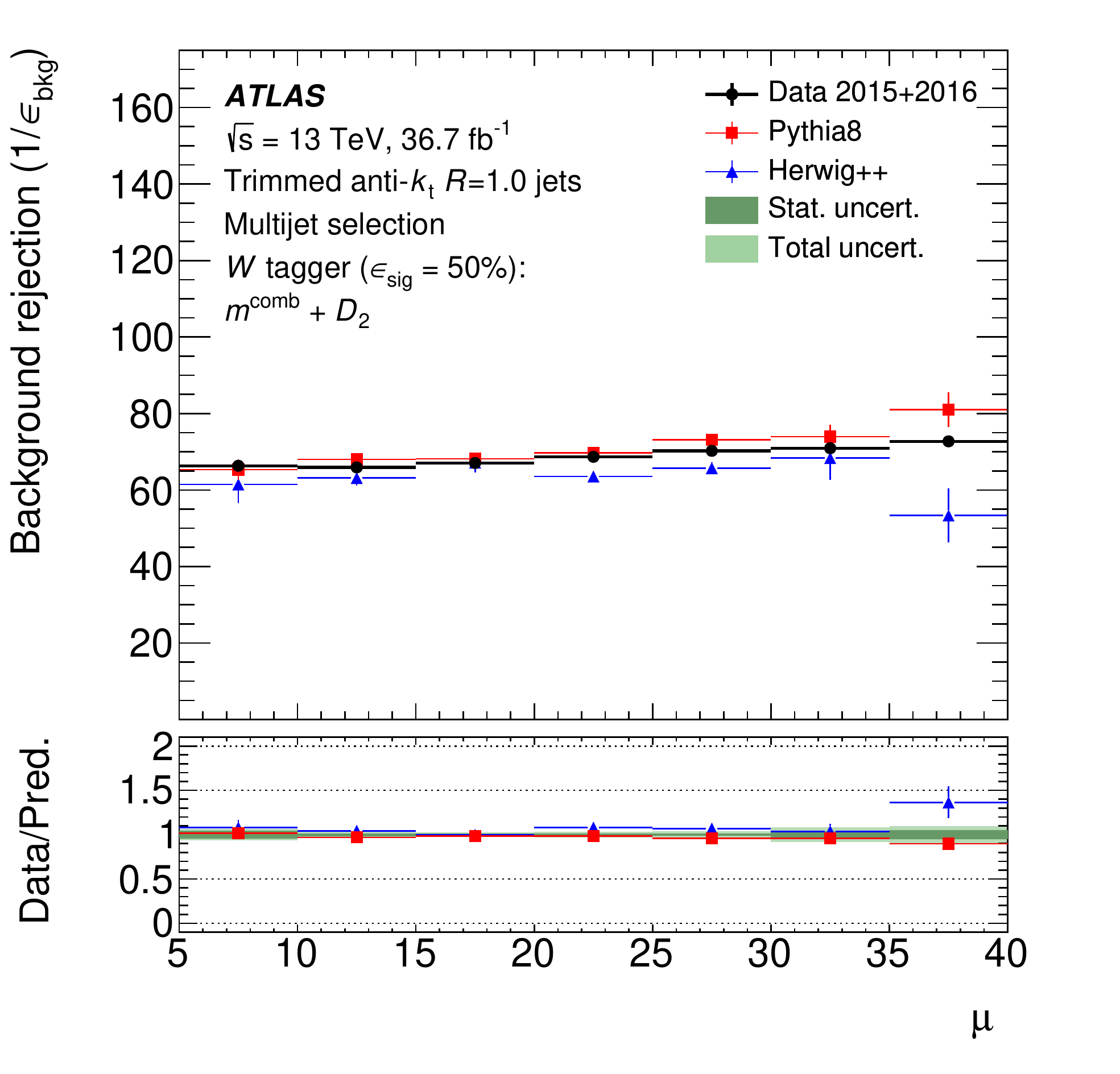}}
\subfigure[][]{  \label{fig:DMC_bkg_rejection_1_d} \includegraphics[width=0.45\textwidth]{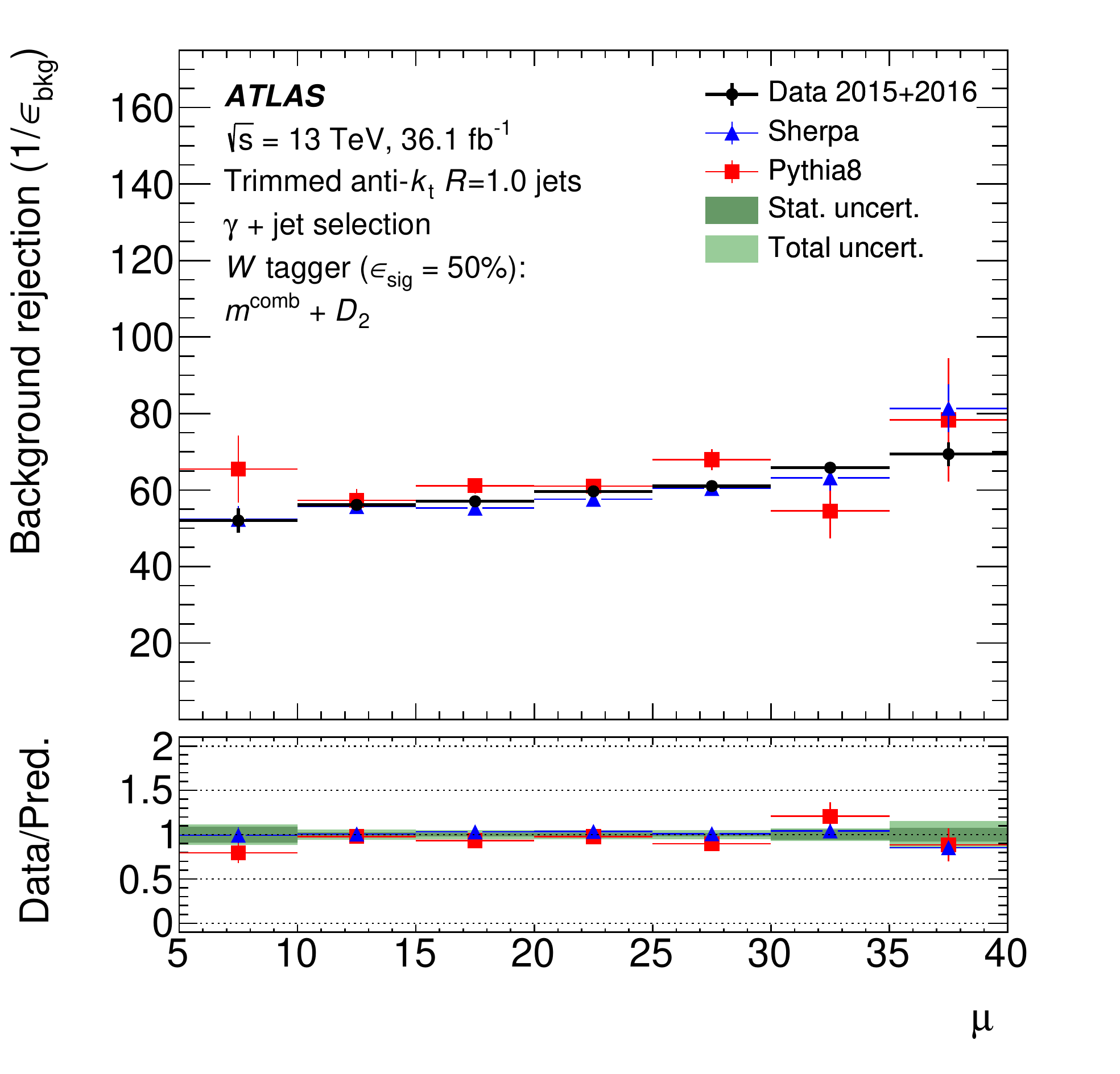}}
\caption{
\label{fig:DMC_bkg_rejection_1} The estimated light-jet rejection
$1/\epsilon_{\mathrm{bkg}}$ as a function of the leading jet \pt and the
average number of interactions per bunch crossing $\mu$ for the two-variable
\Wbosonboson tagger in the
multijet~\subref{fig:DMC_bkg_rejection_1_a}~\subref{fig:DMC_bkg_rejection_1_c}
and
\gammajet~\subref{fig:DMC_bkg_rejection_1_b}~\subref{fig:DMC_bkg_rejection_1_d}
selection.}
\end{figure}
 
\begin{figure}[!h]
\centering
\subfigure[][]{   \label{fig:DMC_bkg_rejection_2_a} \includegraphics[width=0.45\textwidth]{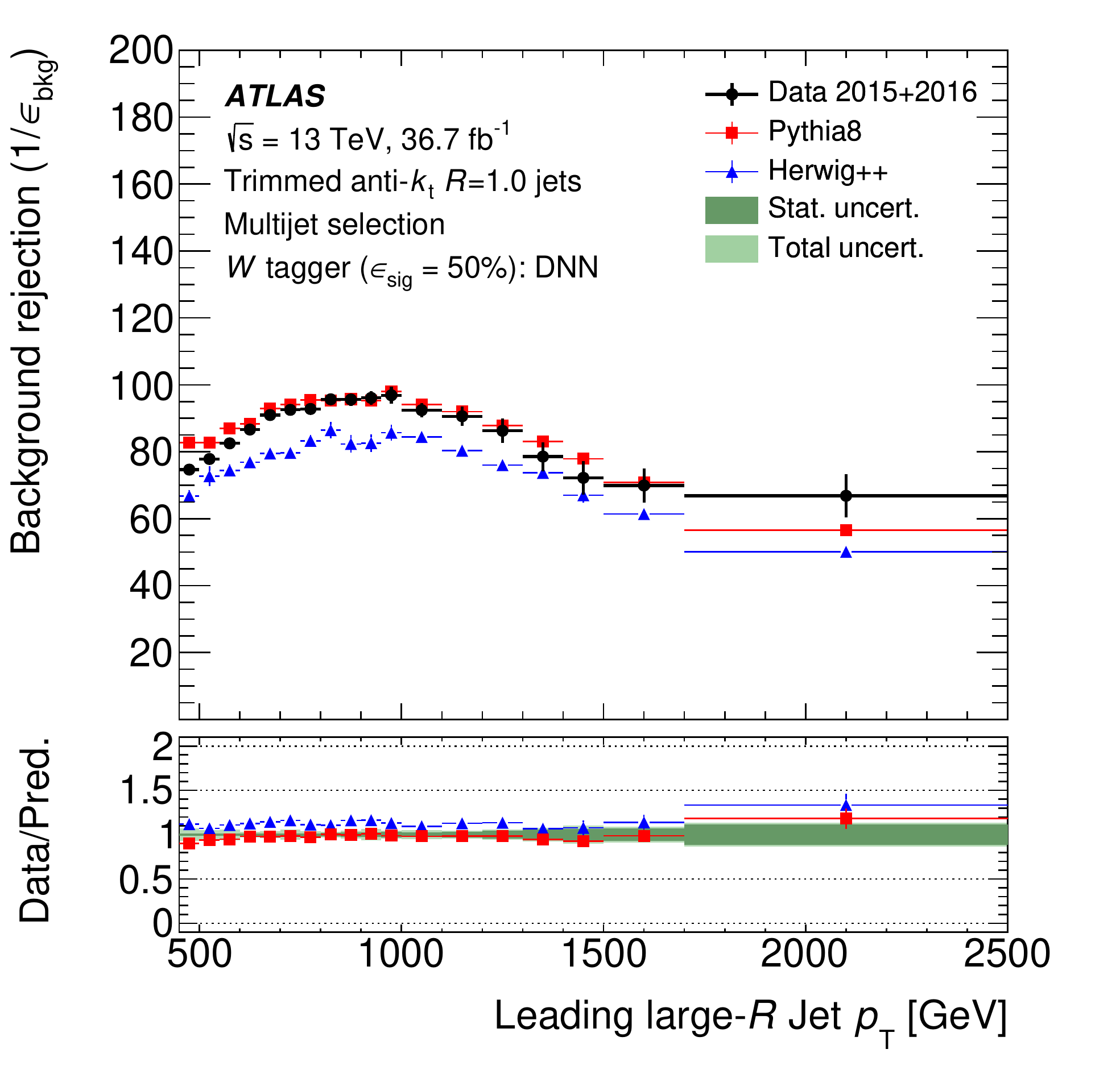}}
\subfigure[][]{  \label{fig:DMC_bkg_rejection_2_b} \includegraphics[width=0.45\textwidth]{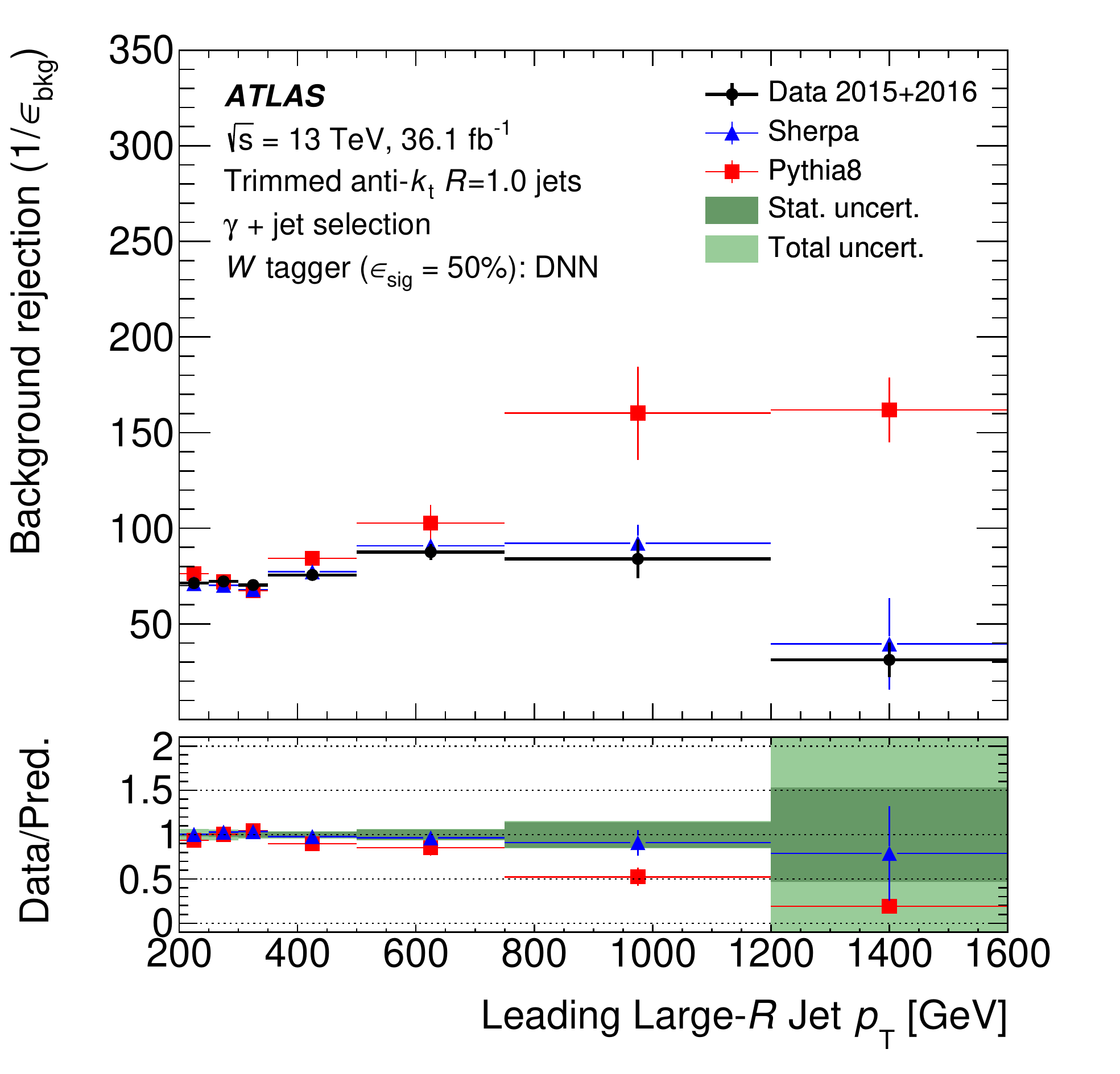}} \\
\subfigure[][]{   \label{fig:DMC_bkg_rejection_2_c} \includegraphics[width=0.45\textwidth]{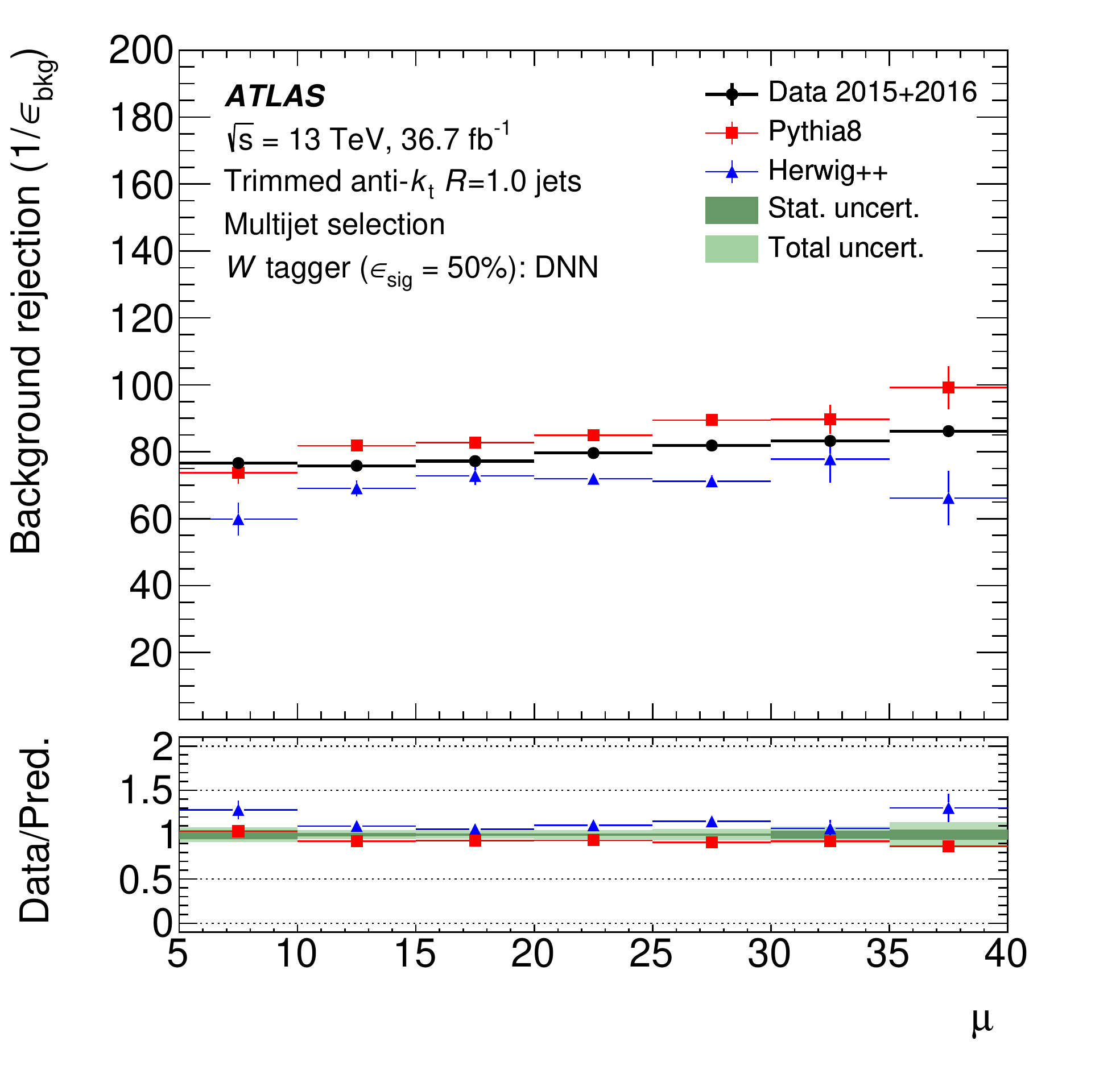}}
\subfigure[][]{  \label{fig:DMC_bkg_rejection_2_d} \includegraphics[width=0.45\textwidth]{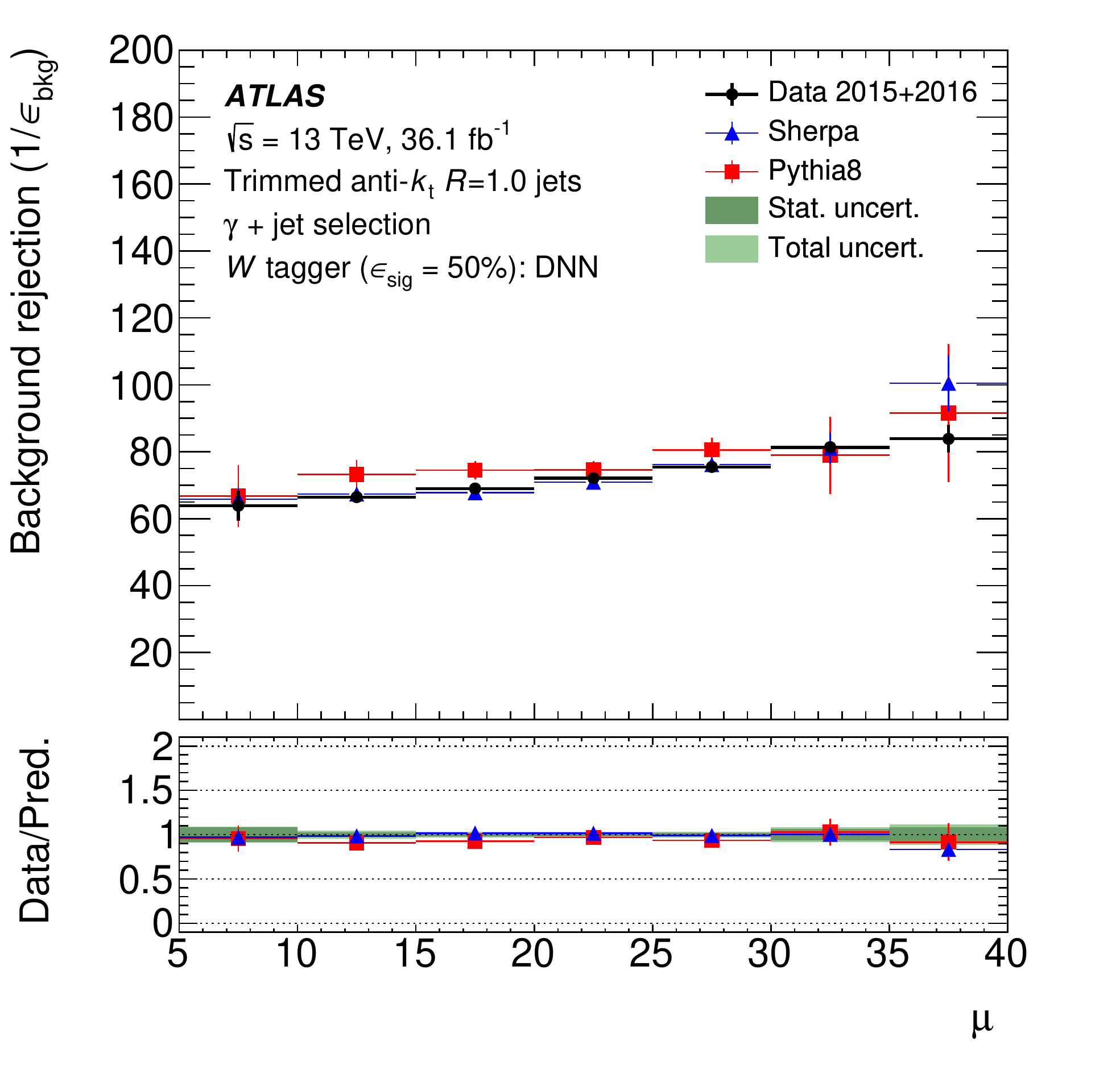}}
\caption{
\label{fig:DMC_bkg_rejection_2} The estimated light-jet rejection
$1/\epsilon_{\mathrm{bkg}}$ as a function of the leading jet \pt and the
average number of interactions per bunch crossing $\mu$ for the DNN
\Wbosonboson tagger in the
multijet~\subref{fig:DMC_bkg_rejection_2_a}~\subref{fig:DMC_bkg_rejection_2_c}
and
\gammajet~\subref{fig:DMC_bkg_rejection_2_b}~\subref{fig:DMC_bkg_rejection_2_d}
selection.}
\end{figure}
 
\begin{figure}[!h]
\centering
\subfigure[][]{   \label{fig:DMC_bkg_rejection_3_a} \includegraphics[width=0.45\textwidth]{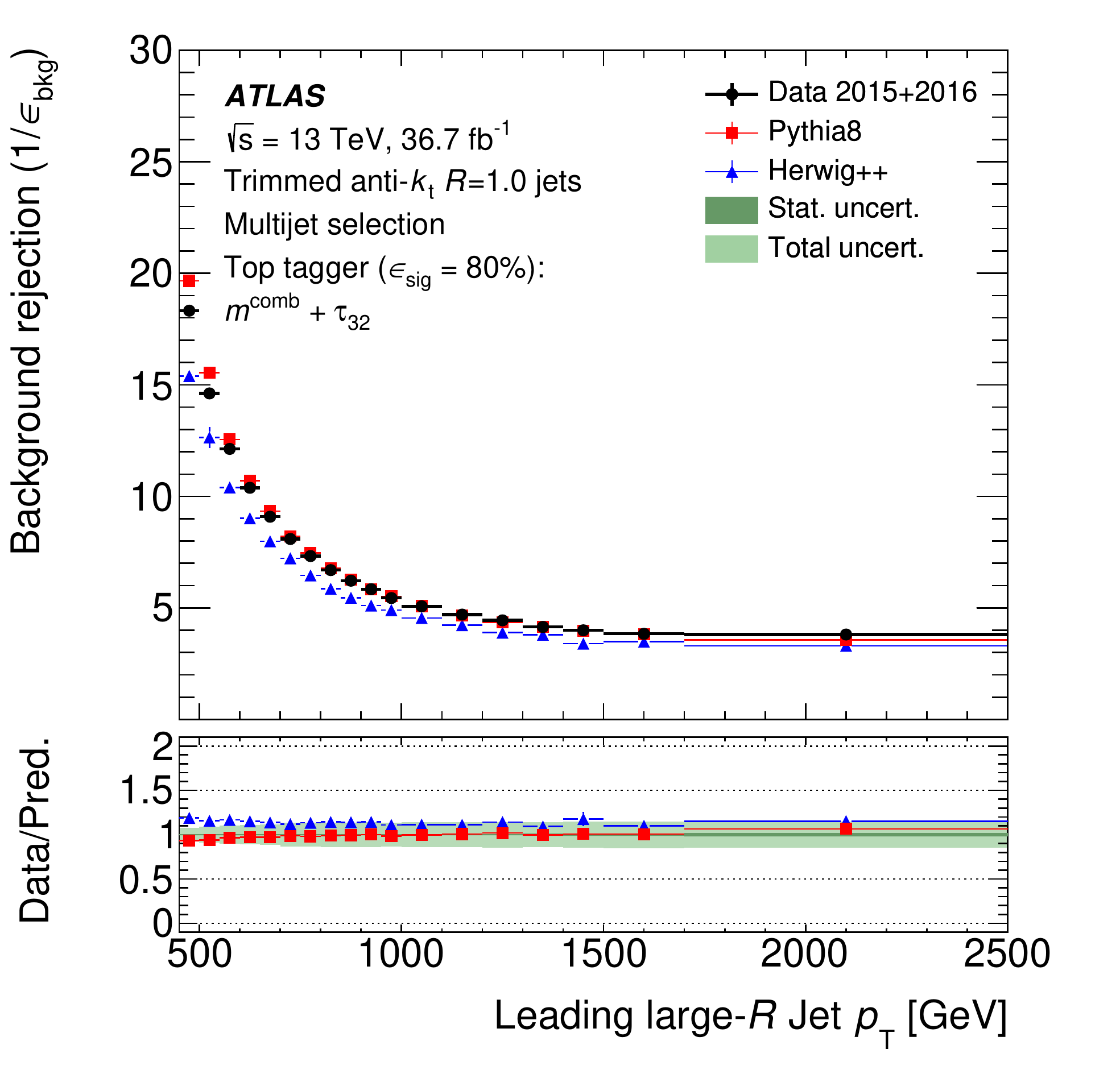}}
\subfigure[][]{  \label{fig:DMC_bkg_rejection_3_b} \includegraphics[width=0.45\textwidth]{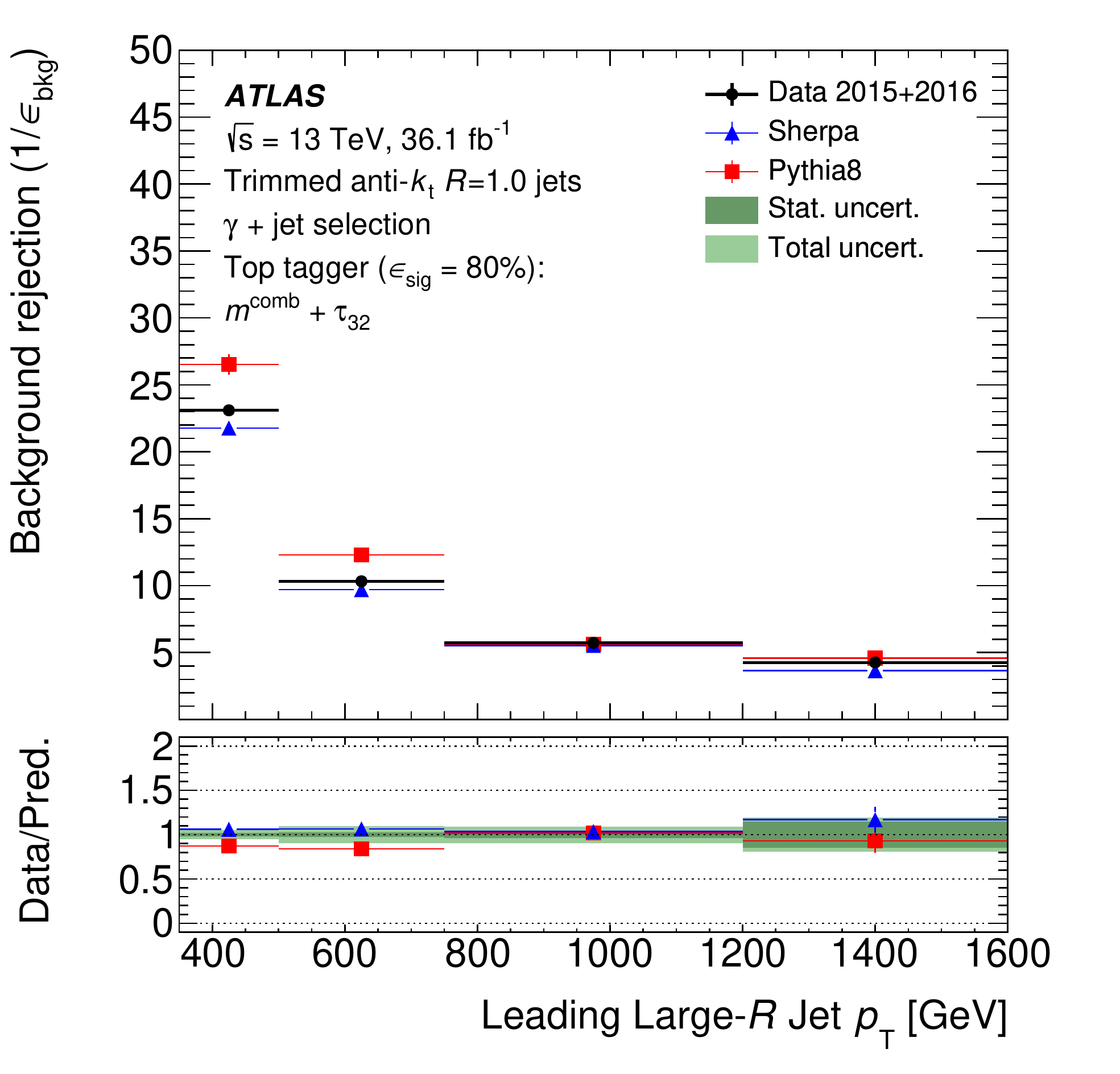}} \\
\subfigure[][]{   \label{fig:DMC_bkg_rejection_3_c} \includegraphics[width=0.45\textwidth]{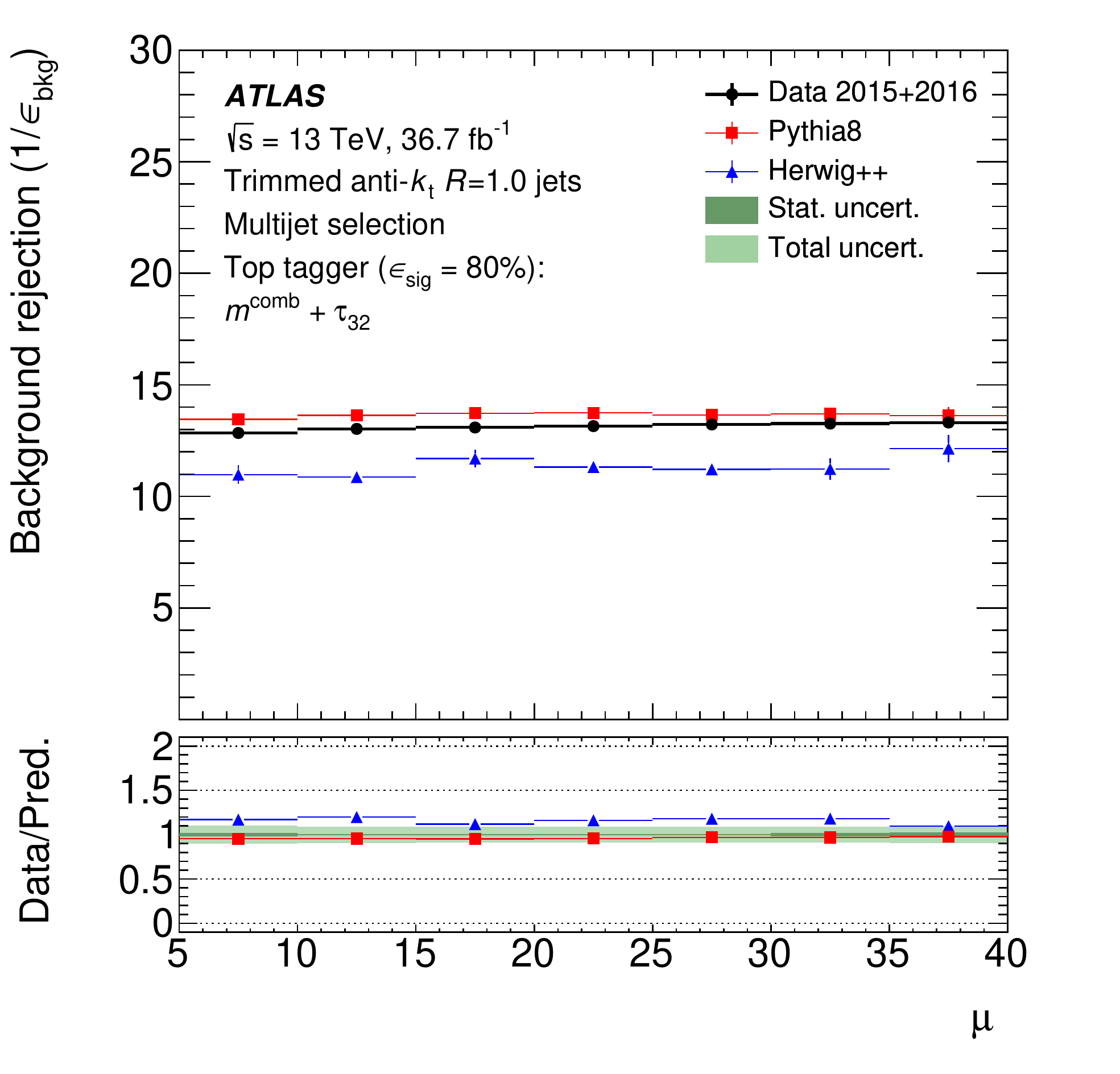}}
\subfigure[][]{  \label{fig:DMC_bkg_rejection_3_d} \includegraphics[width=0.45\textwidth]{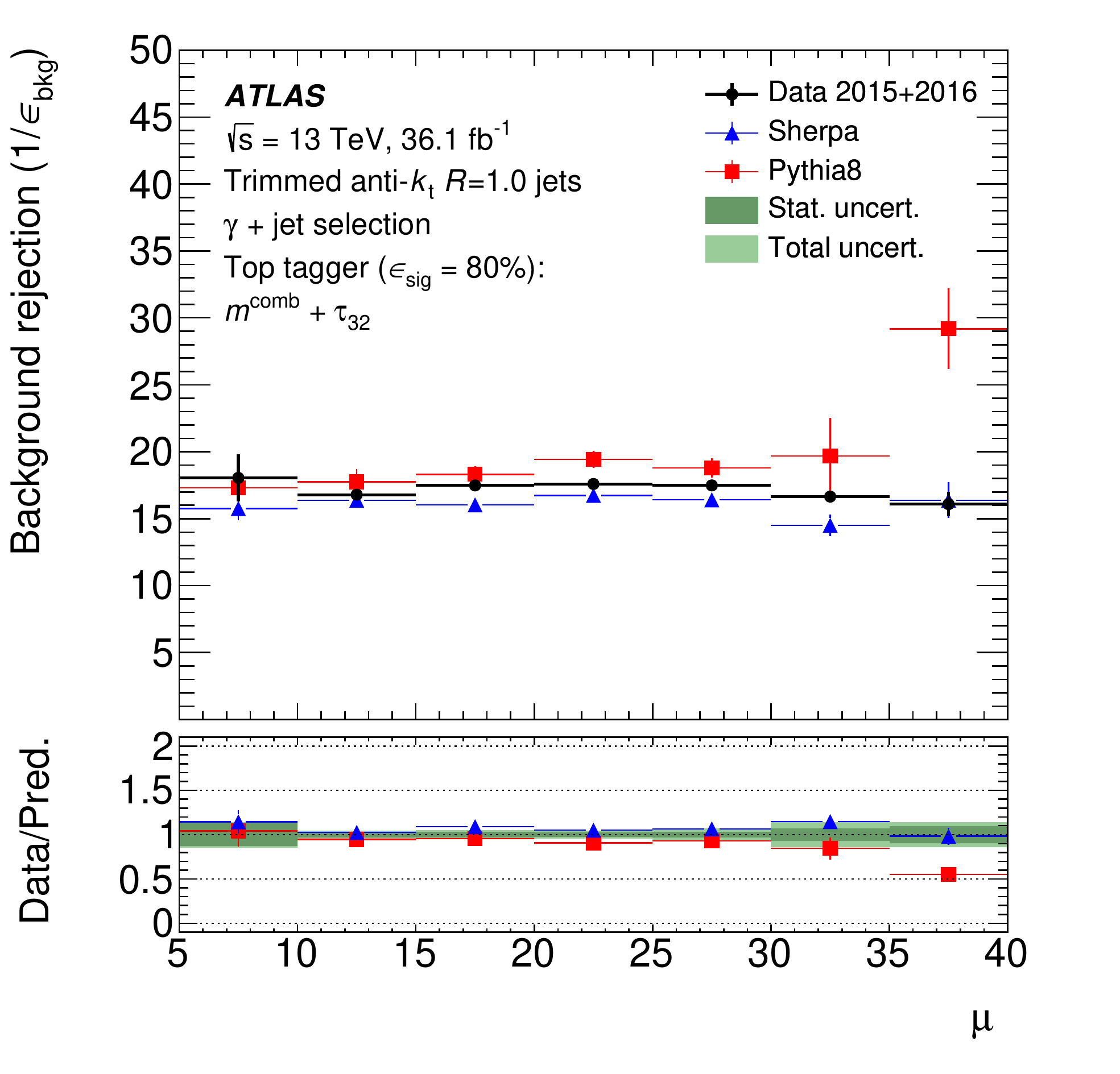}}
\caption{
The estimated light-jet rejection
$1/\epsilon_{\mathrm{bkg}}$ as a function of the leading jet \pt and the
average number of interactions per bunch crossing $\mu$ for the two-variable
\topquark tagger in the
multijet~\subref{fig:DMC_bkg_rejection_3_a}~\subref{fig:DMC_bkg_rejection_3_c}
and
\gammajet~\subref{fig:DMC_bkg_rejection_3_b}~\subref{fig:DMC_bkg_rejection_3_d}
selection.}
\end{figure}
 
\begin{figure}[!h]
\centering
\subfigure[][]{   \label{fig:DMC_bkg_rejection_4_a} \includegraphics[width=0.45\textwidth]{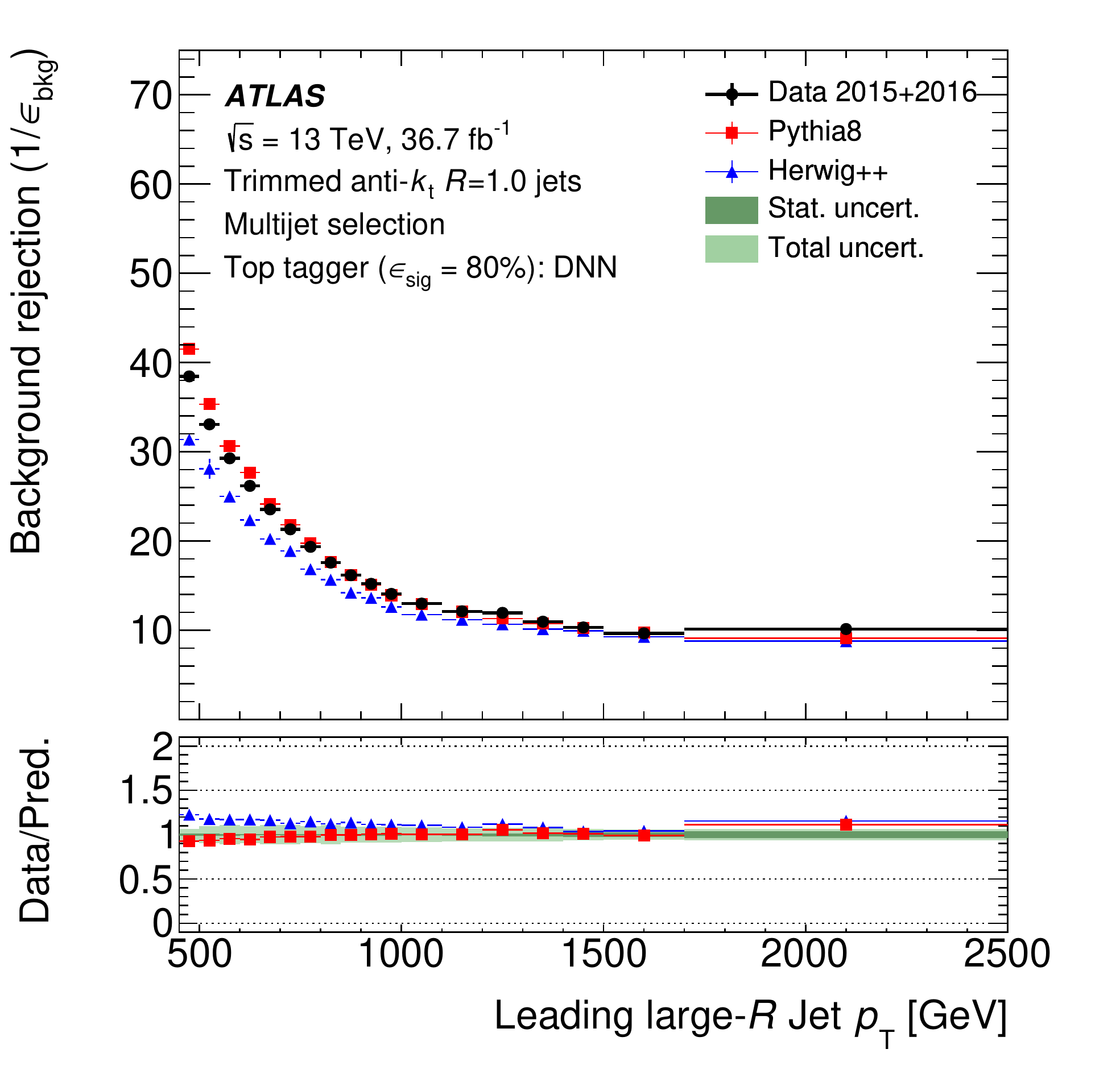}}
\subfigure[][]{  \label{fig:DMC_bkg_rejection_4_b} \includegraphics[width=0.45\textwidth]{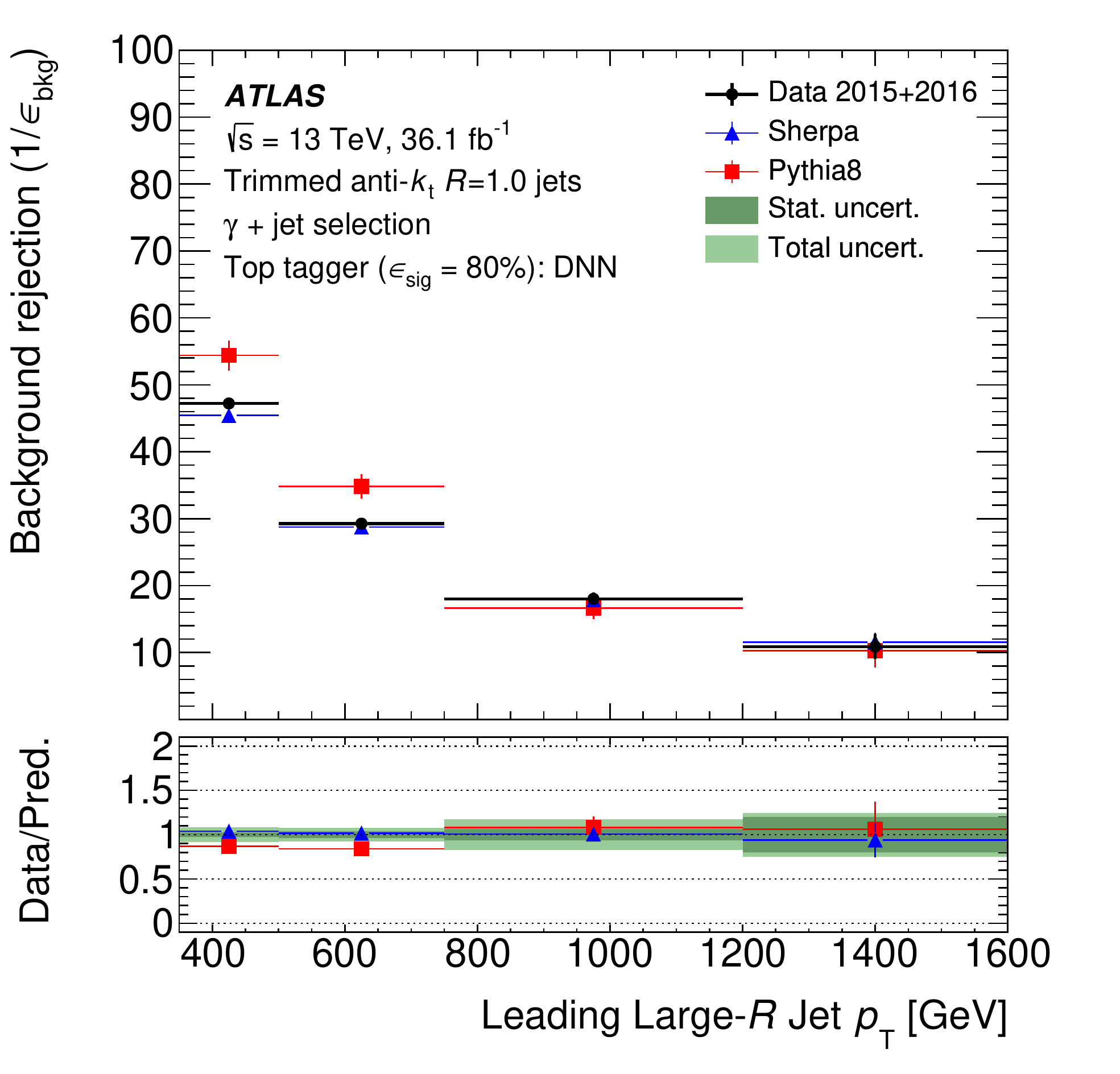}} \\
\subfigure[][]{   \label{fig:DMC_bkg_rejection_4_c} \includegraphics[width=0.45\textwidth]{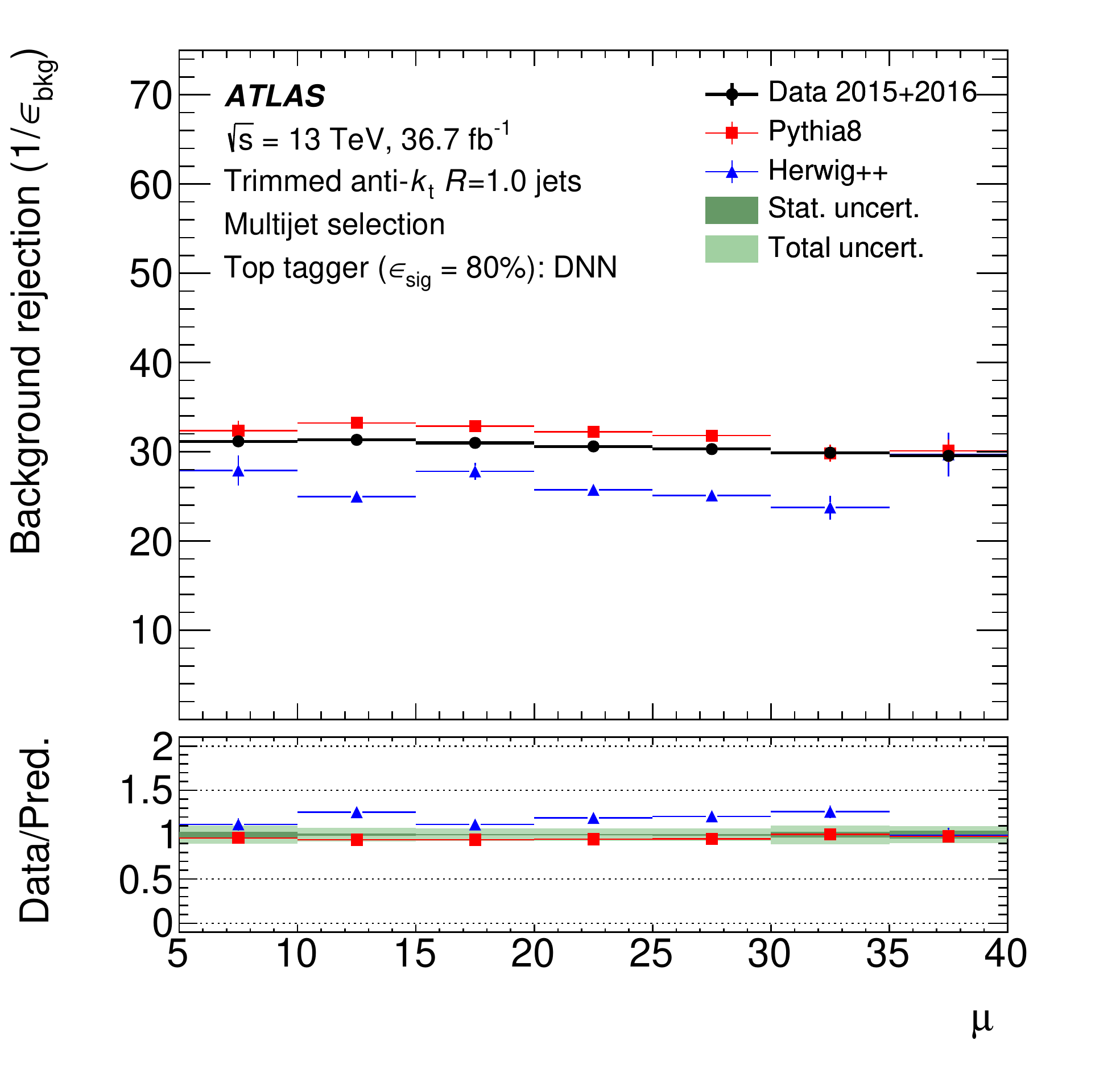}}
\subfigure[][]{  \label{fig:DMC_bkg_rejection_4_d} \includegraphics[width=0.45\textwidth]{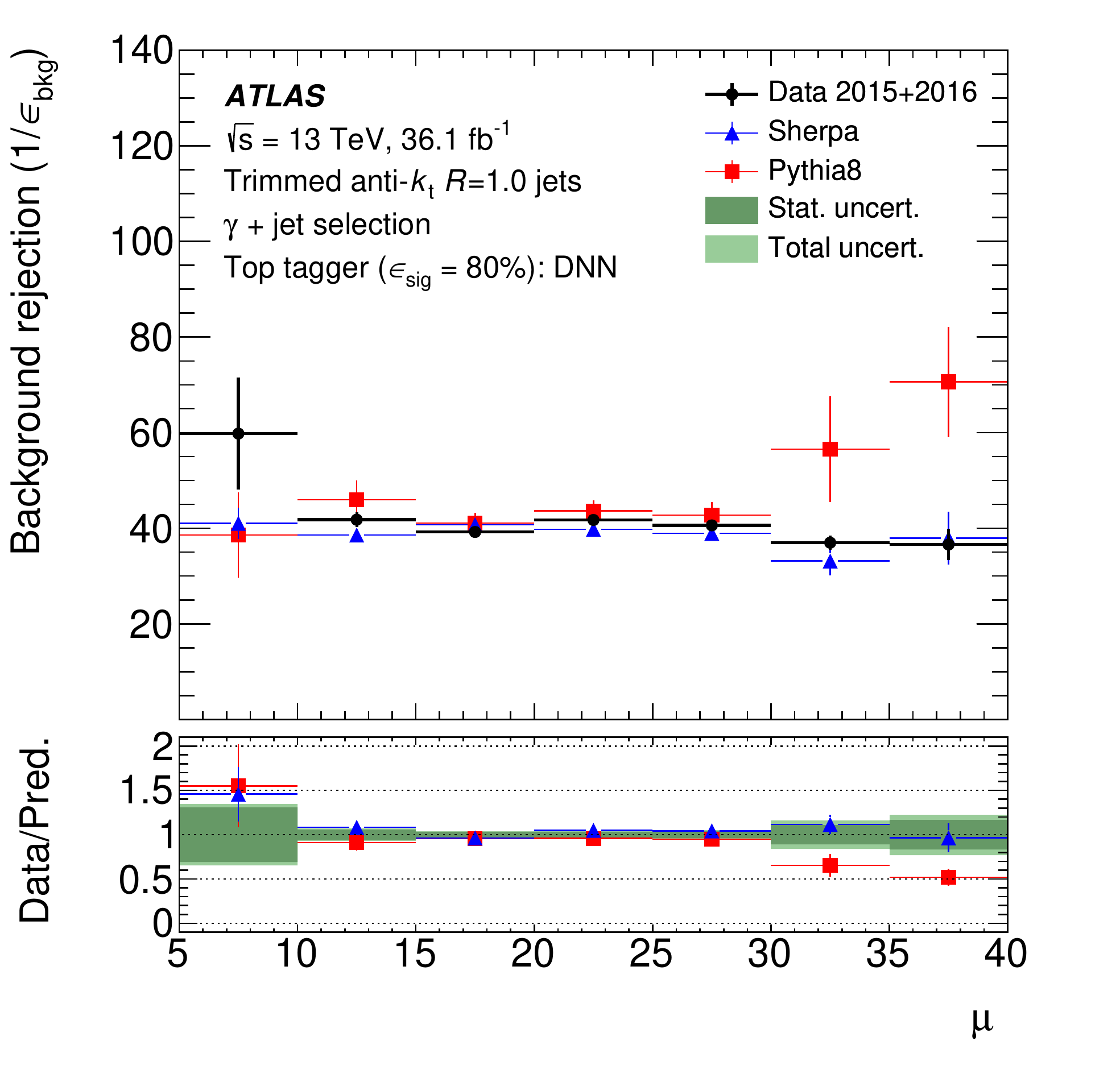}}
\caption{
\label{fig:DMC_bkg_rejection_4} The estimated light-jet rejection
$1/\epsilon_{\mathrm{bkg}}$ as a function of the leading jet \pt and the
average number of interactions per bunch crossing $\mu$ for the DNN \topquark
tagger in the
multijet~\subref{fig:DMC_bkg_rejection_4_a}~\subref{fig:DMC_bkg_rejection_4_c}
and
\gammajet~\subref{fig:DMC_bkg_rejection_4_b}~\subref{fig:DMC_bkg_rejection_4_d}
selection.}
\end{figure}
 
\begin{figure}[!h]
\centering
\subfigure[][]{   \label{fig:DMC_bkg_rejection_5_a} \includegraphics[width=0.45\textwidth]{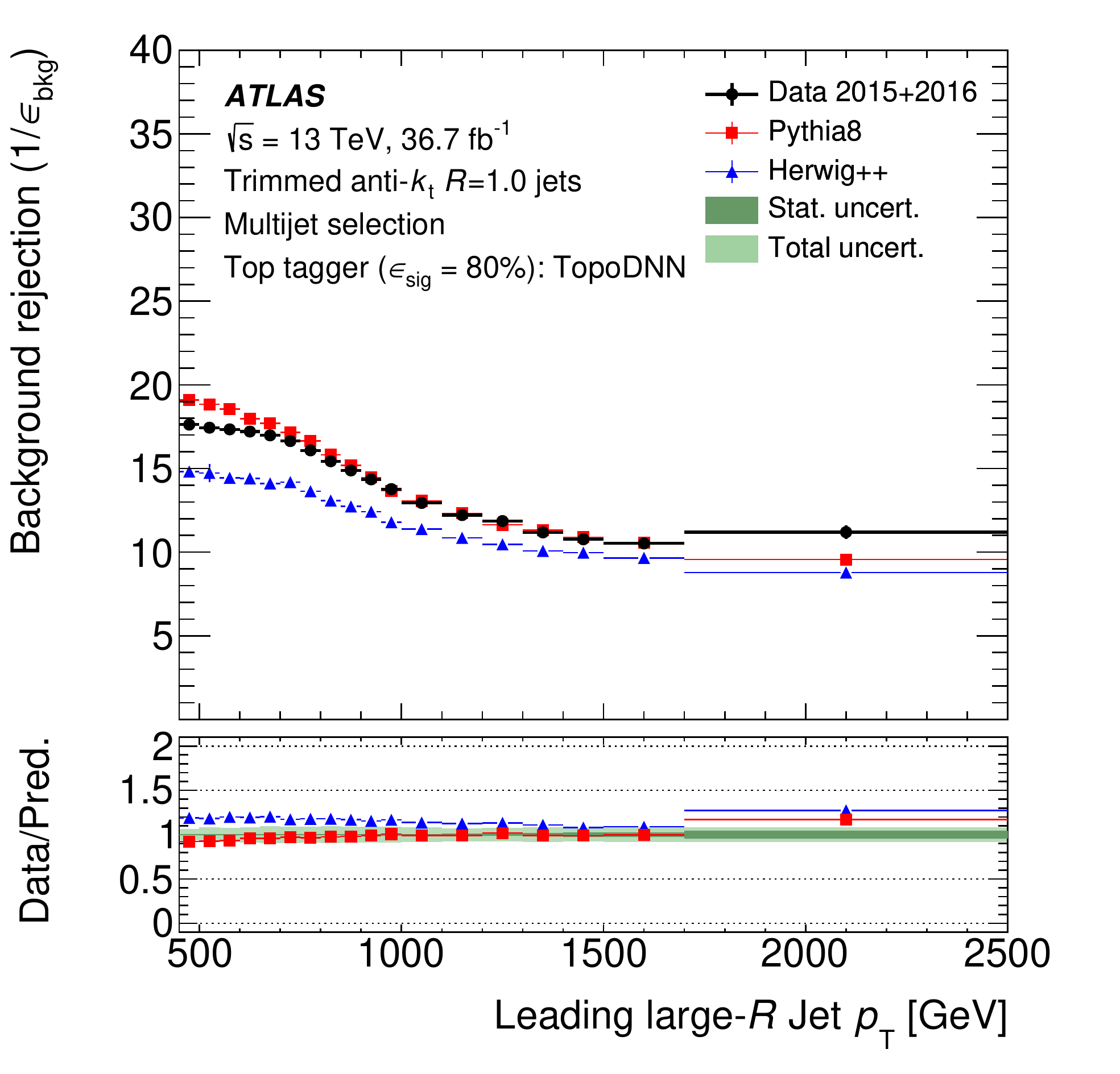}}
\subfigure[][]{  \label{fig:DMC_bkg_rejection_5_b} \includegraphics[width=0.45\textwidth]{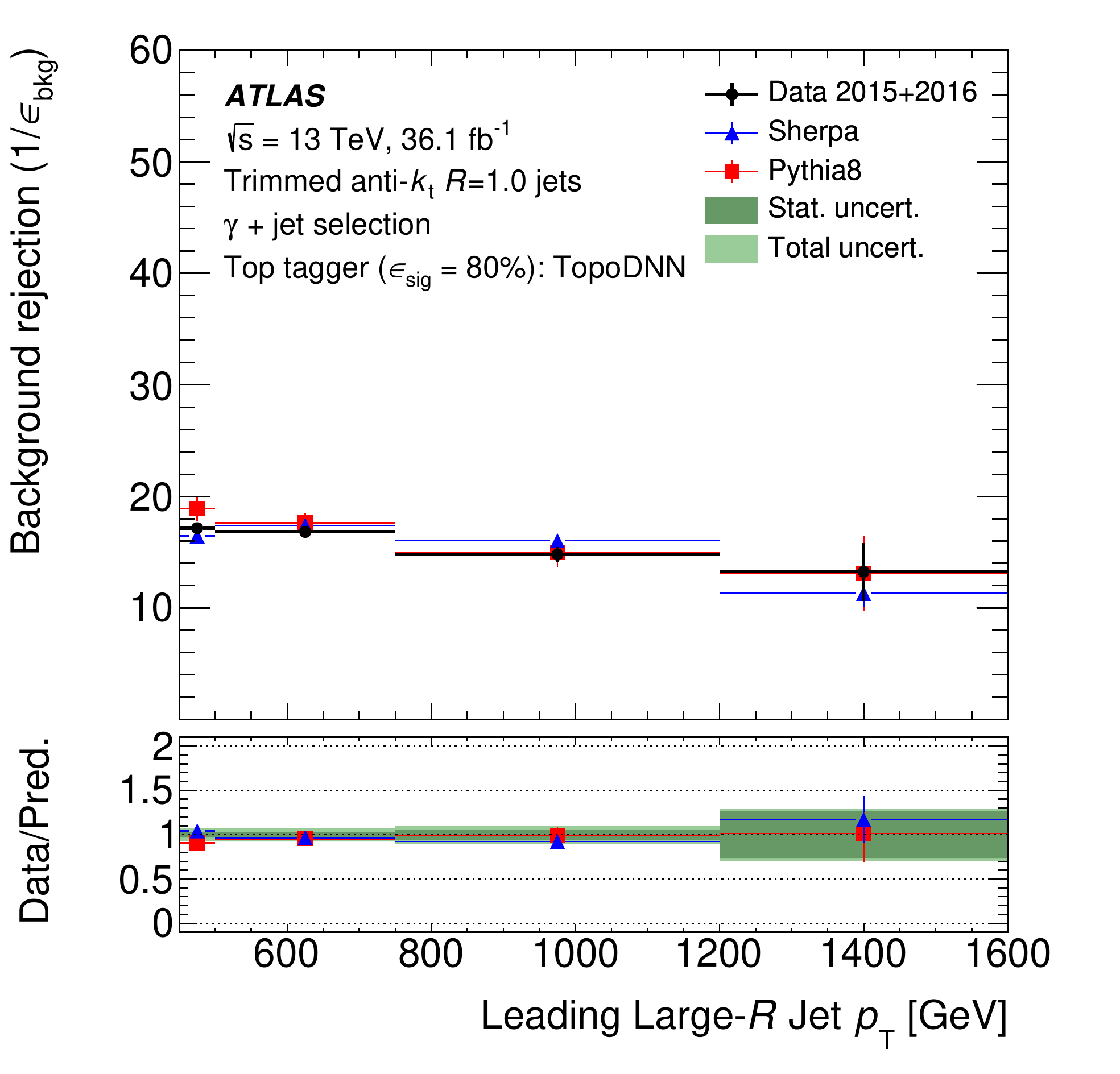}} \\
\subfigure[][]{   \label{fig:DMC_bkg_rejection_5_c} \includegraphics[width=0.45\textwidth]{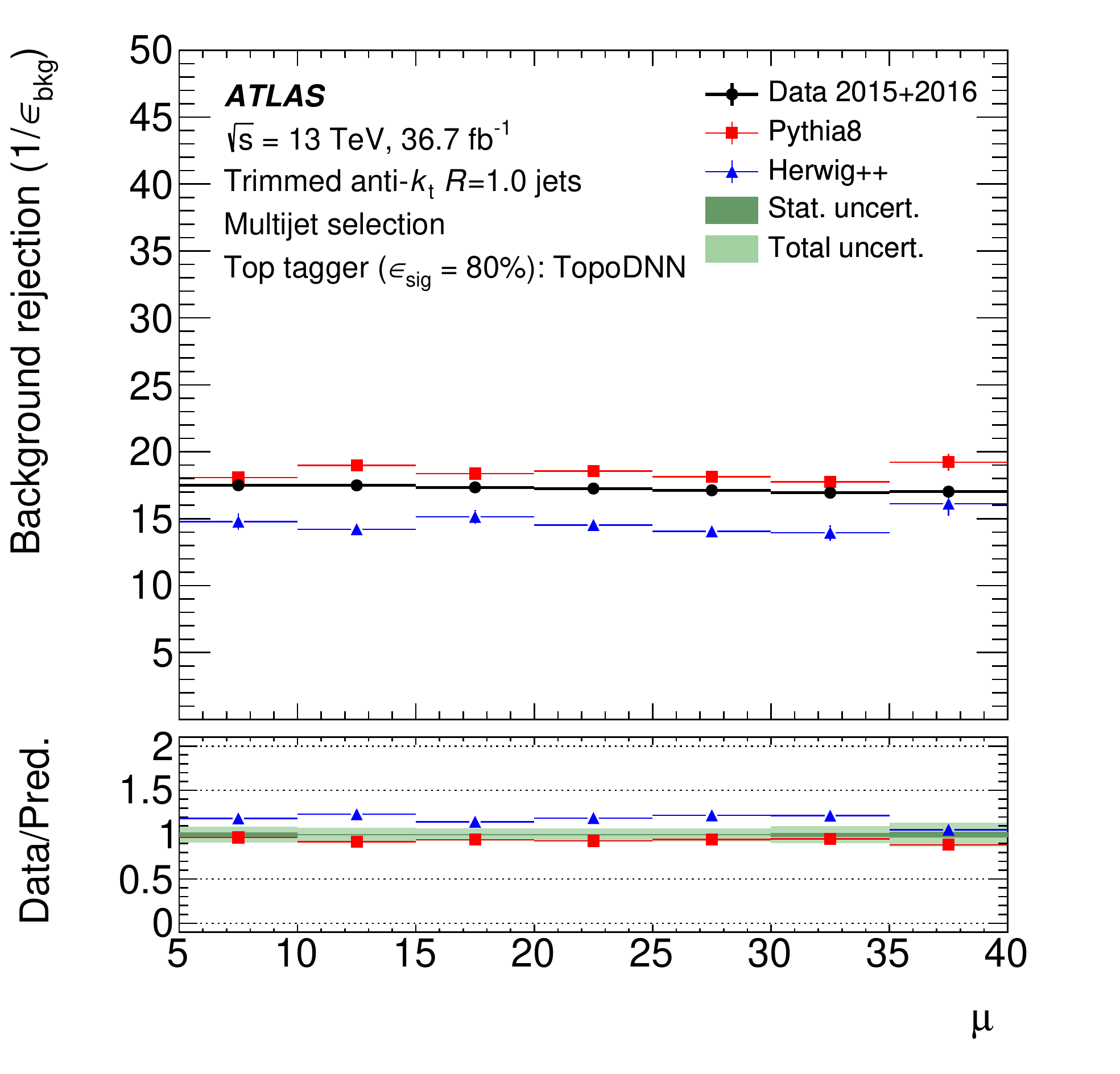}}
\subfigure[][]{  \label{fig:DMC_bkg_rejection_5_d} \includegraphics[width=0.45\textwidth]{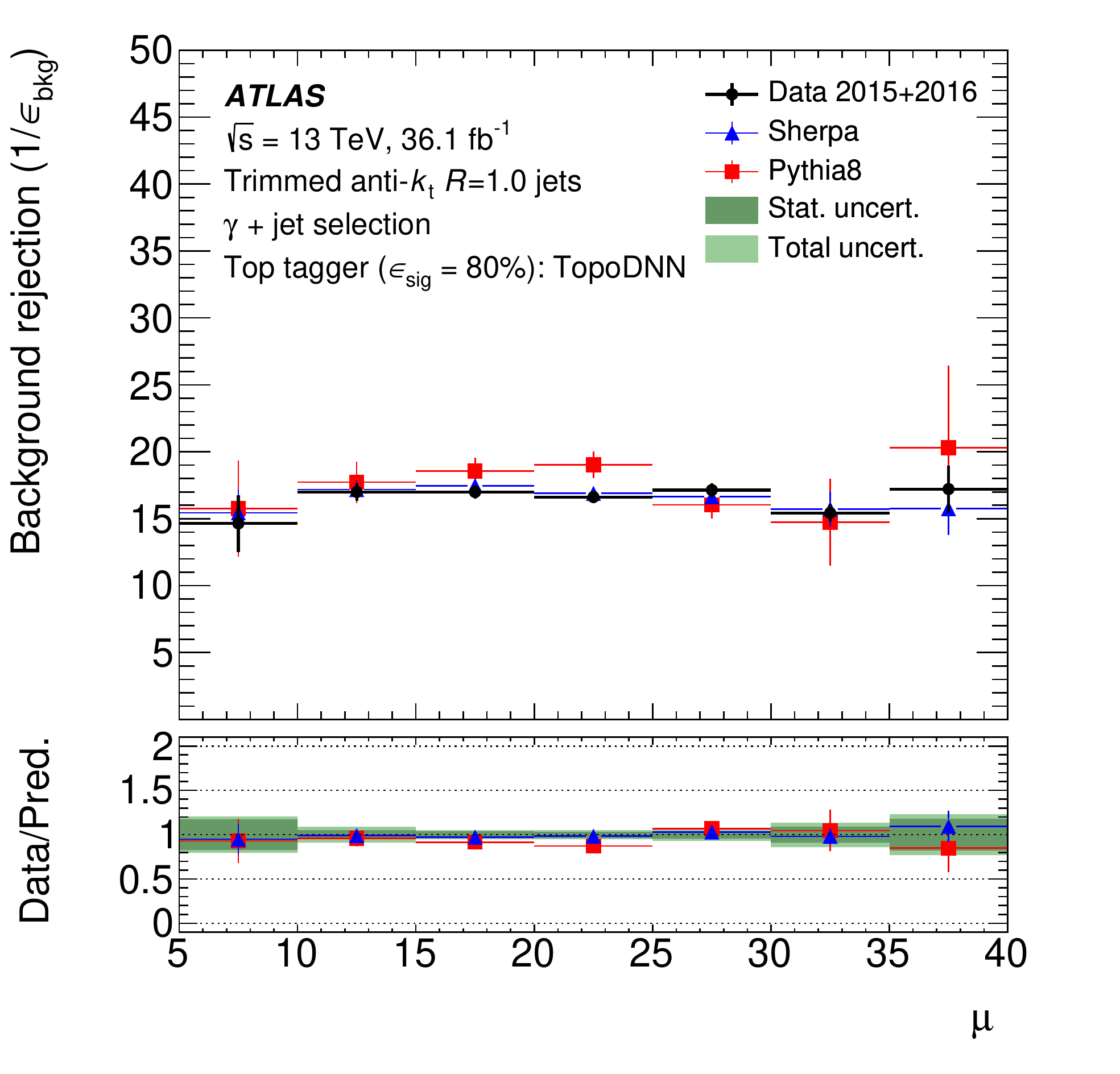}}
\caption{
\label{fig:DMC_bkg_rejection_5} The estimated light-jet rejection
$1/\epsilon_{\mathrm{bkg}}$ as a function of the leading jet \pt and the
average number of interactions per bunch crossing $\mu$ for the
TopoDNN \topquark tagger in the
multijet~\subref{fig:DMC_bkg_rejection_5_a}~\subref{fig:DMC_bkg_rejection_5_c}
and
\gammajet~\subref{fig:DMC_bkg_rejection_5_b}~\subref{fig:DMC_bkg_rejection_5_d}
selection.}
\end{figure}
 
\begin{figure}[!h]
\centering
\subfigure[][]{   \label{fig:DMC_bkg_rejection_6_a} \includegraphics[width=0.45\textwidth]{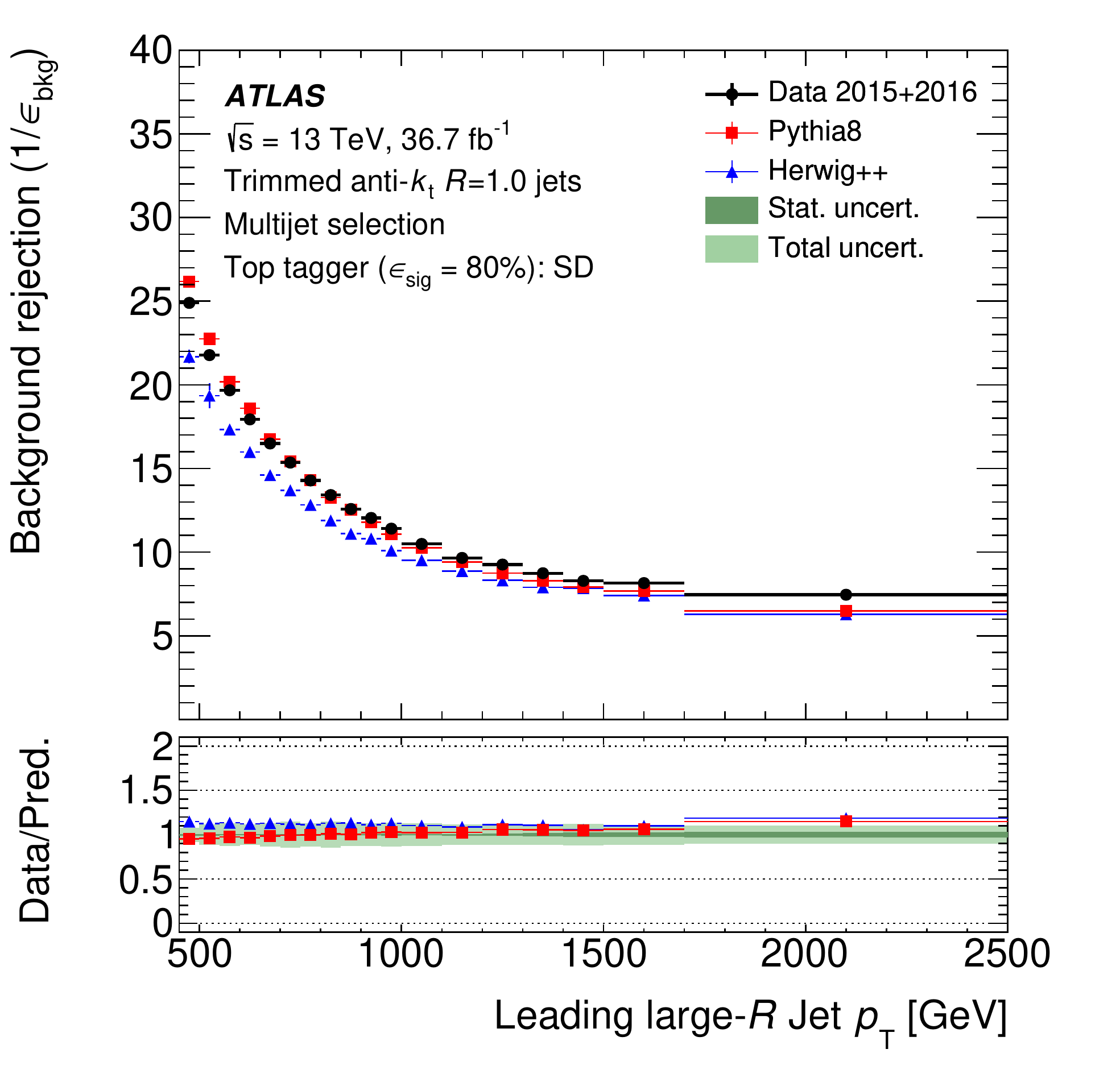}}
\subfigure[][]{  \label{fig:DMC_bkg_rejection_6_b} \includegraphics[width=0.45\textwidth]{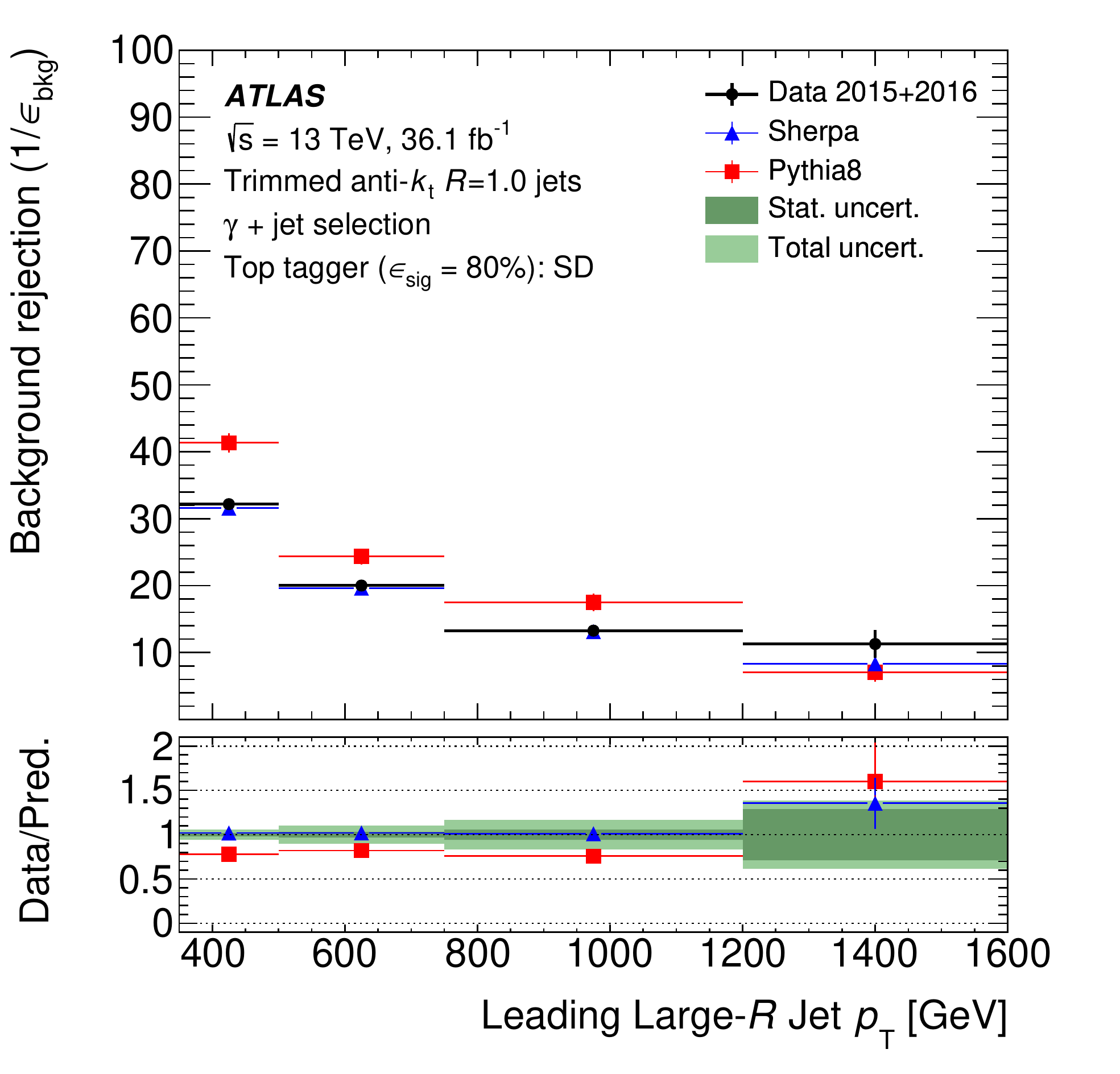}} \\
\subfigure[][]{   \label{fig:DMC_bkg_rejection_6_c} \includegraphics[width=0.45\textwidth]{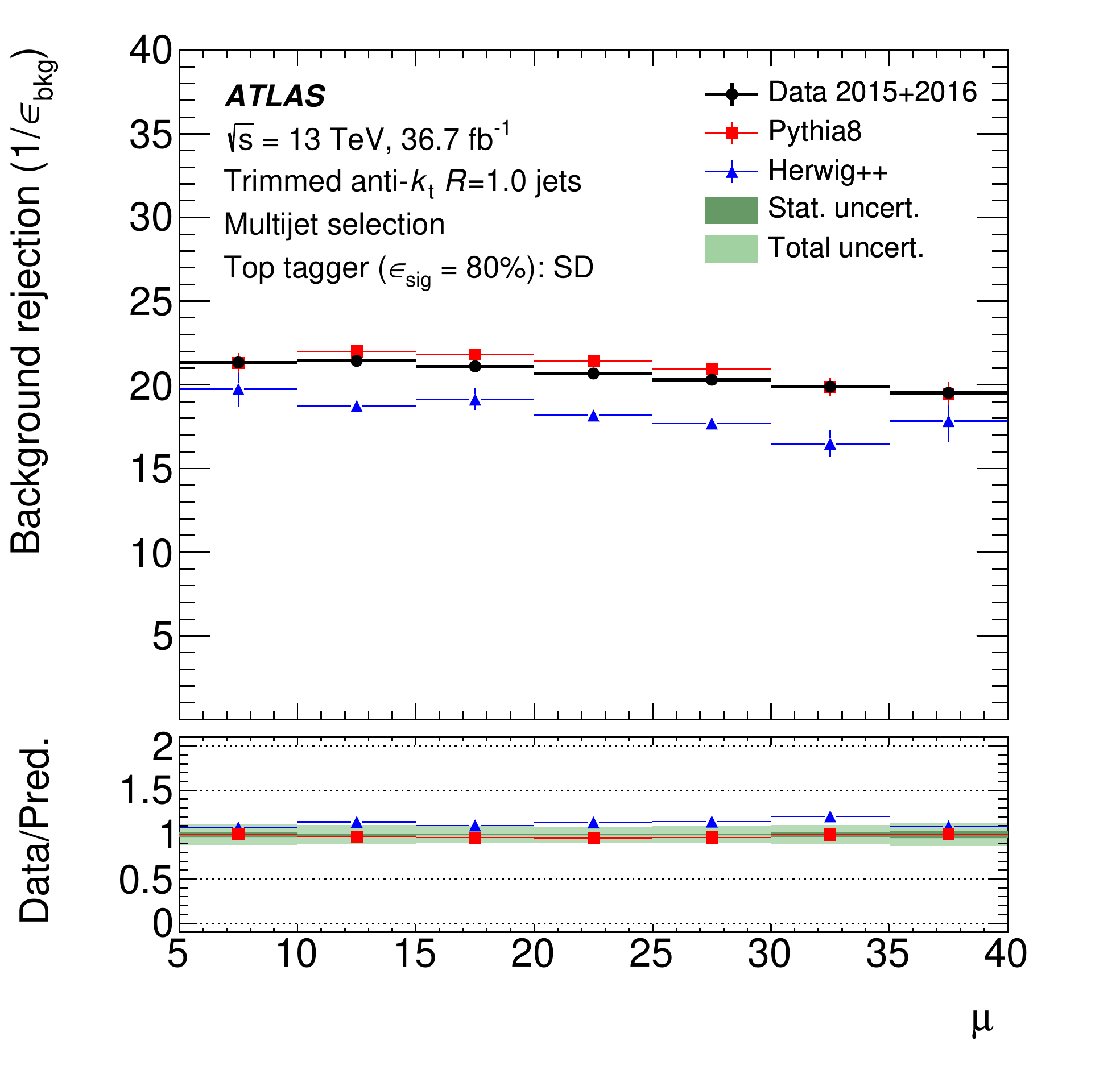}}
\subfigure[][]{  \label{fig:DMC_bkg_rejection_6_d} \includegraphics[width=0.45\textwidth]{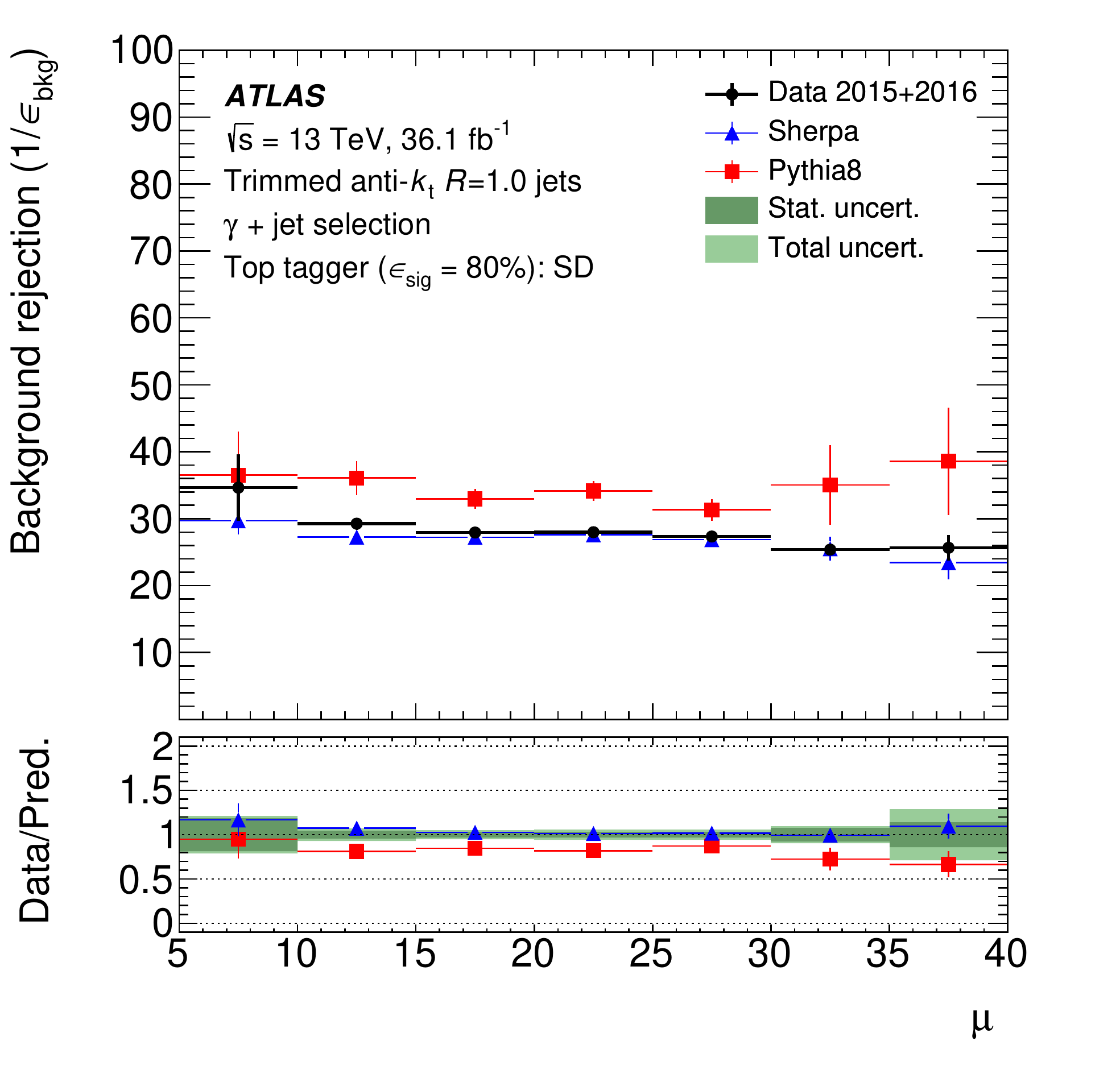}}
\caption{
The estimated light-jet rejection
$1/\epsilon_{\mathrm{bkg}}$ as a function of the leading jet \pt and the
average number of interactions per bunch crossing $\mu$ for the shower
deconstruction \topquark tagger in the
multijet~\subref{fig:DMC_bkg_rejection_6_a}~\subref{fig:DMC_bkg_rejection_6_c}
and
\gammajet~\subref{fig:DMC_bkg_rejection_6_b}~\subref{fig:DMC_bkg_rejection_6_d}
selection.}
\end{figure}
 
\begin{figure}[!h]
\centering
\subfigure[][]{   \label{fig:DMC_bkg_rejection_7_a} \includegraphics[width=0.45\textwidth]{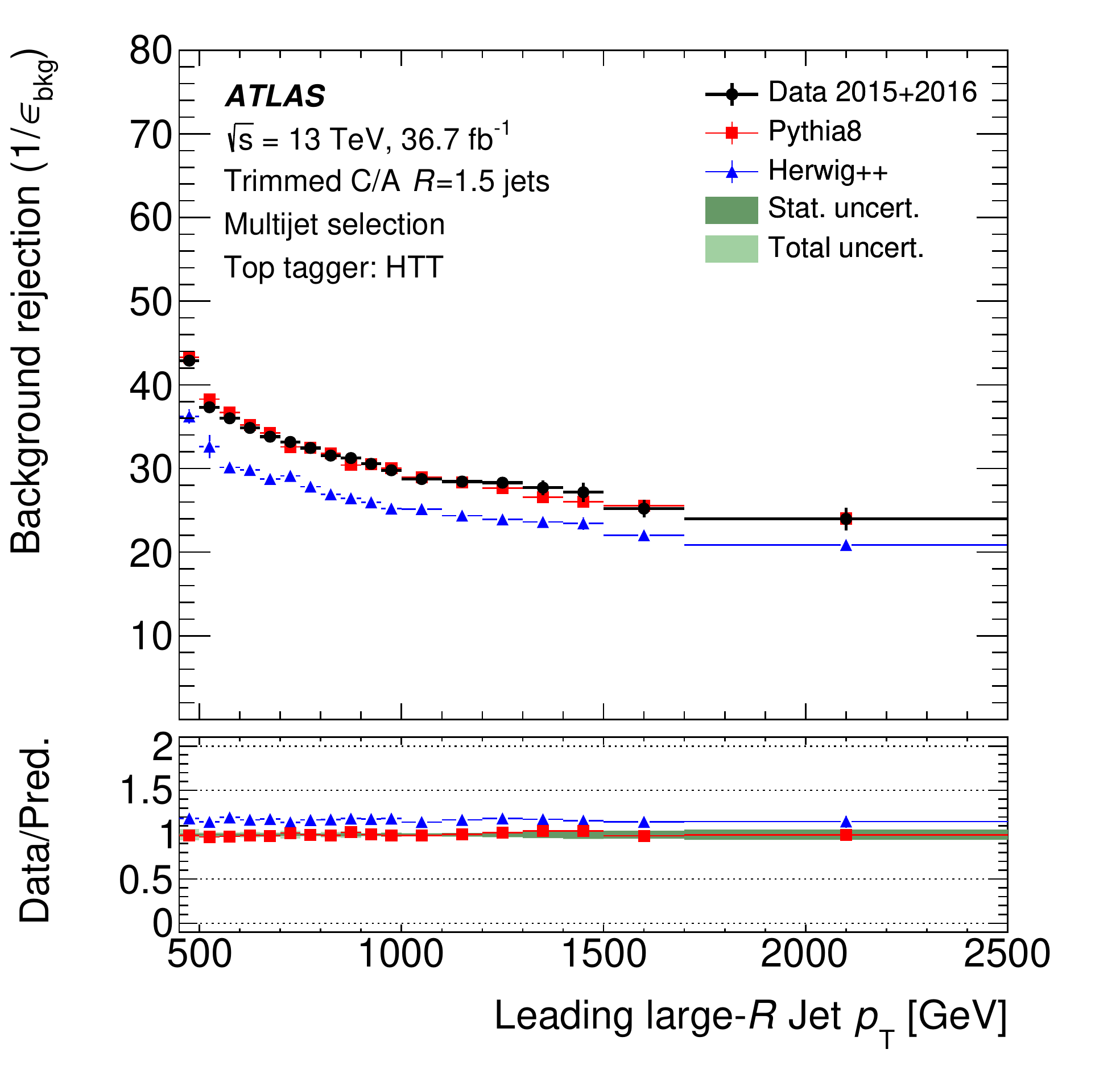}}
\subfigure[][]{  \label{fig:DMC_bkg_rejection_7_b} \includegraphics[width=0.45\textwidth]{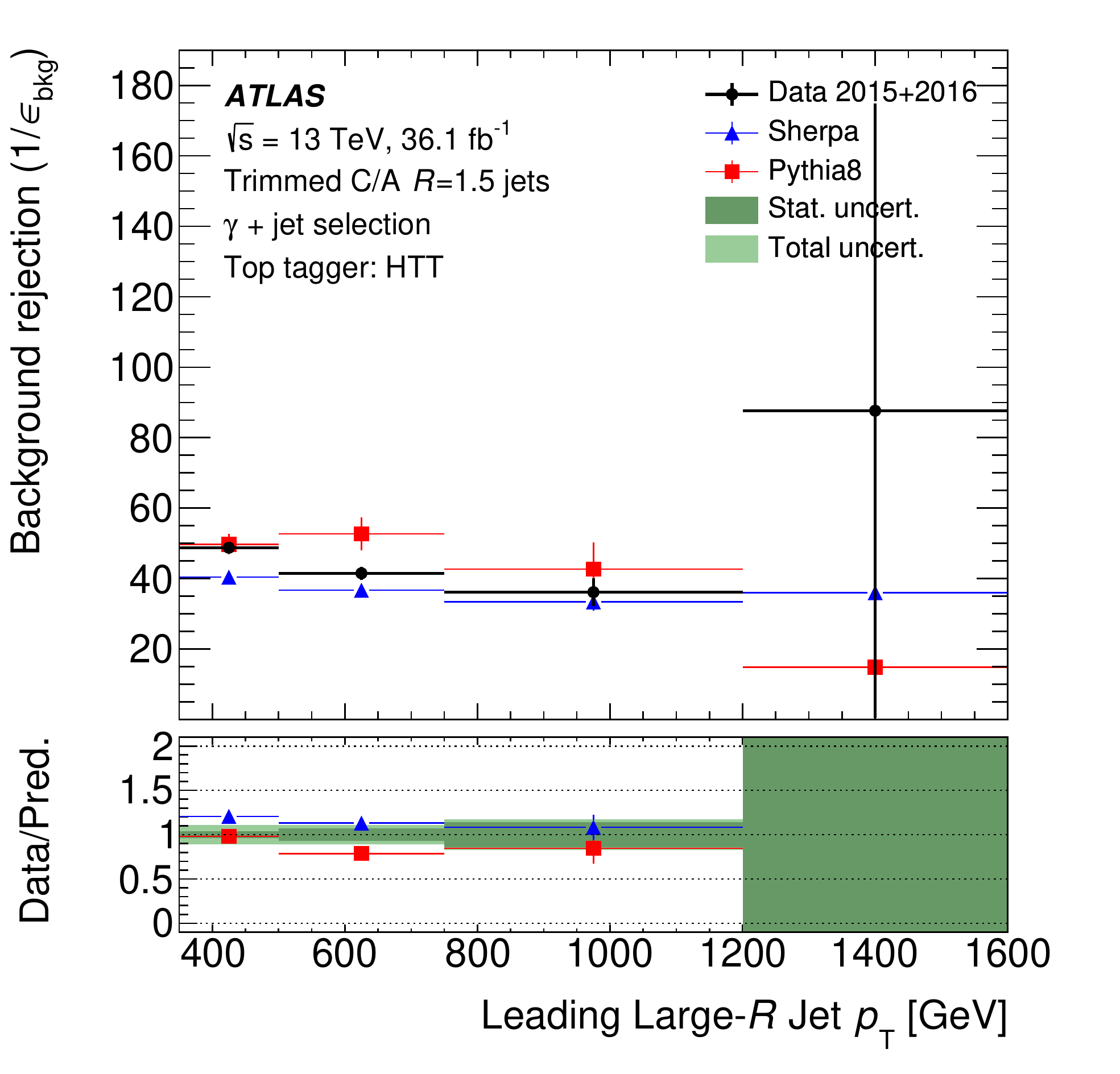}} \\
\subfigure[][]{   \label{fig:DMC_bkg_rejection_7_c} \includegraphics[width=0.45\textwidth]{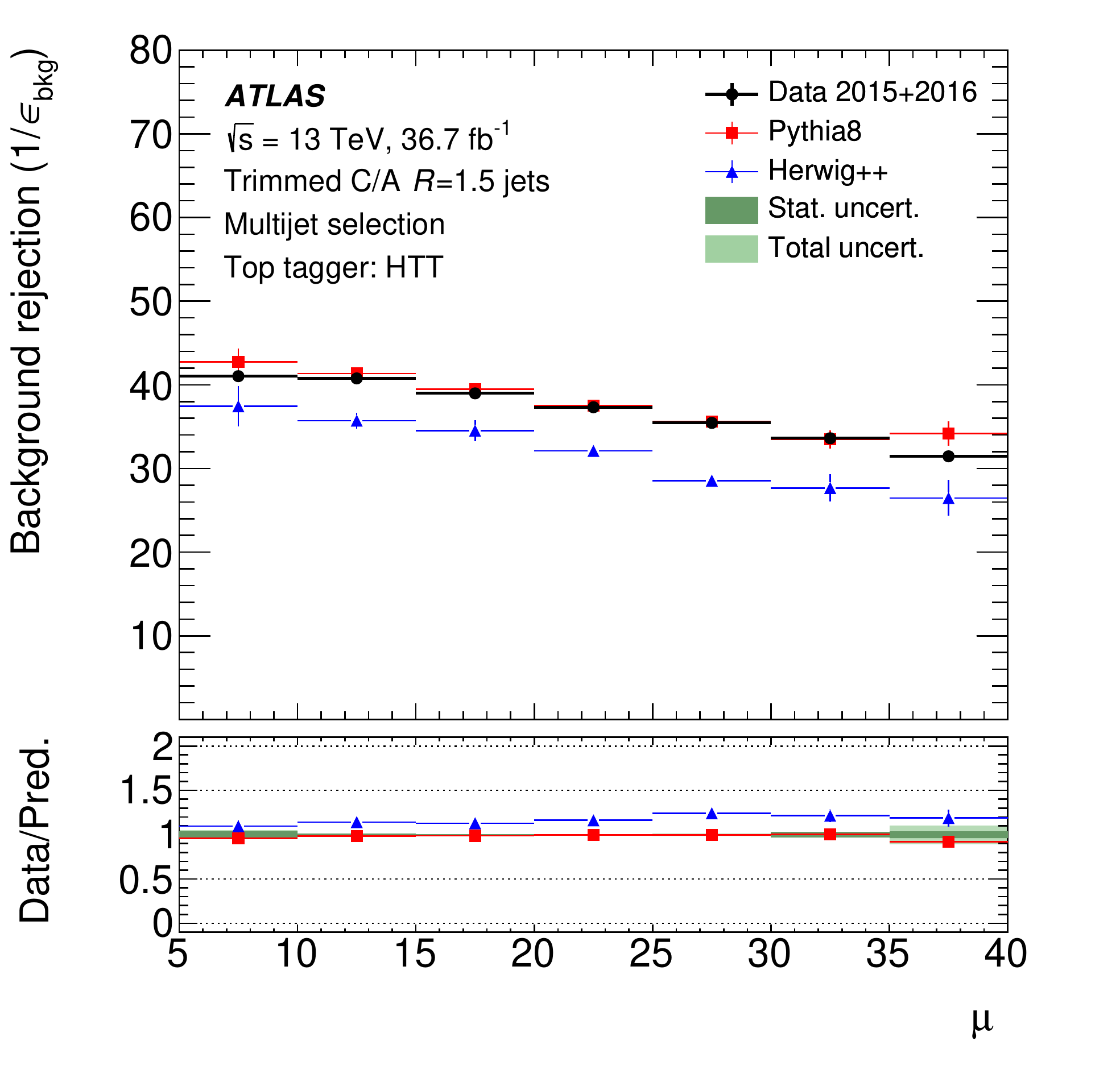}}
\subfigure[][]{  \label{fig:DMC_bkg_rejection_7_d} \includegraphics[width=0.45\textwidth]{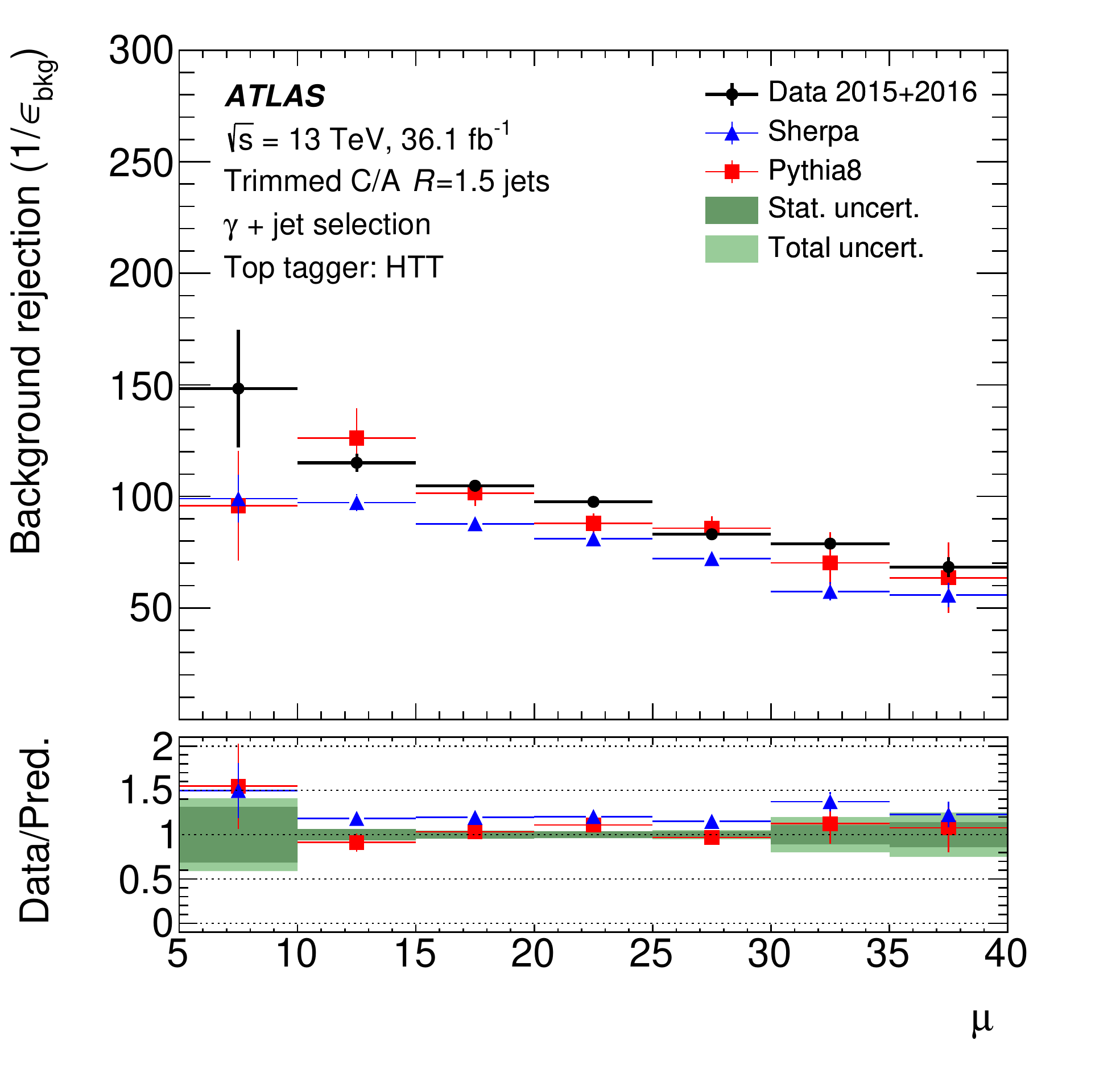}}
\caption{
\label{fig:DMC_bkg_rejection_7} The estimated light-jet rejection
$1/\epsilon_{\mathrm{bkg}}$ as a function of the leading jet \pt and the
average number of interactions per bunch crossing $\mu$ for the \htt in the
multijet~\subref{fig:DMC_bkg_rejection_7_a}~\subref{fig:DMC_bkg_rejection_7_c}
and
\gammajet~\subref{fig:DMC_bkg_rejection_7_b}~\subref{fig:DMC_bkg_rejection_7_d}
selection.}
\end{figure}
\clearpage
\newpage
 
\clearpage
\newpage
 
\subsection{Systematic uncertainties}
\label{subsec:datamc_systematics}
A number of sources of systematic uncertainty enter into the evaluation of the
modelling of data by the Monte Carlo simulation.  These uncertainties derive
both from theoretical assumptions within the Monte Carlo predictions and
from the reconstruction and calibration of the detector response to the physics
objects and therefore affect the three topologies to varying degrees.  These
sources of uncertainty, their effect in this analysis, and the manner in which
they are estimated are summarised in Tables~\ref{tab:systematics_theory}
and~\ref{tab:systematics_experiment}.  Systematic uncertainties are propagated
to the signal efficiency measurement by repeating the fit for varied templates
that correspond to each systematic uncertainty source and comparing the extracted efficiency
for the varied and nominal templates.
 
From this set of uncertainties, those originating from the measurement of
leptons, photons, \akt $R=0.4$ calorimeter jets, and the \MET soft term are
found to be negligible in all cases.  Additionally, the uncertainty related to
the estimation and subsequent subtraction of the multijet background in the
\ttbar analysis in Section~\ref{subsec:datamc_ttbar} and the background with a
real hadronic \Wboson/\Zboson-boson or top-quark decay in the multijet and
\gammajet analysis in Section~\ref{subsec:datamc_dijetgammajet} are found to be
negligible.  The uncertainties due to the application of flavour tagging in the
\ttbar analysis of signal jets are subdominant and affect the yield results with
an impact of the order of 20\% in the region of \mcomb below 100~\GeV. Similarly,
the component of the flavour tagging uncertainties pertaining to the misidentification of light-flavour
jets as $b$-jets tend to have a larger effect at low values of the multivariate
classifier score, in Figures~\ref{fig:DMC_signal_jss2} and~\ref{fig:DMC_signal_jss3} where
non-top-quark jet contributions are more dominant.  However, due to the localization of these
effects, they have a negligible impact on the measurement of the signal efficiency.  The
uncertainties in both the scale and resolution of the observable of interest
(e.g.\ \mcalo, \DTwo and \tauthrtwo) are evaluated by comparing the \largeR jets
formed from calorimeter cell topoclusters to those formed from ID
tracks~\cite{JETM-2018-02}.  These sources of uncertainty generally cause
small (10\%) changes in the yield of events near the most highly populated
regions of the distributions of observables but are generally the dominant uncertainties when
examining both the tails of these distributions and the regions near \MW
and \MTop in Section~\ref{subsec:datamc_ttbar}.  Likewise, in the case of the
\htt, the subjet energy scale uncertainty, which itself is based on Run 1
studies, is a dominant source of systematic uncertainty in the shape of the \htt
mass near \MTop but this uncertainty does not propagate strongly into the final
evaluation of the signal efficiency due to the broad mass window selection
described in Section~\ref{sec:technique_htt}.
 
The dominant systematic uncertainties of these techniques are those related to the
theoretical modelling of the Monte Carlo predictions.  In particular, in
Section~\ref{subsec:datamc_ttbar}, the contribution of the uncertainty in the
modelling of parton shower and hadronisation is dominant in all cases, leading
to variations in the yield of the Monte Carlo when examining the distributions
of \mcomb, \DTwo, and \tauthrtwo of up to 30\%.  This is also true when examining the
modelling of the multivariate classifiers, shown in Figures~\ref{fig:DMC_signal_jss2} and~\ref{fig:DMC_signal_jss3}.
In the tails of these distributions, the uncertainty in the modelling of additional radiation in
\ttbar events yields variations that are comparable in size.  The same behaviour
can be observed in the study of the modelling of light jets, particularly in
Section~\ref{subsubsec:background_analysis}, where predictions from both
\textsc{Pythia8} and \textsc{Herwig++} show shape differences ranging up to
approximately 25\% for certain jet moments as well as for the DNN top tagger.
As seen in Sections~\ref{sec:measured_signal_efficiency}
and~\ref{sec:measured_background_rejection}, these uncertainties manifest
themselves as large variations in the measured signal efficiency and background
rejection.  In the case of the tagging efficiency measurement of top quarks in
particular, the measured signal efficiency is found to be susceptible to both
the truth-level labelling of the top quark and the particular working
point chosen for the tagger.
 
\begin{table}[!b]
\small
\centering
\caption{\label{tab:systematics_theory} Summary of theoretical systematic uncertainties considered in the performance measurements in data.}
\begin{tabular}{>{\flushleft\arraybackslash}m{4.0cm}>{\flushleft\arraybackslash}m{2.5cm}m{8.0cm}}
\toprule
\multicolumn{1}{c}{Source} & \multicolumn{1}{m{1.8cm}}{Affected \mbox{topologies}} & \multicolumn{1}{c}{Description} \\
\midrule
Event generator choice                            & \ttbar                      & Hard-scattering modelling uncertainty estimated as the difference between \textsc{Powheg+Herwig} and \textsc{MC@NLO+Herwig}~\cite{ATL-PHYS-PUB-2016-004}. \\
Showering choice                                  & \ttbar                      & Parton shower and hadronisation modelling uncertainty estimated from the difference between \textsc{Powheg+Pythia6} and \textsc{Powheg+Herwig}~\cite{ATL-PHYS-PUB-2016-004}. \\
Modelling of extra QCD radiation                  & \ttbar                      & Uncertainty in amount of initial/final-state radiation estimated as the difference between the nominal \textsc{Powheg+Pythia6} generator and \emph{radLo} and \emph{radHi} tunes of \textsc{Powheg+Pythia6}. The radiation variations include variation of renormalisation and factorisation scales and the $h_{\text{damp}}$ parameters~\cite{ATL-PHYS-PUB-2016-004}. \\
\ttbar total cross-section uncertainty            & \ttbar                      & Uncertainty in normalisation of \ttbar MC contribution of magnitude $\pm 5.5\%$~\cite{Czakon:2013goa} \\
Single-top total cross-section uncertainty        & \ttbar                      & Uncertainty in normalisation of single-top MC contribution of magnitude $\pm 5.3\%$ (a conservative estimate enveloping the uncertainties on $t$-channel, $s$-channel and $\Wboson t$-channel) \\
$W$+jets total cross-section uncertainty      & \ttbar                      & Uncertainty in normalisation of \Wboson+jets MC contribution of magnitude $\pm 5.0\%$~\cite{ATLAS-CONF-2015-039} \\
$W$+jets theory scale uncertainties           & \ttbar                      & Uncertainty arising from the choice of renormalisation and factorisation scale, CKKW matching scale and QSF scale~\cite{ATL-PHYS-PUB-2017-006}. For the renormalisation and factorisation scale, $\times 0.5$ and $\times 2$ variations are considered and the renormalisation and factorisation scales are varied independently as well as in correlated and anti-correlated ways. The envelope of the variations is considered as the final renormalisation+factorisation scale uncertainty. \\
Signal normalisation                              & multijet, \gammajet         & The uncertainty in the subtraction of processes containing a hadronically decaying top quark or vector boson, conservatively taken to be 25\%. \\
\bottomrule
\end{tabular}
\end{table}
 
\begin{table}[!b]
\small
\centering
\caption{\label{tab:systematics_experiment} Summary of experimental systematic uncertainties considered in the performance measurements in data.}
\begin{tabular}{>{\flushleft\arraybackslash}m{4.0cm}>{\flushleft\arraybackslash}m{2.5cm}m{8.0cm}}
\toprule
\multicolumn{1}{c}{Source} & \multicolumn{1}{m{1.8cm}}{Affected \mbox{topologies}} & \multicolumn{1}{c}{Description} \\
\midrule
anti-$k_t$ $R=$1.0 trimmed jet moment scale       & \ttbar, multijet, \gammajet & The uncertainty in the scale of the detector response for all jet moments derived by comparing the calorimeter quantity with the reference track jet~\cite{PERF-2015-03}.\\
anti-$k_t$ $R=$1.0 trimmed jet moment resolution  & \ttbar, multijet, \gammajet & The uncertainty in the resolution of the detector response conservatively estimated as a 2\% absolute uncertainty in \pt, a 20\% relative uncertainty in jet mass, and a 15\% relative uncertainty in all other jet moments~\cite{JETM-2018-02}.\\
C/A $R=1.5$ subjet energy scale                   & \ttbar, multijet, \gammajet & The uncertainty in the scale of the detector energy response for subjets used in the \htt algorithm conservatively estimated to be 3\% based on Run 1 studies~\cite{PERF-2015-04}\\
\akt $R=0.4$ jet energy scale and resolution      & \ttbar                      & The uncertainty in the scale and resolution of the detector response for the jet \pt derived from simulation and \textit{in situ} calibration measurements~\cite{PERF-2016-04}.\\
\MET track soft term                              & \ttbar                      & The uncertainty on the component of the \MET calculation due to energy flow that is unassigned to a calibrated physics object, estimated in-situ in $Z$+jet events~\cite{PERF-2016-07}. \\
Flavour tagging                                    & \ttbar                      & The uncertainty in the scale factor correcting the efficiency response of the detector to identify heavy-flavour $b$- and $c$-jets as well as light-flavour jets derived \textit{in situ} using \ttbar events~\cite{PERF-2012-04,ATL-PHYS-PUB-2016-012}. \\
Lepton reconstruction and calibration             & \ttbar                      & The uncertainty in the scale factor correcting the efficiency to trigger on, reconstruct, and identify leptons as well as uncertainties in their energy and \pt scale and resolution~\cite{ATLAS-CONF-2016-024,ATL-PHYS-PUB-2016-015,PERF-2015-10}.\\
Photon reconstruction and calibration             & \gammajet                   & The uncertainty in the scale factor correcting the efficiency to trigger on, reconstruct, and identify photons~\cite{PERF-2013-04} as well as uncertainties in their energy scale and resolution~\cite{PERF-2013-05}.\\
Multijet background normalisation                 & \ttbar                      & The uncertainty in the data-driven prediction of the yield of multijet events, conservatively taken to be 50\% based on the estimate in Ref.~\cite{EXOT-2015-04}. \\
Multijet lepton misreconstruction efficiencies    & \ttbar                  & The statistical uncertainty of the real and fake/non-prompt lepton reconstruction efficiencies estimated in Ref.~\cite{EXOT-2015-04} is propagated through the matrix method. \\
Luminosity uncertainty                            & \ttbar, multijet, \gammajet & A \SI{2.1}{\%} relative uncertainty in the MC yield, based on the luminosity uncertainty of the combined 2015+2016 dataset based on~\cite{DAPR-2013-01}. \\
Pile-up uncertainty                               & \ttbar, multijet, \gammajet & Uncertainty in the reweighting of MC pile-up profile to the measured pile-up profile in data based on disagreement between instantaneous luminosity in data and in simulation~\cite{DAPR-2013-01}. \\
\bottomrule
\end{tabular}
\end{table}
\clearpage
\newpage
 
\section{Conclusion}
\label{sec:conclusion}
Various methods to tag boosted, hadronically decaying \Wboson bosons and top
quarks are studied in data and simulation.  A number of techniques, including
the use of physically motivated jet moments, shower deconstruction and the \htt
which were studied in Run 1 are re-optimised for use in LHC Run 2 conditions.
Additionally, the multivariate combination of high-level jet moments using
boosted decision trees and neural networks as well as the combination of
low-level energy flow information in the form of topoclusters using a deep neural network is
studied both in data and Monte Carlo simulation.  The performance of these techniques is
evaluated using Monte Carlo
simulation for jets in the \pT range from $500$ to $2000~\GeV$ and compared in
terms of the central value of the background rejection at fixed signal
efficiency.  This study indicates that a multivariate combination of information
can enhance performance to exceed that of techniques based on more physically motivated
individual features across the full jet \pt range for both \Wboson-boson and
top-quark tagging.
 
The performance of the various tagging techniques is studied using a sample of
36.1\ifb\ of 13~\TeV\ proton--proton collision data collected by the ATLAS detector
at the LHC in 2015 and 2016. A sample of lepton-plus-jets
\ttbar events is used to study the signal \Wboson-boson and top-quark jet
tagging efficiency and compare the predicted efficiency in Monte Carlo simulation
to that in data for a set of working points for the tagging strategies from which
\textit{in situ} calibrations and systematic uncertainties can be derived.
Likewise, background light-jet-enriched event topologies are studied using
multijet and \gammajet samples. We have demonstrated that tagging efficiencies and
the relevant uncertainties for both signal and background can be extracted from data.
This opens opportunities for complex
\Wboson-boson and top-quark taggers using state of the art techniques such as DNNs
and new inputs to be utilized with ATLAS data in the future.
In general, it is found that the inputs to and
the performance of the studied \Wboson-boson and top-quark taggers currently in
use in physics analyses are well-modelled by Monte Carlo simulations.  However,
in all studies, it is found that the primary limiting factor in the description
of the tagging efficiency by the Monte Carlo prediction derives from the
theoretical modelling of the Monte Carlo processes studied, particularly the
parton shower and hadronisation model of the \ttbar process.  Finally, the small
\pileup dependence of each tagger working point is characterised to understand
the relative susceptibility of each strategy to \pileup contamination within the
jet.  In general, the signal efficiency is found to be quite robust against
increased levels of event \pileup whereas the background rejection shows
residual \pileup dependence, particularly in the case of the \Wboson taggers.
In all cases, however, the dependence is well-described by the Monte Carlo
simulation.
\clearpage

\appendix
\part*{Appendix}
\addcontentsline{toc}{part}{Appendix}
 
\section{BDT \& DNN Hyper-parameters}
\label{sec:bdt_&_dnn_hyperparam}
In this section, a description of the tuned hyper-parameters of the BDT and DNN are presented in Tables~\ref{tab:BDTparameter} and \ref{tab:dnn_grid_result_final}.
 
\begin{table}[htb]
\renewcommand{\arraystretch}{1.4}
\centering
\caption{Brief description of the BDT parameters and the chosen parameters.}
\label{tab:BDTparameter}
\begin{tabular}{|l|l|c|}\hline
Setting Name                          & Description                             & Choice  \\  \hline \hline
Software package                      & Package used for training               & TMVA 4.2.1~\cite{Speckmayer:2010zz}    \\ \hline
BoostType                             & Type of boosting technique              & GradientBoost \\  \hline
NTrees                                & Number of trees in the forest           & 500 \\ \hline
MaxDepth                              & Max depth of the decision tree allowed  & 20 \\ \hline
\multirow{1}{*}{MinimumNodeSize}      & \shortstack[l]{Minimum fraction of training events \\ required in a leaf node}  & 1.0$\%$ \\ \hline
Shrinkage                             & Learning rate for GradientBoost algorithm & 0.5 \\ \hline
\multirow{1}{*}{UseBaggedBoost}       & \shortstack[l]{Use only a random (bagged) subsample of all events\\ for growing the trees in each iteration} & True \\ \hline
\multirow{1}{*}{BaggedSampleFraction} & \shortstack[l]{Relative size of bagged event sample\\ to original size of the data sample} & 0.5 \\ \hline
SeparationType                        & Separation criterion for node splitting & GiniIndex \\ \hline
\multirow{1}{*}{nCuts}                & \shortstack[l]{Number of grid points in variable range used \\in finding optimal cut in node splitting} & 500\\ \hline
\end{tabular}
\end{table}
 
\begin{table}
\small
\centering
\caption{Chosen DNN parameters and architecture for shape-based \Wboson-boson and top-quark tagging.}
\label{tab:dnn_grid_result_final}
\begin{tabular}{l|l|l|l|}
\cline{2-4}                                  & \multicolumn{1}{c|}{\Wboson-Boson Tagging} & \multicolumn{1}{c|}{Top-Quark Tagging} & Reference \\      \hline
 
\multicolumn{1}{|l|}{Software package}                    & \multicolumn{2}{c|}{Keras 1.0.8 with Theano backend, lwtnn 2.0}  & \cite{chollet2015keras, 2016arXiv160502688short, daniel_hay_guest_2017_290682}   \\ \hline
\multicolumn{1}{|l|}{Layer type}                    & Dense  & Dense         & \cite{chollet2015keras}   \\
 
\multicolumn{1}{|l|}{Number of hidden layers} &          4     &                     5            & \cite{chollet2015keras}  \\
\multicolumn{1}{|l|}{Architecture} &                 16, 14, 9, 6  &    18, 16, 14, 10, 5      & -  \\
\multicolumn{1}{|l|}{Activation function}           & rectified linear unit (relu)       & rectified linear unit (relu)              & \cite{Goodfellow-et-al-2016}  \\
 
\multicolumn{1}{|l|}{Optimizer}           & Adam       & Adam              & \cite{DBLP:journals/corr/KingmaB14}  \\
\multicolumn{1}{|l|}{Learning rate}                 &  0.0001                          &    0.00005          & \cite{DBLP:journals/corr/KingmaB14}  \\
\multicolumn{1}{|l|}{L1 Regulariser}                &  0.001                              &      0.001             & \cite{Goodfellow-et-al-2016} \\
\multicolumn{1}{|l|}{NN weight initialisation}      & Glorot uniform                       & Glorot uniform                        & \cite{Glorot10understandingthe}  \\
\multicolumn{1}{|l|}{Batch size}                    & 200                                   & 200                                    & \cite{Goodfellow-et-al-2016}  \\
\multicolumn{1}{|l|}{Batch normalisation}           & Yes                                   & Yes                                    & \cite{DBLP:journals/corr/IoffeS15}  \\
 
\multicolumn{1}{|l|}{Number of epochs}                    & 100 with early stopping                                  & 100 with early stopping                                        & \cite{chollet2015keras}  \\
 
\multicolumn{1}{|l|}{Training input group}               & Group 8                                & Group 9                             & - \\ \hline
\end{tabular}
\end{table}
\clearpage

\section*{Acknowledgements}
 
 
We thank CERN for the very successful operation of the LHC, as well as the
support staff from our institutions without whom ATLAS could not be
operated efficiently.
 
We acknowledge the support of ANPCyT, Argentina; YerPhI, Armenia; ARC, Australia; BMWFW and FWF, Austria; ANAS, Azerbaijan; SSTC, Belarus; CNPq and FAPESP, Brazil; NSERC, NRC and CFI, Canada; CERN; CONICYT, Chile; CAS, MOST and NSFC, China; COLCIENCIAS, Colombia; MSMT CR, MPO CR and VSC CR, Czech Republic; DNRF and DNSRC, Denmark; IN2P3-CNRS, CEA-DRF/IRFU, France; SRNSFG, Georgia; BMBF, HGF, and MPG, Germany; GSRT, Greece; RGC, Hong Kong SAR, China; ISF and Benoziyo Center, Israel; INFN, Italy; MEXT and JSPS, Japan; CNRST, Morocco; NWO, Netherlands; RCN, Norway; MNiSW and NCN, Poland; FCT, Portugal; MNE/IFA, Romania; MES of Russia and NRC KI, Russian Federation; JINR; MESTD, Serbia; MSSR, Slovakia; ARRS and MIZ\v{S}, Slovenia; DST/NRF, South Africa; MINECO, Spain; SRC and Wallenberg Foundation, Sweden; SERI, SNSF and Cantons of Bern and Geneva, Switzerland; MOST, Taiwan; TAEK, Turkey; STFC, United Kingdom; DOE and NSF, United States of America. In addition, individual groups and members have received support from BCKDF, CANARIE, CRC and Compute Canada, Canada; COST, ERC, ERDF, Horizon 2020, and Marie Sk{\l}odowska-Curie Actions, European Union; Investissements d' Avenir Labex and Idex, ANR, France; DFG and AvH Foundation, Germany; Herakleitos, Thales and Aristeia programmes co-financed by EU-ESF and the Greek NSRF, Greece; BSF-NSF and GIF, Israel; CERCA Programme Generalitat de Catalunya, Spain; The Royal Society and Leverhulme Trust, United Kingdom.
 
The crucial computing support from all WLCG partners is acknowledged gratefully, in particular from CERN, the ATLAS Tier-1 facilities at TRIUMF (Canada), NDGF (Denmark, Norway, Sweden), CC-IN2P3 (France), KIT/GridKA (Germany), INFN-CNAF (Italy), NL-T1 (Netherlands), PIC (Spain), ASGC (Taiwan), RAL (UK) and BNL (USA), the Tier-2 facilities worldwide and large non-WLCG resource providers. Major contributors of computing resources are listed in Ref.~\cite{ATL-GEN-PUB-2016-002}.
 
\clearpage

\printbibliography
 
\clearpage 
 
\begin{flushleft}
{\Large The ATLAS Collaboration}

\bigskip

M.~Aaboud$^\textrm{\scriptsize 34d}$,    
G.~Aad$^\textrm{\scriptsize 99}$,    
B.~Abbott$^\textrm{\scriptsize 124}$,    
O.~Abdinov$^\textrm{\scriptsize 13,*}$,    
B.~Abeloos$^\textrm{\scriptsize 128}$,    
D.K.~Abhayasinghe$^\textrm{\scriptsize 91}$,    
S.H.~Abidi$^\textrm{\scriptsize 164}$,    
O.S.~AbouZeid$^\textrm{\scriptsize 39}$,    
N.L.~Abraham$^\textrm{\scriptsize 153}$,    
H.~Abramowicz$^\textrm{\scriptsize 158}$
% ...
\bigskip
\\

% ...
$^{13}$Institute of Physics, Azerbaijan Academy of Sciences, Baku; Azerbaijan.\\
% ...
$^{34}$$^{(a)}$Facult\'e des Sciences Ain Chock, R\'eseau Universitaire de Physique des Hautes Energies - Universit\'e Hassan II, Casablanca;$^{(b)}$Centre National de l'Energie des Sciences Techniques Nucleaires (CNESTEN), Rabat;$^{(c)}$Facult\'e des Sciences Semlalia, Universit\'e Cadi Ayyad, LPHEA-Marrakech;$^{(d)}$Facult\'e des Sciences, Universit\'e Mohamed Premier and LPTPM, Oujda;$^{(e)}$Facult\'e des sciences, Universit\'e Mohammed V, Rabat; Morocco.\\
% ...
$^{39}$Niels Bohr Institute, University of Copenhagen, Copenhagen; Denmark.\\
% ...
$^{91}$Department of Physics, Royal Holloway University of London, Egham; United Kingdom.\\
% ...
$^{99}$CPPM, Aix-Marseille Universit\'e, CNRS/IN2P3, Marseille; France.\\
% ...
$^{124}$Homer L. Dodge Department of Physics and Astronomy, University of Oklahoma, Norman OK; United States of America.\\
% ...
$^{128}$LAL, Universit\'e Paris-Sud, CNRS/IN2P3, Universit\'e Paris-Saclay, Orsay; France.\\
% ...
$^{153}$Department of Physics and Astronomy, University of Sussex, Brighton; United Kingdom.\\
% ...
$^{158}$Raymond and Beverly Sackler School of Physics and Astronomy, Tel Aviv University, Tel Aviv; Israel.\\
% ...
$^{164}$Department of Physics, University of Toronto, Toronto ON; Canada.\\
% ...
$^{a}$ Also at Borough of Manhattan Community College, City University of New York, NY; United States of America.\\
$^{b}$ Also at California State University, East Bay; United States of America.\\
$^{c}$ Also at Centre for High Performance Computing, CSIR Campus, Rosebank, Cape Town; South Africa.\\
$^{d}$ Also at CERN, Geneva; Switzerland.\\
$^{e}$ Also at CPPM, Aix-Marseille Universit\'e, CNRS/IN2P3, Marseille; France.\\
% ...
$^{*}$ Deceased

\end{flushleft}


\end{document}